\documentclass[12pt]{article}
\pdfoutput=1
\usepackage{jheppub_modified_green}
\usepackage{epsfig}
\usepackage{amsfonts} 
\usepackage{amssymb}
\usepackage{enumitem}
\usepackage{amsmath}
\usepackage{amsthm}
\usepackage{subfig}
\usepackage{bm}
\usepackage{hyperref}
\usepackage{mathrsfs}
\usepackage{bbm}
\usepackage{bbold}
\usepackage{cancel}
\usepackage{xcolor}
\usepackage{accents}
\usepackage{relsize}
\usepackage{float, slashed, graphicx}
\usepackage{shuffle}
\usepackage{mathtools} %added for coloneqq
\usepackage{hyperref}
\usepackage{todonotes}
\usepackage{yfonts}
\usepackage{comment}
\usepackage{setspace}
\usepackage{placeins}
\usepackage{pgfplots}
\pgfplotsset{width=10cm,compat=1.9}

\usepackage[T1]{fontenc} % if needed

\newcommand{\cA}{\mathcal{A}}
\newcommand{\cI}{\mathcal{I}}
\newcommand{\cM}{\mathcal{M}}
\newcommand{\cC}{\mathcal{C}}
\newcommand{\cJ}{\mathcal{J}}
\newcommand{\cN}{\mathcal{N}}
\newcommand{\cT}{\mathcal{T}}

\newcommand{\cO}{\mathcal{O}}
\newcommand{\cD}{\mathcal{D}}

\newcommand{\eps}{\epsilon}

\newcommand{\mb}{\bar m}

\newcommand{\vb}{\bar v}
\newcommand{\pb}{\bar p}
\newcommand{\yb}{\bar y}
\newcommand{\wb}{\bar w}

\def\nn{\nonumber}

\newmuskip\pFqmuskip

\newcommand*\pFq[6][8]{%
  \begingroup % only local assignments
  \pFqmuskip=#1mu\relax
  \mathcode`\,=\string"8000
  \begingroup\lccode`\~=`\,
  \lowercase{\endgroup\let~}\pFqcomma
  {}_{#2}F_{#3}{\left(\genfrac..{0pt}{}{#4}{#5};#6\right)}%
  \endgroup
}
\newcommand{\pFqcomma}{\mskip\pFqmuskip}

\newcommand{\Cdot}{{\cdot}}

\def\spa#1.#2{\left\langle#1\,#2\right\rangle}
\def\spb#1.#2{\left[#1\,#2\right]}
 
\allowdisplaybreaks

\usepackage{tikz}
\usepackage{tikz-feynman}
\tikzfeynmanset{compat=1.1.0}
\tikzfeynmanset{graviton/.style={circle, draw=green!60, fill=green!5, very thick, minimum size=7mm}}
\tikzfeynmanset{codot/.style={/tikz/shape=circle,/tikz/fill=white,/tikz/minimum size=0.1cm,/tikz/inner sep=1.8pt}}
\tikzfeynmanset{myblob/.style={/tikz/shape=rectangle,/tikz/fill=red,/tikz/minimum size=0.2cm,/tikz/inner sep=1.8pt} }
\tikzfeynmanset{ghc/.style={/tikz/shape=circle,/tikz/fill=white,/tikz/minimum size=0.05cm,} }
\tikzfeynmanset{HV/.style={/tikz/shape=circle,/tikz/fill={rgb:black,1;white,2},/tikz/minimum size=0.1cm,} }
\tikzfeynmanset{GR/.style={/tikz/shape=ellipse,/tikz/fill={rgb:black,1;white,2},/tikz/minimum size=0.3cm,} }
\tikzfeynmanset{GR2/.style={/tikz/shape=ellipse,/tikz/fill={rgb:black,1;white,2},/tikz/minimum width = 1.2cm, 
    /tikz/minimum height = 3.4cm} }
\tikzset{box/.pic={\filldraw[fill=black]  (0,0) circle (2.5pt); \filldraw [fill=black] (0.5,0) circle (2.5pt); \draw [line width=5pt] (0,0) -- (0.5,0);}}
\usetikzlibrary{arrows.meta} % decent arrow in the graphs
\usetikzlibrary{decorations.pathmorphing}
\usetikzlibrary{decorations.pathreplacing}
\usetikzlibrary{decorations.markings}
\tikzset{
   vector2/.style={decorate, decoration={snake, amplitude=1pt, segment length=6pt}, draw,double},
   vector/.style={decorate, decoration={snake, amplitude=1pt, segment length=6pt}, draw},
	provector/.style={decorate, decoration={snake,amplitude=2.5pt}, draw},
	antivector/.style={decorate, decoration={snake,amplitude=-2.5pt}, draw},
    fermion/.style={draw=black, postaction={decorate},
        decoration={markings,mark=at position .55 with {\arrow[draw=black]{>}}}},
    fermionbar/.style={draw=black, postaction={decorate},
        decoration={markings,mark=at position .55 with {\arrow[draw=black]{<}}}},
    fermionnoarrow/.style={draw=black},
    gluon/.style={decorate, draw=black,
        decoration={coil,amplitude=4pt, segment length=5pt}},
    scalar/.style={dashed,draw=black, postaction={decorate},
        decoration={markings,mark=at position .55 with {\arrow[draw=black]{>}}}},
    scalarbar/.style={dashed,draw=black, postaction={decorate},
        decoration={markings,mark=at position .55 with {\arrow[draw=black]{<}}}},
    scalarnoarrow/.style={dashed,draw=black},
    electron/.style={draw=black, postaction={decorate},
        decoration={markings,mark=at position .55 with {\arrow[draw=black]{>}}}},
	bigvector/.style={decorate, decoration={snake,amplitude=4pt}, draw},
}
\tikzset{cross/.style={cross out, draw, 
         minimum size=2*(#1-\pgflinewidth), 
         inner sep=0pt, outer sep=0pt}}

\tikzstyle{block} = [draw, rectangle, 
    minimum height=3em, minimum width=6em]

\title{One-loop   Gravitational Bremsstrahlung and Waveforms from a
 Heavy-Mass Effective Field Theory %and  the Double~Copy 
}
\author{Andreas Brandhuber$\mbox{}^{a,b}$,}
\author{Graham R.~Brown$\mbox{}^{a,b}$,}
\author{Gang Chen$\mbox{}^{c}$,}
\author{Stefano De Angelis$\mbox{}^{d}$,\\}
\author{\hspace{-0.2cm}Joshua Gowdy$\mbox{}^{a,b}$}
\author{and Gabriele Travaglini$\mbox{}^{a,b}$}
\affiliation{$\mbox{}^{a}$Centre for Theoretical Physics, Department of Physics and Astronomy, \\
Queen Mary University of London, Mile End Road, London E1 4NS, United Kingdom}
\affiliation{$\mbox{}^{b}$Kavli Institute for Theoretical Physics, University of California, Santa Barbara, CA~93106,~USA}
\affiliation{$\mbox{}^{c}$Niels Bohr International Academy,
Niels Bohr Institute, University of Copenhagen,\\
Blegdamsvej 17, DK-2100 Copenhagen \O, Denmark} 
\affiliation{$\mbox{}^{d}$Institut de Physique Th\'{e}orique, CEA, CNRS, Universit\'{e} Paris-Saclay,\\ F–91191 Gif-sur-Yvette cedex, France}
\emailAdd{a.brandhuber@qmul.ac.uk}
\emailAdd{graham.brown@qmul.ac.uk}
\emailAdd{gang.chen@nbi.ku.dk}
\emailAdd{stefano.de-angelis@ipht.fr}
\emailAdd{j.k.gowdy@qmul.ac.uk}
\emailAdd{g.travaglini@qmul.ac.uk}
\begin{document}
\begin{flushright}
	QMUL-PH-22-28, 
	SAGEX-22-32-E
\end{flushright}

\abstract{%
Using a  heavy-mass effective field theory (HEFT), we study  gravitational-wave emission  
in the scattering of two spinless black holes or neutron stars of arbitrary masses at next-to-leading order in the Post-Minkowskian expansion.
We compute the contributions to the one-loop scattering amplitude   with four scalars and one graviton  which are relevant to  the calculation of the waveforms, also presenting expressions of 
classical  tree-level amplitudes with four scalars and up to two radiated gravitons. The latter are obtained using a novel on-shell recursion relation for classical amplitudes with four scalars and an arbitrary number of gravitons. Our one-loop five-point amplitude is expressed in terms of a single family of master integrals with  the principal value prescription for linearised massive propagators, which we evaluate using differential equations. In our HEFT approach all hyper-classical iterations and quantum corrections to the amplitude are  dropped at the diagrammatic level, thereby  computing  directly 
contributions to classical physics. Our result exhibits the expected factorisation of infrared divergences,   the correct soft limits, and highly nontrivial cancellations of spurious poles.
Finally,  using our amplitude  result we compute numerically the corresponding next-to-leading corrections to the    spectral waveforms and  the far-field  time-domain waveforms using the Newman-Penrose scalar~$\Psi_4$.  
}

\vspace{-2.6cm}

%\pagenumbering{roman}
\maketitle

\flushbottom
 \tableofcontents
\newpage 
%please don't remove this newpage otherwise the intro has no page number

\section{Introduction}
The extraordinary observation of gravitational waves by the LIGO and Virgo collaborations \cite{LIGOScientific:2016dsl,LIGOScientific:2016aoc,LIGOScientific:2016sjg,LIGOScientific:2017bnn,LIGOScientific:2017vwq}, 100 years after Einstein’s prediction, has initiated a new era of exploration of our universe, with the promise of  major  discoveries in fundamental areas from black holes to particle physics. With the increasing  precision  and scope  of current and future experiments, there is  a pressing  need for accurate theoretical templates for the  gravitational-wave  signal.
A similar demand for ever more precise theoretical predictions has boosted  the development, over the last few decades,  of  
highly efficient methods  to compute scattering amplitudes of elementary particles to high perturbative orders.%
\footnote{For a recent review, also including applications to General Relativity,  see \cite{Travaglini:2022uwo}.} It is then   remarkable that  amplitudes, and  modern methods devised for their computation, have now been put to use in tackling problems in classical gravity. 

The connection with amplitudes was revealed more than 50 years ago in  \cite{Iwasaki:1971iy,Iwasaki:1971vb},  where corrections to the Newtonian potential were computed from one-loop Feynman diagrams. Those papers also appreciated that loop diagrams contribute to classical physics, a point vigorously strengthened  in \cite{Bjerrum-Bohr:2002gqz,Holstein:2004dn}.
An amplitude-based approach was applied at the second Post-Minkowskian (PM) order in
\cite{Neill:2013wsa,Bjerrum-Bohr:2013bxa, Bjerrum-Bohr:2014zsa, Bjerrum-Bohr:2016hpa} to compute corrections to the Newtonian potential using modern amplitude techniques   \cite{Bern:1994zx, Bern:1994cg}.
More recently,  several works have pursued this approach  to compute the conservative part of the potential at  3PM \cite{Bern:2019nnu,Bern:2019crd,Parra-Martinez:2020dzs,Cheung:2020gyp,Bjerrum-Bohr:2021din,DiVecchia:2021bdo,Brandhuber:2021eyq}  and 4PM \cite{Bern:2021dqo,Bern:2021yeh,Bern:2022jvn},    also  including radiation \cite{Luna:2017dtq,Shen:2018ebu,Bautista:2019tdr,Herrmann:2021lqe,Herrmann:2021tct},  in the presence of classical spin \cite{Guevara:2017csg,Arkani-Hamed:2017jhn,Guevara:2018wpp,Chung:2018kqs,Guevara:2019fsj,Arkani-Hamed:2019ymq,Aoude:2020onz,Chung:2020rrz,Guevara:2020xjx,Chen:2021kxt,Kosmopoulos:2021zoq,Chiodaroli:2021eug,Bautista:2021wfy,Cangemi:2022bew,Ochirov_2022,Damgaard:2019lfh,Bern:2020buy,Comberiati:2022ldk,Maybee:2019jus,Haddad:2021znf,Chen:2022clh,Menezes:2022tcs,FebresCordero:2022jts,Alessio:2022kwv,Bern:2022kto,Aoude:2022thd,Aoude:2022trd,Bjerrum-Bohr:2023jau}
and in  theories where Einstein gravity  is modified by higher-derivative interactions 
\cite{Brandhuber:2019qpg, Emond:2019crr, AccettulliHuber:2019jqo,AccettulliHuber:2020oou,AccettulliHuber:2020dal,Carrillo-Gonzalez:2021mqj,Bellazzini:2021shn}.
The fact that gravity is a non-renormalisable theory does not prevent one from making predictions: 
 treating gravity as an effective  theory \cite{Donoghue:1994dn},  non-local/non-analytic effects arising from the low-energy theory can be reliably calculated %without reference 
 and disentangled from a yet-unknown ultraviolet completion. Observables that can be computed in this way include  the deflection  angle between two heavy objects (black holes or neutron stars), the Shapiro time delay,  and waveforms. 
 
Amplitude techniques have thus  emerged as  powerful alternatives to a variety of other approaches,  such as  the effective one-body formulation \cite{Buonanno:1998gg,Damour:2016gwp,Damour:2017zjx,Vines:2017hyw,Vines:2018gqi,Damour:2019lcq}, and the  worldline approach started in  \cite{Goldberger:2004jt, Goldberger:2009qd} and further developed in a relativistic setting in \cite{Kalin:2020mvi, Kalin:2020fhe,Mogull:2020sak,Jakobsen:2021smu,Mougiakakos:2021ckm,Liu:2021zxr,Dlapa:2021npj,Jakobsen:2021lvp,Dlapa:2021vgp,Jakobsen:2022fcj,Riva:2022fru,Jakobsen:2022psy,Dlapa:2022lmu}. 
In these works one performs an expansion in  Newton's constant $G$ while keeping  the dependence on the velocities exact (the PM expansion), which is natural from the quantum field theory viewpoint, as opposed to the Post-Newtonian (PN)  expansion
\cite{Damour:1985mt, Gilmore:2008gq,Damour:2001bu, Bjerrum-Bohr:2002gqz,Blanchet:2003gy, Itoh:2003fy,Foffa:2011ub,Jaranowski:2012eb, Damour:2014jta,Galley:2015kus,Damour:2015isa, Damour:2016abl, Bernard:2015njp, Bernard:2016wrg, Foffa:2012rn, Foffa:2016rgu,Porto:2017dgs,Porto:2017shd,Foffa:2019yfl,Blumlein:2020pog,Foffa:2019hrb,Blumlein:2019zku,Bini:2020wpo,Blumlein:2020pyo,Blumlein:2020znm,Bini:2020uiq,Blumlein:2021txj}, 
also studied for spinning objects
\cite{Porto:2005ac,Steinhoff:2010zz,Levi:2014gsa,Levi:2015msa,Maia:2017yok,Levi:2018nxp,Levi:2020uwu}.
In order to have a sensible perturbative expansion one must require that 
%$\frac{GM}{b}\ll 1$, 
${GM}/{b}\ll 1$, 
where $M$ is the typical   mass of a heavy object and $b$ is the impact parameter. 
Quantum effects  can be discarded since the characteristic Schwarzschild radius of the objects involved  is much larger than their Compton wavelengths, 
%$ GM\gg \frac{\hbar}{M}$, 
$ GM\gg {\hbar}/{M}$,
which combined with the previous relation requires one to work in a regime where 
%$\frac{\hbar}{M}\ll GM\ll b$. 
${\hbar}/{M}\ll GM\ll b$. 

 An ideal  amplitude-based method tailored  to compute classical observables  should possess two features: first, it should easily disentangle quantum corrections from the classical contribution; and second, it should  avoid the subtraction of  the so-called hyper-classical (sometimes called super-classical) terms from complete amplitudes. 
Indeed, in the approximation we are considering $G  M^2$ is not small, and a resummation of higher perturbative orders is mandatory. Remarkably, this is achieved in   impact parameter space  (IPS) \cite{Levy:1969cr,Amati:1987wq,Amati:1987uf,Amati:1990xe,Kabat:1992tb,Bellazzini:2022wzv}, where amplitudes are believed to  exponentiate, and it is precisely the IPS amplitudes that turn out to be relevant  both for computing classical observables such as scattering angles and   waveforms. The extraction of this ``eikonal'' exponent  at a certain perturbative order  requires a delicate subtraction of  terms that reconstruct the exponentiation at lower perturbative  orders; while computing such terms provides a %sanity 
consistency check of the result, an efficient method should  preferably  avoid this.

 An important step in this direction was taken in \cite{Bern:2021dqo}, where  the conservative part of the potential at 4PM was computed from the radial action. Motivated by the WKB formalism, in \cite{Damgaard:2021ipf} an exponential  representation of the $S$-matrix alternative to the eikonal  was proposed, with a clear procedure to compute the matrix elements of the hermitian operator $N$, defined through $S \coloneqq e^{iN}$. Alas, such methods still require the computation of complete amplitudes
and a subsequent subtraction. 

 The  approach   we follow in the present paper,  initiated  in \cite{Brandhuber:2021kpo,Brandhuber:2021eyq},  is based on a Heavy-mass Effective Field Theory (HEFT). Specifically, it was proposed  in \cite{Brandhuber:2021eyq}
that classical observables can be computed entirely  avoiding the subtraction  of iterating terms. This framework was tested in the conservative sector by  deriving the scattering angle for two heavy spinless objects at 3PM, and in this paper we will show how the HEFT approach can  be used  to incorporate radiation emission.  
The key finding of  
\cite{Brandhuber:2021eyq}, which makes the HEFT approach ideally suited for computing classical quantities in gravity, is that   in the elastic case  only  a particular subset of diagrams contributes to the scattering angle up to two loops, namely those that are two massive particle irreducible (2MPI). Conversely, it was found for the scattering angle that diagrams that are two massive particle reducible compute hyper-classical terms, and in the HEFT approach one simply drops them from the get go.

The relevance of a heavy-mass expansion arises from the fact that the momenta exchanged by the heavy particles are much smaller than their masses, thus it is natural to consider an expansion in the  heavy masses. This is precisely the situation one encounters in heavy-quark effective theory \cite{Georgi:1990um,Luke:1992cs,Neubert:1993mb,Manohar:2000dt}.  Such a set-up in gravity was first considered in \cite{Damgaard:2019lfh, Aoude:2020onz}, and in \cite{Brandhuber:2021kpo} some of the present authors were able to combine the heavy-mass expansion   with the colour kinematic/duality \cite{Bern:2008qj,Bern:2010ue,Bern:2019prr}, producing compact expressions for amplitudes where the BCJ numerators are manifestly gauge invariant. 
All-multiplicity expressions for $D$-dimensional amplitudes with two heavy scalars and an arbitrary number of gluons or gravitons    were then presented in \cite{Brandhuber:2021bsf}, where the underlying BCJ kinematic algebra was also related to a quasi-shuffle algebra, further studied in \cite{Chen:2022nei, Brandhuber:2022enp,Cao:2022vou}. Curiously, the number of terms in a numerator with $n{-}2$ massless particles  is the Fubini number $\mathsf{F}_{n-3}$, which counts the number of ordered partitions of $n{-}3$  elements
(or, more mundanely, $\mathsf{F}_{n}$ is the number of possible outcomes of an $n$-horse  race, including ties).
The  HEFT amplitudes enjoy several important properties which make them particularly convenient  as building blocks of loop integrands: in addition to the already mentioned gauge invariance, the BCJ numerators which build these amplitudes are local with respect to the  massless particles,   have  poles corresponding to the  propagation of the heavy particles, and factorise into products of lower-point ones on the massive poles.

We now come to discuss the main observable quantity  of this paper, that is   the gravitational waveforms produced in the scattering process between two spinless heavy objects at one loop in the PM expansion. 
At leading order, the waveforms for spinless objects were computed  in 
\cite{DEath:1976bbo,Kovacs:1977uw,Kovacs:1978eu} and reproduced  recently in  \cite{Jakobsen:2021smu} (and in \cite{Mougiakakos:2021ckm} as a  one-dimensional integral), while in \cite{Jakobsen:2021lvp,Riva:2022fru}
this was generalised to include spin.
Waveforms
in the frequency domain were obtained recently in the zero-frequency limit  \cite{DiVecchia:2022nna} and 
 for generic frequencies \cite{DiVecchia:2021bdo} using amplitudes-based techniques.%
 \footnote{Waveforms have been previously and extensively computed in the PN expansion, see \textit{e.g.}~\cite{Blanchet:1993ec,Ross:2012fc,Galley:2009px, Porto:2010zg,Porto:2012as}.}
 A precise definition of waveforms in terms of amplitudes, which we will employ in this paper,  was %put forward 
 proposed in \cite{Cristofoli:2021vyo}  and further studied in \cite{Cristofoli:2021jas}, based on the KMOC approach  \cite{Kosower:2018adc}. 
 This approach states that the one-loop waveform is computed from two contributions: the first is simply the one-loop five-point amplitude, and the second is a particular unitarity cut of the same amplitude.  In the first version of this paper we incorrectly assumed that, after taking the heavy-mass limit, these pieces combined into a one-loop five-point HEFT amplitude, with the sole purpose of the cut term being to cancel hyper-classical contributions.  We were led to reconsider this after the recent work \cite{Caron-Huot:2023vxl}. There it was shown that the cut contribution not only cancels hyper-classical pieces but also  provides additional classical terms, which we compute in this revised version. The combination of our old and this new contribution leads to the correct one-loop waveform. Furthermore, and remarkably, it was shown in \cite{Caron-Huot:2023vxl} that at one loop, the effect of the cut contribution in the KMOC approach is equivalent to modifying the $i \varepsilon$ prescriptions of the massive propagators from principal value (PV) to retarded propagators.  

In conclusion, the only input required to compute the one-loop waveform is  the one-loop HEFT integrand obtained from 2MPI diagrams but with modified (retarded) propagators for the massive legs. In other words, the second term in KMOC can be interpreted in two ways: either it modifies $i\varepsilon $ prescriptions of massive legs (at one loop) from PV to retarded, or it adds additional classical terms to the one-loop HEFT amplitude evaluated with PV prescription for massive legs. For convenience of our presentation, we will follow the second interpretation.

 The first   goal of this work is then the computation of the one-loop HEFT, or classical,  amplitude.% 
\footnote{The  complete amplitude integrand was computed in \cite{Carrasco:2020ywq, Carrasco:2021bmu}, but for our purposes it is convenient to compute it directly in the HEFT, bypassing a potentially involved extraction of its classical~part.}
The key ingredients in this computation, which enter the unitarity cuts,  are the HEFT amplitudes with two scalars mentioned earlier, but  in addition we find that amplitudes with four scalars and one and two gravitons are also needed in the evaluation of a  particular class of `snail-like' cut diagrams. In order to  compute these four-scalar  amplitudes we devise  a novel incarnation of the BCFW recursion relation, where we shift the momentum {\it transfers}. Such shifts have the advantage of leaving unmodified the linear propagators corresponding to massive particles within the HEFT amplitudes (with two scalars), and lead to the  large-$z$ behaviour that is necessary to avoid boundary terms in the recursion. 

Combining all cuts we first obtain the one-loop HEFT amplitude integrand, which is reduced using LiteRed2 \cite{Lee:2012cn,Lee:2013mka}. There is an additional layer of simplicity introduced by re-parameterising the heavy momenta in the HEFT using what we will refer to as $\pb$- or $\mb$-variables. Indeed, we can reduce to just one  basis of master integrals with a single $i \varepsilon$ prescription, \textit{i.e.}~any linear propagator will appear with a principal value prescription without the need to distinguish between $\pm i \varepsilon$ prescriptions. The expression for our integrand is obtained using $D$-dimensional amplitudes and thus is valid in $D$~dimensions.  We then proceed to evaluate all relevant integrals around $D{=}4$ using the method of differential equations \cite{Kotikov:1990kg,Bern:1993kr,Remiddi:1997ny,Gehrmann:1999as}
in canonical form \cite{Henn:2013pwa}, adapted to  the study of classical gravitational dynamics \cite{Parra-Martinez:2020dzs}.
 The integrated result for our   amplitude is infrared divergent, and  the divergence is in agreement with Weinberg's universal formula \cite{Weinberg:1965nx}. A number of further highly non-trivial checks on our result are also presented, including subtle cancellations of several spurious singularities.   
A similar approach is followed to compute the additional classical contribution arising from the second KMOC contribution mentioned earlier.  
 
Armed with the result for the   integrated amplitude (from the first KMOC term),  augmented by the additional classical piece arising from the second KMOC term,  we then  proceed to   compute the spectral waveform (or waveform in the frequency domain). 
%  
%\begin{align}
%	\label{intwf2intro}
%	\begin{split}
%		W\sim  \int\! d^4q_1 d^4q_2 \, \delta^{(4)} %(q_1 {+} q_2 {-} k) \delta(  {p}_1\Cdot q_1 ) %\delta({p}_2\Cdot q_2 )\, e^{i q_1\Cdot b}\  
 % %\widehat{\mathcal{M}}_{5,\rm HEFT}^{(1)} (q_1, q_2)
%\mathcal{ R}_{5,\rm HEFT}^{(1)} (q_1, q_2; h )
%		\, ,
%	\end{split}
%\end{align}
%where $\mathcal{R}_{5,\rm HEFT}^{(1)} (q_1, q_2; h )$ %is built from two pieces 
%$\widehat{\mathcal{M}}_{5,\rm HEFT}^{(1)} (q_1, q_2)$ 
%related to the finite part of 
%the one-loop HEFT amplitude with the emission of a graviton of momentum $k$ and helicity $h$.  
Interestingly,  this quantity    is infrared divergent, as is the one-loop amplitude; this was already noted in  \cite{Porto:2012as}, where it was observed that  in the time domain  this divergence can be absorbed by a redefinition of the time variable and is thus, reassuringly, unobservable. 
We can then restrict to   its finite part  and move on to evaluate numerically the   two-dimensional integral that gives the spectral waveform. We do this  for several values  of the mass ratios  of the heavy objects   and as a function of the frequency $\omega$ of the emitted gravitational wave.  Finally, we Fourier transform the spectral waveforms to obtain the time-domain waveforms in the far-field region. A convenient quantity to compute is the Newman-Penrose scalar $\Psi_4$ \cite{Newman:1961qr}, which represents the second time derivative of the gravitational strain in the far-field region. 
The results of our numerical evaluations are presented in a number of plots in the frequency and time domains for various mass ratios of the heavy objects.    The interested reader can find {\it Mathematica}  notebooks with expressions for the HEFT amplitudes at tree level with one and two emitted gravitons, and at one loop with one emitted graviton in our   
{\href{https://github.com/QMULAmplitudes/Gravity-Observables-From-Amplitudes}{{\it Gravity Observables from Amplitudes} GitHub repository.}}

The rest of the paper is organised as follows. In Section~\ref{sec:kinematics} we briefly review the five-point kinematics of the process at hand, introducing the  parameterisation employed in subsequent calculations. 
In Section~\ref{sec: HEFTexpansion} we provide a self-contained introduction to the HEFT expansion we use, including a few illustrations thereof: the HEFT expansion of Weinberg's soft factor, and that of the gravitational Compton amplitude. Section~\ref{sec: BCFWtwosources} discusses our new recursion relations, providing expressions of the classical gravitational amplitudes with four scalars and up to two gravitons required in later sections. In Section~\ref{sec: oneloopunitaritycuts} we perform the key computation of the paper: that of the one-loop HEFT amplitude with four scalars and one graviton using unitarity. All cuts are then merged into a single integrand, which is reduced to master integrals using LiteRed2 \cite{Lee:2012cn, Lee:2013mka}. The final result for the integrand is shown in \eqref{finalresultform1} and \eqref{finalresultform2}.
 The relevant family of master integrals is presented in Section~\ref{sec: oneLoopIntegrals}. Section~\ref{sec:finalresult} discusses the final integrated result for the one-loop amplitude, shown in \eqref{eq:OneLoopAmplitude}, along with several consistency checks of our computation. 
 In  Section~\ref{sec: waveforms} we move on to discuss the waveforms. We begin by  reviewing the KMOC approach to such quantities,  and show how to compute waveforms from the HEFT. In this section we also present the computation of the classical contribution arising from the second term in the KMOC expression for the waveforms discussed earlier. 
 We then calculate waveforms  numerically for several mass ratios, in the frequency and time domains, illustrating our results in several plots.  
 Section~\ref{sec: theend} summarises our conclusions and prospects for future work. A few appendices complete the paper: in Appendix~\ref{sec:diffEq} we present a detailed evaluation  of the  integrals listed in Section~\ref{sec: oneLoopIntegrals}
 using the method of differential equations;  in Appendix~\ref{sec:Weinberg-resurrection}  we review Weinberg's classic result for the infrared-divergent part of one-loop amplitudes in gravity and extract its classical part, which we use in the main text as one of the checks on our results; in Appendix~\ref{sec:factorisation} we show in an example that the exponentiation in impact parameter space of two massive particle reducible diagrams occurs also in the presence of radiation;   and finally, in Appendix~\ref{app: c4Details} we give details of one of the unitarity cuts used in our calculation.

%\vspace{0.1cm}

\noindent{\bf Note added:} 
While this paper was in preparation, we became aware of \cite{Herderschee:2023fxh} and 
\cite{Elkhidir:2023dco}, which appear concurrently with our work and partly overlap with it. The key results  of these papers are in agreement.  We thank the authors for communication and for sharing copies of their drafts prior to publication.

\section{Kinematics of the five-point scattering process}
\label{sec:kinematics}
We are interested in computing the scattering amplitudes of two heavy scalars of masses $m_1$ and $m_2,$ accompanied by the emission of a graviton of momentum $k$:
\begin{equation}\label{eq: kinematics}
	\begin{array}{lr}

		\begin{tikzpicture}[scale=15,baseline={([yshift=-1mm]centro.base)}]
			\def\x{0}
			\def\y{0}

			\node at (0+\x,0+\y) (centro) {};
			\node at (-3pt+\x,-3pt+\y) (uno) {$p_1=\bar{p}_1 + \dfrac{q_1}{2}$};
			\node at (-3pt+\x,3pt+\y) (due) {$p_2=\bar{p}_2 + \dfrac{q_2}{2}$};
			\node at (3pt+\x,3pt+\y) (tre) {$p_2^\prime =  \bar{p}_2 - \dfrac{q_2}{2}$};
			\node at (3pt+\x,-3pt+\y) (quattro) {$\ \ p_1^\prime = \bar{p}_1 - \dfrac{q_1}{2}$};
			\node at (5.55pt+\x,\y) (cinque)
			% era 4.25 
			{$k= q_1 + q_2$};

			\draw [thick,double,red] (uno) -- (centro);
			\draw [thick,double,blue] (due) -- (centro);
			\draw [thick,double,blue] (tre) -- (centro);
			\draw [thick,double,red] (quattro) -- (centro);
			\draw [vector,double] (cinque) -- (centro);

			\draw [->] (-2.8pt+\x,-2pt+\y) -- (-1.8pt+\x,-1pt+\y);
			\draw [<-] (2.8pt+\x,-2pt+\y) -- (1.8pt+\x,-1pt+\y);
			\draw [<-] (-1.8pt+\x,1pt+\y) -- (-2.8pt+\x,2pt+\y);
			\draw [->] (1.8pt+\x,1pt+\y) -- (2.8pt+\x,2pt+\y);
			\draw [->] (2.3pt+\x,-0.6pt+\y) -- (3.75pt+\x,-0.6pt+\y);

			\shade [shading=radial] (centro) circle (1.5pt);
		\end{tikzpicture}
		 & \hspace{2cm}
		%	\begin{aligned}
		%& k^\mu = q_1^\mu + q_2^\mu\ ,     \\
		% & p_i^\mu =  m_i v_i^\mu 
		%\ ,       \\
		%& p_i \Cdot q_i \sim 0 
		%  \ .
		%	\end{aligned}
	\end{array}
\end{equation}

Here we have introduced the  convenient  ``barred'' variables \cite{Landshoff:1969yyn,Parra-Martinez:2020dzs}, defined as
\begin{align}
	\label{barredv}
	\begin{split}
		p_1 &= \bar{p}_1 + \frac{q_1}{2}\, , \qquad p_1^\prime = \bar{p}_1 - \frac{q_1}{2}\, , \\
		p_2 &= \bar{p}_2 +  \frac{q_2}{2}\, , \qquad p_2^\prime = \bar{p}_2 - \frac{q_2}{2}\, .
	\end{split}
\end{align}
The  advantage of this parameterisation is that, using the on-shell conditions, one can show that \begin{align}
	\label{eq: HEFTfame}
	\bar{p}_1\Cdot q_1 =\bar{p}_2\Cdot q_2 =0 \, ,
\end{align}
\textit{i.e.}~the momentum transfers $q_{1,2}$ are exactly orthogonal to the barred momenta $\bar{p}_{1,2}$ of the heavy scalars.
It is also useful  to introduce barred masses,
\begin{align}\label{eq: mbardef}
	\bar{m}_i^2 \coloneqq \bar{p}_i^2 = m_i^2 - \frac{q_i^2}{4}\, ,
\end{align}
where $i$ runs from one to the number of heavy particles (two in this case), and $\bar{p}_i {\coloneqq} \bar{m}_i \bar{v}_i$.  As we shall see in Section~\ref{sec: HEFTexpansion}, the HEFT perturbative expansion is organised in powers of the $\bar{m}_i$.
We also mention that it will sometimes  be useful to write the $q_i$ in terms of the radiated momentum $k$ and a single, average momentum transfer $q$ as
\begin{align}\label{qdefs}
	q_1 = q+\frac{k}{2} \ \ , \ \ q_2=-q +\frac{k}{2}  \ , \quad \text{with} \ q \coloneqq \frac{q_1-q_2}{2} \ .
\end{align}
To describe  our five-point scattering process, we need to specify five independent Lorentz-invariant products, which we choose as
\begin{equation}
	\label{fiveiv}
	\begin{split}
		y \coloneqq v_1 \Cdot v_2  \geq 1\ ,\qquad
		q_i^2 \leq 0\ ,\qquad w_i \coloneqq v_i \Cdot k  \geq 0 \, , \qquad i=1,2,
	\end{split}
\end{equation}
where as usual we define the four-velocities using $p_i {=} m_i v_i$, with $v_i^2{=}1$.

To show that $w_i\!\geq\!0$ and $q_i^2\!\leq\!0$, we simply go to a frame where $v_i \!=\! (1,\Vec{0})$  (for fixed $i$)  and $k\!=\!(\omega, 0, 0, \omega)$. Then
$v_i \Cdot k = \omega \geq 0$, and from momentum conservation one finds that $q_i^2 \leq 0$ in the heavy-mass limit (this can be checked  by going again to the rest frame of particle $i$).
%$v_i \Cdot q_i = 0$ imposes $q_i = (0, %\Vec{q}_i)$ } and . 
Furthermore, $y$ is the relativistic factor $\frac{1}{\sqrt{1-\dot{\vec{x}}^2}}$, where $\dot{\vec{x}}$ is the relative velocity of one of the two massive bodies in the rest frame of the other. For example, we can choose the rest frame of particle $1$, where  $v_1^{\mu}=(1,0,0,0)$, and then  $v_2^{\mu}=y(1, \dot{\vec{x}})$. Hence, $y\!\geq\!1$, where $y\!=\!1$ corresponds to the static limit.
We will also regularly use  barred versions of these invariants, namely $\bar{w}_i \coloneqq \bar{v}_i \Cdot k$ and $\bar{y} \coloneqq \bar{v}_1 \Cdot \bar{v}_2$.

In the following, we will denote  the HEFT amplitudes with two or four heavy particles (plus any number of gravitons), as $\cA$  and $\cM$ respectively.  Complete amplitudes will be denoted as $A$ and $M$. Finally, all of the amplitudes in this paper will be matrix elements of $i\, T$ where we write the 
$S$-matrix as $S{=}\mathbb{1}+i\, T$.

%------------------------------------------------------------------------------------------------
\section{Basics of the  HEFT perturbative expansion}\label{sec: HEFTexpansion}

In this section, we give a short introduction and review of the salient features of the HEFT perturbative expansion.
As an invitation to the subject, we illustrate this expansion by applying it to several examples: first, to  the three-point and four-point gravity amplitudes with two heavy particles, and then to the Weinberg soft factor for the emission of a soft graviton in  two-to-two scattering -- the process which is the main focus of this paper.  We then outline the computational strategy used to compute loop corrections to classical quantities in general relativity, following \cite{Brandhuber:2021eyq}.
A final related application will be discussed in Appendix~\ref{sec:Weinberg-resurrection} in connection with the structure of infrared divergences of  gravitational amplitudes at loop level.

\subsection{The three-point amplitude}\label{sec: 3pointHEFT}
Our first simple example is the gravitational three-point tree-level amplitude
\begin{align}
	\label{fig: 3pointFull}
	 & \begin{tikzpicture}[baseline={([yshift=-0.8ex]current bounding box.center)}]\tikzstyle{every node}=[font=\small]
		   \begin{feynman}
			\vertex (a) {\(p_1\)};
			\vertex [right=1.8cm of a] (f2)[HV]{$A_3$};
			\vertex [right=1.8cm of f2] (c){$p_{1'}$};
			\vertex [above=1.8cm of f2] (gm){$q$};
			\vertex [left=0.8cm of gm] (g2){};
			\vertex [right=0.8cm of gm] (g20){};
			\diagram* {
			(a) -- [fermion,thick] (f2) --  [fermion,thick] (c),
			(gm)--[photon,ultra thick,rmomentum'](f2)
			};
		\end{feynman}
	   \end{tikzpicture}\,,
\end{align}
which is given by%
\footnote{In this work we   define
	Newton's constant  as $G={\kappa^2}/ {(32 \pi)}$.}
\begin{equation}
	\label{eq: 3pointFull}
	A_3= -i \kappa (p_1 \Cdot \varepsilon_q)^2\, .
\end{equation}
The two massive scalars carry momenta $p_{1}$ and $p_{1^\prime}$, with
$p_1^2 {=} p_{1'}^2 {=} m^2$ while the graviton has momentum $q$ and polarisation tensor $\varepsilon_q^\mu \varepsilon_q^\nu$. If all of the momenta are real in Minkowski signature then momentum conservation and the on-shell conditions imply that $q{=}0$. However, this amplitude is still non-zero since the polarisation vector $\varepsilon_q$ is well-defined in the limit of zero graviton energy. Additionally, we will frequently use amplitudes like the above in unitary cuts and BCFW diagrams, where it is necessary to make the momentum $q$ complex and non-zero.

We now wish to perform the HEFT expansion of this amplitude and therefore  we use the barred variables \cite{Landshoff:1969yyn,Parra-Martinez:2020dzs} introduced in the previous section
\begin{align}
	p_1 = \bar{p} + \frac{q}{2}\, , \quad
	p_{1'} = \bar{p} - \frac{q}{2}\, ,
\end{align}
which satisfy  $\bar{p}\Cdot q=0$. Furthermore, we define the barred mass and velocity as
\begin{align}
	\bar{p} = \bar{m} \bar{v}\, ,  \qquad \text{with} \ \qquad
	\bar{m} =\sqrt{m^2 - \frac{q^2}{4}}\ .
\end{align}
Expanding the three-point amplitude \eqref{eq: 3pointFull} for large $\mb$, while keeping $q$ fixed, we find that there is only one term, which is of order $\mb^2$
\begin{align}
	\label{eq: 3PointHEFT}
	\cA_{3}(q,\pb)\coloneqq -i\kappa\, \mb^2(\vb\Cdot\varepsilon_q)^2=
	\begin{tikzpicture}[baseline={([yshift=-0.8ex]current bounding box.center)}]\tikzstyle{every node}=[font=\small]
		\begin{feynman}
			\vertex (a) {\(p_1\)};
			\vertex [right=1.8cm of a] (f2)[HV]{H};
			\vertex [right=1.8cm of f2] (c){$p_{1'}$};
			\vertex [above=1.8cm of f2] (gm){$q$};
			\vertex [left=0.8cm of gm] (g2){};
			\vertex [right=0.8cm of gm] (g20){};
			\diagram* {
			(a) -- [fermion,thick] (f2) --  [fermion,thick] (c),
			(gm)--[photon,ultra thick,rmomentum'](f2)
			};
		\end{feynman}
	\end{tikzpicture}\, .
\end{align}
We define this  $\cO(\mb^2)$ term as the three-point HEFT amplitude, and we always label such amplitudes in diagrams with the letter ``H''.

%------------------------------------------------------------------------------------------
\subsection{The gravity Compton amplitude and its HEFT expansion}
\label{sec:4.1}
We  now move on to the tree-level gravitational Compton amplitude, which was derived \textit{e.g.}~in~\cite{Bjerrum-Bohr:2013bxa}.
\begin{align}
	\label{compton}
	 & \begin{tikzpicture}[baseline={([yshift=-0.8ex]current bounding box.center)}]\tikzstyle{every node}=[font=\small]
		   \begin{feynman}
			\vertex (a) {\(p_1\)};
			\vertex [right=1.8cm of a] (f2)[HV]{$A_4$};
			\vertex [right=1.8cm of f2] (c){$p_{1'}$};
			\vertex [above=1.8cm of f2] (gm){};
			\vertex [left=0.8cm of gm] (g2){$\ell_{1}$};
			\vertex [right=0.8cm of gm] (g20){$\ell_{2}$};
			\diagram* {
			(a) -- [fermion,thick] (f2) --  [fermion,thick] (c),
			(g2)--[photon,ultra thick,rmomentum'](f2),(g20)--[photon,ultra thick,rmomentum](f2)
			};
		\end{feynman}
	   \end{tikzpicture}
\end{align}
As before the massive scalars carry momenta $p_{1}$ and $p_{1^\prime}$, with
$p_1^2 {=} p_{1'}^2 {=} m^2$, and the two gravitons have momenta $\ell_{1,2}$, with $\ell_{1,2}^2=0$.
Momentum conservation relates these  as $p_1 {=} p_{1'} {+} \ell_1 {+} \ell_2$, and for later convenience we also introduce $q{\coloneqq}\ell_1 {+} \ell_2$.
The four-point Compton amplitude in the full theory can be written in a way that makes  its double-copy structure manifest:
\begin{align}
	\label{4ptcin}
	A_4 =  \frac{i \kappa^2}{16}  \left( \frac{N_{12}^2}{D_{12} }+ \frac{N_{21}^2}{D_{21} } + \frac{N_{[12]}^2}{D}\right) \, ,
\end{align}
where the denominators and numerators are
\begin{align}
	\begin{split}
		D_{12} &= - 2 (p_1\Cdot \ell_1) + i \varepsilon\, ,\quad
		\\
		D&= q^2 + i \varepsilon\, , \\
		N_{12} &= 2\Big[ 2 (p_1 \Cdot \varepsilon_1)(p_1 - \ell_1)\Cdot \varepsilon_2 + (p_1 \Cdot \ell_1) (\varepsilon_1\Cdot \varepsilon_2)\Big]\, .
	\end{split}
\end{align}
Here $(D_{21}, N_{21}) = (D_{12}, N_{12})_{\ell_1\leftrightarrow \ell_2}$ and $N_{[12]} = N_{12} - N_{21}$. It is also useful to note  that
$(D_{12} + D_{21})_{\varepsilon=0} {=} -q^2 = -D_{\varepsilon=0} $.

Remarkably, one can combine the terms in \eqref{4ptcin} into the compact expression
\begin{align}
	\label{2.3}
	A_4 = \frac{i \kappa^2}{16} \frac{(D_{12}N_{21} + D_{21} N_{12})^2}{D_{12} D_{12}D } =  i \, 4 \kappa^2   \frac{(p_1 \Cdot F_1 \Cdot F_2 \Cdot p_1)^2}{D_{12} D_{21} D}\, ,
\end{align}
where the numerator is the square of the corresponding Compton amplitude in Yang-Mills, and   we introduced the linearised field strength  tensors $F_i^{\mu\nu}=
	\ell_i^\mu \varepsilon_i^\nu - \ell_i^\nu \varepsilon_i^\mu$.

In order to perform the HEFT expansion of the Compton amplitude it is essential to make the Feynman $i\varepsilon$ prescription explicit, as we will see below, and to once again use the barred variables
\begin{align}
	\label{viva-barred}
	p_1 = \bar{p} + \frac{q}{2}\, , \quad
	p_{1'} = \bar{p} - \frac{q}{2}\, \, .
\end{align}
As previously stated these satisfy $\bar{p}\Cdot q=0$ which avoids inconvenient feed-down terms in the expansion arising from dot products of the form
$p_1 \Cdot q = q^2/2$.
Using the barred variables, we can rewrite the denominators as
\begin{align}
	\begin{split}
		D_{12} &= - 2 (\bar{p}\Cdot \ell_1) + i \varepsilon - \frac{q^2}{2}\, , \\
		D_{21} &= - 2 (\bar{p}\Cdot \ell_2) + i \varepsilon - \frac{q^2}{2}\,
		= 2 (\bar{p}\Cdot \ell_1) + i \varepsilon - \frac{q^2}{2}\, .
	\end{split}
\end{align}
Again, we  write $\pb\coloneqq\mb \vb$ and expand  the massive propagators for large $\mb$ keeping the momenta of massless particles fixed:
\begin{align}\label{eq: propexpansion}
	\begin{split}
		\frac{1}{D_{12}}
		&=\frac{1}{- 2 (\bar{p}\Cdot \ell_1) + i \varepsilon} +\frac{q^2}{2 (-2\bar{p}\Cdot \ell_1+i\varepsilon)^2} +\cdots,  %= - \frac{i \pi}{2} \delta(\bar{p}\Cdot \ell_1)  - \frac{1}{2 (\bar{p}\Cdot \ell_1) }+\frac{q^2}{8 (\bar{p}\Cdot \ell_1)^2} +\cdots \, , 
		\\
		\frac{1}{D_{21}}
		&=\frac{1}{ 2 (\bar{p}\Cdot \ell_1) + i \varepsilon} + \frac{q^2}{2 (2\bar{p}\Cdot \ell_1+i\varepsilon)^2} + \cdots ,
		% = - \frac{i \pi}{2} \delta(\bar{p}\Cdot \ell_1)  + \frac{1}{2 (\bar{p}\Cdot \ell_1) }+\frac{q^2}{8 (\bar{p}\Cdot \ell_1)^2}+ \cdots \, , 
	\end{split}
\end{align}
and
\begin{align}
	\begin{split}
		\label{4.8}
		\frac{1}{D_{12}D_{21}D}
		&=-\frac{1}{(q^2+ i \varepsilon)^2 }\Big( \frac{1}{D_{12}} + \frac{1}{D_{21}}\Big)  =
		\frac{1}{(q^2+ i \varepsilon)^2}\Big[ i \pi  \, \delta (\bar{p}\Cdot \ell_1)  - \frac{q^2}{4 (\bar{p}\Cdot \ell_1)^2} \Big]  +\cdots \, .
	\end{split}
\end{align}
One can then use this to expand  the quantity $\frac{(p_1 \Cdot F_1 \Cdot F_2 \Cdot p_1)^2}{D_{12} D_{21} D}$ in
\eqref{2.3}.
The delta-function supported term gives
\begin{align}
	\begin{split}
		\label{primoo}
		i\pi  \frac{\delta (\bar{p}\Cdot \ell_1) (p_1\Cdot F_1 \Cdot F_2 \Cdot p_1)^2}{(q^2+ i \varepsilon)^2} & =
		\frac{i\, \pi}{4}\delta (\bar{p}\Cdot \ell_1)  \left[  (\bar{p}\Cdot \varepsilon_1)(\bar{p}\Cdot \varepsilon_2)-\frac{1}{4} (q\Cdot \varepsilon_1) (q\Cdot \varepsilon_2) + \frac{q^2}{8}(\varepsilon_1\Cdot \varepsilon_2)  \right]^2
		\\
		& = \frac{i\,\pi}{4}\delta ( \bar{p}\Cdot \ell_1) \ (\bar{p}\Cdot \varepsilon_1)^2(\bar{p}\Cdot \varepsilon_2)^2+\cdots\, .
	\end{split}
\end{align}
This is of $\cO (\bar{m}^3)$, while the dots correspond to terms of order
$\cO (\bar{m})$ which are not relevant for classical physics and can be dropped.
The term with the squared linearised propagator, of $\cO (\bar{m}^2)$, is given by
\begin{align}
	\label{secondoo}
	- \frac{(\bar{p} \Cdot F_1 \Cdot F_2 \Cdot \bar{p})^2}{4 (q^2 + i\varepsilon) (\bar{p}\Cdot \ell_1)^2 } \ .
\end{align}
Combining \eqref{primoo} and \eqref{secondoo}, also reinstating the coupling constant dependence,  we arrive at
\begin{align}
	\label{eq: comptonExpansion}
	A_{4}  = A_{4,\bar{m}^3} + A_{4,\bar{m}^2}\, ,
\end{align}
with
\begin{align}
	\label{eq: comptonheft-1}
	%\begin{split}
	A_{4,\bar{m}^3} &= - i \kappa^2 \bar{m}^3  \big[ - i \pi
	\, \delta  ( \bar{v}\Cdot \ell_1) \ (\bar{v}\Cdot \varepsilon_1)^2 (\bar{v}\Cdot \varepsilon_2)^2\big]= \pi \cA_3(\ell_1,\pb)\,\delta(\pb\Cdot \ell_1)\,\cA_3(\ell_2,\pb) \, ,
	\\ \cr
	\label{eq: comptonheft-2}
	A_{4,\bar{m}^2}     & \coloneqq\cA_{4}(\ell_1,\ell_2,\pb)=    - i \kappa^2  \bar{m}^2
	\left(\frac{\bar{v} \Cdot F_1 \Cdot F_2 \Cdot \bar{v}}{\bar{v}\Cdot \ell_1}\right)^2
	\frac{1}{q^2 + i \varepsilon}\ .
	%\end{split}
\end{align}
The normalisation of \eqref{eq: comptonheft-1} and  \eqref{eq: comptonheft-2} is consistent with the three-point HEFT amplitude we found in the previous section.
We now define the  \textit{HEFT amplitude involving two massive scalars} as the term in the HEFT expansion which is homogeneous in  $\mb$ and of $\mathcal{O}(\mb^2)$.
 Hence, the amplitude $\cA_{4}$ in \eqref{eq: comptonheft-2} is the HEFT four-point Compton amplitude. This can also be obtained via the double copy
\cite{Bern:2008qj,Bern:2010ue} from its Yang-Mills counterpart, as shown in \cite{Brandhuber:2021eyq}. The two terms in the expansion of the Compton amplitude in \eqref{eq: comptonExpansion} can be expressed diagrammatically as follows,
\begin{align}
\label{eq: comptonExpansionDiagram}
	\begin{tikzpicture}[baseline={([yshift=-0.8ex]current bounding box.center)}]\tikzstyle{every node}=[font=\small]
		\begin{feynman}
			\vertex (a) {\(p_1\)};
			\vertex [right=1.8cm of a] (f2)[HV]{$A_4$};
			\vertex [right=1.8cm of f2] (c){$p_{1'}$};
			\vertex [above=1.8cm of f2] (gm){};
			\vertex [left=0.8cm of gm] (g2){$\ell_{1}$};
			\vertex [right=0.8cm of gm] (g20){$\ell_{2}$};
			\diagram* {
			(a) -- [fermion,thick] (f2) --  [fermion,thick] (c),
			(g2)--[photon,ultra thick,rmomentum'](f2),(g20)--[photon,ultra thick,rmomentum](f2)
			};
		\end{feynman}
	\end{tikzpicture}
	=
	\begin{tikzpicture}[baseline={([yshift=-0.8ex]current bounding box.center)}]\tikzstyle{every node}=[font=\small]
		\begin{feynman}
			\vertex (a) {\(p_1\)};
			\vertex [right=1.8cm of a] (f2)[HV]{H};
			\vertex [right=1.8cm of f2] (c)[HV]{H};
			\vertex [right=1.8cm of c] (d){\(p_{1'}\)};
			\vertex [above=1.8cm of f2] (g1){$\ell_{1}$};
			\vertex [above=1.8cm of c] (g2){$\ell_{2}$};
			\vertex [right=0.9 cm of f2] (cut);
			\vertex [above=0.3cm of cut] (cutu);
			\vertex [below=0.3cm of cut] (cutb);
			\diagram* {
			(a) -- [fermion,thick] (f2) --  [thick] (c)--  [fermion,thick] (d),
			(g1)--[photon,ultra thick,rmomentum'](f2),
			(g2)--[photon,ultra thick,rmomentum](c),
			(cutu)--[red,thick] (cutb)
			};
		\end{feynman}
	\end{tikzpicture}
	+
	\begin{tikzpicture}[baseline={([yshift=-0.8ex]current bounding box.center)}]\tikzstyle{every node}=[font=\small]
		\begin{feynman}
			\vertex (a) {\(p_1\)};
			\vertex [right=1.8cm of a] (f2)[HV]{H};
			\vertex [right=1.8cm of f2] (c){$p_{1'}$};
			\vertex [above=1.8cm of f2] (gm){};
			\vertex [left=0.8cm of gm] (g2){$\ell_{1}$};
			\vertex [right=0.8cm of gm] (g20){$\ell_{2}$};
			\diagram* {
			(a) -- [fermion,thick] (f2) --  [fermion,thick] (c),
			(g2)--[photon,ultra thick,rmomentum'](f2),(g20)--[photon,ultra thick,rmomentum](f2)
			};
		\end{feynman}
	\end{tikzpicture}
\end{align}
where the  line cut in red   corresponds to the delta function $\pi \delta(\pb\Cdot \ell_1)$ in \eqref{eq: comptonheft-1}. Note that the ordering of the two three-point HEFT amplitudes on either side of the red cut does not matter.

We can now make a few  observations on the general structure of the expansion we have just seen in this example:

{\bf 1.} While the HEFT amplitude is $\mathcal{O}(\mb^2)$, we have also found  a term \eqref{eq: comptonheft-1} with two three-point amplitudes joined by a  ``cut propagator''. This term is of  $\mathcal{O}(\mb^3)$, and  we will refer to it as the  {\it ``hyper-classical term''.}
Note that $\bar{m} = \sqrt{m^2- \frac{q^2}{4}}$, hence this parameter does not have a fixed $\hbar$ scaling, which is however recovered in the large-$m$ limit.%
\footnote{Further comments on the subtle distinction between the $1/\bar{m}$ and the $\hbar$ expansions can be found  in Section~\ref{sec: classicalvsHEFT}.}
In this terminology, the HEFT amplitude is then the {\it classical amplitude.}

	{\bf 2.}
Propagators of massless particles are untouched by the HEFT expansion, and hence are treated with the  standard Feynman $i\varepsilon$ prescription.
However, our  HEFT amplitude contains squared linearised propagators, and it is clear from the preceding derivation (see \textit{e.g.}~\eqref{eq: propexpansion} and \eqref{4.8}) that such propagators appear with the derivative of the principal value prescription%
\footnote{This is  known as the Hadamard's partie finie regularisation.}
\begin{equation}
\label{PVprime}
	\text{PV}'\left(\frac{1}{x}\right)\coloneqq\frac{d}{dx}\text{PV}\left(\frac{1}{x}\right)= -\frac{1}{2}\Big( \frac{1}{(x + i \varepsilon)^2} + \frac{1}{(x - i \varepsilon)^2}\Big)\,.
\end{equation}
For instance, in \eqref{4.8}, by $1 /  (\bar{p}\Cdot\ell_1)^2$ one really  means the combination
\begin{align}\label{eq: hadamard}
	\frac{1}{2} \Big( \frac{1}{(\bar{p}\Cdot\ell_1 + i \varepsilon)^2} + \frac{1}{(\bar{p}\Cdot\ell_1 - i \varepsilon)^2}\Big)\, .
\end{align}
We will see in Section \ref{sec:diag-HEFT-exp}
how these features extend to generic amplitudes. We also comment that when we reduce our integrands using integration by parts identities (IBP), we are left with a basis of master integrals which contain only a single power of the linearised propagators.
These can then be treated with the standard principal value prescription.

\subsection{Weinberg soft factor and its HEFT expansion}
\label{subsec:Weinberg}

An interesting limit of the five-particle process introduced in Section
\ref{sec:kinematics} is the soft limit where the graviton momentum $k \to 0$. In \cite{Weinberg:1965nx} it was shown that in this limit the amplitude factorises into the elastic four-point amplitude (without the graviton) multiplied by the universal Weinberg soft factor. Note that this statement for the leading soft singularity holds at all loop orders in gravity, and will   be used in Section~\ref{sec:finalresult} 
as a consistency check of our one-loop result.

We now want to apply our HEFT expansion to this soft factor, which will    decompose  into  a delta-function supported term and a HEFT term. The kinematics of  our scattering of two heavy bodies with the emission of one graviton has been described earlier in Section~\ref{sec:kinematics}. In terms of the variables introduced there,
Weinberg's factor for the emission of a soft graviton has the form \cite{Weinberg:1965nx}
\begin{align}
	S_{\rm W} & = \frac{\kappa}{2}\,\varepsilon_{\mu\nu}(k) \left[\frac{p_1^{\prime\mu} p_1^{\prime \nu}}{p_1^\prime\Cdot k + i \varepsilon} +  \frac{p_2^{\prime \mu} p_2^{\prime \nu}}{p_2^\prime\Cdot k + i \varepsilon} 
	-\frac{p_1^\mu p_1^\nu}{p_1\Cdot k - i \varepsilon} -  \frac{p_2^\mu p_2^\nu}{p_2\Cdot k - i \varepsilon}  \right]\, ,
\end{align}
where we have kept the Feynman $i \varepsilon$.
Next, we rewrite it using the barred variables
introduced in \eqref{barredv}. 
One  then expands  the denominators using
\begin{align}
	\frac{1}{x+i \varepsilon} =  {\rm PV}\left(\frac{1}{x}\right) - i \pi \delta(x)\, ,
\end{align}
and  $\big[(\pb_i\!-\!{q_i}/{2})\Cdot k\big]^{-1} \to(\pb_i\Cdot k)^{-1} \big[ 1 + \frac{1}{2}(q_i\Cdot k) / (\pb_i \Cdot k)\big] + \cdots$ for large $\pb_i\coloneqq \mb_i\vb_i$, where we retain only terms up to $\mathcal{O}(\mb_i^{-2})$.  We  also set
$q_1 \!=\! - q_2 \!\coloneqq\! q$ where appropriate. 
Doing so one obtains delta-function supported (or hyper-classical) term  
\begin{align}
	\label{SWd}
	S_{\rm W}^{\delta} =  -\kappa\, i\, \pi \varepsilon_{\mu \nu}(k) \Big[ \bar{p}_1^\mu \bar{p}_1^\nu \, \delta (\bar{p}_1\Cdot k) + 1\leftrightarrow 2\Big]\, ,
\end{align}
along with  the HEFT part of the soft factor (effectively derived by setting all the $i \varepsilon$ to zero):
\begin{align}
	\label{SWHEFT}
	S_{\rm W}^{\rm HEFT} = - \frac{\kappa}{2}\,\varepsilon_{\mu \nu}(k) \left[
		\frac{\bar{p}_1^\mu q^\nu +\bar{p}_1^\nu q^\mu}{\bar{p}_1\Cdot k} - \bar{p}_1^\mu \bar{p}_1^\nu \frac{q\Cdot k}{(\bar{p}_1\Cdot k)^2}\, - \, 1\leftrightarrow 2\right]\, .
\end{align}
A few comments are in order here.

	{\bf 1.} First, we observe that the convenience of the barred variable  stems from the fact that $S_{\rm W}$ is neatly decomposes into the sum of the two terms \eqref{SWd} and
\eqref{SWHEFT}. Using unbarred variables, the result would be given by the sum of the unbarred versions of
\eqref{SWd} and \eqref{SWHEFT} and  in addition we would also have the ``feed-down'' term 
\begin{align}
	\label{feed-down-rubbish}
	\Delta S_{\rm W} = \frac{\kappa}{2}\, i \pi
	\varepsilon_{\mu \nu} (k)
	\Big[ (p_1^\mu q^\nu + p_1^\nu q^\mu) \delta(p_1\Cdot k) \ +  \
		p_1^\mu p_1^\nu  (q\Cdot k) \delta^\prime (p_1\Cdot k) \, - \, 1\leftrightarrow 2 \Big]\, .
\end{align}

{\bf 2.}
The expansion of
$S_{\rm W} {\to}  S_{\rm W}^{\delta} + S_{\rm W}^{\rm HEFT} $ parallels  that of the amplitudes, see our previous example \eqref{eq: comptonheft-1}-\eqref{eq: comptonheft-2}
and the general discussion in the next section. In particular, $S_{\rm W}^{\delta}$ and $S_{\rm W}^{\rm HEFT} $ are of $\cO(\mb)$ and $\cO(\mb^0)$, respectively.

	{\bf 3.}  Finally, we note that $S_{\rm W}^{\rm HEFT}$ can  be recast
in the interesting forms% 
\footnote{Note that the  pole in $q\Cdot k$ in is spurious but allows for this compact expression.}
\cite{Bautista:2019evw}
\begin{align}
	S_{\rm W}^{\rm HEFT} = \frac{\kappa}{2}\frac{1}{q\Cdot k}\left[ \frac{\big(\bar{p}_1 \Cdot F \Cdot q\big)^2}{( \bar{p}_1\Cdot k)^2 }  - \frac{\big(\bar{p}_2 \Cdot F \Cdot q\big)^2}{( \bar{p}_2\Cdot k)^2}\right]
	\ ,
\end{align}
and
\begin{align}
	\label{eq:softHEFT}
	S_{\rm W}^{\rm HEFT} =  \frac{\kappa}{2}\frac{\bar{p}_1\Cdot F\Cdot \bar{p}_2}{(\bar{p}_1\Cdot k)(\bar{p}_2\Cdot k)}\left( \frac{\bar{p}_1\Cdot F\Cdot q}{\bar{p}_1\Cdot k}  + \frac{\bar{p}_2\Cdot F \Cdot q}{\bar{p}_2\Cdot k}\right)
	\ ,
\end{align} 
where  $a\Cdot F(k) \Cdot b \!\coloneqq\! (a\Cdot k) (\varepsilon\Cdot b) - (b\Cdot k)(a\Cdot \varepsilon) $ and  $\varepsilon_{\mu\nu} (k) \!\coloneqq\!\varepsilon_{\mu} (k)\varepsilon_{\nu} (k)$ as usual.
Note that $S_{\rm W}^{\rm HEFT}$  is $\mathcal{O}(k^{-1})$ and linear in $q$.
%Similar expressions hold for $S_{\rm W}^{\rm HEFT}$.

\subsection{Diagrammatics of the HEFT expansion with one heavy source}
\label{sec:diag-HEFT-exp}
 
We now summarise some of the key aspects of the HEFT expansion of  tree  amplitudes with two massive scalars of mass $m$ and $n$ gravitons. This section expands on and clarifies the ideas presented in \cite{Brandhuber:2021eyq}.

We first focus in on the expansion of a single scalar propagator. In gravity we must  sum over all possible orderings of gravitons scattering off a heavy particle: for instance, we have contributions coming from the following two schematic Feynman diagrams containing an explicit propagator,
\begin{align}
	\label{join-2-heft}
	 & \begin{tikzpicture}[baseline={([yshift=-0.8ex]current bounding box.center)}]\tikzstyle{every node}=[font=\small]
		   \begin{feynman}
			\vertex (a) {\(p_1\)};
			\vertex [right=1.5cm of a] (f2)[HV]{$~~~$};
			\vertex [right=2.0cm of f2] (f3)[HV]{$~~~$};
			\vertex [right=1.5cm of f3] (c){$p_{1'}$};
			\vertex [below=1.1cm of f2] (gma){$\boldsymbol\cdots$};
			\vertex [below=1.5cm of f2] (gm){};
			\vertex [below=1.1cm of f3] (gm2a){$\boldsymbol\cdots$};
			\vertex [below=1.5cm of f3] (gm2){};
			\vertex [left=0.5cm of gm] (g2){$\ell_{1_L}$};
			\vertex [right=0.5cm of gm] (g20){$\ell_{i_L}$};
			\vertex [left=0.5cm of gm2] (g3){$\ell_{1_R}$};
			\vertex [right=0.5cm of gm2] (g30){$\ell_{j_R}$};
			\diagram* {
			(a) -- [fermion,thick] (f2)-- [fermion,thick] (f3) --  [fermion,thick] (c),
			(g2)--[photon,ultra thick,rmomentum](f2),
			(g20)--[photon,ultra thick,rmomentum'](f2),
			(g3)--[photon,ultra thick,rmomentum](f3),
			(g30)--[photon,ultra thick,rmomentum'](f3),
			};
		\end{feynman}
	   \end{tikzpicture}+\begin{tikzpicture}[baseline={([yshift=-0.8ex]current bounding box.center)}]\tikzstyle{every node}=[font=\small]
		                     \begin{feynman}
			\vertex (a) {\(p_1\)};
			\vertex [right=1.5cm of a] (f2)[HV]{$~~~$};
			\vertex [right=2.0cm of f2] (f3)[HV]{$~~~$};
			\vertex [right=1.5cm of f3] (c){$p_{1'}$};
			\vertex [below=1.1cm of f2] (gma){$\boldsymbol\cdots$};
			\vertex [below=1.5cm of f2] (gm){};
			\vertex [below=1.1cm of f3] (gm2a){$\boldsymbol\cdots$};
			\vertex [below=1.5cm of f3] (gm2){};
			\vertex [left=0.5cm of gm] (g2){$\ell_{1_R}$};
			\vertex [right=0.5cm of gm] (g20){$\ell_{j_R}$};
			\vertex [left=0.5cm of gm2] (g3){$\ell_{1_L}$};
			\vertex [right=0.5cm of gm2] (g30){$\ell_{i_L}$};
			\diagram* {
			(a) -- [fermion,thick] (f2)-- [fermion,thick] (f3) --  [fermion,thick] (c),
			(g2)--[photon,ultra thick,rmomentum](f2),(g20)--[photon,ultra thick,rmomentum'](f2), (g3)--[photon,ultra thick,rmomentum](f3),(g30)--[photon,ultra thick,rmomentum'](f3),
			};
		\end{feynman}
	                     \end{tikzpicture}
\end{align}
where the scalar propagators  in the diagrams above are
$\dfrac{i}{(p_1- Q_L)^2 - m^2 + i \varepsilon}$ and
$\dfrac{i}{(p_1-  Q_R)^2 - m^2 + i \varepsilon}$, with
\begin{align}
	Q_L\coloneqq \ell_{1_L} + \cdots +\ell_{i_L}\, , \qquad
	Q_R\coloneqq \ell_{1_R} + \cdots +\ell_{j_R}\, .
\end{align}
Switching to barred variables \eqref{viva-barred}, as required by the HEFT expansion,
\begin{align}
	p_1 = \bar{p}+ \frac{q}{2}\, , \quad p_{1'} = \bar{p}- \frac{q}{2}\, ,
\end{align}
with $q= Q_L + Q_R$, we can rewrite the propagators  as
\begin{align}
	\begin{split}
		\label{eeeq}
		\frac{i}{-2p_1\Cdot Q_L + Q_L^2 + i \varepsilon} &=\frac{i}{ -2 \bar{p}\Cdot Q_L  - q\Cdot Q_L+ Q_L^2+ i \varepsilon}\\
        &=\pi \delta(-2 \bar{p}\Cdot Q_L  - q\Cdot Q_L+ Q_L^2)+{\rm PV}\frac{i}{-2 \bar{p}\Cdot Q_L  - q\Cdot Q_L+ Q_L^2}, \\
         &=\pi \delta(2 \bar{p}\Cdot Q_L)-{\rm PV}\frac{i}{2 \bar{p}\Cdot Q_L}+(q\Cdot Q_L- Q_L^2)\Big(\pi \delta^{\prime}(2 \bar{p}\Cdot Q_L)-{\rm PV}^{\prime}\frac{1}{2 \bar{p}\Cdot Q_L}\Big)\\
         &+\cdots,\\
		\frac{i}{2p_1 \Cdot Q_L + Q_L^2+ i \varepsilon} &=\frac{i}{ 2 \bar{p}\Cdot Q_L  - q\Cdot Q_L+ Q_L^2+ i \varepsilon} \\
        &=\pi \delta(2 \bar{p}\Cdot Q_L)+{\rm PV}\frac{i}{2 \bar{p}\Cdot Q_L}+(q\Cdot Q_L- Q_L^2)\Big(-\pi \delta^{\prime}(2 \bar{p}\Cdot Q_L)-{\rm PV}^{\prime}\frac{1}{2 \bar{p}\Cdot Q_L}\Big)\\
        &+\cdots,
	\end{split}
\end{align}
where we used   $\bar{p}\Cdot q=0$ and performed a heavy-mass expansion in powers of $\pb$. The notation ${\rm PV}^\prime$ was introduced in \eqref{PVprime}. There may also be contact-like Feynman diagrams where the above propagator is absent, these terms will have their own heavy-mass expansion. 

We perform the above expansion on all propagators and combine pairs of diagrams and their crossed versions in \eqref{join-2-heft} using \eqref{eeeq}. 
The leading-order piece in the $\mb$ expansion comes from Feynman diagrams with the maximal number of scalar propagators.%
\footnote{Contact Feynman diagrams with fewer scalar propagators scale with one fewer power of $\mb$ for each absent propagator, for example a 
two-graviton vertex scales with $\mb^2$ but two one-graviton vertices separated by a propagator scale like $\mb^2\mb^2\mb^{-1}=\mb^3$. See \eqref{4.8} where we carried out this procedure in the specific example of the two-graviton Compton amplitude.} In addition, the scalar propagators in this leading-order piece all combine into delta functions. We denote this leading-order contribution with cut propagators diagrammatically~as
\begin{align}
\label{eq: maximalhypclassicalDiagram}
	\begin{tikzpicture}[baseline={([yshift=-0.8ex]current bounding box.center)}]\tikzstyle{every node}=[font=\small]
		\begin{feynman}
			\vertex (a) {\(p_1\)};
			\vertex [right=1.8cm of a] (f2)[HV]{H};
            \vertex [right=1.8cm of f2] (mid)[]{};
			\vertex [right=1.8cm of mid] (c)[HV]{H};
			\vertex [right=1.8cm of c] (d){\(p_{1'}\)};
			\vertex [above=1.8cm of f2] (g1){$\ell_{1}$};
			\vertex [above=1.8cm of c] (g2){$\ell_{n}$};
            \vertex [above=0.9cm of mid] (dots){\textbf{\Large$\cdots$}};
			\vertex [right=0.9 cm of f2] (cut1);
            \vertex [right=0.9 cm of mid] (cut2);
			\vertex [above=0.3cm of cut1] (cut1u);
			\vertex [below=0.3cm of cut1] (cut1b);
            \vertex [above=0.3cm of cut2] (cut2u);
			\vertex [below=0.3cm of cut2] (cut2b);
			\diagram* {
			(a) -- [fermion,thick] (f2) --  [thick] (mid)--  [thick] (c)--  [fermion,thick] (d),
			(g1)--[photon,ultra thick,rmomentum'](f2),
			(g2)--[photon,ultra thick,rmomentum](c),
			(cut1u)--[red,thick] (cut1b),(cut2u)--[red,thick] (cut2b)
			};
		\end{feynman}
	\end{tikzpicture}
\end{align}
which is explicitly given by \cite{Kabat:1992tb}
\begin{align}
	\begin{split}
		\label{KOargument}
		\Big(\prod_{j=1}^{n-1} \pi \, \delta({ \bar p \Cdot \ell_{j}} )\Big) \cA_{3}(\ell_1,{\pb}) \cdots \cA_{3} (\ell_n,{\pb})\, .
	\end{split}
\end{align}
This amplitude has maximal hyper-classical scaling $\mb^{n+1}$, where a factor of ${\bar m}^{-1}$ comes from each delta function and we have extracted the leading $\mb^2$ part of each three-point vertex which gives a HEFT three-point amplitude.

The first sub-leading contribution scales as $\mb^{n}$ (which may still be hyper-classical); this will involve, amongst other things, the sub-leading terms in the expansion \eqref{eeeq} which feature derivatives of delta functions. For the two-graviton Compton amplitude this sub-leading contribution is of  $\cO(\mb^2)$ and so would be, in fact, classical. However, we saw that the two-graviton HEFT Compton was free of $\delta^{\prime}$ terms --  this is because in the diagram \eqref{join-2-heft} there was just one possible propagator separating three-point amplitudes. As such, there is only one diagram and its crossed version, and from these we simply get the the sum of propagators in \eqref{eeeq} cancelling the $\delta^{\prime}$ terms. This is no longer true beyond the two-graviton case, which we now explore by considering the three-graviton amplitude.%
\footnote{We thank Aidan Herderschee, Radu Roiban and Fei Teng for helpful discussions on this point.}

The three-graviton Compton amplitude
\begin{align}
	\label{5point1SourceHEFT}
	 & \begin{tikzpicture}[baseline={([yshift=-0.8ex]current bounding box.center)}]\tikzstyle{every node}=[font=\small]
		   \begin{feynman}
			\vertex (a)[]{$\pb-q/2$};
			\vertex[left=1.9cm of a] (a0)[HV]{\,\,\,\,\,\,\,\,};
			\vertex[left=1.9cm of a0] (am1){$\pb+q/2$};
			\vertex[above=1.9cm of a0] (b0){$\ell_2$};
			\vertex[left=1.6cm of b0] (bm1){$\ell_1$};
			\vertex[right=1.6cm of b0] (b1){$\ell_3$};
			\diagram*{(a0)--[fermion, very thick](a),(am1)--[fermion, very thick](a0), (a0) -- [photon,ultra thick, momentum] (b0),(a0) -- [photon,ultra thick, momentum] (bm1),(a0) -- [photon,ultra thick, momentum'] (b1)};
		\end{feynman}
	   \end{tikzpicture}\,.
\end{align}
can also be heavy-mass expanded. The leading term will be of the form \eqref{eq: maximalhypclassicalDiagram} and is of  $\cO(\mb^4)$, while the classical amplitude is of  $\cO(\mb^2)$ and so is twice sub-leading. From the expansion of the propagator in \eqref{eeeq} we see that it may contain terms, amongst many others, which have a double derivative of a delta function.
In the remainder of the paper we drop such derivative of delta function contributions in the three-graviton Compton amplitude. In effect the individual HEFT amplitudes are taken with the $i\varepsilon$ prescription set to zero such that only principal value poles remain. We are justified in doing so because when we include (derivatives of) delta function terms in the construction of the one-loop amplitude they give only zero-energy mode contributions to our final result for the waveform. These only contribute a time-independent term to the gravitational strain (defined later in \eqref{strain}) and so do not contribute to the asymptotic Newman-Penrose scalar. This situation will not always be true when one considers higher-loop computations.

Finally, we can see that as we increase the number of gravitons we encounter more and more derivatives of delta functions in the classical amplitude coming from the expansion \eqref{eeeq} of the propagators. This is because the classical amplitude becomes more and more sub-leading compared to the leading $\cO(\mb^{n+1})$ contribution.

\subsection{\texorpdfstring{$\hbar$}{} vs HEFT (or \texorpdfstring{$1/\mb$}{}) expansion}
\label{sec: classicalvsHEFT}

The HEFT expansion is closely related, but subtly distinct from the  $\hbar$ expansion, and here we briefly outline some of the differences, also highlighting the advantages of the HEFT expansion.

In the expansion
around small values of $\hbar$,
one scales  Newton's constant
as $G {\to} G/\hbar$ and the graviton momenta (or sums thereof) as  $k \to \hbar k $  while keeping their wavenumbers fixed \cite{Kosower:2018adc}.
Equivalently, one can take the heavy-mass limit $m {\rightarrow} \infty$: the graviton momenta then scale as $\mathcal{O}(m^0)$ and hence are  subleading compared to any  massive momentum $p {=} m v$ which scales as $\mathcal{O}(m)$. The  expansion in inverse powers of the masses  is therefore equivalent to the expansion in small $\hbar$.
This prescription is conceptually clean
but has  some  unpleasant features.
%For example we may want to declare that the incoming momenta %$p_{1,2}$ are purely classical as they scale with $m_{1,2}$, then %the outgoing momenta $p^\prime_{1,2}=p_{1,2} \pm q$ would be a %mixture of classical and quantum pieces. 
The on-shell conditions for the incoming, $p_i$, and outgoing, $p_{i^\prime} {=} p_i \pm q$ massive momenta
$p_{i}^2 = p_{i'}^2 = m_{i}^2$  imply that 
$p_i \Cdot q = \mp q^2/2\sim \cO (\hbar^2)$. 
Hence,  contracting massless momenta with massive ones  leads to expressions that in general do not have a homogeneous degree in $\hbar$. As a consequence, the hyper-classical part of an amplitude will generate terms that feed down to the classical part of the same amplitude upon use of momentum conservation and on-shell conditions. Going to increasingly higher orders in the loop expansion, one will be faced with a proliferation of such feed-down terms from hyper-classical contributions.%
\footnote{An example of feed-down terms is provided in \eqref{feed-down-rubbish}. }

As we have seen in the preceding sections, the natural expansion parameters in the  HEFT  are the inverse   barred mass variables $1/\bar{m}_i$, and  each term of the expansion has a fixed degree in $\bar{m}_i$.
An advantage of the HEFT expansion is that in terms of the barred variables introduced in Section~\ref{sec:kinematics}, momentum conservation implies the exact statement 
$\bar{p}_i\Cdot q_i{=}0$. 
As a consequence, the HEFT expansion is free of feed-down terms from {\it hyper-classical terms} such as those  appearing in the conventional $\hbar$ expansion. In other words, the expansion in $\mb$ is robust under the use of momentum conservation and on-shell conditions with no mixing of the various perturbative orders. Of course, the HEFT expansion still contains hyper-classical terms, for example those in \eqref{eq: comptonExpansionDiagram}. However, these are subtracted in the classical observables we wish to compute, as we shall see explicitly in Section \ref{sec: waveforms}.

As can be seen from their definition \eqref{eq: mbardef}, the barred masses  do not themselves have a fixed scaling in terms of $\hbar$, therefore each term in the HEFT expansion does not have a fixed scaling in $\hbar$. Of course  an expansion in $\hbar$ can be obtained from the HEFT expansion instantaneously by using \eqref{eq: mbardef}.
In other words, the HEFT expansion is  a reorganisation of the $\hbar$ expansion. Importantly, once we focus on a classical observable and obtain an expression free of hyper-classical terms (of order $\mb^s$ for $s{>}2$) then we can re-expand in $\hbar$ or equivalently $m$. From \eqref{eq: mbardef} it is clear that the re-expansion affects only the quantum terms which are not relevant. At this point we are free to replace $\mb$ with $m$ and we have a bona fide classical object.

\section{Tree-level amplitudes with four heavy scalars from a BCFW recursion}\label{sec: BCFWtwosources}
\subsection{Diagrammatics of the HEFT expansion with four heavy scalars}\label{sec: HEFT4Scalars}
To compute the classical five-point one-loop amplitude we will also require HEFT amplitudes involving four external scalars, in addition to the two-scalar HEFT amplitudes discussed above.
The HEFT expansion for these amplitudes is a natural extension of the two scalar case. We start with an amplitude involving four scalars and $n{-}4$ gravitons, denoted $M_n$, 
\begin{align}\label{eq: 2to2andRgravitonAmp}
     \begin{tikzpicture}[baseline={([yshift=-0.4ex]current bounding box.center)}]\tikzstyle{every node}=[font=\small]
		      \begin{feynman}
			\vertex (p1) {\(p_1\)};
			\vertex [above=2.5cm of p1](p2){$p_2$};
			\vertex [right=1.5cm of p2] (u1) [HV]{$~~~$};
                \vertex [right=1.5cm of u1] (u2) ;
			\vertex [below=1.25cm of u1] (m1) [GR2]{$M_n$};
			\vertex [below=1.25cm of m1] (b0) [HV]{$~~~$};
			\vertex [right=1.5cm of u1] (p3){$p_{2'}$};
			\vertex [right=1.0cm of p1] (b1) []{};
			\vertex [right=1.0cm of b1] (b2) []{};
			\vertex [right=1.0cm of b2] (p4){$p_{1'}$};
   			\vertex [above=2.0cm of p4] (k1){};
                \vertex [above=1.25cm of p4] (k2){$\vdots$};
                \vertex [above=0.4cm of p4] (k3){};
                \vertex [right=0.3cm of k1] (k1l){$k_1$};
                \vertex [right=0.3cm of k3] (k3l){$k_{n-4}$};
			\diagram* {
			(p2) -- [thick] (u1)-- [thick] (p3), (p1) -- [thick] (b0)--  [thick] (p4), (m1)--[photon,ultra thick] (k1),
    (m1)--[photon,ultra thick] (k3)
			};
		\end{feynman}
	      \end{tikzpicture}\, .
\end{align}

To perform the HEFT expansion of the amplitude above amplitude we expand in both the barred masses $\mb_1$ and $\mb_2$ associated with the two massive lines. The analysis of the massive propagators is essentially identical to the two scalar case in Section~\ref{sec:diag-HEFT-exp}, except that there are now two such massive lines in every diagram.

The simplest example is the elastic process, for which we only need to keep the leading order in the large-$\mb_i$ expansion
\begin{equation}
 M_4
=
       \begin{tikzpicture}[baseline={([yshift=-0.4ex]current bounding box.center)}]\tikzstyle{every node}=[font=\small]
		      \begin{feynman}
			\vertex (p1) {\(p_1\)};
			\vertex [above=2.5cm of p1](p2){$p_2$};
			\vertex [right=1.5cm of p2] (u1) [HV]{$~~~$};
                \vertex [right=1.5cm of u1] (u2) ;
			\vertex [below=1.25cm of u1] (m1) [GR2]{H};
			\vertex [below=1.25cm of m1] (b0) [HV]{$~~~$};
			\vertex [right=1.5cm of u1] (p3){$p_{2'}$};
			\vertex [right=1.0cm of p1] (b1) []{};
			\vertex [right=1.0cm of b1] (b2) []{};
			\vertex [right=1.0cm of b2] (p4){$p_{1'}$};
   			\vertex [above=1.25cm of p4] (k1){};
			\diagram* {
			(p2) -- [thick] (u1)-- [thick] (p3), (p1) -- [thick] (b0)--  [thick] (p4)
			};
		\end{feynman}
	      \end{tikzpicture}+\cdots
\end{equation}

where $+\cdots$ are subleading terms which are irrelevant for classical physics.
This leading-order contribution scales like $\mb_1^2\mb_2^2$ and is what we define as the HEFT amplitude with four heavy scalars: $\cM_4$. We will see how to calculate this amplitude in the next section explicitly, but for now let us continue with another example of the four-scalar HEFT expansion. 

The tree-level five-point amplitude, with four scalars and one graviton, when expanded in the large-$\mb_i$ limit, contains iteration terms. These appear since the amplitude contains massive propagators which, when expanded, give delta functions exactly as in \eqref{eeeq}. As usual, we can write the expansion of this amplitude diagrammatically as follows
\begin{align}
\label{eq:5pointHEFTExpansion}
    M_5
=
    \begin{tikzpicture}[baseline={([yshift=-0.4ex]current bounding box.center)}]\tikzstyle{every node}=[font=\small]
		      \begin{feynman}
			\vertex (p1) {\(p_1\)};
			\vertex [above=2.5cm of p1](p2){$p_2$};
			\vertex [right=1.0cm of p2] (u1) [HV]{$~~~$};
			\vertex [below=1.25cm of u1] (m1) [GR2]{H};
			\vertex [below=1.25cm of m1] (b0) [HV]{$~~~$};
			\vertex [right=2.6cm of u1] (p3){$p_{2'}$};
			\vertex [right=1.0cm of p1] (b1) []{};
			\vertex [right=0.8cm of b1] (b2) []{};
			\vertex [right=0.8cm of b2] (b3) [HV]{H};
			\vertex [right=1.0cm of b3](p4){$p_{1'}$};
			\vertex [above=1.25cm of p1] (cutL);
			\vertex [right=3.2cm of cutL] (cutR){$k$};
			\vertex [right=0.8cm of b1] (cut1);
			\vertex [above=0.3cm of cut1] (cut1u);
			\vertex [below=0.3cm of cut1] (cut1b);
			\vertex [above=0.4cm of cut1u] (cutd1);
			\vertex [right=0.4cm of cutd1] (cutd2);
			\vertex [right=0.5cm of b2] (cut2);
			\vertex [above=0.3cm of cut2] (cut2u);
			\vertex [below=0.3cm of cut2] (cut2b);
			\diagram* {
			(p2) -- [thick] (u1) -- [thick] (p3), (p1) -- [thick] (b0)-- [thick] (b3)-- [thick] (p4),  (cut1u)--[ red,thick] (cut1b),(cutR)-- [photon,ultra thick] (b3),
			};
		\end{feynman}
	      \end{tikzpicture}
       +
       \begin{tikzpicture}[baseline={([yshift=-0.4ex]current bounding box.center)}]\tikzstyle{every node}=[font=\small]
		      \begin{feynman}
			\vertex (p1) {\(p_1\)};
			\vertex [above=2.5cm of p1](p2){$p_2$};
			\vertex [right=1.0cm of p2] (u1) [HV]{$~~~$};
                \vertex [right=1.0cm of u1] (u2) ;
			\vertex [below=1.25cm of u1] (m1) [GR2]{H};
			\vertex [below=1.25cm of m1] (b0) [HV]{$~~~$};
			\vertex [right=2.6cm of u1] (p3){$p_{2'}$};
			\vertex [right=1.0cm of p1] (b1) []{};
			\vertex [right=1.0cm of b1] (b2) []{};
			\vertex [right=0.6cm of b2] (b3);
			\vertex [right=0.6cm of b3](p4){$p_{1'}$};
                \vertex [above=2.1cm of b3](u3)[HV]{H};
			\vertex [above=1.25cm of p1] (cutL);
			\vertex [right=3.2cm of cutL] (cutR){$k$};
			\vertex [right=0.8cm of b1] (cut1);
			\vertex [above=0.3cm of cut1] (cut1u);
			\vertex [below=0.3cm of cut1] (cut1b);
                \vertex [above=2.5cm of cut1u] (cut2u);
			\vertex [above=2.5cm of cut1b] (cut2b);
			\vertex [above=0.4cm of cut1u] (cutd1);
			\vertex [right=0.4cm of cutd1] (cutd2);
			\vertex [right=0.5cm of b2] (cut2);
			\diagram* {
			(p2) -- [thick] (u1) -- [thick] (u3)-- [thick] (p3), (p1) -- [thick] (b0)--  [thick] (p4),  (cut2u)--[ red,thick] (cut2b),(cutR)-- [photon,ultra thick] (u3),
			};
		\end{feynman}
	      \end{tikzpicture}
       +
       \begin{tikzpicture}[baseline={([yshift=-0.4ex]current bounding box.center)}]\tikzstyle{every node}=[font=\small]
		      \begin{feynman}
			\vertex (p1) {\(p_1\)};
			\vertex [above=2.5cm of p1](p2){$p_2$};
			\vertex [right=1.5cm of p2] (u1) [HV]{$~~~$};
                \vertex [right=1.5cm of u1] (u2) ;
			\vertex [below=1.25cm of u1] (m1) [GR2]{H};
			\vertex [below=1.25cm of m1] (b0) [HV]{$~~~$};
			\vertex [right=1.5cm of u1] (p3){$p_{2'}$};
			\vertex [right=1.0cm of p1] (b1) []{};
			\vertex [right=1.0cm of b1] (b2) []{};
			\vertex [right=1.0cm of b2] (p4){$p_{1'}$};
   			\vertex [above=1.25cm of p4] (k1){$k$};
			\diagram* {
			(p2) -- [thick] (u1)-- [thick] (p3), (p1) -- [thick] (b0)--  [thick] (p4), (m1)--[photon,ultra thick] (k1)
			};
		\end{feynman}
	      \end{tikzpicture}+\cdots
\end{align}
where the three-point amplitudes are the HEFT amplitudes we found in \eqref{eq: 3PointHEFT}. The red cut line denotes the same delta function as in the two-scalar case: $\pi\delta(\pb_i \Cdot k)$ where $i=1,2$ if the first/second scalar line is cut. The new object here is the four-scalar one-graviton HEFT amplitude, denoted $\cM_5$, which scales as $\mb_1^2\mb_2^2$ and will  be calculated in the next section. 

For amplitudes like those in \eqref{eq: 2to2andRgravitonAmp} with more than one radiated graviton, the HEFT expansion proceeds analogously. The  $\cO(\mb_1^2\mb_2^2)$ term is what we define as the HEFT amplitude with four scalars and $n{-}4$ gravitons, and we denote it as $\cM_n$. Once again, in diagrams these are labelled with the letter ``H''. As was explained in Section \ref{sec:diag-HEFT-exp}, we will drop any $\delta'$ terms which appear in these amplitudes since they only contribute to time-independent terms in the waveform.

\subsection{The HEFT BCFW recursion relation}

Here we present a novel and highly efficient method to construct HEFT amplitudes with two pairs of scalars and any number of gravitons valid in $D$ dimensions. In order to do so we will use a $D$-dimensional version of BCFW on-shell recursion relations \cite{Britto:2004ap, Britto:2005fq}
with a carefully chosen shift that leaves unmodified  the linearised propagators of massive particles,  and only invokes factorisation channels that involve  gravitons. The only required inputs  are the HEFT amplitudes with a single pair of massive scalars  available in any dimension and for any multiplicity \cite{Brandhuber:2021kpo,Brandhuber:2021bsf,Brandhuber:2022enp},
and the well-established factorisation on poles corresponding to massless propagators.

We now describe the shift using the kinematic setup and conventions introduced in \eqref{eq: kinematics} for one radiated graviton. If there are several radiated gravitons we simply replace
\begin{align}
	q_1+q_2 = k \ \ \longrightarrow \ \ q_1+q_2 = \sum_{i=1}^{n-4} k_i \coloneqq K \ ,
\end{align}
where the $k_i$ are the graviton momenta and the corresponding polarisation vectors are denoted by $\varepsilon_i$.

A convenient choice for the $D$-dimensional  shifts turns out to be one where, unlike in the usual BCFW recursion \cite{Britto:2004ap,Britto:2005fq}, one shifts the internal momenta $q_1$ and $q_2$:
\begin{align}\label{bcfw-shift}
	\begin{split}
		q_1 & \rightarrow \ \ \hat{q}_1 = q_1 + z r \ , \\
		q_2 & \rightarrow \ \ \hat{q}_2 = q_2 - z r \ ,
	\end{split}
	\begin{tikzpicture}[baseline={([yshift=-0.4ex]current bounding box.center)}]\tikzstyle{every node}=[font=\small]
		\begin{feynman}
			\vertex (p1) {\(p_1\)};
			\vertex [above=2.0cm of p1](p2){$p_2$};
			\vertex [right=2cm of p2] (u1) [HV]{H};
			\vertex [right=2cm of u1] (p3){$p_{2'}$};
			\vertex [right=2.0cm of p1] (b3) [HV]{H};
			\vertex [right=2.0cm of b3](p4){$p_{1'}$};
			\vertex [above=1.25cm of p1] (cutL);
			\vertex [above=1.1cm of p1] (cutL0);
			\vertex [right=1cm of cutL0] (cutLL);
			\vertex [right=2cm of cutLL] (cutRR);
			\vertex [above right=1.0cm of b3] (gbr);
			\vertex [above left=1.0cm of b3] (gbl);
			\vertex [below right=1.0cm of u1] (gur);
			\vertex [below left=1.0cm of u1] (gul);
			\diagram* {
			(p2) -- [thick] (u1) -- [thick,momentum={[arrow style=red]\(z r\)}] (p3), (b3)-- [photon,ultra thick,rmomentum=\( \),momentum'={[arrow style=red]\(z r\)}] (u1), (p1) -- [thick] (b3), (p4)-- [thick,momentum={[arrow style=red]\(z r\)}] (b3), %(cutLL)--[dashed, red,thick] (cutRR),
			(gbr)-- [photon,ultra thick] (b3),(gbl)-- [photon,ultra thick] (b3), (gur)-- [photon,ultra thick] (u1),(gul)-- [photon,ultra thick] (u1),
			};
		\end{feynman}
	\end{tikzpicture}
\end{align}

where $z\in \mathbb{C}$  and $r$ is a null vector obeying
\begin{align}\label{shift-conditions}
	\begin{split}
		& r^2 = 0\, , \qquad \bar{p}_{1,2} \Cdot r = 0 \ , \qquad
		r \Cdot \varepsilon_i = r \Cdot k_i = 0 \ ,
	\end{split}
\end{align}
for all gravitons $i=1,\ldots , n$.
The first condition makes the shifted propagators scale as $z^{-1}$ for large $z$;  the second and third condition are important to guarantee that the shifted amplitude
$\mathcal{M}(z) \to 0$ as $z\to \infty$ as we show below.
Naively, \eqref{shift-conditions} seems to impose too many constraints on $r$ to allow for a non-trivial solution. The  way out is to demand that $r$ lives in a space whose dimension is larger than the spacetime dimension~$D$. For this to be useful  our knowledge of HEFT amplitudes in any dimension is crucial.
The solution for the shifts involves $q_i\Cdot r$, which need to be nonvanishing, hence also the $q_i$ must live in this larger spacetime, while the null vector $r$ lives only in the extra dimensions. Note that our shifts differ from those employed in \cite{Britto:2021pud}, which also addressed the computation of amplitudes with massive scalars and up to two gravitons, however before taking the classical limit.

Importantly, it is easy to see that the HEFT amplitudes appearing in the  on-shell diagrams are completely unaffected by the shifts
because of their  structure  and the conditions \eqref{shift-conditions}.
This makes these diagrams particularly efficient to compute HEFT amplitudes as we will  demonstrate for  $n {=} 0, 1, 2$ gravitons in Sections~\ref{sec:4ps}, \ref{sec:5ps} and \ref{sec:6ps}.

\subsection{Proof of large-$z$ behaviour}
In the following we want to show
that
HEFT or classical amplitudes with two pairs of scalars vanish as $z \to \infty$ and hence there is no problematic boundary term.

We will infer the  large-$z$ behaviour from general  properties of the Feynman rules  and the special properties of the shift vector $r$ introduced above.
First, these amplitudes scale as $\bar{m}_1^2 \bar{m}_2^2$ and subleading powers in $\bar{m}_i$ will always be dropped.%
\footnote{Note that in a full Feynman diagram computation including the Feynman $i \varepsilon$ prescription we also produce ``hyper-classical'' terms with higher powers of $\bar{m}_i$ and $\delta$-functions, as discussed in Section \ref{sec: HEFT4Scalars}. Such contributions involve products of simpler, lower-point HEFT amplitudes and $\delta$-functions, and hence  we do not discuss them here as they are easily accounted for.}

A generic Feynman diagram contributing to a four-scalar multi-graviton process involves two two-scalar-$m$-graviton vertices,  with masses $\bar{m}_1$ and $\bar{m}_2$, respectively, and up to $n$ multi-graviton vertices connected by graviton propagators.
In order to find the large-$z$ behaviour we only need to trace the momentum flow of the shifted momenta $\hat{q}_i$ through
a given diagram,
where $i$ labels the various propagators that can appear.

Let us consider a Feynman diagram with $s{-}1$ pure multi-graviton vertices and
$s$ graviton propagators with shifted momenta, connected to  two vertices with pairs of massive scalars. Each propagator will contribute a factor of $1/\hat{q}_i^2$ which scales as $1/z$, and hence the propagators produce a total factor of $z^{-s}$. Next there are $s{-}1$ multi-graviton vertices which are quadratic in the  momenta of internal or external gravitons.
Each vertex potentially contains two factors of shifted momenta $\hat{q}_i$. These either contract with $k_i$, $\varepsilon_i$ or $\bar{p}_i$, which removes the
$z$-dependent term in any $\hat{q}_i$ because of \eqref{shift-conditions}; or, the two shifted momenta contract with each other, which gives a factor linear in $z$. Hence, in the worst case one gets an overall scaling of $z^{-s} z^{s-1}= z^{-1}$ for a diagram with $s$ propagators.
Finally, the two-scalar-multi-graviton vertices scale as $\bar{m}_i^2$, from  two powers of $\bar{p}_i$. Even when the shift modifies the $\bar{p}_i$, the associated $z$-dependent  corrections are subleading in the $1/\bar{m}_i$ expansion and cannot contribute to the HEFT amplitude.
In conclusion, the HEFT amplitudes have favourable large-$z$ behaviour under the BCFW shifts
introduced above for any multiplicity. Therefore we can bootstrap any classical  amplitude with two pairs of massive scalars from a BCFW-like recursion of the multi-graviton Compton classical amplitudes.

\subsection{Four-point amplitude: elastic scattering}
\label{sec:4ps}

This corresponds to the classical $2 \rightarrow 2$ amplitude without radiation. In
this case $q_2 = -q_1 \coloneqq -q$ and $\hat{q} = q + z r$. The on-shell condition $\hat{q}^2=0$
is solved by $z = -q^2/(2 q \cdot r)$, however this shift only appears in the polarisation vectors in the BCFW sub-amplitudes due to the judicious choice of shift vector $r$.

There is a single on-shell diagram in the $q^2$-channel and the ingredients are the three-point amplitudes \eqref{eq: 3PointHEFT}
\begin{align}
	\label{3ptampG}
	\cA_3(\hat{q},\bar{p}_i) = -i \kappa (\bar{p}_{i} \Cdot \varepsilon_{\hat{q}} )^2 \ , \quad i=1,2 \ ,
\end{align}
and this diagram can be evaluated instantly as%
\footnote{We remind the reader that in this paper we denote as $\cA$ and $\cM$ the amplitudes containing two or four massive scalars, respectively.}
\begin{align}
	\begin{split}
		\cM_4 & = \sum_{h}
		(-i \kappa) (\bar{p}_{1} \Cdot \varepsilon_{\hat{q}} )^2 \frac{i}{q^2}
		(-i \kappa) (\bar{p}_{2} \Cdot \varepsilon_{-\hat{q}} )^2
		%\\& 
		= \frac{-i \kappa^2 \bar{m}_1^2 \bar{m}_2^2}{q^2}
		\sum_{h} (\bar{v}_{1} \cdot \varepsilon_{\hat{q}} )^2 (\bar{v}_{2} \cdot \varepsilon_{-\hat{q}} )^2 \\
		& = -i \kappa^2 \frac{\bar{m}_1^2 \bar{m}_2^2 (\bar{y}^2 - \frac{1}{D-2})}{q^2} \ ,
	\end{split}
\end{align}
where in the last step the sum over internal graviton polarisations was performed using%
\footnote{For an interesting discussion of completeness relations see \cite{Kosmopoulos:2020pcd}.}
\begin{align}
	\label{simplsum}
	\sum_{h}\varepsilon^{\mu_a}_{-\hat{q}} {\varepsilon}^{\nu_a}_{-\hat{q}}{\varepsilon}^{\mu_b}_{\hat{q}} \varepsilon^{\nu_b}_{\hat{q}}=\frac{1}{2}\Big[ \eta^{\mu_a \mu_b}\eta^{\nu_a \nu_b}+\eta^{\mu_a \nu_b}\eta^{\nu_a \mu_b}-\frac{2}{D-2}\eta^{\mu_a\nu_a}\eta^{\mu_b\nu_b}\Big]
	\ ,
\end{align}
where $d_\phi$  has  the following values for the cases of pure gravity or $\mathcal{N}=0$ supergravity:
\begin{align}
	\label{dphi}
	d_\phi = \begin{cases}
		         \dfrac{1}{D-2} & \qquad \mathrm{gravity,}
		         \\
		         0              & \qquad \mathcal{N}=0\ \mathrm{supergravity.}
	         \end{cases}
\end{align}
From \eqref{simplsum}, it follows that
\begin{align}\label{eq:spinSum}
	\sum_{h_a}(\bar{\varepsilon}_{-\hat{q}}\Cdot v)^2 f(\varepsilon_{\hat{q}}) =f|_{v}-{1\over D-2}f|_{ \eta}\, ,
\end{align}
where by  $f|_{v},f|_{ \eta}$ we denote replacing  $\varepsilon_{\hat{q}}^{\mu}\varepsilon_{\hat{q}}^{\nu}$ inside $f$ by $v^{\mu}v^{\nu}$ and  $\eta^{\mu\nu}$, respectively.

\subsection{Five-point amplitude:   process with one radiated graviton}
\label{sec:5ps}

\begin{align}\label{5point-bcfw}
	D_1=\begin{tikzpicture}[baseline={([yshift=-0.4ex]current bounding box.center)}]\tikzstyle{every node}=[font=\small]
		    \begin{feynman}
			\vertex (p1) {\(p_1\)};
			\vertex [above=2.5cm of p1](p2){$p_2$};
			\vertex [right=2cm of p2] (u1) [HV]{H};
			\vertex [right=2cm of u1] (p3){$p_{2'}$};
			\vertex [right=2.0cm of p1] (b3) [HV]{H};
			\vertex [right=2.0cm of b3](p4){$p_{1'}$};
			\vertex [above=1.25cm of p1] (cutL);
			\vertex [above=0.75cm of p1] (cutL0);
			\vertex [right=1cm of cutL0] (cutLL);
			\vertex [right=2cm of cutLL] (cutRR);
			\vertex [right=4.0cm of cutL] (cutR){$k$};
			\diagram* {
			(p2) -- [thick] (u1) -- [thick] (p3), (u1)-- [photon,ultra thick,rmomentum'=\(\hat{q}_{1}\)] (b3), (p1) -- [thick] (b3)-- [thick] (p4), (cutLL)--[dashed, red,thick] (cutRR),(cutR)-- [photon,ultra thick] (u1),
			};
		\end{feynman}
	    \end{tikzpicture}
	\,, &  &
	D_2=\begin{tikzpicture}[baseline={([yshift=-0.4ex]current bounding box.center)}]\tikzstyle{every node}=[font=\small]
		    \begin{feynman}
			\vertex (p1) {\(p_1\)};
			\vertex [above=2.5cm of p1](p2){$p_2$};
			\vertex [right=2cm of p2] (u1) [HV]{H};
			\vertex [right=2cm of u1] (p3){$p_{2'}$};
			\vertex [right=2.0cm of p1] (b3) [HV]{H};
			\vertex [right=2.0cm of b3](p4){$p_{1'}$};
			\vertex [above=1.25cm of p1] (cutL);
			\vertex [above=1.1cm of p1] (cutL0);
			\vertex [right=1cm of cutL0] (cutLL);
			\vertex [right=2cm of cutLL] (cutRR);
			\vertex [right=4.0cm of cutL] (cutR){$k$};
			\diagram* {
			(p2) -- [thick] (u1) -- [thick] (p3), (b3)-- [photon,ultra thick,rmomentum'=\(\hat{q}_{2}\)] (u1), (p1) -- [thick] (b3)-- [thick] (p4), (cutLL)--[dashed, red,thick] (cutRR),(cutR)-- [photon,ultra thick] (b3),
			};
		\end{feynman}
	    \end{tikzpicture}
	\,.
\end{align}

In Figure \eqref{5point-bcfw} we have drawn the two recursive diagrams that contribute to the five-point HEFT amplitude, $\cM_5$.
The on-shell conditions give $\hat{q}_1^2 = q_1^2 + 2 z_1 q_1 \Cdot r = 0$ and $\hat{q}_2^2 = q_2^2 - 2 z_2 q_2 \Cdot r = 0$ with the solutions
$z_1 = -q_1^2/(2 q_1 \Cdot r)$ and
$z_2 = q_2^2/(2 q_2 \Cdot r)$. The two diagrams
give contributions of the form
\begin{align}
	D_i = i\,  \frac{N_i}{q_i^2} \ , \qquad  i=1, 2 \ ,
\end{align}
where the numerators $N_i$ are obtained
from appropriate products of a three-point and a four-point HEFT amplitude and performing the relevant state sums. As was noted earlier in the elastic case, the three-point amplitudes are not modified by the BCFW shifts of the $q_i$. This is also true for the new ingredient we need in this case, namely the four-point HEFT amplitude. In the first diagram of Figure~\ref{5point-bcfw} this amplitude has the form, referring to \eqref{compton},
\begin{align}
	\cA_4(-\hat{q}_1,k,\bar{p}_2)
	=
	-\frac{i \kappa^2}{(-2 k \Cdot \hat{q}_1)}
	\left( \frac{\bar{p}_2 \Cdot F_k \Cdot F_{-\hat{q}_1} \Cdot \bar{p}_2}{\bar{p}_2 \Cdot k}\right)^2
	=
	-\frac{i \kappa^2}{(-2 k \Cdot q_1)}
	\left( \frac{\bar{p}_2 \Cdot F_k \Cdot F_{-q_1} \Cdot \bar{p}_2}{\bar{p}_2 \Cdot k}\right)^2 \ ,
\end{align}
where we are allowed to replace $\hat{q}_1$ by $q_1$ because $r \Cdot k = r \Cdot \varepsilon_k = r \Cdot \bar{p}_i = 0$. However, note that the polarisation vector in $F_{-q_1}$ remains $\varepsilon_{\hat{q}_1}$.
We can also set
$k \Cdot q_1 = k \Cdot q$, where $q= (q_1-q_2)/2$ is the average momentum transfer defined in \eqref{qdefs}.

With these preliminaries, we can now compute $N_1$:
\begin{align}\label{eq: 5pttreeN1}
	\begin{split}
		N_1 & = -\kappa^3 \sum_{h} (\bar{p}_1 \Cdot \varepsilon_{q_1})^2 \frac{(\bar{p}_2 \Cdot F_k \Cdot F_{-q_1} \Cdot \bar{p}_2)^2}{(-2 k \Cdot q)(\bar{p}_2 \Cdot k)^2}
		\\
		& = \frac{\kappa^3 \bar{m}^2_1 \bar{m}^2_2}{(2 k \Cdot q)\bar{w}_2^2} \left\{
		\Big[  \bar{y} (\bar{v}_2 \Cdot F_k \Cdot q) + \bar{w}_2 (\bar{v}_1 \Cdot F_k \Cdot \bar{v}_2)\Big]^2-
		\frac{1}{D-2} (\bar{v}_2 \Cdot F_k \Cdot q)^2\right\} \ ,
	\end{split}
\end{align}
where we have used \eqref{eq:spinSum} to perform the state sum. Similar manipulations give
\begin{align}
	\begin{split}\label{eq: 5pttreeN2}
		N_2  = -\frac{\kappa^3 \bar{m}^2_1 \bar{m}^2_2}{(2 k \Cdot q)\bar{w}_1^2} \left\{\Big[ \bar{y} (\bar{v}_1 \Cdot F_k \Cdot q) + \bar{w}_1 (\bar{v}_1 \Cdot F_k \Cdot \bar{v}_2)\Big]^2-
		\frac{1}{D-2} (\bar{v}_1 \Cdot F_k \Cdot q)^2\right\} \ ,
	\end{split}
\end{align}
in terms of which the five-point amplitude with one radiated graviton is
\begin{align}\label{eq: 5pttree}
	\cM_5(k,\pb_1,\pb_2) = i \frac{N_1}{q_1^2} + i \frac{N_2}{q_2^2}
	\ .
\end{align}
This result matches the form \cite{Bautista:2021inx} of the  classical five-point tree-level amplitude, first computed  in~\cite{Luna:2017dtq}. Note that both $N_1$ and $N_2$ contain the spurious pole $2 k\Cdot q$ which cancels when we sum the contributions from both BCFW diagrams.

\subsection{Six-point amplitude: process with two radiated gravitons}
\label{sec:6ps}

In the six-point case, four distinct recursive diagrams contribute and they are given by
\begin{align}\label{6point-bcfw}
	D_1=\begin{tikzpicture}[baseline={([yshift=-0.4ex]current bounding box.center)}]\tikzstyle{every node}=[font=\small]
		    \begin{feynman}
			\vertex (p1) {\(p_1\)};
			\vertex [above=2.5cm of p1](p2){$p_2$};
			\vertex [right=2cm of p2] (u1) [HV]{H};
			\vertex [right=2cm of u1] (p3){$p_{2'}$};
			\vertex [right=2.0cm of p1] (b3) [HV]{H};
			\vertex [right=2.0cm of b3](p4){$p_{1'}$};
			\vertex [above=1.25cm of p1] (cutL);
			\vertex [above=0.75cm of p1] (cutL0);
			\vertex [right=1cm of cutL0] (cutLL);
			\vertex [right=2cm of cutLL] (cutRR);
			\vertex [right=4.0cm of cutL] (cutR){$k_1$};
			\vertex [below=0.8cm of cutR] (k2){$k_2$};
			\diagram* {
			(p2) -- [thick] (u1) -- [thick] (p3), (u1)-- [photon,ultra thick,rmomentum'=\(\hat{q}_{1}\)] (b3), (p1) -- [thick] (b3)-- [thick] (p4), (cutLL)--[dashed, red,thick] (cutRR),(cutR)-- [photon,ultra thick] (u1), (k2)-- [photon,ultra thick] (u1)
			};
		\end{feynman}
	    \end{tikzpicture}
	\,, &  &
	D_2=\begin{tikzpicture}[baseline={([yshift=-0.4ex]current bounding box.center)}]\tikzstyle{every node}=[font=\small]
		    \begin{feynman}
			\vertex (p1) {\(p_1\)};
			\vertex [above=2.5cm of p1](p2){$p_2$};
			\vertex [right=2cm of p2] (u1) [HV]{H};
			\vertex [right=2cm of u1] (p3){$p_{2'}$};
			\vertex [right=2.0cm of p1] (b3) [HV]{H};
			\vertex [right=2.0cm of b3](p4){$p_{1'}$};
			\vertex [above=1.25cm of p1] (cutL);
			\vertex [above=1.1cm of p1] (cutL0);
			\vertex [right=1cm of cutL0] (cutLL);
			\vertex [right=2cm of cutLL] (cutRR);
			\vertex [right=4.0cm of cutL] (cutR){$k_1$};
			\vertex [below=0.8cm of cutR] (k2){$k_2$};
			\diagram* {
			(p2) -- [thick] (u1) -- [thick] (p3), (b3)-- [photon,ultra thick,rmomentum'=\(\hat{q}_{2}\)] (u1), (p1) -- [thick] (b3)-- [thick] (p4), (cutLL)--[dashed, red,thick] (cutRR),(cutR)-- [photon,ultra thick] (b3),(k2)-- [photon,ultra thick] (b3),
			};
		\end{feynman}
	    \end{tikzpicture}
	\,.
\end{align}
\begin{align}\label{6point-bcfw2}
	D_3=\begin{tikzpicture}[baseline={([yshift=-0.4ex]current bounding box.center)}]\tikzstyle{every node}=[font=\small]
		    \begin{feynman}
			\vertex (p1) {\(p_1\)};
			\vertex [above=2.5cm of p1](p2){$p_2$};
			\vertex [right=2cm of p2] (u1) [HV]{H};
			\vertex [right=2cm of u1] (p3){$p_{2'}$};
			\vertex [right=2.0cm of p1] (b3) [HV]{H};
			\vertex [right=2.0cm of b3](p4){$p_{1'}$};
			\vertex [above=1.25cm of p1] (cutL);
			\vertex [above=1.1cm of p1] (cutL0);
			\vertex [right=1cm of cutL0] (cutLL);
			\vertex [right=2cm of cutLL] (cutRR);
			\vertex [right=4.0cm of cutL] (cutR){$k_1$};
			\vertex [above=0.5cm of cutR] (k2){$k_2$};
			\diagram* {
			(p2) -- [thick] (u1) -- [thick] (p3), (b3)-- [photon,ultra thick,momentum'=\(\hat{t}_{1}\)] (u1), (p1) -- [thick] (b3)-- [thick] (p4), (cutLL)--[dashed, red,thick] (cutRR),(cutR)-- [photon,ultra thick] (b3),(k2)-- [photon,ultra thick] (u1),
			};
		\end{feynman}
	    \end{tikzpicture}
	\,. &  &
	D_4=\begin{tikzpicture}[baseline={([yshift=-0.4ex]current bounding box.center)}]\tikzstyle{every node}=[font=\small]
		    \begin{feynman}
			\vertex (p1) {\(p_1\)};
			\vertex [above=2.5cm of p1](p2){$p_2$};
			\vertex [right=2cm of p2] (u1) [HV]{H};
			\vertex [right=2cm of u1] (p3){$p_{2'}$};
			\vertex [right=2.0cm of p1] (b3) [HV]{H};
			\vertex [right=2.0cm of b3](p4){$p_{1'}$};
			\vertex [above=1.25cm of p1] (cutL);
			\vertex [above=0.75cm of p1] (cutL0);
			\vertex [right=1cm of cutL0] (cutLL);
			\vertex [right=2cm of cutLL] (cutRR);
			\vertex [right=4.0cm of cutL] (cutR){$k_1$};
			\vertex [below=0.5cm of cutR] (k2){$k_2$};
			\diagram* {
			(p2) -- [thick] (u1) -- [thick] (p3), (u1)-- [photon,ultra thick,rmomentum'=\(\hat{t}_{2}\)] (b3), (p1) -- [thick] (b3)-- [thick] (p4), (cutLL)--[dashed, red,thick] (cutRR),(cutR)-- [photon,ultra thick] (u1),(k2)-- [photon,ultra thick] (b3),
			};
		\end{feynman}
	    \end{tikzpicture}
	\,,
\end{align}
with $t_1=q_1-k_1$, $t_2=q_1-k_2$.
Once again we solve the on-shell conditions for the deformed momenta $\hat q_1^2=\hat q_2^2=\hat t_1^2=\hat t_2^2=0$ to get the value of the $z$-pole for each BCFW diagram. We then calculate each diagram by gluing together the appropriate sub-amplitudes via a state sum. These amplitudes include yet another new ingredient: the five-point single heavy source HEFT amplitude, again derived in \cite{Brandhuber:2021eyq}
\begin{align}
	 & \begin{tikzpicture}[baseline={([yshift=-0.8ex]current bounding box.center)}]\tikzstyle{every node}=[font=\small]
		   \begin{feynman}
			\vertex (a)[]{$\pb-q/2$};
			\vertex[left=1.9cm of a] (a0)[HV]{H};
			\vertex[left=1.9cm of a0] (am1){$\pb+q/2$};
			\vertex[above=1.9cm of a0] (b0){$\ell_2$};
			\vertex[left=1.6cm of b0] (bm1){$\ell_1$};
			\vertex[right=1.6cm of b0] (b1){$\ell_3$};
			\diagram*{(a0)--[fermion, very thick](a),(am1)--[fermion, very thick](a0), (a0) -- [photon,ultra thick, momentum] (b0),(a0) -- [photon,ultra thick, momentum] (bm1),(a0) -- [photon,ultra thick, momentum'] (b1)};
		\end{feynman}
	   \end{tikzpicture}\,.
\end{align}
This five-point HEFT amplitude can be written compactly in terms of a set of BCJ numerators \cite{Brandhuber:2021eyq,Brandhuber:2021kpo,Brandhuber:2021bsf} as follows
\begin{equation}\label{eq: 1source5pt}
	\cA_{5}(\ell_1,\ell_2,\ell_3,\pb)= -i\kappa^3\left( \frac{(\cN([[1,2],3],\pb))^2}{\ell_{12}^2 \ell_{123}^2}+\frac{(\cN([[1,3],2],\pb))^2}{\ell_{13}^2 \ell_{123}^2} +\frac{(\cN([[3,2],1],\pb))^2}{\ell_{23}^2 \ell_{123}^2}\right)
\end{equation}
The first of these numerators is given by
\begin{equation}
	\cN([[1,2],3],\pb)= -\frac{(\pb\Cdot F_{1}\Cdot F_{2}\Cdot \pb) (\ell_{12}\Cdot F_{3}\Cdot \pb)}{\pb\Cdot \ell_1 \pb\Cdot \ell_{12}}
	-\frac{(\pb\Cdot F_{1}\Cdot F_{3}\Cdot \pb) (\ell_{1}\Cdot F_{2}\Cdot \pb)}{\pb\Cdot \ell_1 \pb\Cdot \ell_{13}}
	+\frac{(\pb\Cdot F_{1}\Cdot F_{2}\Cdot  F_{3}\Cdot \pb)}{\pb\Cdot \ell_1 }
\end{equation}
and the rest are related by permuting the massless legs $\ell_1,\ell_2,\ell_3$. Note, we are dropping $\delta'$ contributions as discussed in Section~\ref{sec:diag-HEFT-exp}. Hence the six-point tree-level HEFT amplitude with two heavy sources and two radiated gravitons is
\begin{align}\label{eq: 6pttree}
	\cM_6(k_1,k_2,\pb_1,\pb_2) = i \frac{N_1}{q_1^2} + i \frac{N_2}{q_2^2}+i\frac{N_3}{(q_1-k_1)^2}+i\frac{N_4}{(q_1-k_2)^2}
	\ ,
\end{align}
where we have defined numerators $N_i$ for each BCFW diagram in the same manner as before. 
As was the case for the five-point amplitude, the BCFW shift in $\hat q_1$ simply drops out of the amplitude for the same reasons as before.

\section{One-loop five-point amplitude via unitarity}
\label{sec: oneloopunitaritycuts}

\subsection{Strategy of the calculation}

In this section we construct the one-loop integrand via unitarity cuts, by gluing  tree-level HEFT amplitudes.  The classical amplitude is obtained from two massive particle irreducible  (2MPI) diagrams, which are of  $\cO(\mb_1^3 \mb_2^2)$  and $\cO(\mb_1^2 \mb_2^3)$. These two  terms  are simply related by swapping $1 \leftrightarrow 2$, hence  we  focus here on the  former. We will confirm in Section~\ref{sec: waveforms} that these are precisely the terms needed for the waveforms.

We also mention in passing that the $\cO(\mb_1^3 \mb_2^3)$ hyper-classical diagrams, corresponding to two massive particle reducible HEFT diagrams, factorise when Fourier transformed to impact parameter space. This was seen in the conservative case in \cite{Brandhuber:2021eyq}, and also happens in the presence of radiation, as we show in Appendix~\ref{sec:factorisation}.

The HEFT tree amplitudes that enter the unitarity cuts have either two or four massive scalars plus several gravitons,
and have been described  in Section~\ref{sec:diag-HEFT-exp} and Section~\ref{sec: BCFWtwosources}, respectively. They are all manifestly gauge invariant since the dependence on the graviton polarisations occurs only through the corresponding linearised field strength tensors. The three-point HEFT amplitude \eqref{eq: 3PointHEFT} is an exception which depends directly on the polarisation tensor, but it is nevertheless gauge invariant.

The cut diagrams 
contributing  to the classical amplitude at  $\cO(\mb_1^3 \mb_2^2)$ are
\begin{align}\label{eq: 1loopDiagrams}
	\cC_1=\begin{tikzpicture}[baseline={([yshift=-0.4ex]current bounding box.center)}]\tikzstyle{every node}=[font=\small]
		      \begin{feynman}
			\vertex (p1) {\(p_1\)};
			\vertex [above=2.5cm of p1](p2){$p_2$};
			\vertex [right=2cm of p2] (u1) [HV]{H};
			\vertex [right=2cm of u1] (p3){$p_{2'}$};
			\vertex [right=1.0cm of p1] (b1) [HV]{H};
			\vertex [right=1.0cm of b1] (b2) []{};
			\vertex [right=1.0cm of b2] (b3) [HV]{H};
			\vertex [right=1.0cm of b3](p4){$p_{1'}$};
			\vertex [above=1.25cm of p1] (cutL);
			\vertex [right=4.0cm of cutL] (cutR){$k$};
			\vertex [right=1.0cm of b1] (cut1);
			\vertex [above=0.3cm of cut1] (cut1u);
			\vertex [below=0.3cm of cut1] (cut1b);
			\vertex [right=0.5cm of b2] (cut2);
			\vertex [above=0.3cm of cut2] (cut2u);
			\vertex [below=0.3cm of cut2] (cut2b);
			\diagram* {
			(p2) -- [thick] (u1) -- [thick] (p3),
			(b1)--[photon,ultra thick,rmomentum=\(\ell_{1}\)](u1), (b3)-- [photon,ultra thick,rmomentum'=\(\ell_{2}\)] (u1), (p1) -- [thick] (b1)-- [thick] (b3)-- [thick] (p4), (cutL)--[dashed, red,thick] (cutR), (cut1u)--[ red,thick] (cut1b),(cutR)-- [photon,ultra thick] (b3),
			};
		\end{feynman}
	      \end{tikzpicture}\,, &  &
	\cC_2=\begin{tikzpicture}[baseline={([yshift=-0.4ex]current bounding box.center)}]\tikzstyle{every node}=[font=\small]
		      \begin{feynman}
			\vertex (p1) {\(p_1\)};
			\vertex [above=2.5cm of p1](p2){$p_2$};
			\vertex [right=2cm of p2] (u1) [HV]{H};
			\vertex [right=2cm of u1] (p3){$p_{2'}$};
			\vertex [right=1.0cm of p1] (b1) [HV]{H};
			\vertex [right=1.0cm of b1] (b2) []{};
			\vertex [right=1.0cm of b2] (b3) [HV]{H};
			\vertex [right=1.0cm of b3](p4){$p_{1'}$};
			\vertex [above=1.25cm of p1] (cutL);
			\vertex [right=4.0cm of cutL] (cutR){$k$};
			\vertex [right=1.0cm of b1] (cut1);
			\vertex [above=0.3cm of cut1] (cut1u);
			\vertex [below=0.3cm of cut1] (cut1b);
			\vertex [right=0.5cm of b2] (cut2);
			\vertex [above=0.3cm of cut2] (cut2u);
			\vertex [below=0.3cm of cut2] (cut2b);
			\diagram* {
			(p2) -- [thick] (u1) -- [thick] (p3),
			(b1)--[photon,ultra thick,rmomentum=\(\ell_{1}\)](u1), (b3)-- [photon,ultra thick,rmomentum'=\(\ell_{3}\)] (u1), (p1) -- [thick] (b1)-- [thick] (b3)-- [thick] (p4), (cutL)--[dashed, red,thick] (cutR), (cut1u)--[red,thick] (cut1b),(cutR)-- [photon,ultra thick] (u1),
			};
		\end{feynman}
	      \end{tikzpicture}\,,
\end{align}
and the following cuts which subsume the above cuts
\begin{align}\label{eq: 1loopSnails}
	\cC_3=\begin{tikzpicture}[baseline={([yshift=-0.4ex]current bounding box.center)}]\tikzstyle{every node}=[font=\small]
		      \begin{feynman}
			\vertex (p1) {\(p_1\)};
			\vertex [above=2.5cm of p1](p2){$p_2$};
			\vertex [right=1.0cm of p2] (u1) [HV]{$~~~$};
			\vertex [below=1.25cm of u1] (m1) [GR2]{H};
			\vertex [below=1.25cm of m1] (b0) [HV]{$~~~$};
			\vertex [right=3cm of u1] (p3){$p_{2'}$};
			\vertex [right=1.0cm of p1] (b1) []{};
			\vertex [right=1.0cm of b1] (b2) []{};
			\vertex [right=1.0cm of b2] (b3) [HV]{H};
			\vertex [right=1.0cm of b3](p4){$p_{1'}$};
			\vertex [above=1.25cm of p1] (cutL);
			\vertex [right=4.0cm of cutL] (cutR){$k$};
			\vertex [right=1.0cm of b1] (cut1);
			\vertex [above=0.3cm of cut1] (cut1u);
			\vertex [below=0.3cm of cut1] (cut1b);
			\vertex [above=0.4cm of cut1u] (cutd1);
			\vertex [right=0.4cm of cutd1] (cutd2);
			\vertex [right=0.5cm of b2] (cut2);
			\vertex [above=0.3cm of cut2] (cut2u);
			\vertex [below=0.3cm of cut2] (cut2b);
			\diagram* {
			(p2) -- [thick] (u1) -- [thick] (p3),
			(b3)-- [photon,ultra thick,rmomentum'=\(\ell_{2}\)] (m1), (p1) -- [thick] (b0)-- [thick] (b3)-- [thick] (p4),  (cut1u)--[ red,thick] (cut1b),(cut1u)--[dashed, red,thick] (cutd2),(cutR)-- [photon,ultra thick] (b3),
			};
		\end{feynman}
	      \end{tikzpicture}\, \,, &  &
	\cC_4=\begin{tikzpicture}[baseline={([yshift=-0.4ex]current bounding box.center)}]\tikzstyle{every node}=[font=\small]
		      \begin{feynman}
			\vertex (p1) {\(p_1\)};
			\vertex [above=2.5cm of p1](p2){$p_2$};
			\vertex [right=1.0cm of p2] (u1) [HV]{$~~~$};
			\vertex [below=1.25cm of u1] (m1) [GR2]{H};
			\vertex [below=1.25cm of m1] (b0) [HV]{$~~~$};
			\vertex [right=3cm of u1] (p3){$p_{2'}$};
			\vertex [right=1.0cm of p1] (b1) []{};
			\vertex [right=1.0cm of b1] (b2) []{};
			\vertex [right=1.0cm of b2] (b3) [HV]{H};
			\vertex [right=1.0cm of b3](p4){$p_{1'}$};
			\vertex [above=1.25cm of p1] (cutL);
			\vertex [right=4.0cm of cutL] (cutR);
			\vertex [above=0.5cm of cutR] (cutRu){$k$};
			\vertex [right=1.0cm of b1] (cut1);
			\vertex [above=0.3cm of cut1] (cut1u);
			\vertex [below=0.3cm of cut1] (cut1b);
			\vertex [above=0.4cm of cut1u] (cutd1);
			\vertex [right=0.4cm of cutd1] (cutd2);
			\vertex [right=0.5cm of b2] (cut2);
			\vertex [above=0.3cm of cut2] (cut2u);
			\vertex [below=0.3cm of cut2] (cut2b);
			\diagram* {
			(p2) -- [thick] (u1) -- [thick] (p3),
			(b3)-- [photon,ultra thick,rmomentum'=\(\ell_{3}\)] (m1), (p1) -- [thick] (b0)-- [thick] (b3)-- [thick] (p4),  (cut1u)--[ red,thick] (cut1b),(cut1u)--[dashed, red,thick] (cutd2),(cutRu)-- [photon,ultra thick] (m1),
			};
		\end{feynman}
	      \end{tikzpicture} \,.
\end{align}
There are also HEFT diagrams where the graviton is emitted from an incoming leg, for example the following swapped version of diagram $\cC_1$ (and similarly diagram $\cC_3$)
\begin{align}\label{eq: inoutswappedgraphs}
	\begin{tikzpicture}[baseline={([yshift=-0.4ex]current bounding box.center)}]\tikzstyle{every node}=[font=\small]
		\begin{feynman}
			\vertex (p1) {\(p_1\)};
			\vertex [above=2.5cm of p1](p2){$p_2$};
			\vertex [right=2cm of p2] (u1) [HV]{H};
			\vertex [right=2cm of u1] (p3){$p_{2'}$};
			\vertex [right=1.0cm of p1] (b1) [HV]{H};
			\vertex [right=1.0cm of b1] (b2) []{};
			\vertex [right=1.0cm of b2] (b3) [HV]{H};
			\vertex [right=1.0cm of b3](p4){$p_{1'}$};
			\vertex [above=1.25cm of p1] (cutL){$k$};
			\vertex [right=4.0cm of cutL] (cutR);
			\vertex [right=1.0cm of b1] (cut1);
			\vertex [above=0.3cm of cut1] (cut1u);
			\vertex [below=0.3cm of cut1] (cut1b);
			\vertex [right=0.5cm of b2] (cut2);
			\vertex [above=0.3cm of cut2] (cut2u);
			\vertex [below=0.3cm of cut2] (cut2b);
			\diagram* {
			(p2) -- [thick] (u1) -- [thick] (p3),
			(b1)--[photon,ultra thick](u1), (b3)-- [photon,ultra thick] (u1), (p1) -- [thick] (b1)-- [thick] (b3)-- [thick] (p4), (cutL)--[dashed, red,thick] (cutR), (cut1u)--[ red,thick] (cut1b),(cutL)--[photon,ultra thick](b1),
			};
		\end{feynman}
	\end{tikzpicture}.
\end{align}
However this gives exactly the same contributions as $\cC_1$, which can be seen explicitly using the loop momentum reparameterisation $\ell_1\rightarrow \ell_3=-\ell_1-q_1$ and the following property of the HEFT %massive cut
delta function
\begin{equation}
	\delta(\vb_1\Cdot \ell_1)= \delta(\vb_1 \Cdot \ell_3) \ ,
\end{equation}
where we have used \eqref{eq: HEFTfame}.
Thus we just need to multiply the contributions of diagrams $\cC_1$ and $\cC_3$ by a factor of 2. Note that whenever we cut two gravitons as in $\cC_1$ and $\cC_2$ we also include a symmetry factor of  $S=\frac{1}{2!}$ for the two identical particles crossing the cut.
As noted earlier the contribution of $\mb_1^2 \mb_2^3$ is obtained by swapping $q_1\leftrightarrow q_2$ and $p_1 \leftrightarrow p_2$.

We now have to combine carefully the information from the various cuts to construct the complete integrand.
Naively summing the cut integrands from all the diagrams leads to an over-counting since there are terms in the full integrand detected by more than one of the cuts above. 
%For example, the contributions from the cuts $\cC_1$ and $\cC_2$ are entirely contained in the cuts
%$\cC_3$ and $\cC_4$. 
The correct procedure, called {\it cut merging}, ensures that terms detected in several cuts are only counted once. 

An example of this issue is the particular contribution to the full integrand contained in the overlap of {\it all} cut diagrams. It is easy to see that it corresponds to the following triple cut with all massless propagators present:
\begin{align} \label{eq: overcountingGraph}
	\begin{tikzpicture}[baseline={([yshift=-0.4ex]current bounding box.center)}]\tikzstyle{every node}=[font=\small]
		\begin{feynman}
			\vertex (p1) {\(p_1\)};
			\vertex [above=2.5cm of p1](p2){$p_2$};
			\vertex [right=2cm of p2] (u1) [HV]{H};
			\vertex [right=2cm of u1] (p3){$p_{2'}$};
			\vertex [right=1.0cm of p1] (b1) [HV]{H};
			\vertex [right=1.0cm of b1] (b2) []{};
			\vertex [right=1.0cm of b2] (b3) [HV]{H};
			\vertex [right=1.0cm of b3](p4){$p_{1'}$};
			\vertex [above=1.25cm of p1] (cutL);
			\vertex [right=2.4cm of cutL] (m1)[dot]{};
			\vertex [right=4.0cm of cutL] (cutR){$k$};
			\vertex [right=1.0cm of b1] (cut1);
			\vertex [above=0.3cm of cut1] (cut1u);
			\vertex [above=0.5cm of cut1] (cut1uu);
			\vertex [above=0.9cm of b3] (int);
			\vertex [right=0.2cm of int] (cut1uur);
			\vertex [above=1.4cm of cut1] (cut1uuu);
			\vertex [above=1.7cm of b3] (cut1uuur);
			\vertex [below=0.3cm of cut1] (cut1b);
			\vertex [right=0.5cm of b2] (cut2);
			\vertex [above=0.3cm of cut2] (cut2u);
			\vertex [below=0.3cm of cut2] (cut2b);
			\diagram* {
			(p2) -- [thick] (u1) -- [thick] (p3),
			(b1)--[photon,ultra thick,rmomentum=\(\ell_{1}\)](u1), (b3)-- [photon,ultra thick,rmomentum'=\(\ell_{2}\)] (m1)-- [photon,ultra thick,rmomentum'=\(\ell_{3}\)] (u1)
			, (p1) -- [thick] (b1)-- [thick] (b3)-- [thick] (p4),
			(cutL)--[dashed, red,thick] (cut1uu) ,
			(cut1uu)--[dashed, red,thick] (cut1uur) ,
			(cut1uuu)--[dashed, red,thick] (cut1uuur),
			(cut1u)--[red,thick] (cut1b),
			(cutR)-- [photon,ultra thick] (m1),
			};
		\end{feynman}
	\end{tikzpicture}\,.
\end{align}
There is also a mirror version of this diagram where the emitted graviton appears on the left of the diagram, but this makes an identical contribution.
% Evidently, we must count this contribution only once.

We denote the operation of cut merging as a union
\begin{align}\label{eq: unionofcuts}
	\cC_{\mb_1^3\mb_2^2}=(2*\cC_1)\cup \cC_2 \cup (2*\cC_3)\cup \cC_4\, ,
\end{align}
and we find that the contributions in the overlap of diagrams $2*\cC_1$ and $\cC_2$ are also \emph{exactly} identical to the triple cut diagram \eqref{eq: overcountingGraph}.  Similarly, we found explicitly that the contributions detected by cut $\cC_1$ and $\cC_2$ are exactly contained as a subset of the contributions from cuts $\cC_3$ and $\cC_4$. The exact identification and matching of overlap terms is facilitated by the use of a minimal set of independent scalar products, as we will be explain in more detail in the next section.

The integrand found in this process is given by a linear combination of tensor integrals. As we show in the next sections this {\it bare} integrand can be reduced further
in a two-step process: first, we will convert the tensor integrals into a sum of loop momentum independent coefficients times scalar integrals, second, we will reduce the scalar integrals to a family of master integrals using integration by parts relations (IBP).

The relevant (master) integrals are
scalar one-loop Feyman integrals of the form
\begin{equation}\label{eq: masterIntegral}
	j_{a_1, 1, a_3, a_4, a_5} = \int\!\frac{d^D \ell}{(2 \pi)^{D}} \frac{-i\pi\delta (\, \vb_1 \cdot \ell)}{(\ell^2 + i \varepsilon)^{a_1} [(\ell+q_1)^2+ i \varepsilon]^{a_3} [(\ell-q_2)^2+ i \varepsilon]^{a_4}( \vb_2 \cdot \ell)^{a_5}}\ ,
\end{equation}
with the propagator structure coming from the pentagon master topology
\begin{align}\label{eq: masterGraph}
	\begin{tikzpicture}[baseline={([yshift=-0.4ex]current bounding box.center)}]\tikzstyle{every node}=[font=\small]
		\begin{feynman}
			\vertex (p1) {\(p_1\)};
			\vertex [above=2.5cm of p1](p2){$p_2$};
			\vertex [right=1cm of p2] (u1) [dot]{};
			\vertex [right=3cm of p2] (u2) [dot]{};
			\vertex [right=3cm of u1] (p3){$p_{2}'$};
			\vertex [right=1.0cm of p1] (b1) [dot]{};
			\vertex [right=1.0cm of b1] (b2) []{};
			\vertex [right=1.0cm of b2] (b3) [dot]{};
			\vertex [right=1.0cm of b3](p4){$p_{1}'$};
			\vertex [above=1.25cm of p1] (cutL);
			\vertex [below=1.25cm of u2] (m1)[dot]{};
			\vertex [right=4.0cm of cutL] (cutR){$k$};
			\vertex [right=1.0cm of b1] (cut1);
			\vertex [above=0.3cm of cut1] (cut1u);
			\vertex [below=0.3cm of cut1] (cut1b);
			\vertex [right=0.5cm of b2] (cut2);
			\vertex [above=0.3cm of cut2] (cut2u);
			\vertex [below=0.3cm of cut2] (cut2b);
			\diagram* {
			(p2) -- [thick] (u1) -- [thick] (p3),
			(b1)--[boson,rmomentum=\(\ell_{1}\)](u1), (b3)-- [boson,rmomentum=\(\ell_{3}\)] (m1)-- [boson,rmomentum=\(\ell_{2}\)] (u2), (p1) -- [thick] (b1)-- [thick] (b3)-- [thick] (p4), (cut1u)--[red,thick] (cut1b),(cutR)-- [boson] (m1),
			};
		\end{feynman}
	\end{tikzpicture}\,,
\end{align}
where the top line corresponds to a linearised massive propagator taken to some integer power%
\footnote{In the family of master integrals the linearised massive propagator $\cD_5$ corresponding to the top line always appears with power $a_5=1$ and is regulated with the principal value prescription. If $a_5 > 1$, it is regulated according to \eqref{eq: hadamard} and its generalisation to higher powers, however such integrals can always be reduced to master integrals with $a_5=1$.}, while the bottom line is a HEFT delta function which is present in all diagrams. In the following we will  refer to the propagators by the labels $\cD_i$ as defined in Table~\ref{tab:propagatorbasis} 
\begin{table}[H]
	\centering
	\begin{tabular}{ | c | c | c | c | c | }
		\hline
		$\cD_1$  & $\cD_2$          & $\cD_3$        & $\cD_4$         & $\cD_5$          \\
		\hline
		$\ell^2$ & $\vb_1 \Cdot \ell$ & $(\ell+q_1)^2$ & $ (\ell-q_2)^2$ & $\vb_2 \Cdot \ell$ \\
		\hline
	\end{tabular}
	\caption{\it Propagator basis}
	\label{tab:propagatorbasis}
\end{table}
\noindent
and in this notation the master topology is given by
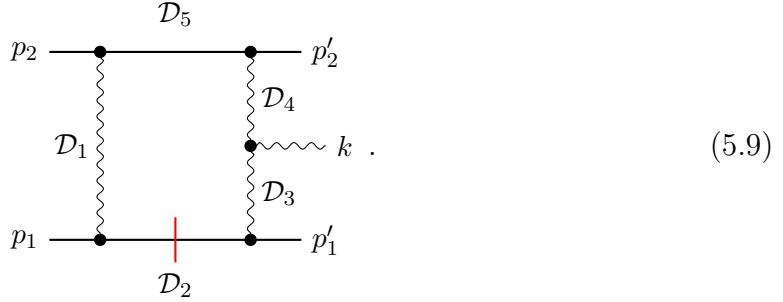
\begin{align}\label{eq: masterGraphSimple}
	\begin{tikzpicture}[baseline={([yshift=-0.4ex]current bounding box.center)}]\tikzstyle{every node}=[font=\small]
		\begin{feynman}
			\vertex (p1) {\(p_1\)};
			\vertex [above=2.5cm of p1](p2){$p_2$};
			\vertex [right=1cm of p2] (u1) [dot]{};
			\vertex [right=3cm of p2] (u2) [dot]{};
			\vertex [right=3cm of u1] (p3){$p_{2}'$};
			\vertex [right=1.0cm of p1] (b1) [dot]{};
			\vertex [right=1.0cm of b1] (b2) []{};
			\vertex [right=1.0cm of b2] (b3) [dot]{};
			\vertex [right=1.0cm of b3](p4){$p_{1}'$};
			\vertex [above=1.25cm of p1] (cutL);
			\vertex [below=1.25cm of u2] (m1)[dot]{};
			\vertex [right=4.0cm of cutL] (cutR){$k$};
			\vertex [right=1.0cm of b1] (cut1);
			\vertex [above=0.3cm of cut1] (cut1u);
			\vertex [below=0.3cm of cut1] (cut1b){$\cD_2$};
			\vertex [right=1cm of u1] (cut2);
			\vertex [above=0.2cm of cut2] (cut2u){$\cD_5$};
			\diagram* {
			(p2) -- [thick] (u1) -- [thick] (p3),
			(b1)--[boson, edge label=$\cD_1$](u1), (b3)-- [boson,edge label'=$\cD_3$] (m1)-- [boson,edge label'=$\cD_4$] (u2), (p1) -- [thick] (b1)-- [thick] (b3)-- [thick] (p4), (cut1u)--[red,thick] (cut1b),(cutR)-- [boson] (m1),
			};
		\end{feynman}
	\end{tikzpicture}\,.
\end{align}
 By using a minimal basis of scalar products and using all possible identities, we can merge integrands in different   cuts before performing IBP reductions.

We will study the various cuts in more detail momentarily, but already at this stage we can identify the topologies that occur in the various overlaps of  cut diagrams. The overlap terms between cut diagrams $\cC_1$ and $\cC_2$ are precisely those with the pentagon master topology \eqref{eq: masterGraphSimple},  where all the massless propagators are present  (possibly with higher powers of some propagators) and also the box topology where we collapse the massive propagator $\cD_5$. The overlap terms between cut diagrams $\cC_3$ and $\cC_4$ include in addition the box topology where we collapse just propagator $\cD_1$.

The final step is the reduction to a  basis of master integrals using IBP relations. We will describe  our  basis in full in Section~\ref{sec: oneLoopIntegrals}, where we also present explicit results for each  integral.

%-------------------------------------------------------
\subsection{Cut one}
For the first diagram \eqref{eq: 1loopDiagrams}, which we have denoted by $\cC_1$, the integrand is
\begin{align}
	 & \cC_1=\begin{tikzpicture}[baseline={([yshift=-0.4ex]current bounding box.center)}]\tikzstyle{every node}=[font=\small]
		         \begin{feynman}
			\vertex (p1) {\(p_1\)};
			\vertex [above=2.5cm of p1](p2){$p_2$};
			\vertex [right=2cm of p2] (u1) [HV]{H};
			\vertex [right=2cm of u1] (p3){$p_{2'}$};
			\vertex [right=1.0cm of p1] (b1) [HV]{H};
			\vertex [right=1.0cm of b1] (b2) []{};
			\vertex [right=1.0cm of b2] (b3) [HV]{H};
			\vertex [right=1.0cm of b3](p4){$p_{1'}$};
			\vertex [above=1.25cm of p1] (cutL);
			\vertex [right=4.0cm of cutL] (cutR){$k$};
			\vertex [right=1.0cm of b1] (cut1);
			\vertex [above=0.3cm of cut1] (cut1u);
			\vertex [below=0.3cm of cut1] (cut1b);
			\vertex [right=0.5cm of b2] (cut2);
			\vertex [above=0.3cm of cut2] (cut2u);
			\vertex [below=0.3cm of cut2] (cut2b);
			\diagram*{
			(p2) -- [thick] (u1) -- [thick] (p3),
			(b1)--[photon,ultra thick,rmomentum=\(\ell_{1}\)](u1), (b3)-- [photon,ultra thick,rmomentum'=\(\ell_{2}\)] (u1), (p1) -- [thick] (b1)-- [thick] (b3)-- [thick] (p4), (cutL)--[dashed, red,thick] (cutR), (cut1u)--[ red,thick] (cut1b),(cutR)-- [photon,ultra thick] (b3),
			};
		\end{feynman}
	         \end{tikzpicture}\nn                            \\
	 & =\int\! \frac{d^D \ell_1}{(2 \pi)^{D}} \delta(\vb_1\Cdot \ell_1)\sum_{h_1,h_2} \frac{\cA_4^{h_1,h_2}(\ell_1, \ell_2,  \vb_2)
		\cA_3^{-h_1}(-\ell_1,\vb_1)\cA_4^{-h_2}(-\ell_2,k,\vb_1)}{ \ell_1^2 \ell_2^2  }\, .
\end{align}
In the above and for the remainder of this section we suppress the explicit Feynman $i\varepsilon$ and principal value prescriptions as they can be reinstated unambigously. The three and four-point amplitudes with two scalars and one or two radiated  gravitons are given in \eqref{3ptampG} and  \eqref{eq: comptonheft-2}.

First we perform the sum over intermediate states $\ell_1$ and $\ell_2$ in $D$ dimensions using
\begin{align}
\label{statesum}
	\begin{split}
		\sum_{h} \varepsilon^{\mu_a}_{-p}\varepsilon^{\nu_a}_{-p}\varepsilon^{\mu_b}_{p}\varepsilon^{\nu_b}_{p} &= \frac{1}{2}\left( P^{\mu_a\mu_b}  P^{\nu_a \nu_b}+P^{\mu_a\nu_b} P^{\nu_a \mu_b}\right)-d_\phi P^{\mu_a\nu_a} P^{\mu_b \nu_b}\, ,
	\end{split}
\end{align}
where
\begin{align}
	\label{projector}
	P^{\alpha\beta} & = \eta^{\alpha\beta}-\frac{p^\alpha q^\beta + p^\beta q^\alpha}{p\Cdot q}\, ,
\end{align}
for some reference momentum $q$,  and with  $d_\phi$ defined in \eqref{dphi}.
Note that if the dilaton is included in the state sum, then the polarisation tensors $\varepsilon^{\alpha\beta}=\varepsilon^{\alpha}\varepsilon^{\beta}$ are no longer traceless. Since the HEFT amplitudes are manifestly gauge-invariant, we are entitled to make the simplification $P^{\alpha\beta}\rightarrow \eta^{\alpha \beta}$ \cite{Kosmopoulos:2020pcd} in \eqref{statesum}. This corresponds to the state sum shown and used already earlier in \eqref{simplsum}.  
 Using the full projector \eqref{projector} with reference momentum $q$, however, also allows an intermediate check since the reference momentum can be seen to drop out of the result explicitly. In addition, some care is needed when dealing with diagrams involving  a three-point graviton amplitude, which is not written in terms of field strengths -- an example of this situation is the diagram
\eqref{eq: overcountingGraph}, and in such cases we need to use the full projector $P^{\alpha\beta}$ given in \eqref{projector}.

Once we have performed the state sum, the integrand is a function of 
\begin{align}
	A\Cdot F_k\Cdot B, \qquad A\Cdot B,
\end{align}
where $A,B$ can be any of the vectors $\ell_1, q_1, q_2, \vb_1, \vb_2$.
The scalar products involving  field-strength tensors $F_k$ are not all independent due to the identities \cite{Feng:2020jck}
\begin{align}
	\begin{split}
		A\Cdot F_k\Cdot B \ k\Cdot C + C\Cdot F_k\Cdot A \ k\Cdot &B +B\Cdot F_k\Cdot C \  A\Cdot k = 0\, ,\\
		A\Cdot F_k\Cdot A&=0\, ,\\
		k\Cdot F_k\Cdot B&=0\,,
	\end{split}
\end{align}
where $A, B, C$ can be any vector. The first of these relations is simply the Bianchi identity in momentum space, and the last two follow trivially from the antisymmetry of $F_k$ and the fact that $k$ is on shell. Using these relations we can write the integrand in terms of the independent tensor structures which involve products of a pair of field strengths taken from the following list, where we have fixed the second vector contracted into $F_k$ to always be $\bar{v}_2$,
\begin{align}
	\ell_1\Cdot F_k\Cdot \vb_2, &  & q_1\Cdot F_k\Cdot \vb_2, &  & \vb_1\Cdot F_k\Cdot \vb_2\, .
\end{align}
In all of our calculations the internal cut lines are in $D$ dimensions, while  external momenta are kept in four dimensions. Four-dimensional external kinematics allows us to rewrite the tensor structure
$q_2\Cdot F_k\Cdot \ell_1$ by expanding  $F_k$ in terms of a  basis formed by   taking anti-symmetric products of the vectors $\vb_1,\vb_2,q_1,q_2$:
\begin{equation}
	\label{Fexp}
	F_{k}^{\alpha\beta}= a \,\vb_1^{[\alpha} \vb_2^{\beta]}+b\, \vb_1^{[\alpha} q_1^{\beta]}+c\, \vb_1^{[\alpha} q_2^{\beta]}+ d\, \vb_2^{[\alpha} q_1^{\beta]} + e\, \vb_2^{[\alpha} q_2^{\beta]}
	+ f\, q_1^{[\alpha} q_2^{\beta]}\,,
\end{equation}
and then solving the linear system for the coefficients $a,b,c,d,e,f$. The coefficients  are then written in terms of the traces
\begin{align}
	q_1\Cdot F_k\Cdot \vb_2\, ,\qquad  \vb_1\Cdot F_k\Cdot \vb_2\,,
\end{align}
and hence we will be left with an integrand made of products of these structures. 
%(although this may not yield the most compact expression for this quantity).

The fact that the external kinematics is restricted to four dimensions implies even more identities due to the vanishing of the Gram determinant $G(v_1,v_2,q_1,q_2, \varepsilon_k)$. This is equivalent to the fact that any fully anti-symmetrised tensor with more than four Lorentz indices  must vanish in four dimensions. This gives for example the following identity involving $F_k$
\begin{equation}
	F_{k}^{[\mu\nu} \vb_1^{\rho} \vb_{2}^\sigma q_{1}^{\tau]}=0\, ,
\end{equation}
and contracting relations like this with the $F_{k}, \vb_1,\vb_2,q_1,q_2$ gives us an additional relation between tensor structures that we use to simplify the integrand further. The additional relation we generate is quadratic in the field strengths and allows us to reduce the possible combinations of field strengths that appear in the integrand to two such traces,
\begin{align}
	q_1\Cdot F_k\Cdot \vb_2\,\vb_1\Cdot F_k\Cdot \vb_2, &  & (\vb_1\Cdot F_k\Cdot \vb_2)^2\,.
\end{align}
Now one can rewrite all  scalar products in the numerator depending on the loop momentum $\ell_1$
in terms of inverse powers of propagators (assuming the cut conditions). This gives a fully tensor-reduced integrand which can be written in terms of loop momentum independent coefficients $c_i$ and scalar integrals which are sub-topologies of the master topology \eqref{eq: masterGraphSimple}, possibly with higher powers of propagators.

We then perform IBP reduction using LiteRed2 \cite{Lee:2012cn,Lee:2013mka}
and assuming the cut conditions of cut diagram $\cC_1$. We find the following four master integrals,
\begin{align}
	\label{eq: cut1MIs}
	\cI_1\coloneqq j_{11010}= \begin{tikzpicture}[scale=0.5, transform shape, baseline={([yshift=-0.4ex]current bounding box.center)}]\tikzstyle{every node}=[font=\small]
		                          \begin{feynman}
			\vertex (p1) {};
			\vertex [above=2.5cm of p1](p2){};
			\vertex [right=2cm of p2] (u1) [dot]{};
			\vertex [right=2.5cm of p2] (ur){};
			\vertex [right=4cm of p2] (p3){};
			\vertex [right=1.0cm of p1] (b1) [dot]{};
			\vertex [right=1.0cm of b1] (b2) []{};
			\vertex [right=1.0cm of b2] (b3) [dot]{};
			\vertex [right=1.0cm of b3](p4){};
			\vertex [above=1.25cm of p1] (cutL);
			\vertex [below=1.25cm of ur] (m1);
			\vertex [right=4.0cm of cutL] (cutR){};
			\vertex [right=1.0cm of b1] (cut1);
			\vertex [above=0.3cm of cut1] (cut1u);
			\vertex [below=0.3cm of cut1] (cut1b){};
			\vertex [right=1cm of u1] (cut2);
			\vertex [above=0.2cm of cut2] (cut2u){};
			\diagram* {
			(p2) -- [thick] (u1) -- [thick] (p3),
			(b1)--[boson](u1), (b3)-- [boson] (u1), (p1) -- [thick] (b1)-- [thick] (b3)-- [thick] (p4), (cut1u)--[red,thick] (cut1b),(cutR)-- [boson] (b3),
			};
		\end{feynman}
	                          \end{tikzpicture}\,,
	 &  & \cI_3\coloneqq j_{11011}= \begin{tikzpicture}[scale=0.5, transform shape, baseline={([yshift=-0.4ex]current bounding box.center)}]\tikzstyle{every node}=[font=\small]
		                                \begin{feynman}
			\vertex (p1) {};
			\vertex [above=2.5cm of p1](p2){};
			\vertex [right=1cm of p2] (u1) [dot]{};
			\vertex [right=3cm of p2] (u2) [dot]{};
			\vertex [right=3cm of u1] (p3){};
			\vertex [right=1.0cm of p1] (b1) [dot]{};
			\vertex [right=1.0cm of b1] (b2) []{};
			\vertex [right=1.0cm of b2] (b3) [dot]{};
			\vertex [right=1.0cm of b3](p4){};
			\vertex [above=1.25cm of p1] (cutL);
			\vertex [below=1.25cm of u2] (m1)[]{};
			\vertex [right=4.0cm of cutL] (cutR){};
			\vertex [right=1.0cm of b1] (cut1);
			\vertex [above=0.3cm of cut1] (cut1u);
			\vertex [below=0.3cm of cut1] (cut1b){};
			\vertex [right=1cm of u1] (cut2);
			\vertex [above=0.2cm of cut2] (cut2u){};
			\diagram* {
			(p2) -- [thick] (u1) -- [thick] (p3),
			(b1)--[boson](u1), (b3)-- [boson] (u2), (p1) -- [thick] (b1)-- [thick] (b3)-- [thick] (p4), (cut1u)--[red,thick] (cut1b),(cutR)-- [boson] (b3),
			};
		\end{feynman}
	                                \end{tikzpicture}\,, \\
	\cI_5\coloneqq j_{11110}= \begin{tikzpicture}[scale=0.5, transform shape, baseline={([yshift=-0.4ex]current bounding box.center)}]\tikzstyle{every node}=[font=\small]
		                          \begin{feynman}
			\vertex (p1) {};
			\vertex [above=2.5cm of p1](p2){};
			\vertex [right=2cm of p2] (u1) [dot]{};
			\vertex [right=2.5cm of p2] (ur){};
			\vertex [right=4cm of p2] (p3){};
			\vertex [right=1.0cm of p1] (b1) [dot]{};
			\vertex [right=1.0cm of b1] (b2) []{};
			\vertex [right=1.0cm of b2] (b3) [dot]{};
			\vertex [right=1.0cm of b3](p4){};
			\vertex [above=1.25cm of p1] (cutL);
			\vertex [below=1.25cm of ur] (m1)[dot]{};
			\vertex [right=4.0cm of cutL] (cutR){};
			\vertex [right=1.0cm of b1] (cut1);
			\vertex [above=0.3cm of cut1] (cut1u);
			\vertex [below=0.3cm of cut1] (cut1b){};
			\vertex [right=1cm of u1] (cut2);
			\vertex [above=0.2cm of cut2] (cut2u){};
			\diagram* {
			(p2) -- [thick] (u1) -- [thick] (p3),
			(b1)--[boson](u1), (b3)-- [boson] (u1), (p1) -- [thick] (b1)-- [thick] (b3)-- [thick] (p4), (cut1u)--[red,thick] (cut1b),(cutR)-- [boson] (m1),
			};
		\end{feynman}
	                          \end{tikzpicture}\,,
	 &  & \cI_6\coloneqq j_{11111}=
	\begin{tikzpicture}[scale=0.5, transform shape, baseline={([yshift=-0.4ex]current bounding box.center)}]\tikzstyle{every node}=[font=\small]
		\begin{feynman}
			\vertex (p1) {};
			\vertex [above=2.5cm of p1](p2){};
			\vertex [right=1cm of p2] (u1) [dot]{};
			\vertex [right=3cm of p2] (u2) [dot]{};
			\vertex [right=3cm of u1] (p3){};
			\vertex [right=1.0cm of p1] (b1) [dot]{};
			\vertex [right=1.0cm of b1] (b2) []{};
			\vertex [right=1.0cm of b2] (b3) [dot]{};
			\vertex [right=1.0cm of b3](p4){};
			\vertex [above=1.25cm of p1] (cutL);
			\vertex [below=1.25cm of u2] (m1)[dot]{};
			\vertex [right=4.0cm of cutL] (cutR){};
			\vertex [right=1.0cm of b1] (cut1);
			\vertex [above=0.3cm of cut1] (cut1u);
			\vertex [below=0.3cm of cut1] (cut1b){};
			\vertex [right=1cm of u1] (cut2);
			\vertex [above=0.2cm of cut2] (cut2u){};
			\diagram* {
			(p2) -- [thick] (u1) -- [thick] (p3),
			(b1)--[boson](u1), (b3)-- [boson] (m1)-- [boson] (u2), (p1) -- [thick] (b1)-- [thick] (b3)-- [thick] (p4), (cut1u)--[red,thick] (cut1b),(cutR)-- [boson] (m1),
			};
		\end{feynman}
	\end{tikzpicture}\,,\nn
\end{align}
and therefore  the contributions are given by a sum of these integrals multiplied by their coefficients,  which we write as
\begin{equation}
	\cC_1= \frac{c_1}{2}\, \cI_1 + \frac{c_3}{2}\, \cI_3+\frac{c_5}{2}\, \cI_5+\frac{c_6}{2}\, \cI_6,
\end{equation}
where the overall factor of $1/2$ was introduced for convenience since this contribution is doubled up when we include the swapped graph \ref{eq: inoutswappedgraphs}.

\subsection{Cut two}
For the second diagram, the integrand $\cC_2$ is given by
\begin{align}
	 & \cC_2=\begin{tikzpicture}[baseline={([yshift=-0.4ex]current bounding box.center)}]\tikzstyle{every node}=[font=\small]
		         \begin{feynman}
			\vertex (p1) {\(p_1\)};
			\vertex [above=2.5cm of p1](p2){$p_2$};
			\vertex [right=2cm of p2] (u1) [HV]{H};
			\vertex [right=2cm of u1] (p3){$p_{2'}$};
			\vertex [right=1.0cm of p1] (b1) [HV]{H};
			\vertex [right=1.0cm of b1] (b2) []{};
			\vertex [right=1.0cm of b2] (b3) [HV]{H};
			\vertex [right=1.0cm of b3](p4){$p_{1'}$};
			\vertex [above=1.25cm of p1] (cutL);
			\vertex [right=4.0cm of cutL] (cutR){$k$};
			\vertex [right=1.0cm of b1] (cut1);
			\vertex [above=0.3cm of cut1] (cut1u);
			\vertex [below=0.3cm of cut1] (cut1b);
			\vertex [right=0.5cm of b2] (cut2);
			\vertex [above=0.3cm of cut2] (cut2u);
			\vertex [below=0.3cm of cut2] (cut2b);
			\diagram* {
			(p2) -- [thick] (u1) -- [thick] (p3),
			(b1)--[photon,ultra thick,rmomentum=\(\ell_{1}\)](u1), (b3)-- [photon,ultra thick,rmomentum'=\(\ell_{3}\)] (u1), (p1) -- [thick] (b1)-- [thick] (b3)-- [thick] (p4), (cutL)--[dashed, red,thick] (cutR), (cut1u)--[red,thick] (cut1b),(cutR)-- [photon,ultra thick] (u1),
			};
		\end{feynman}
	         \end{tikzpicture}                             \\
	 & =\int\! \frac{d^{D}\ell_1}{ (2\pi)^{D}} \delta(\vb_1\Cdot \ell_1)\sum_{h_1,h_3} \frac{\cA_5^{h_1,h_3}(\ell_1, \ell_3,k, \vb_2)
		\cA_3^{-h_1}(-\ell_1,\vb_1)\cA_3^{-h_3}(-\ell_3,\vb_1)}{ \ell_1^2 \ell_3^2  }\, .
\end{align}
In addition to the three and four-point HEFT amplitudes required up until now, we now also need the five-point tree-level HEFT amplitude which is given in \eqref{eq: 1source5pt}.

The analysis and simplification of this diagram follows the same steps as for cut $\cC_1$, however now there seem to emerge new graph topologies which are distinct to those contained in \eqref{eq: masterGraph}. Explicitly, propagator structures like the following can appear
\begin{equation}
	\cT_1=
	\begin{tikzpicture}[baseline={([yshift=-0.4ex]current bounding box.center)}]\tikzstyle{every node}=[font=\small]
		\begin{feynman}
			\vertex (p1) {\(p_1\)};
			\vertex [above=2.5cm of p1](p2){$p_2$};
			\vertex [right=1cm of p2] (u1) [dot]{};
			\vertex [right=3cm of p2] (u2) [dot]{};
			\vertex [right=3cm of u1] (p3){$p_{2}'$};
			\vertex [right=1.0cm of p1] (b1) [dot]{};
			\vertex [above=1.25cm of b1] (lm1) [dot]{};
			\vertex [right=1.0cm of b1] (b2) []{};
			\vertex [right=1.0cm of b2] (b3) [dot]{};
			\vertex [right=1.0cm of b3](p4){$p_{1}'$};
			\vertex [above=1.25cm of p1] (cutL){$k$};
			\vertex [right=2.9cm of cutL] (m1)[]{};
			\vertex [right=4.0cm of cutL] (cutR);
			\vertex [right=1.0cm of b1] (cut1);
			\vertex [above=0.3cm of cut1] (cut1u);
			\vertex [below=0.3cm of cut1] (cut1b);
			\vertex [right=0.5cm of b2] (cut2);
			\vertex [above=0.3cm of cut2] (cut2u);
			\vertex [below=0.3cm of cut2] (cut2b);
			\diagram* {
			(p2) -- [thick] (u1) -- [thick] (p3),
			(b1)--[boson,rmomentum'=\(\ell_{1}\)](lm1)--[boson,rmomentum'=\(\ell_{4}\)](u1), (b3)-- [ boson,rmomentum'=\(\ell_{3}\)] (u2), (p1) -- [thick] (b1)-- [thick] (b3)-- [thick] (p4), (cut1u)--[red,thick] (cut1b),(cutL)-- [boson%, rmomentum'
			] (lm1),
			};
		\end{feynman}
	\end{tikzpicture}
	,\quad \cT_2=
	\begin{tikzpicture}[baseline={([yshift=-2.4ex]current bounding box.center)}]\tikzstyle{every node}=[font=\small]
		\begin{feynman}
			\vertex (p1) {\(p_1\)};
			\vertex [above=2.5cm of p1](p2){$p_2$};
			\vertex [right=1cm of p2] (u1) [dot]{};
			\vertex [right=2cm of p2] (u2) [dot]{};
			\vertex [right=3cm of p2] (u3) [dot]{};
			\vertex [above=1cm of u2] (kout) []{$k$};
			\vertex [right=3cm of u1] (p3){$p_{2}'$};
			\vertex [right=1.0cm of p1] (b1) [dot]{};
			\vertex [above=1.25cm of b1] (lm1) []{};
			\vertex [right=1.0cm of b1] (b2) []{};
			\vertex [right=1.0cm of b2] (b3) [dot]{};
			\vertex [right=1.0cm of b3](p4){$p_{1}'$};
			\vertex [above=1.25cm of p1] (cutL);
			\vertex [right=2.9cm of cutL] (m1)[]{};
			\vertex [right=4.0cm of cutL] (cutR);
			\vertex [right=1.0cm of b1] (cut1);
			\vertex [above=0.3cm of cut1] (cut1u);
			\vertex [below=0.3cm of cut1] (cut1b);
			\vertex [right=0.5cm of b2] (cut2);
			\vertex [above=0.3cm of cut2] (cut2u);
			\vertex [below=0.3cm of cut2] (cut2b);
			\diagram* {
			(p2) -- [thick] (u1) -- [thick] (p3),
			(b1)--[ boson,rmomentum=\(\ell_{1}\)](u1), (b3)-- [ boson,rmomentum'=\(\ell_{3}\)] (u3), (p1) -- [thick] (b1)-- [thick] (b3)-- [thick] (p4), (cut1u)--[red,thick] (cut1b),(u2)-- [boson%, momentum'
			] (kout),
			};
		\end{feynman}
	\end{tikzpicture}\,,
\end{equation}
where $\ell_4=\ell_1+k$ and all massive propagators are linearised. In fact, in the $\mb$ expansion of the HEFT, both of these topologies can be rewritten into the form of \eqref{eq: masterGraph} as we now shall demonstrate. 

The origin of topology $\cT_1$ can be seen by expanding the five-point HEFT amplitude in terms of BCJ numerators using \eqref{eq: 1source5pt}
\begin{align}
    \cC_2=\int\! \frac{d^{D}\ell_1}{ (2\pi)^{D}}& \delta(\vb_1\Cdot \ell_1)\sum_{h_1,h_3} -i\kappa^3 \frac{
		\cA_3^{-h_1}(-\ell_1,\vb_1)\cA_3^{-h_3}(-\ell_3,\vb_1)}{ \ell_1^2 \ell_3^2  }\times\nn\\ 
    &\left( \frac{(\cN([[\ell_1,\ell_3],k],\vb_2))^2}{q_1^2 q_2^2} + \frac{(\cN([[\ell_1,k],\ell_3],\vb_2))^2}{\ell_4^2 q_2^2} +\frac{(\cN([[\ell_3,k],\ell_1],\vb_2))^2}{\ell_2^2 q_2^2}\right)
     \,.
\end{align}
The BCJ numerators $\cN$ themselves never contain any massless propagators, hence all terms in the topology  $\cT_1$ must come from the denominator $1/l_4^2 q_2^2$ associated with the second BCJ numerator above. Thus, we can eliminate the topology $\cT_2$ by reparameterising the loop momentum \textit{for this particular (second) term} as $\ell_1\rightarrow-\ell_1-q_1$, which at the level of the 
integrand is equivalent to the replacements $\ell_1\leftrightarrow \ell_3$, $\ell_4\leftrightarrow \ell_2$.
This leaves us with the expression
\begin{align}
       \cC_2=\int\! \frac{d^{D}\ell_1}{ (2\pi)^{D}}\delta(\vb_1\Cdot \ell_1)\sum_{h_1,h_3} -i\kappa^3& \frac{
		\cA_3^{-h_1}(-\ell_1,\vb_1)\cA_3^{-h_3}(-\ell_3,\vb_1)}{ \ell_1^2 \ell_3^2  }\times\nn\\ 
    &\left( \frac{(\cN([[\ell_1,\ell_3],k],\vb_2))^2}{q_1^2 q_2^2} +2\frac{(\cN([[\ell_3,k],\ell_1],\vb_2))^2}{\ell_2^2 q_2^2}\right)
    \, .
\end{align}
We would like to note that the rewriting above relies on the HEFT specific condition $\vb_1\Cdot q_1{=}0$
imposed by the delta function
and the principal value prescription for the linearised massive propagators in the HEFT. The delta function (regularised linear propagators) are even (odd) when the sign of the momentum is flipped, and this allows manipulations which otherwise would change the $i\varepsilon$ prescription.

Next, we consider the terms in cut $\cC_2$ belonging to the topology $\cT_2$ which can appear in either of the two remaining BCJ numerators
\begin{equation}
	\cC_2\lvert_{\cT_2} = \int\! \frac{d^D \ell_1}{(2 \pi)^{D}} \frac{\delta ( \vb_1 \Cdot \ell_1)}{\ell_1^2 \ell_3^2 (\vb_2 \Cdot \ell_1) (\vb_2 \Cdot (\ell_1+q_1))} g(\ell_1)\,,
\end{equation}
where $g(\ell_1)$ is shorthand for the rest of the integrand. The topology $\cT_2$ can contain two powers of linearised massive propagators, for example,  $1/(\vb_2 \Cdot \ell_1)^2$. However, in what follows we will assume for simplicity only single powers as the analysis in either case is the same. The first step is to perform partial fractions on the two linearised propagators to yield
\begin{equation}
	\cC_2\lvert_{\cT_2} =\int\! \frac{d^D \ell_1}{(2 \pi)^{D}} \frac{\delta ( \vb_1 \Cdot \ell_1)}{\ell_1^2 \ell_3^2 (\vb_2 \Cdot \ell_1) } \frac{g(\ell_1)}{\vb_2 \Cdot q_1}-
	\int\! \frac{d^D \ell_1}{(2 \pi)^{D}} \frac{\delta ( \vb_1 \Cdot \ell_1)}{\ell_1^2 \ell_3^2 (\vb_2 \Cdot (\ell_1+q_1))} \frac{g(\ell_1)}{\vb_2 \Cdot q_1}
	\,,
\end{equation}
Next we again re-parameterise loop momentum by $\ell_1\rightarrow -\ell_1-q_1$, but only in the second term above,
\begin{equation}
	\cC_2\lvert_{\cT_2} =\int\! \frac{d^D \ell_1}{(2 \pi)^{D}} \frac{\delta ( \vb_1 \Cdot \ell_1)}{\ell_1^2 \ell_3^2 (\vb_2 \Cdot \ell_1) } \frac{g(\ell_1)+g(\ell_3)}{\vb_2 \Cdot q_1}\, , 
\end{equation}
which is a subtopology of \eqref{eq: masterGraph}. Graphically this process can be written as
\begin{equation}
	\begin{split}
		\mkern-89mu
		\begin{tikzpicture}[baseline={([yshift=-2.4ex]current bounding box.center)}]\tikzstyle{every node}=[font=\small]
			\begin{feynman}
				\vertex (p1) {\(p_1\)};
				\vertex [above=2.5cm of p1](p2){$p_2$};
				\vertex [right=1cm of p2] (u1) [dot]{};
				\vertex [right=2cm of p2] (u2) [dot]{};
				\vertex [right=3cm of p2] (u3) [dot]{};
				\vertex [above=1cm of u2] (kout) []{$k$};
				\vertex [right=3cm of u1] (p3){$p_{2}'$};
				\vertex [right=1.0cm of p1] (b1) [dot]{};
				\vertex [above=1.25cm of b1] (lm1) []{};
				\vertex [right=1.0cm of b1] (b2) []{};
				\vertex [right=1.0cm of b2] (b3) [dot]{};
				\vertex [right=1.0cm of b3](p4){$p_{1}'$};
				\vertex [above=1.25cm of p1] (cutL);
				\vertex [right=2.9cm of cutL] (m1)[]{};
				\vertex [right=4.0cm of cutL] (cutR);
				\vertex [right=1.0cm of b1] (cut1);
				\vertex [above=0.3cm of cut1] (cut1u);
				\vertex [below=0.3cm of cut1] (cut1b);
				\vertex [right=0.5cm of b2] (cut2);
				\vertex [above=0.3cm of cut2] (cut2u);
				\vertex [below=0.3cm of cut2] (cut2b);
				\diagram* {
				(p2) -- [thick] (u1) -- [thick] (p3),
				(b1)--[ boson,rmomentum=\(\ell_{1}\)](u1), (b3)-- [ boson,rmomentum'=\(\ell_{3}\)] (u3), (p1) -- [thick] (b1)-- [thick] (b3)-- [thick] (p4), (cut1u)--[red,thick] (cut1b),(u2)-- [boson] (kout),
				};
			\end{feynman}
		\end{tikzpicture} \xrightarrow[{\rm fraction}]{\rm partial}
		\begin{aligned}
			\begin{tikzpicture}[baseline={([yshift=-2.4ex]current bounding box.center)}]\tikzstyle{every node}=[font=\small]
				\begin{feynman}
					\vertex (p1) {\(p_1\)};
					\vertex [above=2.5cm of p1](p2){$p_2$};
					\vertex [right=1cm of p2] (u1) [dot]{};
					\vertex [right=2cm of p2] (u2)[]{};
					\vertex [right=3cm of p2] (u3) [dot]{};
					\vertex [above=1cm of u1] (kout) []{$k$};
					\vertex [right=3cm of u1] (p3){$p_{2}'$};
					\vertex [right=1.0cm of p1] (b1) [dot]{};
					\vertex [above=1.25cm of b1] (lm1) []{};
					\vertex [right=1.0cm of b1] (b2) []{};
					\vertex [right=1.0cm of b2] (b3) [dot]{};
					\vertex [right=1.0cm of b3](p4){$p_{1}'$};
					\vertex [above=1.25cm of p1] (cutL);
					\vertex [right=2.9cm of cutL] (m1)[]{};
					\vertex [right=4.0cm of cutL] (cutR);
					\vertex [right=1.0cm of b1] (cut1);
					\vertex [above=0.3cm of cut1] (cut1u);
					\vertex [below=0.3cm of cut1] (cut1b);
					\vertex [right=0.5cm of b2] (cut2);
					\vertex [above=0.3cm of cut2] (cut2u);
					\vertex [below=0.3cm of cut2] (cut2b);
					\diagram* {
					(p2) -- [thick] (u1) -- [thick] (p3),
					(b1)--[ boson,rmomentum=\(\ell_{1}\)](u1), (b3)-- [ boson,rmomentum'=\(\ell_{3}\)] (u3), (p1) -- [thick] (b1)-- [thick] (b3)-- [thick] (p4), (cut1u)--[red,thick] (cut1b),(u1)-- [boson] (kout),
					};
				\end{feynman}
			\end{tikzpicture} \\
			+	\begin{tikzpicture}[baseline={([yshift=-2.4ex]current bounding box.center)}]\tikzstyle{every node}=[font=\small]
				  \begin{feynman}
					\vertex (p1) {\(p_1\)};
					\vertex [above=2.5cm of p1](p2){$p_2$};
					\vertex [right=1cm of p2] (u1) [dot]{};
					\vertex [right=2cm of p2] (u2)[]{};
					\vertex [right=3cm of p2] (u3) [dot]{};
					\vertex [above=1cm of u3] (kout) []{$k$};
					\vertex [right=3cm of u1] (p3){$p_{2}'$};
					\vertex [right=1.0cm of p1] (b1) [dot]{};
					\vertex [above=1.25cm of b1] (lm1) []{};
					\vertex [right=1.0cm of b1] (b2) []{};
					\vertex [right=1.0cm of b2] (b3) [dot]{};
					\vertex [right=1.0cm of b3](p4){$p_{1}'$};
					\vertex [above=1.25cm of p1] (cutL);
					\vertex [right=2.9cm of cutL] (m1)[]{};
					\vertex [right=4.0cm of cutL] (cutR);
					\vertex [right=1.0cm of b1] (cut1);
					\vertex [above=0.3cm of cut1] (cut1u);
					\vertex [below=0.3cm of cut1] (cut1b);
					\vertex [right=0.5cm of b2] (cut2);
					\vertex [above=0.3cm of cut2] (cut2u);
					\vertex [below=0.3cm of cut2] (cut2b);
					\diagram* {
					(p2) -- [thick] (u1) -- [thick] (p3),
					(b1)--[ boson,rmomentum=\(\ell_{1}\)](u1), (b3)-- [ boson,rmomentum'=\(\ell_{3}\)] (u3), (p1) -- [thick] (b1)-- [thick] (b3)-- [thick] (p4), (cut1u)--[red,thick] (cut1b),(u3)-- [boson] (kout),
					};
				\end{feynman}
			  \end{tikzpicture}
		\end{aligned}
		\begin{tikzpicture}[baseline={([yshift=-2.4ex]current bounding box.center)}]\tikzstyle{every node}=[font=\small]
			\begin{feynman}
				\vertex (p0) {};
				\vertex [below=0.5cm of p0](p1) {};
				\vertex [below=2.5cm of p1](p2){};
				\diagram* {
				(p1) -- [fermion, thick, out=-30, in=30] (p2),
				};
			\end{feynman}
		\end{tikzpicture}
		%\begin{tikzpicture}[->,auto,node distance=3cm,
		% thick]
		% \node[] (a1) {};
		% \node[] (a4) [below=3cm of a1] {};
		% \node[] (a5) [below=3cm of a4] {};
		% \path[every node]
		%   (a1) edge[bend left] node [left] {} (a4) node [near end, fill=white] {};
		%\end{tikzpicture},
		\rotatebox[origin=c]{270}{$\ell_1\rightarrow -\ell_1-q_1$}
	\end{split}
\end{equation}
and, as has been explained above, such manipulations are possible because we have principal valued and/or delta function cut linear massive propagators.

Hence, as was the case for cut $\cC_1$, the integrand from cut $\cC_2$ contains the same basis of propagators $\cD_i$ given in Table~\ref{tab:propagatorbasis} and corresponding to the master topology \eqref{eq: masterGraphSimple}. We then  follow the same method: first, tensor reduce to scalar integrals in this master topology, and second, perform IBP reduction assuming the cut conditions of diagram $\cC_2$ to get a set of master integrals. In this case these we get the box $\cI_5$, the pentagon $\cI_6$ previously found in cut $\cC_1$, and two additional topologies
\begin{align}
	\label{eq: cut2MIs}
	\cI_2\coloneqq j_{11100}= \begin{tikzpicture}[scale=0.5, transform shape, baseline={([yshift=-0.4ex]current bounding box.center)}]\tikzstyle{every node}=[font=\small]
		                          \begin{feynman}
			\vertex (p1) {};
			\vertex [above=2.5cm of p1](p2){};
			\vertex [right=2cm of p2] (u1) [dot]{};
			\vertex [right=2.5cm of p2] (ur){};
			\vertex [right=4cm of p2] (p3){};
			\vertex [right=1.0cm of p1] (b1) [dot]{};
			\vertex [right=1.0cm of b1] (b2) []{};
			\vertex [right=1.0cm of b2] (b3) [dot]{};
			\vertex [right=1.0cm of b3](p4){};
			\vertex [above=1.25cm of p1] (cutL);
			\vertex [below=1.25cm of ur] (m1);
			\vertex [right=4.0cm of cutL] (cutR){};
			\vertex [right=1.0cm of b1] (cut1);
			\vertex [above=0.3cm of cut1] (cut1u);
			\vertex [below=0.3cm of cut1] (cut1b){};
			\vertex [right=1cm of u1] (cut2);
			\vertex [above=0.2cm of cut2] (cut2u){};
			\diagram* {
			(p2) -- [thick] (u1) -- [thick] (p3),
			(b1)--[boson](u1), (b3)-- [boson] (u1), (p1) -- [thick] (b1)-- [thick] (b3)-- [thick] (p4), (cut1u)--[red,thick] (cut1b),(cutR)-- [boson] (u1),
			};
		\end{feynman}
	                          \end{tikzpicture}\,, &  &
	\cI_4\coloneqq j_{11101}= \begin{tikzpicture}[scale=0.5, transform shape, baseline={([yshift=-0.4ex]current bounding box.center)}]\tikzstyle{every node}=[font=\small]
		                          \begin{feynman}
			\vertex (p1) {};
			\vertex [above=2.5cm of p1](p2){};
			\vertex [right=1cm of p2] (u1) [dot]{};
			\vertex [right=3cm of p2] (u2) [dot]{};
			\vertex [right=3cm of u1] (p3){};
			\vertex [right=1.0cm of p1] (b1) [dot]{};
			\vertex [right=1.0cm of b1] (b2) []{};
			\vertex [right=1.0cm of b2] (b3) [dot]{};
			\vertex [right=1.0cm of b3](p4){};
			\vertex [above=1.25cm of p1] (cutL);
			\vertex [below=1.25cm of u2] (m1)[]{};
			\vertex [right=4.0cm of cutL] (cutR){};
			\vertex [right=1.0cm of b1] (cut1);
			\vertex [above=0.3cm of cut1] (cut1u);
			\vertex [below=0.3cm of cut1] (cut1b){};
			\vertex [right=1cm of u1] (cut2);
			\vertex [above=0.2cm of cut2] (cut2u){};
			\diagram* {
			(p2) -- [thick] (u1) -- [thick] (p3),
			(b1)--[boson](u1), (b3)-- [boson] (u2), (p1) -- [thick] (b1)-- [thick] (b3)-- [thick] (p4), (cut1u)--[red,thick] (cut1b),(cutR)-- [boson] (u2),
			};
		\end{feynman}
	                          \end{tikzpicture}\,.
\end{align}
To summarise, the contributions from cut $\cC_2$ are
\begin{equation}
	\cC_2= c_2\, \cI_2 + c_4\, \cI_4+ c_5\, \cI_5+ c_6\, \cI_6\, ,
\end{equation}
where the coefficients $c_5$ and $c_6$ exactly match those found from cut $\cC_1$.

%--------------------------------------------------------

\subsection{Cut three -- the first ``snail'' diagram}

In the above cuts, which gave contributions $\cC_1$ and $\cC_2$, we always cut the massless propagator $\cD_1$ with momenta $\ell_1$. There are, however, other unitarity cut diagrams relevant for classical physics that are $\cO(\mb_1^3 \mb_2^2)$ in the HEFT expansion. These are the diagrams $\cC_3$ and $\cC_4$ which only involve a single cut massless propagator and a single HEFT cut of the massive scalar on the bottom line of \eqref{eq: cut3}. These new diagrams probe all the master integrals found in cuts one and two but also involve four new master integrals with the collapsed massless propagator $\cD_1$. We find these new contributions to be crucial for capturing the full infrared divergence of the classical five-point one-loop amplitude. That is, with these contributions the infrared-divergent part of our amplitude is given by a five-point tree HEFT amplitude multiplied by a infrared phase, which exactly matches  Weinberg's prediction \cite{Weinberg:1965nx}. We explain this matching  in detail in Appendix \ref{sec:Weinberg-resurrection}, and now move on to computing the contributions from these new diagrams.

The first new cut we consider is $\cC_3$ in \eqref{eq: 1loopSnails} which we replicate below for convenience 
\begin{align}\label{eq: cut3}
	\cC_3=\begin{tikzpicture}[baseline={([yshift=-0.4ex]current bounding box.center)}]\tikzstyle{every node}=[font=\small]
		      \begin{feynman}
			\vertex (p1) {\(p_1\)};
			\vertex [above=2.5cm of p1](p2){$p_2$};
			\vertex [right=1.0cm of p2] (u1) [HV]{$~~~$};
			\vertex [below=1.25cm of u1] (m1) [GR2]{H};
			\vertex [below=1.25cm of m1] (b0) [HV]{$~~~$};
			\vertex [right=3cm of u1] (p3){$p_{2'}$};
			\vertex [right=1.0cm of p1] (b1) []{};
			\vertex [right=1.0cm of b1] (b2) []{};
			\vertex [right=1.0cm of b2] (b3) [HV]{H};
			\vertex [right=1.0cm of b3](p4){$p_{1'}$};
			\vertex [above=1.25cm of p1] (cutL);
			\vertex [right=4.0cm of cutL] (cutR){$k$};
			\vertex [right=1.0cm of b1] (cut1);
			\vertex [above=0.3cm of cut1] (cut1u);
			\vertex [below=0.3cm of cut1] (cut1b);
			\vertex [above=0.4cm of cut1u] (cutd1);
			\vertex [right=0.4cm of cutd1] (cutd2);
			\vertex [right=0.5cm of b2] (cut2);
			\vertex [above=0.3cm of cut2] (cut2u);
			\vertex [below=0.3cm of cut2] (cut2b);
			\diagram* {
			(p2) -- [thick] (u1) -- [thick] (p3),
			(b3)-- [photon,ultra thick,rmomentum'=\(\ell_{2}\)] (m1), (p1) -- [thick] (b0)-- [thick] (b3)-- [thick] (p4),  (cut1u)--[ red,thick] (cut1b),(cut1u)--[dashed, red,thick] (cutd2),(cutR)-- [photon,ultra thick] (b3),
			};
		\end{feynman}
	      \end{tikzpicture}\,,
\end{align}
which features the gluing of a four-point tree-level HEFT amplitude with two scalars into a new ingredient, the five-point tree level HEFT amplitude with \emph{four} scalars. This five-point amplitude was derived in Section \ref{sec: BCFWtwosources} using BCFW recursion relations and is given in  \eqref{eq: 5pttreeN1}, \eqref{eq: 5pttreeN2} and \eqref{eq: 5pttree}. The contribution from cut $\cC_3$ is then given by
\begin{equation}
	\cC_3=\int\! \frac{d^{D}\ell_1}{ (2\pi)^{D}} \delta(\vb_1\Cdot \ell_1)\sum_{h_2} \frac{\cM_5^{h_2}(\ell_2, \vb_1, \vb_2)
		\cA_4^{-h_2}(-\ell_2,k,\vb_1)}{ \ell_2^2  }\, .
\end{equation}
Next we perform the state sum, tensor reduce the result and finally perform IBP reduction to a set of MIs. In addition to the master integrals found in cut diagram $\cC_1$,  we also have to include the following master integrals
\begin{equation}
	\begin{split}\label{eq: cut3MIs}
		&\widetilde \cI_1\coloneqq j_{01010}= \begin{tikzpicture}[scale=0.5, transform shape, baseline={([yshift=-0.4ex]current bounding box.center)}]\tikzstyle{every node}=[font=\small]
			\begin{feynman}
				\vertex (p1) {};
				\vertex [above=2.5cm of p1](p2){};
				\vertex [right=1cm of p2] (u1) {};
				\vertex [right=3cm of p2] (u2);
				\vertex [right=2.4cm of u1] (p3){};
				\vertex [right=1.0cm of p1] (b1) {};
				\vertex [right=1.0cm of b1] (b2) []{};
				\vertex [right=0.9cm of b2] (b3) [dot]{};
				\vertex [left=1.2cm of b3] (x1) []{};
				\vertex [above=0.3cm of x1] (x2) []{};
				\vertex [above=0.8cm of x2] (y) []{};
				\vertex [right=0.65cm of y] (x3) []{};
				\vertex [right=1.0cm of b3](p4){};
				\vertex [above=1.25cm of p1] (cutL);
				\vertex [right=0.4cm of cutL] (cutL3);
				\vertex [right=1.00cm of cutL3] (cutL2)[dot]{};
				\vertex [below=1.25cm of u2] (m1){};
				\vertex [right=4.0cm of cutL] (cutR){};
				\vertex [right=1.0cm of b1] (cut1);
				\vertex [above=0.3cm of cut1] (cut1u);
				\vertex [below=0.3cm of cut1] (cut1b){};
				\vertex [right=1cm of u1] (cut2);
				\vertex [above=0.2cm of cut2] (cut2u){};
				\diagram* {
				(p2) -- [thick] (cutL2), (cutL2) -- [thick] (p3),(b3)--[boson, half right] (cutL2), (p1) -- [thick] (cutL2)-- [thick] (b3)-- [thick] (p4),(cutR)-- [boson] (b3),(x2)--[red,thick](x3)
				};
			\end{feynman}
		\end{tikzpicture}\,,\\
		& \widetilde \cI_2\coloneqq j_{01011}=\begin{tikzpicture}[scale=0.5, transform shape, baseline={([yshift=-0.4ex]current bounding box.center)}]\tikzstyle{every node}=[font=\small]
			\begin{feynman}
				\vertex (p1) {};
				\vertex [above=2.5cm of p1](p2){};
				\vertex [right=1cm of p2] (u1) {};
				\vertex [right=3cm of p2] (u2) [dot]{};
				\vertex [right=3cm of u1] (p3){};
				\vertex [right=1.0cm of p1] (b1) {};
				\vertex [right=1.0cm of b1] (b2) []{};
				\vertex [right=1.0cm of b2] (b3) [dot]{};
				\vertex [right=1.0cm of b3](p4){};
				\vertex [above=1.25cm of p1] (cutL);
				\vertex [right=1.00cm of cutL] (cutL2)[dot]{};
				\vertex [below=1.25cm of u2] (m1){};
				\vertex [left=1.2cm of b3] (x1) []{};
				\vertex [above=0.15cm of x1] (x2) []{};
				\vertex [above=0.8cm of x2] (y) []{};
				\vertex [right=0.65cm of y] (x3) []{};
				\vertex [right=4.0cm of cutL] (cutR){};
				\vertex [right=1.0cm of b1] (cut1);
				\vertex [above=0.3cm of cut1] (cut1u);
				\vertex [below=0.3cm of cut1] (cut1b){};
				\vertex [right=1cm of u1] (cut2);
				\vertex [above=0.2cm of cut2] (cut2u){};
				\diagram* {
				(p2) -- [thick] (cutL2) -- [thick] (u2), (u2) -- [thick] (p3),(b3)--[boson] (u2), (p1) -- [thick] (cutL2)-- [thick] (b3)-- [thick] (p4),(cutR)-- [boson] (b3),(x2)--[red,thick](x3)
				};
			\end{feynman}
		\end{tikzpicture}\,,\\
		& \widetilde \cI_4\coloneqq j_{01111}= \begin{tikzpicture}[scale=0.5, transform shape, baseline={([yshift=-0.4ex]current bounding box.center)}]\tikzstyle{every node}=[font=\small]
			\begin{feynman}
				\vertex (p1) {};
				\vertex [above=2.5cm of p1](p2){};
				\vertex [right=1cm of p2] (u1) {};
				\vertex [right=3cm of p2] (u2) [dot]{};
				\vertex [right=3cm of u1] (p3){};
				\vertex [right=1.0cm of p1] (b1) {};
				\vertex [right=1.0cm of b1] (b2) []{};
				\vertex [right=1.0cm of b2] (b3) [dot]{};
				\vertex [right=1.0cm of b3](p4){};
				\vertex [above=1.25cm of p1] (cutL);
				\vertex [right=1.00cm of cutL] (cutL2)[dot]{};
				\vertex [below=1.25cm of u2] (m1)[dot]{};
				\vertex [left=1.2cm of b3] (x1) []{};
				\vertex [above=0.15cm of x1] (x2) []{};
				\vertex [above=0.8cm of x2] (y) []{};
				\vertex [right=0.65cm of y] (x3) []{};
				\vertex [right=4.0cm of cutL] (cutR){};
				\vertex [right=1.0cm of b1] (cut1);
				\vertex [above=0.3cm of cut1] (cut1u);
				\vertex [below=0.3cm of cut1] (cut1b){};
				\vertex [right=1cm of u1] (cut2);
				\vertex [above=0.2cm of cut2] (cut2u){};
				\diagram* {
				(p2) -- [thick] (cutL2) -- [thick] (u2), (u2) -- [thick] (p3),(b3)-- [boson] (m1)-- [boson] (u2), (p1) -- [thick] (cutL2)-- [thick] (b3)-- [thick] (p4),(cutR)-- [boson] (m1),(x2)--[red,thick](x3)
				};
			\end{feynman}
		\end{tikzpicture}\,.
	\end{split}
\end{equation}
The contributions from this cut are then given by
\begin{equation}
	\cC_3= \cC_1+\frac{\widetilde c_1}{2}\, \widetilde \cI_1 + \frac{\widetilde c_2}{2}\, \widetilde \cI_2+\frac{\widetilde c_4}{2}\, \widetilde\cI_4,
\end{equation}
which include the contributions from cut $\cC_1$ as expected. Once again these contributions will be doubled up when we include also the swapped graphs in \eqref{eq: inoutswappedgraphs}.

\subsection{Cut four -- the second ``snail'' diagram}
\label{sec:second-snail}

In addition to cut $\cC_3$ we also have contributions coming from cut diagram $\cC_4$ which involves gluing in the six-point tree-level HEFT amplitude. The cut diagram has the form
\begin{align}\label{eq: cutfour}
	\cC_4=\begin{tikzpicture}[baseline={([yshift=-0.4ex]current bounding box.center)}]\tikzstyle{every node}=[font=\small]
		      \begin{feynman}
			\vertex (p1) {\(p_1\)};
			\vertex [above=2.5cm of p1](p2){$p_2$};
			\vertex [right=1.0cm of p2] (u1) [HV]{$~~~$};
			\vertex [below=1.25cm of u1] (m1) [GR2]{H};
			\vertex [below=1.25cm of m1] (b0) [HV]{$~~~$};
			\vertex [right=3cm of u1] (p3){$p_{2'}$};
			\vertex [right=1.0cm of p1] (b1) []{};
			\vertex [right=1.0cm of b1] (b2) []{};
			\vertex [right=1.0cm of b2] (b3) [HV]{H};
			\vertex [right=1.0cm of b3](p4){$p_{1'}$};
			\vertex [above=1.25cm of p1] (cutL);
			\vertex [right=4.0cm of cutL] (cutR);
			\vertex [above=0.5cm of cutR] (cutRu){$k$};
			\vertex [right=1.0cm of b1] (cut1);
			\vertex [above=0.3cm of cut1] (cut1u);
			\vertex [below=0.3cm of cut1] (cut1b);
			\vertex [above=0.4cm of cut1u] (cutd1);
			\vertex [right=0.4cm of cutd1] (cutd2);
			\vertex [right=0.5cm of b2] (cut2);
			\vertex [above=0.3cm of cut2] (cut2u);
			\vertex [below=0.3cm of cut2] (cut2b);
			\diagram* {
			(p2) -- [thick] (u1) -- [thick] (p3),
			(b3)-- [photon,ultra thick,rmomentum'=\(\ell_{3}\)] (m1), (p1) -- [thick] (b0)-- [thick] (b3)-- [thick] (p4),  (cut1u)--[ red,thick] (cut1b),(cut1u)--[dashed, red,thick] (cutd2),(cutRu)-- [photon,ultra thick] (m1),
			};
		\end{feynman}
	      \end{tikzpicture} \,.
\end{align}
The six-point tree-level amplitude with four scalars and two radiated gravitons was derived in Section~\ref{sec: BCFWtwosources}.
Computing this contribution is more involved than the previous cut diagrams, hence we relegate the details to Appendix~\ref{app: c4Details}.

Cut diagram $\cC_4$ contains those master integrals probed by cut $\cC_2$ but  in addition the box $\widetilde \cI_4$ previously found from cut $\cC_3$ and the following new master integral
\begin{equation}
	\begin{split}\label{eq: cut4MIs}
		&\widetilde \cI_3\coloneqq j_{01101}= \begin{tikzpicture}[scale=0.5, transform shape, baseline={([yshift=-0.4ex]current bounding box.center)}]\tikzstyle{every node}=[font=\small]
			\begin{feynman}
				\vertex (p1) {};
				\vertex [above=2.5cm of p1](p2){};
				\vertex [right=1cm of p2] (u1) {};
				\vertex [right=3cm of p2] (u2) [dot]{};
				\vertex [right=3cm of u1] (p3){};
				\vertex [right=1.0cm of p1] (b1) {};
				\vertex [right=1.0cm of b1] (b2) []{};
				\vertex [right=1.0cm of b2] (b3) [dot]{};
				\vertex [right=1.0cm of b3](p4){};
				\vertex [above=1.25cm of p1] (cutL);
				\vertex [right=1.00cm of cutL] (cutL2)[dot]{};
				\vertex [below=1.25cm of u2] (m1){};
				\vertex [left=1.2cm of b3] (x1) []{};
				\vertex [above=0.15cm of x1] (x2) []{};
				\vertex [above=0.8cm of x2] (y) []{};
				\vertex [right=0.65cm of y] (x3) []{};
				\vertex [right=4.0cm of cutL] (cutR){};
				\vertex [right=1.0cm of b1] (cut1);
				\vertex [above=0.3cm of cut1] (cut1u);
				\vertex [below=0.3cm of cut1] (cut1b){};
				\vertex [right=1cm of u1] (cut2);
				\vertex [above=0.2cm of cut2] (cut2u){};
				\diagram* {
				(p2) -- [thick] (cutL2) -- [thick] (u2), (u2) -- [thick] (p3),(b3)--[boson] (u2), (p1) -- [thick] (cutL2)-- [thick] (b3)-- [thick] (p4),(cutR)-- [boson] (u2),(x2)--[red,thick](x3)
				};
			\end{feynman}
		\end{tikzpicture}\,.
	\end{split}
\end{equation}
The contributions from cut diagram $\cC_4$ are then given by
\begin{equation}
	\cC_4= \cC_2+\widetilde c_3\, \widetilde \cI_3+ \widetilde c_4\, \widetilde\cI_4,
\end{equation}
and contain those contributions coming from cut $\cC_2$ as expected and where $\widetilde c_4$ matches the same coefficient appearing in cut $\cC_3$.

\subsection{Final result before integration}

We can now merge the contributions from all the cuts according to \eqref{eq: unionofcuts} and present the one-loop five-point amplitude at order $\mb_1^3\mb_2^2$ in the HEFT expansion in terms of the master integrals and their coefficients:
\begin{align}
	\begin{split}
		\label{finalresultform1}
		\cM^{(1)}_{5,\bar{m}_1^3 \bar{m}_2^2} &=(2*\cC_1)\cup \cC_2 \cup (2*\cC_3)\cup \cC_4
		% \\ &
		=  \sum_{i=1}^4 \tilde{c}_i \widetilde{\cI}_i + \sum_{i=1}^6 c_i \cI_i\ .
	\end{split}
\end{align} 
The contribution at order $\mb_1^2\mb_2^3$ can be found by swapping labels of the massive lines $1 \leftrightarrow 2$ in the above amplitude and together these contributions  completely determine the classical one-loop five-point
amplitude,% 
\footnote{The $\mb_1^3\mb_2^3$  (or hyper-classical)
contribution will not be needed but can be computed as discussed in Appendix~\ref{sec:factorisation}.}
\begin{equation}
	\label{finalresultform2}
	\mathcal{M}_{5}^{(1)} = (\cM_{5,\mb_1^3\mb_2^2}^{(1)}+\cM_{5,\mb_1^2\mb_2^3}^{(1)})\vert_{\mb_i \rightarrow m_i}\ .
\end{equation}
In the next section, we outline our strategy used to evaluate the master integral topologies \eqref{figurineintegrali} and present complete analytic results in dimensional regularisation (around $D{=}4$). In Section \ref{sec:finalresult} we will then discuss
the full, integrated result of the one-loop five-point amplitude.

\section{The one-loop integrals from differential equations}
\label{sec: oneLoopIntegrals}

\subsection{The structure of the integrals}
The integrals we are considering have the form%
\footnote{The definitions of the integrals $\cI_1$ and $\cI_2$ in the {\href{https://github.com/QMULAmplitudes/Gravity-Observables-From-Amplitudes}{\it ancillary files }} differ from those in the main text (in 
\eqref{figurineintegrali} and below in \eqref{eq: integralsList})  by a factor of $1/2$. The coefficients of these integrals will also change accordingly so that the complete expression for the amplitude is the same here as in the paper.}
\begin{equation}
	j_{a_1, 1, a_3, a_4, a_5} = \mu_{\rm IR}^{4-D} \int\! \frac{d^D \ell}{(2 \pi)^{D}} \frac{- i \pi\, \delta (\vb_1 \cdot \ell)}{(\ell^2+i \varepsilon)^{a_1} [(\ell+q_1)^2+i \varepsilon]^{a_3} [(\ell-q_2)^2+i \varepsilon]^{a_4} } \ \mathrm{PV} \frac{1}{
		(\vb_2 \Cdot \ell)^{a_5}}\ ,
\end{equation}
where
\begin{equation}
	\label{eq::deltaPVdef}
	- i \pi \, \delta (x) = \frac{1}{2} \left(\frac{1}{x+i \varepsilon} + \frac{1}{-x+i \varepsilon}\right)\ , \qquad {\rm PV} \frac{1}{x} = \frac{1}{2} \left(\frac{1}{x+i \varepsilon} - \frac{1}{-x+i \varepsilon}\right)\ .
\end{equation}
We compute the master integral (MI) basis using LiteRed2 \cite{Lee:2012cn,Lee:2013mka}. The full list of MIs from the merger of the contributions from the cuts in Section \ref{sec: oneloopunitaritycuts} are
 \begin{equation}
 	\begin{split}
 		\label{figurineintegrali}
 		\begin{aligned}
 			\widetilde{\mathcal{I}}_1=j_{01010}=\begin{tikzpicture}[scale=0.5, transform shape, baseline={([yshift=-0.4ex]current bounding box.center)}]\tikzstyle{every node}=[font=\small]
 				                                    \begin{feynman}
 					\vertex (p1) {};
 					\vertex [above=2.5cm of p1](p2){};
 					\vertex [right=1cm of p2] (u1) {};
 					\vertex [right=3cm of p2] (u2);
 					\vertex [right=2.4cm of u1] (p3){};
 					\vertex [right=1.0cm of p1] (b1) {};
 					\vertex [right=1.0cm of b1] (b2) []{};
 					\vertex [right=0.9cm of b2] (b3) [dot]{};
 					\vertex [left=1.2cm of b3] (x1) []{};
 					\vertex [above=0.3cm of x1] (x2) []{};
 					\vertex [above=0.8cm of x2] (y) []{};
 					\vertex [right=0.65cm of y] (x3) []{};
 					\vertex [right=1.0cm of b3](p4){};
 					\vertex [above=1.25cm of p1] (cutL);
 					\vertex [right=0.4cm of cutL] (cutL3);
 					\vertex [right=1.00cm of cutL3] (cutL2)[dot]{};
 					\vertex [below=1.25cm of u2] (m1){};
 					\vertex [right=4.0cm of cutL] (cutR){};
 					\vertex [right=1.0cm of b1] (cut1);
 					\vertex [above=0.3cm of cut1] (cut1u);
 					\vertex [below=0.3cm of cut1] (cut1b){};
 					\vertex [right=1cm of u1] (cut2);
 					\vertex [above=0.2cm of cut2] (cut2u){};
 					\diagram* {
 					(p2) -- [thick] (cutL2), (cutL2) -- [thick] (p3),(b3)--[boson, half right] (cutL2), (p1) -- [thick] (cutL2)-- [thick] (b3)-- [thick] (p4),(cutR)-- [boson] (b3),(x2)--[red,thick](x3)
 					};
 				\end{feynman}
 			                                    \end{tikzpicture}\,,
 			 &  & \widetilde{\mathcal{I}}_2=j_{01011}=\begin{tikzpicture}[scale=0.5, transform shape, baseline={([yshift=-0.4ex]current bounding box.center)}]\tikzstyle{every node}=[font=\small]
 				                                          \begin{feynman}
 					\vertex (p1) {};
 					\vertex [above=2.5cm of p1](p2){};
 					\vertex [right=1cm of p2] (u1) {};
 					\vertex [right=3cm of p2] (u2) [dot]{};
 					\vertex [right=3cm of u1] (p3){};
 					\vertex [right=1.0cm of p1] (b1) {};
 					\vertex [right=1.0cm of b1] (b2) []{};
 					\vertex [right=1.0cm of b2] (b3) [dot]{};
 					\vertex [right=1.0cm of b3](p4){};
 					\vertex [above=1.25cm of p1] (cutL);
 					\vertex [right=1.00cm of cutL] (cutL2)[dot]{};
 					\vertex [below=1.25cm of u2] (m1){};
 					\vertex [right=4.0cm of cutL] (cutR){};
 					\vertex [left=1.2cm of b3] (x1) []{};
 					\vertex [above=0.15cm of x1] (x2) []{};
 					\vertex [above=0.8cm of x2] (y) []{};
 					\vertex [right=0.65cm of y] (x3) []{};
 					\vertex [right=1.0cm of b1] (cut1);
 					\vertex [above=0.3cm of cut1] (cut1u);
 					\vertex [below=0.3cm of cut1] (cut1b){};
 					\vertex [right=1cm of u1] (cut2);
 					\vertex [above=0.2cm of cut2] (cut2u){};
 					\diagram* {
 					(p2) -- [thick] (cutL2) -- [thick] (u2), (u2) -- [thick] (p3),(b3)--[boson] (u2), (p1) -- [thick] (cutL2)-- [thick] (b3)-- [thick] (p4),(cutR)-- [boson] (b3),(x2)--[red,thick](x3)
 					};
 				\end{feynman}
 			                                          \end{tikzpicture}\,,  \\
 			\widetilde{\mathcal{I}}_3=j_{01101}= \begin{tikzpicture}[scale=0.5, transform shape, baseline={([yshift=-0.4ex]current bounding box.center)}]\tikzstyle{every node}=[font=\small]
 				                                     \begin{feynman}
 					\vertex (p1) {};
 					\vertex [above=2.5cm of p1](p2){};
 					\vertex [right=1cm of p2] (u1) {};
 					\vertex [right=3cm of p2] (u2) [dot]{};
 					\vertex [right=3cm of u1] (p3){};
 					\vertex [right=1.0cm of p1] (b1) {};
 					\vertex [right=1.0cm of b1] (b2) []{};
 					\vertex [right=1.0cm of b2] (b3) [dot]{};
 					\vertex [right=1.0cm of b3](p4){};
					\vertex [left=1.2cm of b3] (x1) []{};
					\vertex [above=0.15cm of x1] (x2) []{};
					\vertex [above=0.8cm of x2] (y) []{};
					\vertex [right=0.65cm of y] (x3) []{};
					\vertex [above=1.25cm of p1] (cutL);
					\vertex [right=1.00cm of cutL] (cutL2)[dot]{};
					\vertex [below=1.25cm of u2] (m1){};
					\vertex [right=4.0cm of cutL] (cutR){};
					\vertex [right=1.0cm of b1] (cut1);
					\vertex [above=0.3cm of cut1] (cut1u);
					\vertex [below=0.3cm of cut1] (cut1b){};
					\vertex [right=1cm of u1] (cut2);
					\vertex [above=0.2cm of cut2] (cut2u){};
					\diagram* {
					(p2) -- [thick] (cutL2) -- [thick] (u2), (u2) -- [thick] (p3),(b3)--[boson] (u2), (p1) -- [thick] (cutL2)-- [thick] (b3)-- [thick] (p4),(cutR)-- [boson] (u2),(x2)--[red,thick](x3)
					};
				\end{feynman}
			                                     \end{tikzpicture}\, ,
			 &  & \widetilde{\mathcal{I}}_4=j_{01111}= \begin{tikzpicture}[scale=0.5, transform shape, baseline={([yshift=-0.4ex]current bounding box.center)}]\tikzstyle{every node}=[font=\small]
				                                           \begin{feynman}
					\vertex (p1) {};
					\vertex [above=2.5cm of p1](p2){};
					\vertex [right=1cm of p2] (u1) {};
					\vertex [right=3cm of p2] (u2) [dot]{};
					\vertex [right=3cm of u1] (p3){};
					\vertex [right=1.0cm of p1] (b1) {};
					\vertex [right=1.0cm of b1] (b2) []{};
					\vertex [right=1.0cm of b2] (b3) [dot]{};
					\vertex [right=1.0cm of b3](p4){};
					\vertex [above=1.25cm of p1] (cutL);
					\vertex [right=1.00cm of cutL] (cutL2)[dot]{};
					\vertex [below=1.25cm of u2] (m1)[dot]{};
					\vertex [right=4.0cm of cutL] (cutR){};
					\vertex [left=1.2cm of b3] (x1) []{};
					\vertex [above=0.15cm of x1] (x2) []{};
					\vertex [above=0.8cm of x2] (y) []{};
					\vertex [right=0.65cm of y] (x3) []{};
					\vertex [right=1.0cm of b1] (cut1);
					\vertex [above=0.3cm of cut1] (cut1u);
					\vertex [below=0.3cm of cut1] (cut1b){};
					\vertex [right=1cm of u1] (cut2);
					\vertex [above=0.2cm of cut2] (cut2u){};
					\diagram* {
					(p2) -- [thick] (cutL2) -- [thick] (u2), (u2) -- [thick] (p3),(b3)-- [boson] (m1)-- [boson] (u2), (p1) -- [thick] (cutL2)-- [thick] (b3)-- [thick] (p4),(cutR)-- [boson] (m1),(x2)--[red,thick](x3)
					};
				\end{feynman}
			                                           \end{tikzpicture}\,, \\
			\cI_1\coloneqq j_{11010}= \begin{tikzpicture}[scale=0.5, transform shape, baseline={([yshift=-0.4ex]current bounding box.center)}]\tikzstyle{every node}=[font=\small]
				                          \begin{feynman}
					\vertex (p1) {};
					\vertex [above=2.5cm of p1](p2){};
					\vertex [right=2cm of p2] (u1) [dot]{};
					\vertex [right=2.5cm of p2] (ur){};
					\vertex [right=4cm of p2] (p3){};
					\vertex [right=1.0cm of p1] (b1) [dot]{};
					\vertex [right=1.0cm of b1] (b2) []{};
					\vertex [right=1.0cm of b2] (b3) [dot]{};
					\vertex [right=1.0cm of b3](p4){};
					\vertex [above=1.25cm of p1] (cutL);
					\vertex [below=1.25cm of ur] (m1);
					\vertex [right=4.0cm of cutL] (cutR){};
					\vertex [right=1.0cm of b1] (cut1);
					\vertex [above=0.3cm of cut1] (cut1u);
					\vertex [below=0.3cm of cut1] (cut1b){};
					\vertex [right=1cm of u1] (cut2);
					\vertex [above=0.2cm of cut2] (cut2u){};
					\diagram* {
					(p2) -- [thick] (u1) -- [thick] (p3),
					(b1)--[boson](u1), (b3)-- [boson] (u1), (p1) -- [thick] (b1)-- [thick] (b3)-- [thick] (p4), (cut1u)--[red,thick] (cut1b),(cutR)-- [boson] (b3),
					};
				\end{feynman}
			                          \end{tikzpicture}\,,
			 &  & \mathcal{I}_2=j_{11100}=\begin{tikzpicture}[scale=0.5, transform shape, baseline={([yshift=-0.4ex]current bounding box.center)}]\tikzstyle{every node}=[font=\small]
				                              \begin{feynman}
					\vertex (p1) {};
					\vertex [above=2.5cm of p1](p2){};
					\vertex [right=2cm of p2] (u1) [dot]{};
					\vertex [right=2.5cm of p2] (ur){};
					\vertex [right=4cm of p2] (p3){};
					\vertex [right=1.0cm of p1] (b1) [dot]{};
					\vertex [right=1.0cm of b1] (b2) []{};
					\vertex [right=1.0cm of b2] (b3) [dot]{};
					\vertex [right=1.0cm of b3](p4){};
					\vertex [above=1.25cm of p1] (cutL);
					\vertex [below=1.25cm of ur] (m1);
					\vertex [right=4.0cm of cutL] (cutR){};
					\vertex [right=1.0cm of b1] (cut1);
					\vertex [above=0.3cm of cut1] (cut1u);
					\vertex [below=0.3cm of cut1] (cut1b){};
					\vertex [right=1cm of u1] (cut2);
					\vertex [above=0.2cm of cut2] (cut2u){};
					\diagram* {
					(p2) -- [thick] (u1) -- [thick] (p3),
					(b1)--[boson](u1), (b3)-- [boson] (u1), (p1) -- [thick] (b1)-- [thick] (b3)-- [thick] (p4), (cut1u)--[red,thick] (cut1b),(cutR)-- [boson] (u1),
					};
				\end{feynman}
			                              \end{tikzpicture}\, ,              \\
			\cI_3\coloneqq j_{11011}= \begin{tikzpicture}[scale=0.5, transform shape, baseline={([yshift=-0.4ex]current bounding box.center)}]\tikzstyle{every node}=[font=\small]
				                          \begin{feynman}
					\vertex (p1) {};
					\vertex [above=2.5cm of p1](p2){};
					\vertex [right=1cm of p2] (u1) [dot]{};
					\vertex [right=3cm of p2] (u2) [dot]{};
					\vertex [right=3cm of u1] (p3){};
					\vertex [right=1.0cm of p1] (b1) [dot]{};
					\vertex [right=1.0cm of b1] (b2) []{};
					\vertex [right=1.0cm of b2] (b3) [dot]{};
					\vertex [right=1.0cm of b3](p4){};
					\vertex [above=1.25cm of p1] (cutL);
					\vertex [below=1.25cm of u2] (m1)[]{};
					\vertex [right=4.0cm of cutL] (cutR){};
					\vertex [right=1.0cm of b1] (cut1);
					\vertex [above=0.3cm of cut1] (cut1u);
					\vertex [below=0.3cm of cut1] (cut1b){};
					\vertex [right=1cm of u1] (cut2);
					\vertex [above=0.2cm of cut2] (cut2u){};
					\diagram* {
					(p2) -- [thick] (u1) -- [thick] (p3),
					(b1)--[boson](u1), (b3)-- [boson] (u2), (p1) -- [thick] (b1)-- [thick] (b3)-- [thick] (p4), (cut1u)--[red,thick] (cut1b),(cutR)-- [boson] (b3),
					};
				\end{feynman}
			                          \end{tikzpicture}\,,
			 &  & \mathcal{I}_4=j_{11101}= \begin{tikzpicture}[scale=0.5, transform shape, baseline={([yshift=-0.4ex]current bounding box.center)}]\tikzstyle{every node}=[font=\small]
				                               \begin{feynman}
					\vertex (p1) {};
					\vertex [above=2.5cm of p1](p2){};
					\vertex [right=1cm of p2] (u1) [dot]{};
					\vertex [right=3cm of p2] (u2) [dot]{};
					\vertex [right=3cm of u1] (p3){};
					\vertex [right=1.0cm of p1] (b1) [dot]{};
					\vertex [right=1.0cm of b1] (b2) []{};
					\vertex [right=1.0cm of b2] (b3) [dot]{};
					\vertex [right=1.0cm of b3](p4){};
					\vertex [above=1.25cm of p1] (cutL);
					\vertex [below=1.25cm of u2] (m1)[]{};
					\vertex [right=4.0cm of cutL] (cutR){};
					\vertex [right=1.0cm of b1] (cut1);
					\vertex [above=0.3cm of cut1] (cut1u);
					\vertex [below=0.3cm of cut1] (cut1b){};
					\vertex [right=1cm of u1] (cut2);
					\vertex [above=0.2cm of cut2] (cut2u){};
					\diagram* {
					(p2) -- [thick] (u1) -- [thick] (p3),
					(b1)--[boson](u1), (b3)-- [boson] (u2), (p1) -- [thick] (b1)-- [thick] (b3)-- [thick] (p4), (cut1u)--[red,thick] (cut1b),(cutR)-- [boson] (u2),
					};
				\end{feynman}
			                               \end{tikzpicture}\, ,             \\
			\cI_5\coloneqq j_{11110}= \begin{tikzpicture}[scale=0.5, transform shape, baseline={([yshift=-0.4ex]current bounding box.center)}]\tikzstyle{every node}=[font=\small]
				                          \begin{feynman}
					\vertex (p1) {};
					\vertex [above=2.5cm of p1](p2){};
					\vertex [right=2cm of p2] (u1) [dot]{};
					\vertex [right=2.5cm of p2] (ur){};
					\vertex [right=4cm of p2] (p3){};
					\vertex [right=1.0cm of p1] (b1) [dot]{};
					\vertex [right=1.0cm of b1] (b2) []{};
					\vertex [right=1.0cm of b2] (b3) [dot]{};
					\vertex [right=1.0cm of b3](p4){};
					\vertex [above=1.25cm of p1] (cutL);
					\vertex [below=1.25cm of ur] (m1)[dot]{};
					\vertex [right=4.0cm of cutL] (cutR){};
					\vertex [right=1.0cm of b1] (cut1);
					\vertex [above=0.3cm of cut1] (cut1u);
					\vertex [below=0.3cm of cut1] (cut1b){};
					\vertex [right=1cm of u1] (cut2);
					\vertex [above=0.2cm of cut2] (cut2u){};
					\diagram* {
					(p2) -- [thick] (u1) -- [thick] (p3),
					(b1)--[boson](u1), (b3)-- [boson] (u1), (p1) -- [thick] (b1)-- [thick] (b3)-- [thick] (p4), (cut1u)--[red,thick] (cut1b),(cutR)-- [boson] (m1),
					};
				\end{feynman}
			                          \end{tikzpicture}\,,
			 &  & \cI_6\coloneqq j_{11111}=
			\begin{tikzpicture}[scale=0.5, transform shape, baseline={([yshift=-0.4ex]current bounding box.center)}]\tikzstyle{every node}=[font=\small]
				\begin{feynman}
					\vertex (p1) {};
					\vertex [above=2.5cm of p1](p2){};
					\vertex [right=1cm of p2] (u1) [dot]{};
					\vertex [right=3cm of p2] (u2) [dot]{};
					\vertex [right=3cm of u1] (p3){};
					\vertex [right=1.0cm of p1] (b1) [dot]{};
					\vertex [right=1.0cm of b1] (b2) []{};
					\vertex [right=1.0cm of b2] (b3) [dot]{};
					\vertex [right=1.0cm of b3](p4){};
					\vertex [above=1.25cm of p1] (cutL);
					\vertex [below=1.25cm of u2] (m1)[dot]{};
					\vertex [right=4.0cm of cutL] (cutR){};
					\vertex [right=1.0cm of b1] (cut1);
					\vertex [above=0.3cm of cut1] (cut1u);
					\vertex [below=0.3cm of cut1] (cut1b){};
					\vertex [right=1cm of u1] (cut2);
					\vertex [above=0.2cm of cut2] (cut2u){};
					\diagram* {
					(p2) -- [thick] (u1) -- [thick] (p3),
					(b1)--[boson](u1), (b3)-- [boson] (m1)-- [boson] (u2), (p1) -- [thick] (b1)-- [thick] (b3)-- [thick] (p4), (cut1u)--[red,thick] (cut1b),(cutR)-- [boson] (m1),
					};
				\end{feynman}
			\end{tikzpicture}\,,
		\end{aligned}
	\end{split}
\end{equation}
where we dubbed with a tilde the integrals which have pinched the massless propagator $\mathcal{D}_1$ in \eqref{eq: masterGraphSimple}. These are new types of integrals that appear in the bremsstrahlung process, and their contribution is fundamental for the infrared behaviour of the amplitude, as will be discussed in Appendix~\ref{sec:Weinberg-resurrection}.

In general, these integrals  depend on the five kinematic variables $(\yb,\wb_1,\wb_2,-q_1^2,-q_2^2)$ introduced in \eqref{fiveiv}. The strategy to compute the integrals is the following:
\begin{itemize}
	\item[{\bf 1.}] The delta function and the principal value make most of the integrals finite. Then it is easy to evaluate the integrals by direct integration of Feynman parameters, except for $\widetilde{\mathcal{I}}_4$, $\mathcal{I}_5$ and $\mathcal{I}_6$.
	\item[{\bf 2.}] We consider the two remaining box MIs, $\widetilde{\mathcal{I}}_4$, $\mathcal{I}_5$, and their sub-topologies separately setting up the differential equations for a single kinematic variable for each of them ($\yb$ and $\wb_1$, respectively).
	\item[{\bf 3.}] We find the $\epsilon$-canonical form \cite{Henn:2013pwa} for each of these two linear systems of differential equations and solve them (details can be found in Appendix~\ref{sec:diffEq}).
	\item[{\bf 4.}] We then compute the asymptotic behaviour (boundary value) near a codimension-one surface in the space of the kinematic variables  
 ($\yb {\sim} 1$ and $\wb_1 {\sim} 0$, respectively), using the method of regions by \cite{Beneke:1997zp,Pak:2010pt,Jantzen:2012mw}.
	\item[{\bf 5.}] Finally, since we are interested in the scattering amplitude to order $\cO(\epsilon^0)$, we can write the pentagon $\mathcal{I}_6$ as a linear combination of the four boxes \cite{Bern:1993kr}, as we  show later in this section.
\end{itemize}

\subsection{The analytic form of the master integrals}

In the following, we present the  analytic expression of the MIs in dimensional regularisation up to  $\cO(\epsilon)$:%
\begin{equation}\label{eq: integralsList}
	\begin{split}
		\begin{aligned}
			\widetilde{\mathcal{I}}_1 & = - \frac{\wb_1}{8 \pi}\Big[1+ \left(i \pi + 2 - \gamma_E + \log\pi - \log {\wb_1^2 \over \mu_{\rm IR}^2 } \right) \tilde{\epsilon}\Big]\ ,                                        &
			\widetilde{\mathcal{I}}_2 & =-\frac{i\, \pi - 2\log\left(\sqrt{\yb^2-1}+\yb\right)}{16\pi \sqrt{\yb^2-1}} \ ,                                                                                \\
			\widetilde{\mathcal{I}}_3 & =-\frac{i}{16 \sqrt{\yb^2-1}}\ ,                                                                                                                           &
			\widetilde{\mathcal{I}}_4 & =\frac{1}{32 \pi \wb_1 \wb_2 \tilde{\epsilon} }+\frac{i\, \pi - \log {\wb_2^2 \over \mu_{\rm IR}^2}}{32 \pi \wb_1 \wb_2}\ ,                                                                 \\
			\mathcal{I}_1             & =-\frac{i\, \pi -2 \log \frac{\sqrt{-q_2^2+\wb_1^2}+\wb_1}{\sqrt{-q_2^2}}}{16 \pi  \sqrt{-q_2^2+\wb_1^2}}\ ,                                                   &
			\mathcal{I}_2             & =-\frac{i}{16 \sqrt{-q_1^2}}\ ,                                                                                                                            \\
			\mathcal{I}_3             & =\frac{ i\, \pi - 2\log\left(\sqrt{\yb^2-1}+\yb\right)}{16\pi (-q_2^2) \sqrt{\yb^2-1}}\ ,                                                                      &
			\mathcal{I}_4             & =\frac{i}{16 (-q_1^2) \sqrt{\yb^2-1}}\ ,                                                                                                                     \\
			\mathcal{I}_5             & =\frac{1}{32 \pi  \left(-q_1^2\right) \wb_1 \tilde{\epsilon} }+\frac{i\, \pi -2 \log \frac{q_1^2}{q_2^2} - \log {\wb_1^2 \over \mu_{\rm IR}^2}}{32 \pi  \left(-q_1^2\right) \wb_1}\ ,
		\end{aligned}
	\end{split}
\end{equation}
where, for convenience, we defined $\tilde{\epsilon} =  \epsilon\,e^{(\gamma_E - \log \pi)\epsilon}$, $\gamma_E$ is the Euler-Mascheroni constant
and $\mu_{\rm IR}$ is the infrared scale introduced in dimensional regularisation. We have expanded the $\widetilde{\mathcal{I}}_1$ integral up to and including $\cO(\eps)$ since its coefficient arising from IBP reduction is infrared divergent.
The final integral to be computed is the pentagon $\mathcal{I}_6$, which can be decomposed in terms of the boxes $\mathcal{I}_{3,4,5}$ and $\widetilde{\mathcal{I}}_4$ \cite{vanNeerven:1983vr,Ellis:2011cr}:
\begin{equation}
\label{eq:pentagon_dec}
	\begin{split}
		\mathcal{I}_6 &= \frac{- 2 q_2^2 \wb_1 \wb_2 \yb+2 q_1^2 \wb_1^2-q_2^4 \yb^2+q_1^2 q_2^2 \yb^2+q_2^4-q_1^2 q_2^2}{2 \left(-2 q_1^2 q_2^2 \wb_2 \wb_1 \yb+q_1^4 \wb_1^2+q_2^4 \wb_2^2\right)} \mathcal{I}_3\\
		&+\frac{-2 q_1^2 \wb_1 \wb_2 \yb+2 q_2^2 \wb_2^2-q_1^4 \yb^2+q_1^2 q_2^2 \yb^2+q_1^4-q_1^2 q_2^2}{2 \left(-2 q_1^2 q_2^2 \wb_2 \wb_1 \yb+q_1^4 \wb_1^2+q_2^4 \wb_2^2\right)}\mathcal{I}_4\\
		&+\frac{-q_1^2 \wb_2 \wb_1 \yb-q_2^2 \wb_2 \wb_1 \yb+q_1^2 \wb_1^2+q_2^2 \wb_2^2-2 \wb_2^2 \wb_1^2}{-2 q_1^2 q_2^2 \wb_2 \wb_1 \yb+q_1^4 \wb_1^2+q_2^4 \wb_2^2}\widetilde{\mathcal{I}}_4\\
		&+\frac{-q_1^4 \wb_1 \yb+q_1^2 q_2^2 \wb_1 \yb-2 q_1^2
			\wb_1^2 \wb_2+q_1^2 q_2^2 \wb_2-q_2^4 \wb_2}{-2 q_1^2 q_2^2 \wb_2 \wb_1 \yb+q_1^4 \wb_1^2+q_2^4 \wb_2^2} \mathcal{I}_5\ .
	\end{split}
\end{equation}
The infrared-divergent part of $\mathcal{I}_6$ and its imaginary part take a particularly simple form:
\begin{align}
	\mathcal{I}_6            & = -\frac{1}{32 \pi (-q_1^2) \wb_1 \wb_2 \tilde{\epsilon}} + \cO(\tilde{\epsilon}^{\, 0})\ ,              \\
	{\rm Im}\, \mathcal{I}_6 & = -\frac{1}{32 (-q_1^2) \wb_1 \wb_2} -\frac{1}{16\, q_1^2 q_2^2 \sqrt{\yb^2-1}} + \cO(\tilde{\epsilon})\ .
\end{align}

%\subsection{Feynman integrals -- differential equations %and region expansion}

% The analytic expression of the bubble and triangle integrals can be easily obtained by direct Feynman parameter integration:

% \SDA{The following two functions are to be checked. I am not sure about $i\pi$ terms and overall $i$ factors.}

% \begin{equation}
%     \begin{split}
%         j_{01101} &= \frac{(-1)^{\frac{d-6}{2}} i\, \Gamma\left[3-\frac{d}{2}\right]}{16 M_1 M_2 (4\pi)^{d/2}} \int_{-\infty}^{+\infty}\! d x_2 \int_0^{+\infty} \!d x_3 d x_5\, \delta \left(1-\sum_l x_l\right)\, x_3^{3-d} \left(w_2 x_3 x_5+\frac{y}{2} x_5 x_2+\frac{x_2^2}{4}+\frac{x_5^2}{4}\right)^{\frac{d-6}{2}}\ ,\\
%         & = \frac{1}{128 \pi  M_1 M_2 \sqrt{y^2-1} \epsilon }+\frac{\log \frac{4 \pi  \left(y^2-1\right)}{w_2^2}\SDA{-4 i \pi} -\gamma_E }{128 \pi M_1 M_2 \sqrt{y^2-1}}+\cO\left(\epsilon\right)\ .
%     \end{split}
% \end{equation}

% \begin{equation}
%     \begin{split}
%         j_{01011} & = \frac{i \, \Gamma\left[3-\frac{d}{2}\right]}{16 M_1 M_2 (4\pi)^{d/2}} \int_{-\infty}^{+\infty}\! d x_2 \int_0^{+\infty} \!d x_4 d x_5\,\delta \left(1-\sum_l x_l\right)\, x_4^{3-d} \left(w_1 x_4 x_2-\frac{y}{2} x_5 x_2-\frac{x_2^2}{4}-\frac{x_5^2}{4} \right)^{\frac{d-6}{2}} \\
%         & = \frac{1}{128 \pi  M_1 M_2 \sqrt{y^2-1} \epsilon }+\frac{\log \frac{4 \pi  \left(y^2-1\right)}{w_1^2}-2 i \arcsin(y)-3 i \pi -\gamma }{128 \pi M_1 M_2 \sqrt{y^2-1}}+\cO\left(\epsilon\right)\ .
%     \end{split}
% \end{equation}

\section{Final result after integration and checks}
\label{sec:finalresult}

Our final result before integration was written in \eqref{finalresultform1} and \eqref{finalresultform2}. The  expression for the one-loop amplitude in
\eqref{finalresultform1}
is expanded in our basis of master integrals, with the
ten coefficients $c_i$ and $\tilde{c}_i$ potentially  having spurious Gram determinant singularities at $\epsilon_{\mu \nu \rho \sigma} \vb_{1}^\mu \vb_2^\nu q_1^\rho q_2^\sigma = 0$.
Instead, we chose to present the result in terms of the functions that appear in the integrals, which makes the analytic structures more transparent:
\begin{equation}
	\label{eq:OneLoopAmplitude}
	\begin{split}
		\cM_{5,\bar{m}_1^3 \bar{m}_2^2}^{(1)} &= \frac{\mathfrak{d}_{\rm IR}}{\epsilon} + \mathfrak{R}+ i\pi\, \mathfrak{i}_1 + \frac{i \pi}{\sqrt{\yb^2-1}} \mathfrak{i}_2 + c_{1,0}\, \cI_1 + c_{2,0}\, \cI_2\\
		&+\mathfrak{l}_{\wb_1} \log \frac{\wb_1^2}{\mu_{\rm IR}^2} +\mathfrak{l}_{\wb_2} \log \frac{\wb_2^2}{\mu_{\rm IR}^2}+ \mathfrak{l}_{q} \log \frac{q_1^2}{q_2^2} + \mathfrak{l}_{\yb} \frac{\log\left(\sqrt{\yb^2-1}+\yb\right)}{\sqrt{\yb^2-1}} + \mathcal{O}(\epsilon^0 )\ ,
	\end{split}
\end{equation}
where the coefficients $\mathfrak{d}_{\rm IR}$, $\mathfrak{R}$, $\mathfrak{i}_i$, $\mathfrak{l}_i$ are rational functions of the kinematics variables and are homogeneous in the linearised field strength $F_k^{\mu \nu}$, and $\mu_{\rm IR}$ is an infrared scale. In particular, each of the coefficients is a linear combination of the various $c_i$s and $\tilde{c}_i$s and we notice these combinations are always manifestly free of spurious Gram determinant singularities. Here we are considering the amplitude in dimensional regularisation up to terms $\mathcal{O}(\epsilon^0 )$ and the second subscript of the coefficients specify the power of $\epsilon$ in their  Laurent expansion.

A first non-trivial check of our amplitude is to confirm  that the  coefficients of:
\begin{itemize}
	\item[{\bf 1.}] the infrared divergences $\mathfrak{d}_{\rm IR}$,
	\item[{\bf 2.}] the logarithms of the infrared scale $-(\mathfrak{l}_{\wb_1}+\mathfrak{l}_{\wb_2})$,
	\item[{\bf 3.}] the associated imaginary part $\mathfrak{i}_{1}$
\end{itemize}
are all the same, as dictated by unitarity and the Callan-Symanzik equation (for a recent discussion, see \cite{Caron-Huot:2016cwu}):
\begin{equation}
	\begin{split}
		\mathfrak{d}_{\rm IR} = \mathfrak{i}_{1} = -\mathfrak{l}_{\wb_1}-\mathfrak{l}_{\wb_2} &= \frac{8 q_1^2 \wb_2 \wb_1^2 \tilde{c}_{1,-1}+q_1^2 \tilde{c}_{4,0}+2 \wb_2 c_{5,0}+c_{6,0}}{128 \pi  \left(-q_1^2\right) \wb_1 \wb_2 } \\
		&=- \frac{i \kappa^2}{32 \pi} \bar{m}_1 \wb_1 \cM_{5,\bar{m}_1^2 \bar{m}_2^2}^{(0)}\ .
	\end{split}
\end{equation}
The fact that the infrared-divergent part is proportional to the tree-level amplitude can be also seen as a direct consequence of Weinberg's exponentiation of infrared divergences, which is reviewed in Appendix~\ref{sec:Weinberg-resurrection}:
\begin{equation}
\label{eq:classicalWeinberg}
	\left.  \cM_{2\to 3} \right|_{\text{IR-div}}  =  e^{i\frac{G}{\epsilon}(-\bar{m}_1 \bar{m}_2 \frac{(2\yb^2-1)}{\sqrt{\yb^2-1}} - \bar{m}_1 \wb_1 - \bar{m}_2 \wb_2)} \left. \cM_{2\to 3} \right|_{\rm finite}\ .
\end{equation}
The second imaginary contribution $\mathfrak{i}_{2}$ is also proportional to the tree-level amplitude:
\begin{equation}
	\begin{split}
		\label{7.4}
		\mathfrak{i}_{2}&=\frac{q_1^2 q_2^2 \left(\tilde{c}_{2,0}+\tilde{c}_{3,0}\right)+q_1^2 c_{3,0}+q_2^2 c_{4,0}+c_{6,0}}{64 q_1^2 q_2^2}\\
		& = \frac{i \kappa^2}{64 \pi} \frac{\, \bar{m}_1 \wb_1 \yb\, (2 \yb^2-3)}{(\yb^2 -1)} \cM_{5,\bar{m}_1^2 \bar{m}_2^2}^{(0)}\ ,
	\end{split}
\end{equation}
but is not accompanied by corresponding infrared-divergent or scale dependent terms.
To understand the difference compared to the previous case, we need to take into account the $\bar{m}$ expansion. Indeed, we have shown in Section~\ref{sec: HEFTexpansion} that this expansion implies the use of the principal value prescription for the linear propagators, which in turn has the effect to make the otherwise divergent integrals $\widetilde{\mathcal{I}}_2$, $\widetilde{\mathcal{I}}_3$, $\mathcal{I}_3$ and $\mathcal{I}_4$ finite, without altering their imaginary part. Then,  $\mathfrak{i}_{2}$ in \eqref{7.4} should be thought of as the imaginary part corresponding to  infrared divergent/scale dependent contributions that have been moved to iteration terms.
This can be checked explicitly by expanding Weinberg's soft phase in terms of the  $p_i$ variables, instead of $\pb_i$, to order $m_1^1 m_2^0$ and $m_1^0 m_2^1$:
\begin{equation}
    \left.  \cM_{2\to 3} \right|_{\text{IR-div}}  =  e^{i\frac{G}{\epsilon}\left[-m_1 m_2 \frac{(2y^2-1)}{\sqrt{y^2-1}} - y\frac{2y^2-3}{(y^2-1)^{3/2}} (m_1 w_1 + m_2 w_2) - m_1 w_1 - m_2 w_2\right]} \left. \cM_{2\to 3} \right|_{\rm finite}\, .
\end{equation}
Moreover, the amplitude \eqref{eq:OneLoopAmplitude} has spurious singularities in the physical region which have to cancel when we combine different logarithmic contributions near the poles. First of all, let us emphasise that the argument of the logarithm and the square roots appearing in $\mathcal{I}_1$ and $\mathcal{I}_2$ are positive in the physical region, as discussed in Section~\ref{sec:kinematics}.

At leading order in the soft limit, \textit{i.e.}~$\cO(\omega^{-1})$, only the two-massless triangles contribute,  and  they reproduce exactly Weinberg's soft factor in the HEFT \eqref{eq:softHEFT}:
\begin{equation}
\label{eq:leadingWeinberg}
	c_1 \mathcal{I}_1 + c_2 \mathcal{I}_2 =
	\frac{\kappa}{2}
	\frac{\vb_1 \Cdot F_k \Cdot \vb_2}{\wb_1\, \wb_2} \left( \frac{\vb_1 \Cdot F_k \Cdot q}{\wb_1} + \frac{\vb_2 \Cdot F_k \Cdot q}{\wb_2} \right) \mathcal{M}_{4, \bar{m}_1^3 \bar{m}_2^2}^{(1)} + \cO(\omega^0)\ ,
\end{equation}
$q \simeq  q_1 \simeq -q_2$ 
as in Section~\ref{subsec:Weinberg} and $\mathcal{M}_{4, \bar{m}_1^3 \bar{m}_2^2}^{(1)}$ is the classical four-point one-loop amplitude:
\begin{equation}
	\mathcal{M}_{4, \bar{m}_1^3 \bar{m}_2^2}^{(1)} = i\, G^2\, \bar{m}_1^3 \bar{m}_2^2\,  \frac{6 \pi^2 (5 \yb^2 -1)}{\sqrt{-q^2}}.
\end{equation}
Finally, we notice that:
\begin{itemize}
	\item[{\bf 1.}] $\mathfrak{l}_{\wb_2}$ and $\mathfrak{l}_{\yb}$ have a double pole at  $\hat{\yb}_1=\frac{\wb_1^2+\wb_2^2}{2 \wb_1 \wb_2} \geq 1$\ ,

	\item[{\bf 2.}] $\mathfrak{l}_{\wb_2}$, $\mathfrak{l}_{q}$ and $\mathfrak{l}_{\yb}$ have a double pole at $\hat{\yb}_2=\frac{(q_1^2 \wb_1)^2+(q_2^2 \wb_2)^2}{2 q_1^2 q_2^2 \wb_1 \wb_2}\geq 1$\ ,

	\item[{\bf 3.}] $c_1$, $c_2$ and $\mathfrak{l}_q$ have a pole of order four at%
		\footnote{The combination $- 4 q_1^2 \wb_1^2- \left(q_1^2-q_2^2\right)^2$ is non-negative. Indeed, if we choose the frame where $\vb_1=(1,0,0,0)$ and $k=\omega (1,0,0,1)$, then this becomes $4 \,\omega^2 (q_{1,x}^2 + q_{1,y}^2)$. This means that the poles sit in the physical configuration for which $\vec{k}$, $\vec{q}_1$ and $\vec{q}_2$ are taken to be aligned (modulo boosts),  which is expected to be smooth. 
  Likewise, one can show that $\hat{y}_1$ corresponds to $\vec{k}$ being orthogonal to $\vec{q}_2$ in the rest frame of particle 1.}
		$\hat{\wb}_1=\frac{\left|q_1^2-q_2^2\right|}{2\sqrt{-q_1^2}}$,

	\item[{\bf 4.}] the rational terms $\mathfrak{R}$ have single poles in both $\hat{\yb}_1$ and $\hat{\yb}_2$ and a pole of order three in $\hat{\wb}_1$.
\end{itemize}
For example,  near $\yb\simeq \hat{\yb}_1$, the logarithms simplify as
\begin{equation}
	\frac{\log\left(\sqrt{\yb^2-1}+\yb\right)}{\sqrt{\yb^2-1}}= \frac{2 \wb_1 \wb_2 \log \frac{\wb_1}{\wb_2}}{\wb_1^2-\wb_2^2}+4 \wb_1^2 \wb_2^2 \frac{\wb_1^2-\wb_2^2+\left(\wb_1^2+\wb_2^2\right) \log \frac{\wb_2}{\wb_1}}{\left(\wb_1^2-\wb_2^2\right){}^3}\left(\yb-\hat{\yb}_1\right)+\cdots\ ,
\end{equation}
and one can show that the double poles and the logarithms cancel when we combine $\mathfrak{l}_{\wb_2}$ and $\mathfrak{l}_{\yb}$. Nevertheless, the leftover term has still a simple pole in $\yb\simeq \hat{\yb}_1$. This  pole is only cancelled once we take into account the rational contribution $\mathfrak{R}$, which is needed to restore locality and cancel spurious poles \cite{Bern:2007dw,Huang:2013vha}. The spurious singularities in $\yb\simeq \hat{\yb}_2$ and $\wb_1\simeq \hat{\wb}_1$ share the same fate, even though showing it explicitly is more involved because the terms to be combined are more complicated.

As a final check, the present authors and those of \cite{Herderschee:2023fxh} have performed independent comparisons of their two results for the one-loop amplitude, finding perfect agreement.

\section{Waveforms from the HEFT}\label{sec: waveforms}

\subsection{Blitz review of the KMOC approach}

In this section we review the connection between waveforms and amplitudes following the KMOC approach \cite{Kosower:2018adc,Cristofoli:2021vyo,Cristofoli:2021jas}. The two heavy objects are taken to be in an initial state represented as 
\begin{align}
	\label{psi}
	|\psi\rangle_{\rm in} \coloneqq \int\!d\Phi(p_1) d\Phi(p_2) e^{i (p_1 \Cdot b_1 + p_2 \Cdot b_2)} \phi(p_1) \phi(p_2) |p_1  p_2\rangle_{\rm in}
	\ ,
\end{align}
where the wavefunctions $\phi(p_1)$ and $\phi(p_2)$ are peaked around the classical values of the momenta.
We use the same conventions as \cite{Cristofoli:2021vyo}, 
\begin{align}
\label{dphik}
	\begin{split}
		d\Phi(p) & \coloneqq \frac{d^Dp}{(2\pi)^{D{-}1} }\delta^{(+)} (p^2 - m^2)\, , \qquad 
		|p\rangle  \coloneqq a^\dagger (\vec{p})  |0\rangle\, , \end{split}
\end{align}
with $[a (\vec{p}) , a^\dagger (\vec{p}^{\, \prime})] = (2\pi)^{D{-}1} (2 E_p) \delta^{(D{-}1)}(\vec{p} - \vec{p}^{\, \prime})$, 
for the massive objects. Similarly, for gravitons we choose
\begin{align}
	\begin{split}
		d\Phi(k)  \coloneqq \frac{d^Dk}{(2\pi)^{D{-}1} }\delta^{(+)} (k^2)\, , \qquad 
		|k^h\rangle \coloneqq a^\dagger_h (\vec{k})  |0\rangle\, ,
  \end{split}
\end{align}
with $\big[a_h (\vec{k}) , a^\dagger_{h^\prime} (\vec{k}^{\, \prime})\big] = (2\pi)^{D{-}1} (2 E_k) \delta^{(D{-}1)}(\vec{k} - \vec{k}^{\, \prime})\delta_{h h^\prime}$,
	where $h$ denotes the helicity.
 
The waveform we are interested in is related to the expectation value of the Riemann tensor \cite{Kosower:2018adc,Cristofoli:2021vyo,Cristofoli:2021jas}, 
\begin{align}
\begin{split}
\label{KMOCW}
	\langle R^{\rm out}_{\mu \nu \rho \lambda} (x) \rangle_{\rm \psi}  & \coloneqq \mbox{}_{\rm out}\langle \psi | \mathbb{R}_{\mu \nu \rho \lambda} (x)  | \psi \rangle_{\rm out}\, = \mbox{}_{\rm in}\langle \psi | S^\dagger \mathbb{R}_{\mu \nu \rho \lambda} (x) S | \psi\rangle_{\rm in}\\
		& = \mbox{}_{\rm in}\langle \psi |  \mathbb{R}_{\mu \nu \rho \lambda} (x)  | \psi\rangle_{\rm in} + 2 {\rm Re} \ i \,
		\mbox{}_{\rm in}\langle \psi |  \mathbb{R}_{\mu \nu \rho \lambda} (x) \, T  | \psi\rangle_{\rm in} +
		\mbox{}_{\rm in}\langle \psi | T^\dagger \mathbb{R}_{\mu \nu \rho \lambda} (x) T | \psi\rangle_{\rm in}\, ,
	\end{split}
\end{align}
where  $S=\mathbb{1} +iT$ and $\mathbb{R}_{\mu \nu \rho \lambda} (x)$ is the Riemann tensor, evaluated at the position $x$ of the observer, in the far future of the event. 
Expanding $\mathbb{R}_{\mu \nu \rho \lambda} (x)$ as 
\begin{align}
\label{erreR}
	\mathbb{R}_{\mu \nu \rho \lambda} (x)  =  \frac{\kappa}{2} \, \sum_h \int\!d\Phi(k) \,
	\Big[
	a_h (\vec{k}) e^{-i k\Cdot x} k_{[\mu}\varepsilon^{(h)\ast}_{\nu]} (\vec{k}) k_{[\rho}\varepsilon^{(h)\ast }_{\lambda]} (\vec{k}) \ + \ {\rm h.c.}
	\Big] \, ,
\end{align}
one finds 
that $\mbox{}_{\rm in}\langle \psi |  \mathbb{R}_{\mu \nu \rho \lambda} (x)  | \psi\rangle_{\rm in}=0$,
  and  \cite{Kosower:2018adc,Cristofoli:2021vyo,Cristofoli:2021jas}
\begin{align}\begin{split}
		&\mbox{}_{\rm in}\langle \psi |  \mathbb{R}_{\mu \nu \rho \lambda} (x) \, T  | \psi\rangle_{\rm in} \\
		& = \frac{\kappa}{2}\sum_h\int\!d\Phi (p_1) d\Phi (p_2) d\Phi (p_1^\prime) d\Phi (p_2^\prime) d\Phi (k) \, \phi^\ast (p_1^\prime) \phi^\ast (p_2^\prime) \phi(p_1) \phi (p_2) \\
		& e^{-i k \Cdot x + i (p_1 - p_1^\prime)\Cdot b_1 + i (p_2 - p_2^\prime)\Cdot b_2 }\,
		k_{[\mu}\varepsilon^{(h)\ast}_{\nu]} (\vec{k}) k_{[\rho}\varepsilon^{(h)\ast}_{\lambda]} (\vec{k}) \,
		\langle p_1^\prime p_2^\prime k^h  |  T | p_1 p_2\rangle \\
		& =  \frac{\kappa}{2} \sum_h \int\!d\Phi (k) \,  e^{- i k\Cdot x}k_{[\mu}\varepsilon^{(h)\ast}_{\nu]} (\vec{k}) k_{[\rho}\varepsilon^{(h)\ast}_{\lambda]} (\vec{k}) \,\int\! d\Phi (p_1) d\Phi (p_2)  \phi (p_1) \phi (p_2)  \\
		& \int\!\frac{d^D q_1}{(2\pi)^{D{-}1}} \frac{d^D q_2}{(2\pi)^{D{-}1}} \delta( 2\bar{p}_1\Cdot q_1 ) \delta( 2\bar{p}_2\Cdot q_2 ) e^{i(q_1\Cdot b_1 + q_2 \Cdot b_2)}
		\langle p_1^\prime p_2^\prime k^h  |  T | p_1 p_2\rangle \\
		& \phi^\ast (p_1 {-}  q_1)\phi^\ast (p_2 {-}  q_2)\, ,
	\end{split}
\end{align}
where we introduced barred variables
in the delta functions as in \eqref{barredv}. In the last equality we have also changed integration variables from $(p_1^\prime, p_2^\prime){\to} (q_1,  q_2)$.
Approximating $\phi (p_i {-} q_i) {\to} \phi (p_i)$, with $i=1,2$, we obtain
\begin{align}\begin{split}
		&\mbox{}_{\rm in}\langle \psi |  \mathbb{R}_{\mu \nu \rho \lambda} (x) \, T  | \psi\rangle_{\rm in}
		 =  \frac{\kappa}{2} \int\!\prod_{j=1}^2 d\Phi (p_j)  \, |\phi(p_1)|^2 |\phi (p_2)|^2  \\ &
		\sum_h \int\!d\Phi(k) e^{-i k \Cdot x} \,
		k_{[\mu}\varepsilon^{(h)\ast}_{\nu]} (\vec{k}) k_{[\rho}\varepsilon^{(h)\ast}_{\lambda]} (\vec{k}) \
  \\ & 
		\int\!\frac{d^D q_1}{(2\pi)^{D{-}1}} \frac{d^D q_2}{(2\pi)^{D{-}1}}  \delta( 2\bar{p}_1\Cdot q_1 ) \delta( 2\bar{p}_2\Cdot q_2 ) e^{i(q_1\Cdot b_1 + q_2 \Cdot b_2)}
		\langle p_1^\prime p_2^\prime k^h  |   T | p_1 p_2\rangle 
		\, .
	\end{split}
\end{align}
Next we consider the term $\mbox{}_{\rm in}\langle \psi | T^\dagger \mathbb{R}_{\mu \nu \rho \lambda} (x) T | \psi\rangle_{\rm in}$ in \eqref{KMOCW}. It can  be rewritten    in a similar form to the previous one noting that
\begin{align}
	\mbox{}_{\rm in}\langle \psi | T^\dagger \mathbb{R}_{\mu \nu \rho \lambda} (x) T | \psi\rangle_{\rm in} = \kappa\,  {\rm Re} \,
	\sum_h \int\!d\Phi(k) e^{-i k \Cdot x} \,
	k_{[\mu}\varepsilon^{(h)\ast}_{\nu]} (\vec{k}) k_{[\rho}\varepsilon^{(h)\ast}_{\lambda]} (\vec{k}) \
	\mbox{}_{\rm in}\langle \psi | T^\dagger a_h (\vec{k})  T | \psi\rangle_{\rm in}   \, .
\end{align}
Following identical manipulations as before and combining the two non-vanishing  contributions from \eqref{KMOCW}, we arrive at the following expression for the expectation value of the Riemann tensor:  
\begin{align}
	\label{KMOCsub}
	\begin{split}
		&\langle R^{\rm out}_{\mu \nu \rho \lambda} (x) \rangle_{\rm \psi}  \!=\!    \kappa \,  {\rm Re} \, \Big\{ 
		i  \int\!\prod_{j=1}^2 d\Phi (p_j)   \, |\phi(p_1)|^2 |\phi (p_2)|^2
		%\\ & 
		\sum_h\!\int\!d\Phi(k) e^{-i k \Cdot x} \,
		k_{[\mu}\varepsilon^{(h)\ast}_{\nu]} (\vec{k}) k_{[\rho}\varepsilon^{(h)\ast}_{\lambda]} (\vec{k})
		\\ &
		\int\!\prod_{j=1}^2\frac{d^D q_j}{(2\pi)^{D{-}1}} 
		%(2\pi)^D \delta^{(D)}  (q_1 + q_2 - k) 
		\delta( 2\bar{p}_1\Cdot q_1 ) \delta( 2\bar{p}_2\Cdot q_2 ) e^{i(q_1\Cdot b_1 + q_2 \Cdot b_2)}%\\ &
		\Big[
			\langle p_1^\prime p_2^\prime k^h  |  T | p_1 p_2\rangle  {-}  i \, \langle p_1^\prime p_2^\prime   |  T^\dagger a_h (\vec{k})  T| p_1 p_2\rangle
			\Big] \Big\}
		\, .
	\end{split}
\end{align}
Of course one can also follow the same procedure for $h_{\mu\nu}(x)$. With the free-field expansion
\begin{align}
    \mathbb{h}_{\mu \nu} (x)  =  \kappa \, \sum_h \int\!d\Phi(k) \,
	\Big[
	a_h (\vec{k}) \varepsilon^{(h)\ast}_{\mu} (\vec{k}) \varepsilon^{(h)\ast}_{\nu}  (\vec{k}) e^{-i k\Cdot x}   \ + \ {\rm h.c.}
	\Big]\, , 
\end{align}
one quickly arrives  at%
\footnote{In our conventions the linearised Riemann tensor is 
$R_{\mu \nu \rho \lambda} {=} \frac{1}{2} \big( \partial_\rho \partial_\nu  h_{\mu \lambda} {+} \partial_\lambda \partial_\mu  h_{\nu \rho} {-} 
\partial_\lambda \partial_\nu  h_{\mu \rho} {-} \partial_\rho \partial_\mu  h_{\nu \lambda}
\big)$.}
\begin{align}
	\label{KMOCsubhmunu}
	\begin{split}
		&\langle h^{\rm out}_{\mu \nu } (x) \rangle_{\rm \psi}  \!=\!    2\kappa \,  {\rm Re} \, \Big\{ 
		i \int\!\prod_{j=1}^2 d\Phi (p_j)   \, |\phi(p_1)|^2 |\phi (p_2)|^2
		%\\ & 
		\sum_h \int\!d\Phi(k) e^{-i k \Cdot x} \,
		\varepsilon^{(h)\ast}_{\mu} (\vec{k}) \varepsilon^{(h)\ast}_{\nu} (\vec{k})
		\\ &
		\int\!\prod_{j=1}^2\frac{d^D q_j}{(2\pi)^{D{-}1}} 
		%(2\pi)^D \delta^{(D)}  (q_1 + q_2 - k) 
		\delta( 2\bar{p}_1\Cdot q_1 ) \delta( 2\bar{p}_2\Cdot q_2 ) e^{i(q_1\Cdot b_1 + q_2 \Cdot b_2)}%\\ &
		\Big[
			\langle p_1^\prime p_2^\prime k^h  |  T | p_1 p_2\rangle  {-}  i \, \langle p_1^\prime p_2^\prime   |  T^\dagger a_h (\vec{k})  T| p_1 p_2\rangle
			\Big] \Big\}
		\, .
	\end{split}
\end{align}
Note that   $\langle R^{\rm out}_{\mu \nu \rho \lambda} (x) \rangle_{\rm \psi} $ or  $\langle h^{\rm out}_{\mu \nu} (x) \rangle_{\rm \psi} $ are effectively computed using 
\begin{align}
\begin{split}
\langle \psi | S^\dagger a_h(\vec{k}) S |\psi\rangle 
&= \langle \psi | i a_h(\vec{k}) T + T^\dagger a_h(\vec{k}) T|\psi\rangle\, , \\ 
\langle \psi |S^\dagger a_h^\dagger (\vec{k}) S |\psi \rangle&= \langle \psi |  - i T^\dagger a_h^\dagger(\vec{k}) + T^\dagger a_h^\dagger (\vec{k})T|\psi\rangle\, ,  
   \end{split}
    \end{align}
where from now on we will drop the subscript ``in'' in the state $|\psi\rangle$.

\subsection{From KMOC to  HEFT}
\label{sec:1lWFHEFT}

\subsubsection{General structure}

Our  next task is  to compute the quantity that appears in \eqref{KMOCsub},
\begin{align}
	\label{KMOCsub2}
	\langle p_1^\prime p_2^\prime k^h  |  T | p_1 p_2\rangle  -  i \langle p_1^\prime p_2^\prime   |  T^\dagger a_h (\vec{k})  T| p_1 p_2\rangle \, ,
\end{align}
 up to one loop in the PM expansion. 
The first term is the complete amplitude, where we note  that in our conventions 
\begin{align}
\label{defampli}
	\langle p_1^\prime p_2^\prime k^h  | i\,  T | p_1 p_2\rangle  \coloneqq (2\pi )^D\delta^{(D)}  (q_1 + q_2 - k) M_5(q_1, q_2; h) \ , 
\end{align}
where we have also indicated the dependence on the polarisation $h$ of the graviton. To relate the second term in \eqref{KMOCsub2} to scattering amplitudes we insert a complete set of states 
\begin{align}
\label{nonamesofar}
	\langle p_1^\prime p_2^\prime | T^\dagger a_h(\vec{k})  T | p_1 p_2\rangle 
 %= \sum_n \langle p_1^\prime p_2^\prime | T^\dagger|n\rangle \langle n |  a_h(\vec{k})  T | p_1 p_2\rangle
 % =\sum_n \langle p_1^\prime p_2^\prime | T^\dagger|n\rangle \langle n, k^h |    T | p_1 p_2\rangle\,
 =\sum_{r_1,r_2,s} \langle p_1^\prime p_2^\prime | T^\dagger|r_1 r_2,s\rangle \langle r_1 r_2,s, k^h |    T | p_1 p_2\rangle\,.
\end{align}
where $r_i$ are intermediate scalars and $s$ can be any number of intermediate gravitons.
We now examine the two contributions \eqref{defampli} and \eqref{nonamesofar} in perturbation theory up to and including next-to-leading order identifying the possible amplitudes involved and explaining how hyper-classical contributions are removed.

\textbf{Tree level:}
At leading order, $\cO(\kappa^3)$,  the first term in \eqref{KMOCsub2} after a heavy-mass expansion is 
\begin{equation}
    \langle p_1^\prime p_2^\prime k^h  | i\,  T | p_1 p_2\rangle^{(0)}  \coloneqq (2\pi )^D\delta^{(D)}  (q_1 + q_2 - k) \cM_5^{(0)}(q_1, q_2; h) +\cdots
\end{equation}
where the superscript denotes the loop order,  $\cM_5^{(0)}$ is the classical five-point amplitude we computed in \eqref{eq: 5pttree}, and $+\cdots$ are the hyper-classical terms appearing in \eqref{eq:5pointHEFTExpansion}.
At leading order, for the second term in \eqref{KMOCsub2}  the intermediate state  is just  $|n\rangle = |r_1  r_2\rangle$, and the only piece which can contribute involves a disconnected five-point amplitude, 
\begin{align}
	\label{firstintst-treelevel}
	\sum_{r_1, r_2} \langle p_1^\prime p_2^\prime | T^\dagger|r_1 r_2\rangle^{(0)} \langle r_1 r_2 k^h |    T | p_1 p_2\rangle^{\rm (Disc)}\, .
\end{align}
This contribution is proportional to an on-shell three-point amplitude with an emitted graviton and so gives a static contribution containing $\delta(p_i\Cdot k){\sim}\delta(\omega)$, where $\omega$ is the frequency of the emitted graviton. This gives an overall time-independent shift in $ \langle h^{\rm out}_{\mu \nu } (x)\rangle_{\rm \psi}$.  Such a time-independent shift does not contribute to $\langle R^{\rm out}_{\mu \nu \rho \lambda} (x) \rangle_{\rm \psi}$ and can be fixed by requiring, for example, that $ \langle h^{\rm out}_{\mu \nu } (x)\rangle_{\rm \psi}$ is zero in the far past. For this reason we will neglect static contributions like these at all times. Note, however, that these static contributions also contain a hyper-classical term after expanding in the heavy-mass limit, which pleasingly cancels an identical hyper-classical term in the full amplitude, described in 
\eqref{eq:5pointHEFTExpansion}. 
% By counting powers of the masses, we see that this contribution is hyper-classical and  simply serves to cancel an identical hyper-classical term in the full amplitude, described in \eqref{eq:5pointHEFTExpansion}. It is for this reason that we do not compute such terms. 

% Furthermore, \eqref{firstintst-treelevel} is proportional to an entirely on-shell three-point amplitude with an emitted graviton and gives a static contribution proportional to $\delta(\omega)$, where $\omega$ is the frequency of the emitted graviton. This gives an overall constant-in-time shift of the tree-level waveform, which we ignore. Such an overall shift can be fixed by requiring that the waveform has no radiation in the distant past and furthermore does not contribute to $\langle R^{\rm out}_{\mu \nu \rho \lambda} (x) \rangle_{\rm \psi}$.
% %on which we focus. 

\textbf{One loop:}
At one loop, $\cO(\kappa^5)$, the first type of classical contribution is
\begin{align}
\label{defampliHEFT}
	\langle p_1^\prime p_2^\prime k^h  | i\,  T | p_1 p_2\rangle^{(1)}  = (2\pi )^D\delta^{(D)}  (q_1 + q_2 - k) \cM_{5, \rm HEFT}^{(1)}(q_1, q_2; h) + \cdots \ , 
\end{align}
where the dots represent two types of terms:  hyper-classical terms which  will cancel against the hyper-classical contribution from the second KMOC term in \eqref{KMOCsub2}; and   one-loop static contributions which, much like in the tree-level case, only contribute an overall time-independent shift of the waveform.

Next we  consider the second KMOC term at one loop. Discarding  static contributions as we did at tree level, this term is equal to the  on-shell integration of the product of tree-level four- and five-point amplitudes%
\footnote{One could also consider the case where
$|n\rangle = |r_1  r_2 \tilde{k}^{\tilde{h}}  \rangle$, with an additional intermediate graviton
%as well as further cases involving three-point amplitudes,
but these diagrams all give a vanishing contribution because of the kinematics.}
\begin{align}
	\label{firstintst}
	\sum_{r_1, r_2} \langle p_1^\prime p_2^\prime | T^\dagger|r_1 r_2\rangle^{(0)} \langle r_1 r_2 k^h |    T | p_1 p_2\rangle^{(0)}\, .
\end{align}
The contribution in  \eqref{firstintst} is represented diagrammatically as  
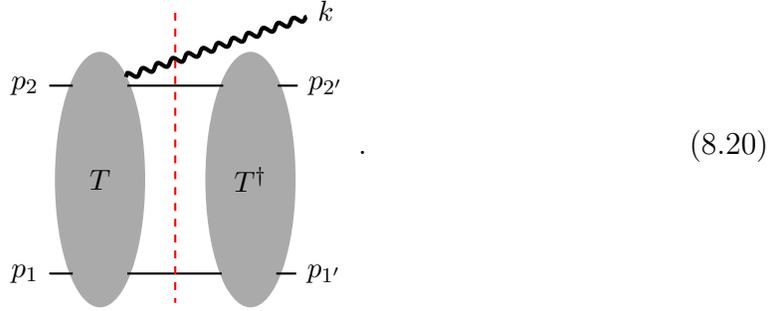
\begin{align}\label{eq:fourSubstract-wf}
	\begin{tikzpicture}[baseline={([yshift=-0.4ex]current bounding box.center)}]\tikzstyle{every node}=[font=\small]
		\begin{feynman}
			\vertex (p1){$p_{1}$};
			\vertex [above=2.5cm of p1](p2){$p_2$};
			\vertex [right=1.0cm of p2] (u1) [HV]{$~~~$};
			\vertex [below=1.25cm of u1] (m1) [GR2]{$T$};
			\vertex [below=1.25cm of m1] (b0) [HV]{$~~~$};
			\vertex [right=3cm of u1] (p3){$p_{2'}$};
			\vertex[left=1cm of p3](u2)[HV]{$~~~$};
			\vertex [below=1.25cm of u2] (m2) [GR2]{$T^\dagger$};
			\vertex [below=1.25cm of m2] (bx) [HV]{$~~~$};
			\vertex [right=1.0cm of p1] (b1) []{};
			\vertex [right=1.0cm of b1] (b2) []{};
			\vertex [right=0.98cm of b2] (b3) [HV]{$~~~$};
			\vertex [right=1.0cm of b3](p4){$p_{1'}$};
			\vertex [above=1.25cm of p1] (cutL);
			\vertex [above=1.0cm of p3] (cutR){$k$};
			\vertex [right=1.0cm of b1] (cut1);
			\vertex [above=0.19cm of cut1] (cut1u);
			\vertex [below=0.39cm of cut1] (cut1b);
			\vertex [right=1.0cm of u1] (cut3);
			\vertex [above=0.17cm of cut3] (cut3u);
			\vertex [above=0.97cm of cut3] (cut3k);
			\vertex [below=0.18cm of cut3] (cut3b);
			\vertex [above=0.4cm of cut1u] (cutd1);
			\vertex [right=0.4cm of cutd1] (cutd2);
			\vertex [right=0.5cm of b2] (cut2);
			\vertex [above=0.19cm of cut2] (cut2u);
			\vertex [below=0.19cm of cut2] (cut2b);
			\diagram* {
			(p2) -- [thick] (u1) -- [thick] (u2)--[thick] (p3),
			(p1) -- [thick] (b0)-- [thick] (b3)-- [thick] (p4),  (cutR)-- [photon,ultra thick] (u1),(cut3k)--[dashed, red,thick] (cut1b),
			};
		\end{feynman}
	\end{tikzpicture}\, .
\end{align}
Unlike at tree level, this term will contribute non-trivially in the classical limit \cite{Caron-Huot:2023vxl}, as we shall now see. In terms of on-shell phase-space integrals, we have
\begin{equation}
\label{eq:oneloop_sub}
\begin{split}
	\sum_{r_1, r_2} \langle p_1^\prime p_2^\prime | T^\dagger|r_1 r_2\rangle^{(0)} \langle r_1 r_2 k^h |    T | p_1 p_2\rangle^{(0)} =\!\! \int \!\!\! \prod_{i=1,2} \! d\Phi(r_i)\,  \, (2\pi)^D\delta^{D} (r_1 + r_2 + k - p_1 - p_2) \, \\
	\times M_{5}^{(0)}(p_1 p_2 \to r_1 r_2 k^h) M_{4}^{(0)\, *}(r_1 r_2 \to p^\prime_1 p^\prime_2)\ .
\end{split}
\end{equation}
We   parameterise the loop momenta as 
\begin{equation}
	r_i = p_i - q_i + l_i\ ,
\end{equation}
such that the delta function in \eqref{eq:oneloop_sub} becomes $\delta^D (l_1+l_2)$, and we have 
\begin{equation}\label{eq: KMOC2Expanded}
\begin{split}
	&\sum_{r_1, r_2} \langle p_1^\prime p_2^\prime | T^\dagger|r_1 r_2\rangle^{(0)}  \langle r_1 r_2 k^h |    T | p_1 p_2\rangle^{(0)} \\
 &=\!\! \int \!\!\! \prod_{i=1,2} \frac{d^D l_i}{(2\pi)^{D-1}} \delta((p_i - q_i + l_i)^2 - m_i^2) \, (2\pi)^D\delta^{D} (l_1 + l_2)\\&
	\times M_{5}^{(0)}(p_1, p_2 \to p_1^\prime+l_1, p_2^\prime+l_2, k^h) M_{4}^{(0)\, *}(p^\prime_1 p^\prime_2 \to p_1^\prime+l_1, p_2^\prime+l_2)\ .
\end{split}
\end{equation}
Expanding in the heavy-mass limit around the $\pb$ variables, the on-shell delta functions become
\begin{equation}
\begin{split}
\label{eq: expand on shell deltas}
	\delta\big((p_i - q_i + l_i)^2 - m_i^2\big) &= \delta((\pb_i - \frac{q_i}{2} + l_i)^2 - m_i^2)\\
	& =\frac{\delta \left(l_i \Cdot \vb_i\right)}{2 \mb_i}+\frac{l_i\Cdot\left(l_i - q_i\right) \delta^\prime\!\left(l_i \Cdot \vb_i\right)}{4 \mb_i^2}+O\left(\mb_i^{-3}\right)\ .
\end{split}
\end{equation}
The same heavy-mass expansion has to be performed at the level of the four- and five-point tree-level amplitudes appearing in \eqref{eq:oneloop_sub}.
Combining the leading contributions to these amplitudes with the delta function in \eqref{eq: expand on shell deltas}, one gets the hyper-classical terms of order $\cO (\bar{m}_1^3 \bar{m}_2^3)$. As expected, subtracting this particular contribution   from  the first KMOC term \eqref{KMOCsub2} has   the effect of  peeling  the hyper-classical contribution off  the complete one-loop five-point matrix element. In our  HEFT approach,
the subtraction of such terms is achieved by simply dropping all the $\cO (\bar{m}_1^3 \bar{m}_2^3)$ terms by power counting. On the other hand, one can obtain a classical contribution by combining the delta-prime term in \eqref{eq: expand on shell deltas} with the leading terms in the scattering amplitude and by taking the leading delta term in \eqref{firstintst} with sub-leading terms in the amplitude. It is this particular contribution that was predicted in \cite{Caron-Huot:2023vxl} and was missed in the first version of this paper.
Such classical contributions from the second KMOC term in \eqref{firstintst} will be computed separately in the next section.

Incidentally, we   note that the
hyper-classical contribution to the second KMOC term \eqref{firstintst} exponentiates in impact parameter space, which we  show explicitly in Appendix~\ref{sec:factorisation}. 
A similar one-loop cancellation of hyper-classical pieces in the expression of the  waveforms at that order was advocated in \cite{Cristofoli:2021jas}, which studied  generalisations of the eikonal  in the presence of radiation \cite{Ciafaloni:2018uwe,DiVecchia:2022nna,DiVecchia:2022piu}.

 Summarising, the quantity of interest is the combination of the one-loop matrix element computed in the HEFT plus an additional classical contribution coming from the second term in the KMOC expression.  For this quantity we introduce the notation 
\begin{align}
\begin{split}
	\label{KMOCfinal}
	&
 i \Big(  \langle p_1^\prime p_2^\prime k^h  | T | p_1 p_2\rangle^{(1)}  {-}  i \sum_{r_1, r_2} \langle p_1^\prime p_2^\prime | T^\dagger|r_1 r_2\rangle^{(0)} \langle r_1 r_2 k^h |    T | p_1 p_2\rangle^{(0)} \Big)  \\ &\coloneqq
 (2\pi)^D \delta^{(D)} (q_1 {+} q_2 {-} k) \ \mathcal{R}_{5}^{(1)} \, .
\end{split}
\end{align} 
Finally, we  mention that in \cite{Caron-Huot:2023vxl} it was pointed out that $\mathcal{R}_{5}^{(1)}$ can be reinterpreted as the one-loop HEFT amplitude where the integrations over massive propagators are done using retarded propagators.  This points to an interesting potential connection  with the approach of \cite{Jakobsen:2021smu,Jakobsen:2021lvp}.

\subsubsection{Explicit computation of the classical contribution from the second KMOC term}

We now turn to the computation of the classical contribution to the second KMOC term given in \eqref{eq:fourSubstract-wf}. As mentioned in the previous section, the classical integrand in \eqref{eq: KMOC2Expanded} contains both $\delta$ and $\delta'$ terms. The relevant integrals we must compute are of the form

\begin{equation}
	\hat j_{a_1, a_2, a_3, a_4, a_5} = \mu^{4-D} \int\! \frac{d^D \ell}{(2 \pi)^{D}} \frac{(- i \pi)^2 \, \delta^{[a_2]} (\bar v_1 \cdot \ell) \delta^{[a_5]} (\bar v_2 \cdot \ell)}{(\ell^2)^{a_1} [(\ell-q_1)^2]^{a_3} [(\ell+q_2)^2]^{a_4}}\ ,
\end{equation}
where
\begin{equation}
	\delta^{[a]}(x) = \frac{(-1)^{a-1}}{(a-1)!} \delta^{(a-1)}(x)\ .
\end{equation}
After applying IBP reduction to the integrand, we obtain the basis integrals are presented in  \eqref{fig:integral_basis} below. These basis integrals only involve delta functions with no derivatives, which are denoted by the red lines on the diagrams.
 \begin{align}
        \widetilde{\mathcal{J}}_1=\hat j_{01011}=\begin{tikzpicture}[scale=0.5, transform shape, baseline={([yshift=-0.4ex]current bounding box.center)}]\tikzstyle{every node}=[font=\small]
        \begin{feynman}
                \vertex (p1) {};
                \vertex [above=2.5cm of p1](p2){};
                \vertex [right=1cm of p2] (u1) {};
                \vertex [right=3cm of p2] (u2) [dot]{};
                \vertex [right=3cm of u1] (p3){};
                \vertex [right=1.0cm of p1] (b1) {};
                \vertex [right=1.0cm of b1] (b2) []{};
                \vertex [right=1.0cm of b2] (b3) [dot]{};
                \vertex [left=1.2cm of u2] (z1) []{};
                \vertex [below=0.15cm of z1] (z2) []{};
                \vertex [below=0.8cm of z2] (w) []{};
                \vertex [right=0.65cm of w] (z3) []{};
                \vertex [right=1.0cm of b3](p4){};
                \vertex [above=1.25cm of p1] (cutL);
                \vertex [right=1.00cm of cutL] (cutL2)[dot]{};
                \vertex [below=1.25cm of u2] (m1){};
                \vertex [right=4.0cm of cutL] (cutR){};
                \vertex [left=1.2cm of b3] (x1) []{};
                \vertex [above=0.15cm of x1] (x2) []{};
                \vertex [above=0.8cm of x2] (y) []{};
                \vertex [right=0.65cm of y] (x3) []{};
                \vertex [right=1.0cm of b1] (cut1);
                \vertex [above=0.3cm of cut1] (cut1u);
                \vertex [below=0.3cm of cut1] (cut1b){};
                \vertex [right=1cm of u1] (cut2);
                \vertex [above=0.2cm of cut2] (cut2u){};
                \diagram* {
                (p2) -- [thick] (cutL2) -- [thick] (u2), (u2) -- [thick] (p3),(b3)--[boson] (u2), (p1) -- [thick] (cutL2)-- [thick] (b3)-- [thick] (p4),(cutR)-- [boson] (b3),(x2)--[red,thick](x3), (z2)--[red,thick](z3)
                };
            \end{feynman}
            \end{tikzpicture}\,,  
            &  & \widetilde{\mathcal{J}}_2=\hat j_{01101}= \begin{tikzpicture}[scale=0.5, transform shape, baseline={([yshift=-0.4ex]current bounding box.center)}]\tikzstyle{every node}=[font=\small]
                                                 \begin{feynman}
                \vertex (p1) {};
                \vertex [above=2.5cm of p1](p2){};
                \vertex [right=1cm of p2] (u1) {};
                \vertex [right=3cm of p2] (u2) [dot]{};
                \vertex [right=3cm of u1] (p3){};
                \vertex [right=1.0cm of p1] (b1) {};
                \vertex [right=1.0cm of b1] (b2) []{};
                \vertex [right=1.0cm of b2] (b3) [dot]{};
                \vertex [left=1.2cm of u2] (z1) []{};
                \vertex [below=0.15cm of z1] (z2) []{};
                \vertex [below=0.8cm of z2] (w) []{};
                \vertex [right=0.65cm of w] (z3) []{};
                \vertex [right=1.0cm of b3](p4){};
                \vertex [left=1.2cm of b3] (x1) []{};
                \vertex [above=0.15cm of x1] (x2) []{};
                \vertex [above=0.8cm of x2] (y) []{};
                \vertex [right=0.65cm of y] (x3) []{};
                \vertex [above=1.25cm of p1] (cutL);
                \vertex [right=1.00cm of cutL] (cutL2)[dot]{};
                \vertex [below=1.25cm of u2] (m1){};
                \vertex [right=4.0cm of cutL] (cutR){};
                \vertex [right=1.0cm of b1] (cut1);
                \vertex [above=0.3cm of cut1] (cut1u);
                \vertex [below=0.3cm of cut1] (cut1b){};
                \vertex [right=1cm of u1] (cut2);
                \vertex [above=0.2cm of cut2] (cut2u){};
                \diagram* {
                (p2) -- [thick] (cutL2) -- [thick] (u2), (u2) -- [thick] (p3),(b3)--[boson] (u2), (p1) -- [thick] (cutL2)-- [thick] (b3)-- [thick] (p4),(cutR)-- [boson] (u2),(x2)--[red,thick](x3), (z2)--[red,thick](z3)
                };
            \end{feynman}
                                             \end{tikzpicture}\, ,
         &  & \widetilde{\mathcal{J}}_3=\hat j_{01111}= \begin{tikzpicture}[scale=0.5, transform shape, baseline={([yshift=-0.4ex]current bounding box.center)}]\tikzstyle{every node}=[font=\small]
                                                       \begin{feynman}
                \vertex (p1) {};
                \vertex [above=2.5cm of p1](p2){};
                \vertex [right=1cm of p2] (u1) {};
                \vertex [right=3cm of p2] (u2) [dot]{};
                \vertex [right=3cm of u1] (p3){};
                \vertex [right=1.0cm of p1] (b1) {};
                \vertex [right=1.0cm of b1] (b2) []{};
                \vertex [right=1.0cm of b2] (b3) [dot]{};
                \vertex [left=1.2cm of u2] (z1) []{};
                \vertex [below=0.15cm of z1] (z2) []{};
                \vertex [below=0.8cm of z2] (w) []{};
                \vertex [right=0.65cm of w] (z3) []{};
                \vertex [right=1.0cm of b3](p4){};
                \vertex [above=1.25cm of p1] (cutL);
                \vertex [right=1.00cm of cutL] (cutL2)[dot]{};
                \vertex [below=1.25cm of u2] (m1)[dot]{};
                \vertex [right=4.0cm of cutL] (cutR){};
                \vertex [left=1.2cm of b3] (x1) []{};
                \vertex [above=0.15cm of x1] (x2) []{};
                \vertex [above=0.8cm of x2] (y) []{};
                \vertex [right=0.65cm of y] (x3) []{};
                \vertex [right=1.0cm of b1] (cut1);
                \vertex [above=0.3cm of cut1] (cut1u);
                \vertex [below=0.3cm of cut1] (cut1b){};
                \vertex [right=1cm of u1] (cut2);
                \vertex [above=0.2cm of cut2] (cut2u){};
                \diagram* {
                (p2) -- [thick] (cutL2) -- [thick] (u2), (u2) -- [thick] (p3),(b3)-- [boson] (m1)-- [boson] (u2), (p1) -- [thick] (cutL2)-- [thick] (b3)-- [thick] (p4),(cutR)-- [boson] (m1),(x2)--[red,thick](x3), (z2)--[red,thick](z3)
                };
            \end{feynman}
        \end{tikzpicture}\,, \nonumber \\
        \mathcal{J}_1= \hat j_{11011}= \begin{tikzpicture}[scale=0.5, transform shape, baseline={([yshift=-0.4ex]current bounding box.center)}]\tikzstyle{every node}=[font=\small]
                                      \begin{feynman}
                \vertex (p1) {};
                \vertex [above=2.5cm of p1](p2){};
                \vertex [right=1cm of p2] (u1) [dot]{};
                \vertex [right=3cm of p2] (u2) [dot]{};
                \vertex [right=3cm of u1] (p3){};
                \vertex [right=1.0cm of p1] (b1) [dot]{};
                \vertex [right=1.0cm of b1] (b2) []{};
                \vertex [right=1.0cm of b2] (b3) [dot]{};
                \vertex [right=1.0cm of b3](p4){};
                \vertex [above=1.25cm of p1] (cutL);
                \vertex [below=1.25cm of u2] (m1)[]{};
                \vertex [right=4.0cm of cutL] (cutR){};
                \vertex [right=1.0cm of b1] (cut1);
                \vertex [above=0.3cm of cut1] (cut1u);
                \vertex [below=0.3cm of cut1] (cut1b){};
                \vertex [right=1cm of u1] (cut2);
                \vertex [above=0.3cm of cut2] (cut2u);
                \vertex [below=0.3cm of cut2] (cut2b);
                \diagram* {
                (p2) -- [thick] (u1) -- [thick] (p3),
                (b1)--[boson](u1), (b3)-- [boson] (u2), (p1) -- [thick] (b1)-- [thick] (b3)-- [thick] (p4), (cut1u)--[red,thick] (cut1b), (cut2u)--[red,thick] (cut2b), (cutR)-- [boson] (b3),
                };
            \end{feynman}
                                  \end{tikzpicture}\,,
         &  & \mathcal{J}_2=\hat j_{11101}= \begin{tikzpicture}[scale=0.5, transform shape, baseline={([yshift=-0.4ex]current bounding box.center)}]\tikzstyle{every node}=[font=\small]
                                           \begin{feynman}
                \vertex (p1) {};
                \vertex [above=2.5cm of p1](p2){};
                \vertex [right=1cm of p2] (u1) [dot]{};
                \vertex [right=3cm of p2] (u2) [dot]{};
                \vertex [right=3cm of u1] (p3){};
                \vertex [right=1.0cm of p1] (b1) [dot]{};
                \vertex [right=1.0cm of b1] (b2) []{};
                \vertex [right=1.0cm of b2] (b3) [dot]{};
                \vertex [right=1.0cm of b3](p4){};
                \vertex [above=1.25cm of p1] (cutL);
                \vertex [below=1.25cm of u2] (m1)[]{};
                \vertex [right=4.0cm of cutL] (cutR){};
                \vertex [right=1.0cm of b1] (cut1);
                \vertex [above=0.3cm of cut1] (cut1u);
                \vertex [below=0.3cm of cut1] (cut1b){};
                \vertex [right=1cm of u1] (cut2);
                \vertex [above=0.3cm of cut2] (cut2u);
                \vertex [below=0.3cm of cut2] (cut2b);
                \diagram* {
                (p2) -- [thick] (u1) -- [thick] (p3),
                (b1)--[boson](u1), (b3)-- [boson] (u2), (p1) -- [thick] (b1)-- [thick] (b3)-- [thick] (p4), (cut1u)--[red,thick] (cut1b), (cut2u)--[red,thick] (cut2b), (cutR)-- [boson] (u2),
                };
            \end{feynman}
                                       \end{tikzpicture}\, ,
         &  & \mathcal{J}_3= \hat j_{11111}=
        \begin{tikzpicture}[scale=0.5, transform shape, baseline={([yshift=-0.4ex]current bounding box.center)}]\tikzstyle{every node}=[font=\small]
            \begin{feynman}
                \vertex (p1) {};
                \vertex [above=2.5cm of p1](p2){};
                \vertex [right=1cm of p2] (u1) [dot]{};
                \vertex [right=3cm of p2] (u2) [dot]{};
                \vertex [right=3cm of u1] (p3){};
                \vertex [right=1.0cm of p1] (b1) [dot]{};
                \vertex [right=1.0cm of b1] (b2) []{};
                \vertex [right=1.0cm of b2] (b3) [dot]{};
                \vertex [right=1.0cm of b3](p4){};
                \vertex [above=1.25cm of p1] (cutL);
                \vertex [below=1.25cm of u2] (m1)[dot]{};
                \vertex [right=4.0cm of cutL] (cutR){};
                \vertex [right=1.0cm of b1] (cut1);
                \vertex [above=0.3cm of cut1] (cut1u);
                \vertex [below=0.3cm of cut1] (cut1b){};
                \vertex [right=1cm of u1] (cut2);
                \vertex [above=0.3cm of cut2] (cut2u);
                \vertex [below=0.3cm of cut2] (cut2b);
                \diagram* {
                (p2) -- [thick] (u1) -- [thick] (p3),
                (b1)--[boson](u1), (b3)-- [boson] (m1)-- [boson] (u2), (p1) -- [thick] (b1)-- [thick] (b3)-- [thick] (p4), (cut1u)--[red,thick] (cut1b), (cut2u)--[red,thick] (cut2b), (cutR)-- [boson] (m1),
                };
            \end{feynman}
        \end{tikzpicture}\,.
\label{fig:integral_basis}
\end{align}
	% \caption{The Feynman integral basis for computing the second term in KMOC. The red lines on the above diagram denote cut propagators consisting of a linearised delta function.}
%
The triangle and box master integrals are computed directly using Feynman parameterisation and are given by
\begin{align}
	\widetilde{\cJ}_1 & = \frac{1}{16 \pi  \sqrt{y ^2-1} \check{\epsilon}}+\frac{\log (y ^2-1)-\log \frac{w_1^2}{\mu^2}}{16 \pi  \sqrt{y ^2-1}}\ ,                                                \\
	\widetilde{\cJ}_2 & = \left. \cJ_1 \right|_{w_1 \rightarrow w_2}\ ,                                                                                                                                                   \\
	\widetilde{\cJ}_3 & = -\frac{\log \left(\sqrt{y ^2-1}+y \right)}{16 \pi  w_1 w_2}\ ,                                                                                                                         \\
	\cJ_1             & = \frac{1}{16 \pi  \sqrt{y ^2-1} (-q_2^2) \check{\epsilon} }-\frac{\log \left(y ^2-1\right)+2 \log \frac{-q_2^2}{\mu^2}-\log \frac{w_1^2}{\mu^2}}{16 \pi  \sqrt{y ^2-1} (-q_2^2)}\ , \\
	\cJ_2             & = \left. \cJ_1 \right|_{w_1 \rightarrow w_2,q_2 \rightarrow q_1}\ ,
\end{align}
where we have defined $\check{\epsilon} =  \epsilon\,e^{(\gamma_E - \log 4\pi)\epsilon}$,  and $\gamma_E$ is the Euler-Mascheroni constant.
The pentagon $\cJ_3$ can then be written in terms of the three boxes up to terms $\cO(\epsilon^0)$. This is also true for the previous pentagon integral $\cI_6$ encountered in the computation of the amplitude, see the relation~\eqref{eq:pentagon_dec}. After setting $\cI_6 \to \cJ_3$, $\widetilde{\cI}_4\to \widetilde{\cJ}_3$, $\cI_3 \to \cJ_1$, $\cI_4 \to \cJ_2$ and $\cI_5 \to 0$, \eqref{eq:pentagon_dec} gives the corresponding relation for the pentagon integral $\cJ_3$ for the case at hand. The divergent part of the pentagon integral is
\begin{equation}
	\left. \cJ_3 \right|_{\text{div}} = - \frac{1}{16 \pi  \sqrt{y ^2-1} q_1^2 q_2^2 \check{\epsilon} } \ .
\end{equation}
Having computed the integrals we now give the expression of the classical $\cO(\mb_1^3\mb_2^2)$ contribution coming from the second KMOC term \footnote{The cut contribution is more naturally written in terms of $\check{\epsilon} $ rather than $\tilde{\epsilon}$. However the difference between the two only results in a finite shift of retarded time in the final waveform.}: 
\begin{equation}\label{secondKMOCtermResult}
\begin{split}
	\cM_{5,\bar{m}_1^3 \bar{m}_2^2, \rm cut}^{(1)} &= \frac{d_{\rm IR}}{\sqrt{\yb^2-1}} \times \frac{1}{\check{\epsilon}} + c_{q_1} \frac{\log\frac{-q_1^2}{\mu^2}}{\sqrt{\yb^2-1}} + c_{q_2} \frac{\log\frac{-q_2^2}{\mu^2}}{\sqrt{\yb^2-1}} + c_{w_1} \frac{\log\frac{w_1^2}{\mu^2}}{\sqrt{\yb^2-1}}\\ &+ c_{w_2} \frac{\log\frac{w_2^2}{\mu^2}}{\sqrt{\yb^2-1}}
	+ c^{1}_{y} \frac{\log \left(\yb^2-1\right)}{\sqrt{\yb^2-1}} + c^{2}_{y} \log \left(\sqrt{\yb^2-1} + \yb\right) + \frac{R}{\sqrt{\yb^2-1}} + \cO(\epsilon),
\end{split}
\end{equation}
where
\begin{equation}
	c_{q_1} + c_{q_2} + c_{w_1} + c_{w_2} = - d_{\rm IR}\ .
\end{equation}
As for the amplitude, there is also an order $\cO(\mb_1^2\mb_2^3)$ piece which can be obtained from \eqref{secondKMOCtermResult} by swapping particles 1 and 2.
We wish  to highlight the following points:
\begin{itemize}
	\item[{\bf 1.}] The coefficient of $-\log \mu^2$ in the amplitude is identical to the coefficient of the infrared divergences, as they are identical in each single Feynman integral.
	\item[{\bf 2.}] We find a new letter $\frac{\log(y^2-1)}{\sqrt{y^2-1}}$ and all the old logarithms are multiplied by a factor of $\frac{1}{\sqrt{y^2-1}}$.
	\item[{\bf 3.}] In the five-point amplitude the logarithms of $q_1^2$ and $q_2^2$ appear always together as $\log\frac{q_1^2}{q_2^2}$ (they appear only in $\cI_5$ and $\cI_6$) and the running logarithms come from a combination of the $w_i$ logarithms. For the cut terms this is no longer true and only the sum of all the logarithms will be proportional to the tree level.
        \item[{\bf 4.}] There is a new infrared divergence, correctly predicted in \cite{Caron-Huot:2023vxl}, associated with the initial conditions of the two massive particles. Moreover, $d_{\rm IR}$ contains also an ultraviolet divergence which is completely local in $q_1^2$ and $q_2^2$ and integrates to delta functions of the impact parameter. Then, such ultraviolet divergence does not affect the waveform.
\end{itemize}

\subsubsection{Final expression for $\mathcal{R}_{5}^{(1)}$}

Combining  \eqref{secondKMOCtermResult}  with \eqref{eq:OneLoopAmplitude} (and the terms with $m_1$ swapped with $m_2$) gives the final expression for the KMOC kernel $\mathcal{R}_{5}^{(1)}$ \eqref{KMOCfinal}:
\begin{align}\label{eq: KMOCkernel}
    \mathcal{R}_{5}^{(1)}= \cM_5^{(1)} +\cM_{5,\rm cut}^{(1)}\, , 
\end{align}
where $\cM_5^{(1)}$ was defined in \eqref{finalresultform2},  and  
\begin{equation}
	\label{extrapiece}
	\mathcal{M}_{5, \rm cut}^{(1)} = (\cM_{5, \mb_1^3\mb_2^2,{\rm cut}}^{(1)}+\cM_{5,  \mb_1^2\mb_2^3,{\rm cut }}^{(1)})\vert_{\mb_i \rightarrow m_i}\ , 
\end{equation}
with  $\cM_{5, \mb_1^3\mb_2^2,{\rm cut}}^{(1)}$   given in \eqref{secondKMOCtermResult}. 
We can then use this result to evaluate the earlier expressions \eqref{KMOCsub} and \eqref{KMOCsubhmunu} for the expectation value of the Riemann tensor and the gravitational field.

The results for the rational coefficients multiplying the integral functions (logarithms, square roots etc.) in \eqref{eq:OneLoopAmplitude} and \eqref{secondKMOCtermResult} can be found in the {\href{https://github.com/QMULAmplitudes/Gravity-Observables-From-Amplitudes}{{\it Gravity Observables from Amplitudes} GitHub repository.}}

\subsection{From HEFT to  waveforms}
\label{sec:8.3}

Combining \eqref{KMOCsub} and \eqref{KMOCsubhmunu} with \eqref{KMOCfinal}, we can write the 
one-loop expectation value of the Riemann tensor or the metric in terms of the KMOC kernel \eqref{eq: KMOCkernel}, 
\begin{align}
	\label{KMOCsubfinal}
	\begin{split}
		\langle R^{\rm out}_{\mu \nu \rho \lambda} (x) \rangle_{\rm \psi}  &=    \kappa\,  {\rm Re} \, \Big[
	i \int\!\prod_{j=1}^2d\Phi (p_j)  \, |\phi(p_1)|^2 |\phi (p_2)|^2
		\\ & 
		\sum_h\!\int\!d\Phi(k) e^{-i k \Cdot x} \,
		k_{[\mu}\varepsilon^{(h)\ast}_{\nu]} (\vec{k}) k_{[\rho}\varepsilon^{(h)\ast}_{\lambda]} (\vec{k})
		\, 
		\widetilde {W}\Big]
		, 
	\end{split}
\end{align}
and 
\begin{align}
	\label{KMOCsubfinalforhmunu}
	\begin{split}
		\langle h^{\rm out}_{\mu \nu} (x) \rangle_{\rm \psi}  \!=\!    2 \, \kappa\,  {\rm Re} \, \Big[
	i \int\!\prod_{j=1}^2d\Phi (p_j)  \, |\phi(p_1)|^2 |\phi (p_2)|^2
		%\\ & 
		\sum_h \int\!d\Phi(k) e^{-i k \Cdot x} \,
		\varepsilon^{(h)\ast}_{\mu} (\vec{k}) \varepsilon^{(h)\ast}_{\nu} (\vec{k})
		\, 
		\widetilde {W}\Big]
		, 
	\end{split}
\end{align}
where $\widetilde{W}=\widetilde{W}(\vec{b}, k^h)$ is given by%
\footnote{The factor of $-i$ cancels the $i$ which is present in our definition of amplitudes as matrix elements of $i\, T$, see  \eqref{defampli}.} 
\begin{equation}
    \widetilde{W}(\vec{b},k^h)\coloneqq -i\int \frac{d^D q_1}{(2\pi)^{D{-}1}} \frac{d^D q_2}{(2\pi)^{D{-}1}} \delta(2  {p}_1\Cdot q_1 ) \delta(2  {p}_2\Cdot q_2 ) e^{i(q_1\Cdot b_1 + q_2 \Cdot b_2)} \langle p_1^\prime p_2^\prime  |  S^{\dagger}a_h(\vec{k}) S | p_1 p_2\rangle \,.
\end{equation}%
At one loop, this expression reduces to 
\begin{align}
\label{KMOCsubfinalbis}
\widetilde{W}^{(1)}(\vec{b}, k^h) \coloneqq  -i \int\!d\mu^{(D)}\  e^{i(q_1\Cdot b_1 + q_2 \Cdot b_2)} \ \mathcal{R}_{5}^{(1)}(q_1, q_2; h) \, , 
\end{align}
and we have defined 
\begin{align}
\label{measureinnn}
	d\mu^{(D)} \coloneqq \frac{d^Dq_1}{(2\pi)^{D{-}1}} \frac{d^Dq_2}{(2\pi)^{D{-}1}}
	\, (2\pi)^D  \delta^{(D)} (q_1 + q_2 - k) \delta(2  {p}_1\Cdot q_1 ) \delta(2  {p}_2\Cdot q_2 )\, .
\end{align}
As usual, $k^h$ denotes a graviton with helicity $h{=}\pm$. Note that  having  eliminated any hyper-classical terms from the waveform, we are now free to express the KMOC kernel in terms of unbarred variables since any feed-down terms will be quantum.

$\widetilde{W}(\vec{b}, k^h)$ is directly related to the waveforms, but before making this connection more precise we would like to pause and make a few comments: 

\noindent
{\bf 1. Dependence on $\vec{b}$}\\
First, the dependence  of \eqref{KMOCsubfinalbis}  on $\vec{b}$ can be made more explicit: changing variables from  $(q_1, q_2)$ to the variables $(q, k)$ introduced in \eqref{qdefs}, we can rewrite \eqref{KMOCsubfinalbis} as
%$k= q_1 + q_2$ and $q\coloneqq \frac{1}{2} (q_1 - q_2)$
%we an rewrite $\cM_5(\vec{b})$ as
\begin{align}
	\label{5pIPSbis}
	\hspace{-0.4cm}-i e^{i \frac{b_1 + b_2}{2} \Cdot k }\int\!\frac{d^Dq}{(2\pi)^{D{-}2}}  \,  \delta \Big(2{p}_{1}\Cdot \Big(q {+} \frac{k}{2}  \Big)\Big) \delta \Big(2{p}_{2}\Cdot \Big({-}q {+} \frac{k}{2}  \Big)\Big)\, \, e^{iq\Cdot (  b_1 -  b_2) }{\mathcal{R}}_{5}^{(1)} \Big(q{+}\frac{k}{2}, {-}q{+}\frac{k}{2}; h\Big),
\end{align}
showing a non-trivial dependence on  $b {\coloneqq} b_1 - b_2$ and a simple phase dependence on the average impact parameter  $(b_1 + b_2)/2$.
Alternatively, one may observe that under a translation $(b_1, b_2){\to} (b_1{+}a, b_2{+}a)$,  \eqref{KMOCsubfinalbis} picks  a factor of $e^{ik\Cdot a}$, hence it is sufficient to compute the quantity  
\begin{align}
  -i \int\!d\mu^{(D)}\  e^{iq_1\Cdot b} \ \mathcal{R}_{5}^{(1)}(q_1, q_2; h) \, .
\end{align}
We will henceforth set $b_2{=}0$ and  $b_1{=}b$ and drop the overall phase in \eqref{5pIPSbis}.

\noindent{\bf 2. Infrared (in)finiteness of the  gravitational waveform}\\ The one-loop HEFT amplitude $ \mathcal{M}_{5}^{(1)}$  in \eqref{KMOCsubfinalbis} contains infrared divergences, which in gravity give a non-vanishing contribution to the waveform. This is in agreement with earlier computations performed in the PN expansion \cite{Goldberger:2009qd}. While such divergent phases drop out of quantities such as cross sections, they are still present in the waveform, which is linear in the classical (or HEFT) amplitude. The question then arises as to what is their fate. The answer was suggested  in
\cite{Porto:2012as}, where the waveform in the time-domain was considered and it was noted  that the divergent phase simply shifts time (in the exponential $e^{-ik\Cdot x}$ in \eqref{KMOCsub})  by an amount proportional to $1/\epsilon$. Using the classical limit of Weinberg's formula computed in \eqref{Wphase}, we see that this shift has the form
\begin{align}
\label{shapiroshift}
	t\to t - \frac{G(p_1+p_2)\Cdot n}{\eps}\, ,
\end{align}
with $k^\mu {=} \omega\, n^\mu$. Physically this time shift,  a Shapiro time delay, is not relevant as ultimately we only deal with time differences --  we measure  time with respect to when an  experimenter begins tracking the wave signal.
As a consequence, we can safely drop the infrared divergences in the one-loop amplitude.

It is also interesting to note that,  by contrast, the electromagnetic analogue of the classical  amplitude  is free of this infrared divergence. Indeed,  the infrared-divergent Coulomb phase in QED \cite{Weinberg:1965nx}, evaluated in the 
HEFT expansion is  $e^{\frac{i}{4 \pi \epsilon}e_1 e_2 \frac{y}{\sqrt{y^2-1}} + \mathcal{O}(m_i^{-2})}$, and   only has hyper-classical contributions. The absence of classical infrared divergences in QED has a very clear physical interpretation: the photon does not interact electromagnetically with the system at large distances, while the graviton is influenced by the total (ADM) mass of the binary system. For example, this difference is crucial when computing the waveshape of \cite{Cristofoli:2021jas} at next-to-leading order, which is finite in electrodynamics but  infrared-divergent in gravity. 

In addition, there is a separate infrared divergence arising from the second KMOC term. This appears because the incoming particles are not truly free: they feel each other's $1/r$ potential and this becomes relevant for inclusive observables such as the waveform. This leads to an additional time shift along the trajectory of the incoming particles prior to their collision. This shift is of the form
\cite{Caron-Huot:2023vxl}
\begin{align}
t\to t -\frac{G(p_1+p_2)\Cdot n}{\eps} \ \frac{y\, (y^2 -\frac{3}{2})}{(y^2-1)^{\frac{3}{2}}}\, ,
\end{align}
and similarly to the shift \eqref{shapiroshift}, it can be discarded in the waveform when we consider only time-differences between measurements. 

In conclusion, we can just focus  on the simpler  four-dimensional integral  
\begin{align}
	\label{intwf2}
	\begin{split}
		W^{(1)} (b, k^h)\coloneqq -i \int\! d\mu^{(4)} \
		e^{i q_1\Cdot b}\  {\mathcal{R}}_{5, \rm fin}^{(1)} (q_1, q_2; h)
		\, , 
	\end{split}
\end{align}
where $ {\mathcal{R}}_{5,\rm  fin}^{(1)}$ is the infrared-finite part of the expression given in \eqref{KMOCfinal}, and  
$d\mu^{(4)}$ is the measure introduced in \eqref{measureinnn} evaluated for $D{=}4$. We will then safely   use    $W$ in \eqref{intwf2} within \eqref{KMOCsubfinal} and 
\eqref{KMOCsubfinalforhmunu} instead of $\widetilde{W}$.

\subsection{Waveforms and Newman-Penrose scalar from the HEFT} 

In the previous section we saw that $\langle R^{\rm out}_{\mu \nu \rho \lambda}\rangle_\psi$ can be written as in \eqref{KMOCsubfinal}, which for convenience we recast here as%
\footnote{From now on we drop the integrations $\int\!\prod_{j=1}^2d\Phi (p_j)  \, |\phi(p_1)|^2 |\phi (p_2)|^2$ as we are assuming that the wavefunctions $\phi(p_i)$ are peaked around the classical value of the momenta of the heavy objects, and are furthermore simply spectators in the evaluations.} 
\begin{align}
	\label{KMOCsubfinal-interm}
	\begin{split}
		\langle R^{\rm out}_{\mu \nu \rho \lambda} (x) \rangle_{\rm \psi}  =   i \,  \frac{\kappa}{2} \,   
		\sum_h\!\int\! & d\Phi(k)  \Big[ e^{-i k \Cdot x} \,
		k_{[\mu}\varepsilon^{(h)\ast}_{\nu]} (\vec{k}) k_{[\rho}\varepsilon^{(h)\ast}_{\lambda]} (\vec{k})
		 \ {W}(b, k^h)
  \\ &
 \quad \ \  \, - \,
  e^{i k \Cdot x} \,\, 
		k_{[\mu}\varepsilon^{(h)}_{\nu]} (\vec{k}) k_{[\rho}\varepsilon^{(h)}_{\lambda]} (\vec{k})
		\  {W}^\ast (b, k^h)
  \Big]
		\, . 
	\end{split}
\end{align}
At large observer's distance $r{\coloneqq }|\vec{x}|$,  the exponentials in \eqref{KMOCsubfinal-interm} oscillate very fast. Introducing 
the retarded time $u{\coloneqq} t{-}r$, one can   rewrite the plane waves $e^{\mp i k\Cdot x}$ using 
$k\Cdot x {=}\omega (t {-}  r \hat{\mathbf{x}}\Cdot \hat{\mathbf{n}}) {=}   \omega u {+} \omega r ( 1 - \cos \theta)$.  A well-known  stationary phase approximation argument \cite{He:2014laa} then gives
\begin{align}
\begin{split}
    \int\!d\Phi (k) e^{\mp i k\Cdot x} f (\omega, \omega \hat{\mathbf{n}}) &\to 
    %\pm  \frac{1}{(2\pi)^{3}}\int_{0}^{+\infty} \!
%\frac{d\omega }{2\omega} \omega^{2} \frac{2\pi}{i\omega r} f (\omega, %\omega \hat{\mathbf{x}}) e^{\mp i \omega u}\\ &  
%
\mp \frac{i}{4 \pi r  }\int_{0}^{+\infty}\!\frac{d\omega}{2\pi}  \ e^{\mp i \omega u} f (\omega, \omega \hat{\mathbf{x}})\, ,  
\end{split}
\end{align}
where $f(\vec{k}) {=} f(\omega, \omega \hat{\mathbf{n}})$ is a function of the graviton momentum $\vec{k}$; note that after the minimisation, the direction $\hat{\mathbf{n}}$ of $\vec{k}$ is aligned to that of  $\vec{x}$. Using this result we can then rewrite 
\begin{align}
	\label{KMOCsubfinal-afterlocal}
	\begin{split}
		\langle R^{\rm out}_{\mu \nu \rho \lambda} (x) \rangle_{\rm \psi} \stackrel{r\rightarrow \infty }{=}     \frac{\kappa}{8 \pi r } \,  \sum_h \int_{0}^{+\infty}\!&\frac{d\omega}{2\pi}\, \Big[
		 k_{[\mu}\varepsilon^{(h)\ast}_{\nu]} (\vec{k}) k_{[\rho}\varepsilon^{(h)\ast}_{\lambda]} (\vec{k})
		\, 
		{W}(b, k^h) e^{-i \omega u}\, 
  \\ &\ \  + \, k_{[\mu}\varepsilon^{(h)}_{\nu]} (\vec{k}) k_{[\rho}\varepsilon^{(h)}_{\lambda]} (\vec{k})\, 
		\ {W}^\ast (b,  k^h) 
  e^{i \omega u} \Big]_{ k = \omega(1, \hat{\mathbf{x}})} \, . 
	\end{split}
\end{align}
Several quantities can now be introduced to describe the waveforms. 
One that is commonly  used in the study of gravitational waves is the  Newman-Penrose scalar~\cite{Newman:1961qr}
\begin{align}
\label{psi4}
    \Psi_4 (x) \coloneqq N^\mu M^{\nu \, \ast}N^\rho M^{\sigma\, \ast}  \langle W_{\mu \nu \rho \sigma}^{\rm out}  (x) \rangle\, .  
\end{align}
Here $W$ is the Weyl tensor, in our case  equal to  the Riemann tensor, and  
\begin{align}
\label{dotting}
N_\mu = \zeta_\mu\, , \qquad M_\mu = \varepsilon^{(+)}_\mu\, , \qquad M_{\mu}^{\ast} = \varepsilon^{(-)}_\mu\, , 
\end{align}
where $\zeta$ is a reference vector chosen such that $\zeta\Cdot \varepsilon^{(\pm)} {=} 0$, and  $\zeta\Cdot n {=} 1$.     
$\Psi_4 (x)$ is often used to illustrate the gravitational
waveform \cite{Pretorius:2005gq,Pretorius:2007nq,CalderonBustillo:2022dph}, 
and  is  also the quantity considered  in the open-source numerical relativity code GRCHombo \cite{GRCHombo,Clough:2015sqa}.

Starting from \eqref{KMOCsubfinal-afterlocal}, we can now compute   $\Psi_4 (x)$. In the far-field domain it has the form  
 \begin{align}
\Psi_4 (x) \stackrel{r\rightarrow \infty }{\longrightarrow}   \frac{\Psi_4^0 (x)}{|\vec{x}\, |}
\, , 
\end{align}
where
    \begin{align}
      \begin{split}
        \label{tdwf2}  
     \Psi_4^0(x) =   
  \frac{\kappa}{8 \pi} \int_{0}^{+\infty}\!\frac{d\omega}{2\pi}\  \omega^2  \Big[   W(b; k^-)  e^{-i \omega u }  +   \big[ W(b; k^+) ]^{\ast}  e^{i \omega u} \Big]_{ k = \omega(1, \hat{\mathbf{x}})}   \, ,
   \end{split}
    \end{align}
    where  $u$ is the retarded time, and 
    the  $\varepsilon_\mu^{(\pm)}$ vectors satisfy 
\begin{align}
\varepsilon_\mu^{(+)\, \ast } = \varepsilon_\mu^{(-)}\, , \qquad \varepsilon^{(\pm)}\Cdot \varepsilon^{(
\pm)\, \ast} = -1\, , \qquad \varepsilon^{(\pm)}\Cdot \varepsilon^{(
\mp)\, \ast} = 0\, . 
\end{align}    
    This is the result we will use to compute waveforms in the time domain. 
    
    Three final comments are in order here. 

\begin{itemize}
    \item[{\bf 1.}] We recall that $W$ was defined in \eqref{intwf2}. It is constructed out of the finite part of the 2MPI HEFT amplitude, which contains only classical physics. 
    \item[{\bf 2.}] Furthermore, we observe that at tree level   
  \begin{align}
    \label{Wpm}
        W^{(0)} (b, k^{\pm}) =\big[ W^{(0)} (-b, k^{\mp})\big]^\ast\, , 
    \end{align}
    which follows from the form of the tree-level  five-point amplitude and the definition of $W$ in  \eqref{intwf2}.

\item[{\bf 3.}] Finally we comment that a quantity widely used to characterise the waveforms is the gravitational strain $h$, of which $\Psi_4$ is the second derivative with respect to the retarded time $u$, $\Psi_4 {=} d^2 h  / d u^2$. This can also be obtained from our previous formulae: 
\begin{align}
      \begin{split}
        \label{strain}  
     h ( x)  =   
 -  \frac{\kappa}{8 \pi |\vec{x}\, | } \int_{0}^{+\infty}\!\frac{d\omega}{2\pi}\    \Big[   W(b; k^-)  e^{-i \omega u }  +   \big[ W(b; k^+) ]^{\ast}  e^{i \omega u} \Big]_{ k = \omega(1, \hat{\mathbf{x}})}   \, .
   \end{split}
    \end{align}
\end{itemize}
In the next section we will perform numerical integrations and will present various plots of $W(b, k^{\pm})$ and $ \omega^2 W(b, k^{\pm})$ which will  illustrate the waveforms in the frequency domain;  we will then move on to show the waveforms in the time domain as obtained from \eqref{tdwf2}. 
Note that from now on we
 will refer to $W(b, k^{\pm})$ simply as the spectral waveform.  

\subsection{Set-up of the  integration for waveforms}

In this section we address the computation of the   one-loop waveform   introduced in \eqref{intwf2},  which enters the Newman-Penrose scalar $\Psi_4^0$ \eqref{tdwf2}. 
A convenient way to perform the integrations in \eqref{intwf2}  was discussed in \cite{Cristofoli:2021vyo}. After integrating out $q_2$ using the delta function, one can parameterise the remaining integration variable $q_1$ as (renaming $q_1\to q$ for notational simplicity),
\begin{align}
	\label{qparam}
	q= z_1 v_1 + z_2 v_2 + z_v \tilde{v} + z_b \tilde{b}\, ,
\end{align}
where
\begin{align}
	\begin{split}
		v_1= \frac{p_1}{m_1}\, ,\quad  v_2= \frac{p_2}{m_2}\, , \quad\tilde{v} = \frac{v}{\sqrt{-v^2}}\, , \quad\tilde{b} = \frac{b}{\sqrt{-b^2}}\, ,
	\end{split}
\end{align}
and
\begin{align}
	v = \epsilon (\bullet\,  v_1\,  v_2\,  b)\, ,\qquad  \text{with} \quad  v^2 =  b^2 (y^2-1)\, .
 \end{align}
Choosing $b$ to be the asymptotic impact parameter, we also have  that $b\Cdot v_1 = b\Cdot v_2 =0$.
The Jacobian is then $d^4q = \sqrt{y^2-1} \prod_{a=1,2,v,b} d z_a$, so that
\begin{align}
	d\mu^{(4)}  \to \frac{1}{(4\pi)^2}\frac{\sqrt{y^2-1}}{m_1 m_2}  \prod_{a=1,2,v,b} d z_a\ \delta \big(z_1 + y z_2\big) \delta \big(z_2 (y^2-1) + w_2\big)\, .
\end{align}
The delta functions set
\begin{align}
	z_1 =  \frac{y}{y^2-1} w_2\, , \qquad z_2 =
	-  \frac{w_2}{y^2-1} \ ,
\end{align}
hence
\eqref{intwf2} becomes%
\footnote{The $z_b$ integration can, in principle, be performed analytically by closing the integration contour in the lower-half plane. Therefore, this integration can be rewritten as a sum over residues on poles and integrals over discontinuities on branch cuts using Cauchy's theorem. %The integration on $z_v$ would represent a challenge in the computation.
To compute the waveform in the time domain, it is convenient to perform the $\omega$ integration first, which evaluates to (derivatives of) a delta function and a PV of $u-z_b\sqrt{-b^2}$ for the real and imaginary part of the amplitude, respectively.}
\begin{align}
	\label{intwf3}
	\begin{split}
		W = \frac{-i }{(4\pi)^2m_1 m_2\sqrt{y^2-1}} \int\! dz_v dz_b \ e^{ - i z_b \sqrt{-b^2} }\,
		\left. \mathcal{R}_{5,\rm  fin}^{(1)}\right|_{ z_1 =  \frac{y}{y^2-1} w_2\, , \ z_2 =                   -  \frac{w_2}{y^2-1} }
		\, .
	\end{split}
\end{align}
We also note that
\begin{align}
	q^2 = - \frac{w_2^2}{y^2-1} - z_v^2 - z_b^2\, .
\end{align}

\subsection{Waveform for  binary scattering}

We now move on to evaluate the waveforms numerically using our one-loop result for the HEFT amplitude. For completeness we will first briefly review the tree-level waveforms, before considering the   one-loop case.
In the following we   parameterise the kinematic data as 
\begin{align}
\begin{split}
v_1&=(1,0,0,0)\, , \\
v_2&=(y,\sqrt{y^2-1},0,0)\,  ,\\
k&=\omega  (1,\sin \theta  \cos \phi ,\sin \theta  \sin \phi ,\cos \theta )\, , \\
    \varepsilon^{(\pm)}&=\frac{1}{\sqrt{2}}(0,\cos \theta  \cos \phi \mp i \sin \phi ,\cos \theta  \sin \phi \pm i \cos \phi ,-\sin \theta )\, , 
\end{split}
\end{align}
therefore working in the rest frame of the first heavy object. We also  choose the impact parameter as $b{=} \sqrt{-b^2} (0, 0, 1, 0)$.
The polarisation vectors $\varepsilon^{(\pm)}$ are related to the graviton polarisation tensors corresponding to positive/negative helicity~as
\begin{align}
\varepsilon^{(\pm \pm)} = \varepsilon^{(\pm)} \otimes \varepsilon^{(\pm)} = 
\varepsilon^{\text{TT},+} \pm i \, 
\varepsilon^{\text{TT},\times} \ ,
\end{align}
where we have also written the relation of the polarisation tensors of $\pm$ helicity to the two standard transverse-traceless (TT) polarisation tensors plus $(+)$ and cross $(\times)$.

\subsubsection{Tree level}
At  tree level the  waveform  is given by   
\begin{align}
\label{eq: treeLevelWaveform}
\begin{split}
    {W}^{(0)}_{\pm} &= -i \int\!d\mu^{(4)} \ e^{i q_1\Cdot b}\  \cM_{\bar{m}_1^2 \bar{m}_2^2}^{(0)}
    \coloneqq m_1 m_2 W_{\pm,m_1 m_2}^{(0)}\\
    &=(m_1+m_2)^2\chi(1-\chi) W_{\pm,m_1 m_2}^{(0)}\, , 
\end{split}
\end{align}
where $\cM_{\bar{m}_1^2 \bar{m}_2^2}^{(0)}$ is the classical tree-level five-point amplitude, obtained by taking the $\bar{m}_1^2 \bar{m}_2^2$  term in the HEFT expansion which was computed in Section \ref{sec:5ps}. $W_{\pm}$ is shorthand for  $W(b, h^{\pm})$, and the subscript $m_1 m_2$ indicates the mass dependence of the corresponding term.  We have also introduced
\begin{align}
\label{chii}
   %\totalmass=m_1+m_2,\quad
   \chi\coloneqq \frac{m_2}{
    %\totalmass
    m_1 + m_2}\, , 
\end{align}
which parameterises the relative mass ratio of the two massive objects. 
The hyper-classical terms at tree-level in \eqref{eq:5pointHEFTExpansion} are subtracted in the calculation of the waveform, as explained in Section \ref{sec:1lWFHEFT}. This has also been noted in \cite{Manohar:2022dea}. As mentioned previously, this allows us to work with $m_i$ and $v_i$ instead of their barred versions, since any feed-down  terms  we generate will be quantum.

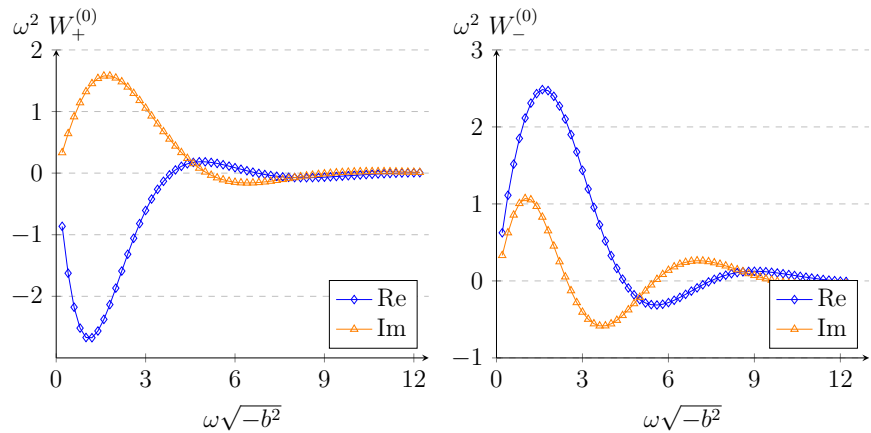
\begin{figure}[t]
\centering
\scalebox{0.8}{
\begin{tikzpicture}
	\begin{axis}[width=0.5\textwidth,
			title={},
			xmin=0, xmax=12.5,
			ymin=-3, ymax=2,
			axis lines=left,
			xtick={-12,-9,-6,-3,0,3,6,9,12},
			ytick={-2,-1,0,1,2},
			compat=newest,
			xlabel=$\omega \sqrt{-b^2}$, %xlabel style={at={(1,0)}, anchor=west},
			ylabel=$\omega^2$ ${W}_{+}^{(0)}$, ylabel style={rotate=-90,at={(0,1)},anchor=south},
			legend pos=south east,
			ymajorgrids=true,
			grid style=dashed,
		]
		\addplot[
			color=blue,
			mark=diamond,
		]
		coordinates {(0.2, -0.8638783333333333)(0.4, -1.6249083333333332)(0.6, -2.1768666666666663)(0.8, -2.51585)(1., -2.66885)(1.2, -2.67225)(1.4, -2.5629)(1.6, -2.3741499999999998)(1.8, -2.134433333333333)(2., -1.8668833333333332)(2.2, -1.5897099999999997)(2.4, -1.3166366666666667)(2.6, -1.0576283333333332)(2.8, -0.8194883333333333)(3., -0.6064633333333334)(3.2, -0.42076499999999994)(3.4, -0.2630283333333333)(3.6, -0.13270099999999999)(3.8, -0.028365166666666667)(4., 0.05199933333333333)(4.2, 0.11079666666666667)(4.4, 0.15064316666666666)(4.6, 0.17423)(4.8, 0.18422166666666664)(5., 0.18316666666666664)(5.2, 0.17345333333333335)(5.4, 0.15725899999999998)(5.6, 0.13652716666666664)(5.8, 0.11295866666666667)(6., 0.08800549999999999)(6.2, 0.06287583333333333)(6.4, 0.038546)(6.6, 0.0157752)(6.8, -0.0048756)(7., -0.0230225)(7.2, -0.03843733333333333)(7.4, -0.05102416666666666)(7.6, -0.060799)(7.8, -0.067867)(8., -0.07240333333333332)(8.2, -0.0746355)(8.4, -0.07482583333333333)(8.6, -0.0732575)(8.8, -0.07022216666666667)(9., -0.06600966666666666)(9.2, -0.0608985)(9.4, -0.0551505)(9.6, -0.049005)(9.8, -0.04267533333333333)(10., -0.036347166666666667)(10.2, -0.030177666666666665)(10.4, -0.024294833333333335)(10.6, -0.01879983333333333)(10.8, -0.013767066666666668)(11., -0.009247483333333332)(11.2, -0.005270383333333333)(11.4, -0.0018462833333333332)(11.6, 0.0010303783333333334)(11.8, 0.00337845)(12., 0.0052274)(12.2, 0.006614816666666666)};
    \addplot[
			color=orange,
			mark=triangle,
		]
		coordinates {(0.2, 0.33222166666666664)(0.4, 0.6419116666666667)(0.6, 0.9149266666666667)(0.8, 1.1431783333333332)(1., 1.3229483333333332)(1.2, 1.4537383333333331)(1.4, 1.5373766666666664)(1.6, 1.5773183333333334)(1.8, 1.5780816666666664)(2., 1.5448216666666665)(2.2, 1.482975)(2.4, 1.3980166666666665)(2.6, 1.2952516666666667)(2.8, 1.1796766666666665)(3., 1.05588)(3.2, 0.9279766666666666)(3.4, 0.7995683333333333)(3.6, 0.6737299999999999)(3.8, 0.5530083333333333)(4., 0.4394416666666666)(4.2, 0.33458166666666667)(4.4, 0.239535)(4.6, 0.15500033333333332)(4.8, 0.08131183333333333)(5., 0.0184895)(5.2, -0.033714666666666664)(5.4, -0.07576983333333333)(5.6, -0.10832)(5.8, -0.13214199999999998)(6., -0.14810416666666665)(6.2, -0.157132)(6.4, -0.16017483333333332)(6.6, -0.15817883333333332)(6.8, -0.152063)(7., -0.1427005)(7.2, -0.13090366666666664)(7.4, -0.11741266666666667)(7.6, -0.102888)(7.8, -0.0879065)(8., -0.07295983333333333)(8.2, -0.0584555)(8.4, -0.04472016666666666)(8.6, -0.032004)(8.8, -0.020487166666666667)(9., -0.010286116666666666)(9.2, -0.0014618849999999998)(9.4, 0.0059729499999999994)(9.6, 0.012046099999999999)(9.8, 0.016818166666666665)(10., 0.020375166666666666)(10.2, 0.022822333333333333)(10.4, 0.024277499999999997)(10.6, 0.02486583333333333)(10.8, 0.0247155)(11., 0.023953166666666668)(11.2, 0.022700666666666664)(11.4, 0.021073333333333333)(11.6, 0.0191765)(11.8, 0.017105833333333334)(12., 0.0149451)(12.2, 0.012766483333333332)};
 
\legend{{\rm Re},{\rm Im}}
	\end{axis}
\end{tikzpicture} 
\begin{tikzpicture}
	\begin{axis}[width=0.5\textwidth,
			title={},
			xmin=0, xmax=13,
			ymin=-1, ymax=3,
			axis lines=left,
			xtick={0,3,6,9,12},
			ytick={-1,-0,1,2,3},
			compat=newest,
			xlabel=$\omega \sqrt{-b^2}$, %xlabel style={at={(1,0)}, anchor=west},
			ylabel=$\omega^2$ ${W}_{-}^{(0)}$, ylabel style={rotate=-90,at={(0,1)},anchor=south},
			legend pos=south east,
			ymajorgrids=true,
			grid style=dashed,
		]
		\addplot[
			color=blue,
			mark=diamond,
		]
		coordinates {(0.2, 0.6250566666666666)(0.4, 1.1120616666666665)(0.6, 1.516725)(0.8, 1.8507)(1., 2.1152666666666664)(1.2, 2.3091999999999997)(1.4, 2.4318833333333334)(1.6, 2.484633333333333)(1.8, 2.4711833333333333)(2., 2.3975166666666663)(2.2, 2.2714999999999996)(2.4, 2.102333333333333)(2.6, 1.9)(2.8, 1.6746666666666665)(3., 1.436175)(3.2, 1.19372)(3.4, 0.9554483333333332)(3.6, 0.7283)(3.8, 0.5178799999999999)(4., 0.3284183333333333)(4.2, 0.16280683333333332)(4.4, 0.022683166666666664)(4.6, -0.09144116666666666)(4.8, -0.18002999999999997)(5., -0.24435166666666666)(5.2, -0.28631166666666663)(5.4, -0.3082733333333333)(5.6, -0.31290833333333334)(5.8, -0.3030483333333333)(6., -0.281555)(6.2, -0.25121666666666664)(6.4, -0.214665)(6.6, -0.174305)(6.8, -0.13226966666666667)(7., -0.090391)(7.2, -0.05018883333333333)(7.4, -0.0128694)(7.6, 0.020663)(7.8, 0.049786)(8., 0.07413349999999999)(8.2, 0.09356433333333333)(8.4, 0.10812883333333334)(8.6, 0.11803583333333334)(8.8, 0.12361833333333333)(9., 0.125301)(9.2, 0.12357166666666666)(9.4, 0.11895366666666665)(9.6, 0.1119825)(9.8, 0.10318683333333334)(10., 0.09307149999999999)(10.2, 0.08210566666666666)(10.4, 0.07071266666666666)(10.6, 0.05926466666666667)(10.8, 0.048078666666666665)(11., 0.03741583333333333)(11.2, 0.027481666666666668)(11.4, 0.018429333333333332)(11.6, 0.01036315)(11.8, 0.0033431)(12., -0.0026096333333333332)(12.2, -0.007507149999999999)};
  
    \addplot[
			color=orange,
			mark=triangle,
		]
		coordinates {(0.2, 0.3306533333333333)(0.4, 0.6255716666666666)(0.6, 0.8560733333333334)(0.8, 1.00579)(1., 1.0701533333333333)(1.2, 1.0539533333333333)(1.4, 0.9684666666666667)(1.6, 0.8287883333333333)(1.8, 0.6515983333333333)(2., 0.45342666666666664)(2.2, 0.24942333333333333)(2.4, 0.052554)(2.6, -0.12682483333333333)(2.8, -0.2811166666666667)(3., -0.4054066666666667)(3.2, -0.497235)(3.4, -0.5562716666666666)(3.6, -0.5839366666666667)(3.8, -0.5830033333333333)(4., -0.5572083333333333)(4.2, -0.5108933333333333)(4.4, -0.4486866666666667)(4.6, -0.3752333333333333)(4.8, -0.29497166666666663)(5., -0.21197499999999997)(5.2, -0.129825)(5.4, -0.051539833333333326)(5.6, 0.020459166666666667)(5.8, 0.08435233333333332)(6., 0.138898)(6.2, 0.18338833333333332)(6.4, 0.21759333333333333)(6.6, 0.2416883333333333)(6.8, 0.25619)(7., 0.2618783333333333)(7.2, 0.2597233333333333)(7.4, 0.25082666666666664)(7.6, 0.23635666666666666)(7.8, 0.217495)(8., 0.19539833333333334)(8.2, 0.17115666666666665)(8.4, 0.14577166666666666)(8.6, 0.12013283333333333)(8.8, 0.09500633333333333)(9., 0.0710295)(9.2, 0.04870983333333333)(9.4, 0.028428666666666665)(9.6, 0.010449166666666666)(9.8, -0.005073783333333333)(10., -0.01808133333333333)(10.2, -0.02859733333333333)(10.4, -0.036714666666666666)(10.6, -0.04258133333333333)(10.8, -0.046387)(11., -0.04835099999999999)(11.2, -0.048711000000000004)(11.4, -0.04771216666666667)(11.6, -0.04559966666666666)(11.8, -0.04261083333333333)(12., -0.0389695)(12.2, -0.034881833333333334)};
 
\legend{{\rm Re},{\rm Im}}
	\end{axis}
\end{tikzpicture} 
}
\caption{\it Spectral version of the Newman-Penrose  scalar at tree level. The two plots show different circular polarisations of the graviton.}
\label{Figure-1}
\end{figure}

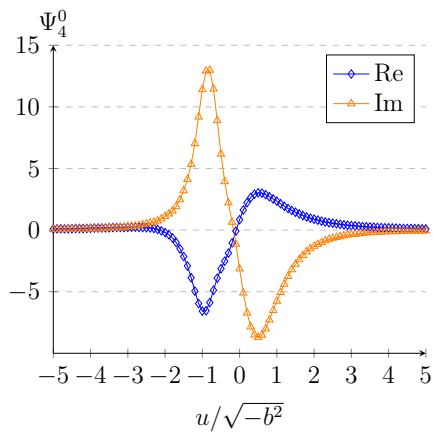
\begin{figure}[t]
\centering
\scalebox{0.8}{
\begin{tikzpicture}
	\begin{axis}[width=0.5\textwidth,
			title={},
			xmin=-5, xmax=5,
			ymin=-10, ymax=15,
			axis lines=left,
			xtick={-5,...,5},
			ytick={-5,0,5,10,15},
			compat=newest,
			xlabel=$u/\sqrt{-b^2}$, %xlabel style={at={(1,0)}, anchor=west},
			ylabel=$\Psi_4^0$, ylabel style={rotate=-90,at={(0,1)},anchor=south},
			legend pos=north east,
			ymajorgrids=true,
			grid style=dashed,
		]
		\addplot[
			color=blue,
			mark=diamond,
		]
		coordinates {(-5., 0.091358)(-4.9, 0.100133)(-4.8, 0.102375)(-4.7, 0.10132116666666666)(-4.6, 0.10524033333333332)(-4.5, 0.116475)(-4.4, 0.12796733333333332)(-4.3, 0.132376)(-4.2, 0.13207066666666667)(-4.1, 0.136458)(-4., 0.14955916666666666)(-3.9, 0.16418566666666667)(-3.8, 0.17105166666666666)(-3.7, 0.1713083333333333)(-3.6, 0.17559166666666665)(-3.5, 0.18996666666666664)(-3.4, 0.207125)(-3.3, 0.21524499999999996)(-3.2, 0.21365666666666666)(-3.1, 0.21423999999999999)(-3., 0.22546)(-2.9, 0.23954999999999999)(-2.8, 0.2408033333333333)(-2.7, 0.22530666666666666)(-2.6, 0.20565166666666668)(-2.5, 0.19238833333333333)(-2.4, 0.17539166666666664)(-2.3, 0.13097116666666667)(-2.2, 0.047387)(-2.1, -0.06552849999999999)(-2., -0.199195)(-1.9, -0.37455499999999997)(-1.8, -0.6368499999999999)(-1.7, -1.019345)(-1.6, -1.5232299999999999)(-1.5, -2.1447833333333333)(-1.4, -2.9129666666666663)(-1.3, -3.865966666666666)(-1.2, -4.955016666666666)(-1.1, -5.96665)(-1., -6.577183333333332)(-0.9, -6.542883333333333)(-0.8, -5.8824000000000005)(-0.7, -4.882516666666667)(-0.6, -3.9027999999999996)(-0.5, -3.135933333333333)(-0.4, -2.5210833333333333)(-0.3, -1.8651166666666665)(-0.2, -1.0414666666666665)(-0.1, -0.0896915)(0., 0.8434166666666667)(0.1, 1.62974)(0.2, 2.22665)(0.3, 2.6497166666666665)(0.4, 2.915683333333333)(0.5, 3.0269666666666666)(0.6, 2.9981)(0.7, 2.873616666666667)(0.8, 2.7078166666666665)(0.9, 2.5312666666666663)(1., 2.34455)(1.1, 2.1429166666666664)(1.2, 1.9381833333333331)(1.3, 1.7519)(1.4, 1.5931433333333331)(1.5, 1.4513116666666666)(1.6, 1.3127533333333332)(1.7, 1.1784866666666667)(1.8, 1.0610283333333332)(1.9, 0.9664283333333332)(2., 0.88576)(2.1, 0.8063083333333334)(2.2, 0.7267933333333333)(2.3, 0.6569633333333333)(2.4, 0.6031716666666667)(2.5, 0.5593766666666666)(2.6, 0.514705)(2.7, 0.46686833333333333)(2.8, 0.4238383333333333)(2.9, 0.39231166666666667)(3., 0.36854666666666663)(3.1, 0.34321499999999994)(3.2, 0.3132716666666666)(3.3, 0.2851433333333333)(3.4, 0.2656033333333333)(3.5, 0.25266666666666665)(3.6, 0.23829166666666668)(3.7, 0.21881666666666666)(3.8, 0.19927666666666666)(3.9, 0.18631166666666665)(4., 0.17931333333333332)(4.1, 0.17136833333333334)(4.2, 0.15834183333333332)(4.3, 0.14401233333333333)(4.4, 0.13474516666666667)(4.5, 0.1310325)(4.6, 0.12698916666666665)(4.7, 0.11815666666666666)(4.8, 0.10717633333333333)(4.9, 0.10004066666666667)(5., 0.09812716666666667)};
    \addplot[
			color=orange,
			mark=triangle,
		]
		coordinates {(-5., 0.054309833333333335)(-4.9, 0.04627333333333333)(-4.8, 0.03955633333333333)(-4.7, 0.048403666666666664)(-4.6, 0.067119)(-4.5, 0.076861)(-4.4, 0.07064633333333332)(-4.3, 0.06355566666666666)(-4.2, 0.073492)(-4.1, 0.09704333333333333)(-4., 0.11302883333333333)(-3.9, 0.11050566666666667)(-3.8, 0.10451983333333333)(-3.7, 0.11734916666666667)(-3.6, 0.148672)(-3.5, 0.1747433333333333)(-3.4, 0.179695)(-3.3, 0.17832833333333334)(-3.2, 0.19846999999999998)(-3.1, 0.24410833333333332)(-3., 0.28858833333333334)(-2.9, 0.3097816666666666)(-2.8, 0.3222733333333333)(-2.7, 0.36173833333333333)(-2.6, 0.43875166666666665)(-2.5, 0.5234583333333334)(-2.4, 0.5859383333333332)(-2.3, 0.6421366666666666)(-2.2, 0.7417400000000001)(-2.1, 0.9083916666666667)(-2., 1.1115866666666667)(-1.9, 1.3162616666666667)(-1.8, 1.5527916666666668)(-1.7, 1.9121499999999998)(-1.6, 2.4631999999999996)(-1.5, 3.20345)(-1.4, 4.129099999999999)(-1.3, 5.3449833333333325)(-1.2, 7.028833333333333)(-1.1, 9.188099999999999)(-1., 11.410733333333333)(-0.9, 12.916166666666665)(-0.8, 12.983799999999999)(-0.7, 11.458416666666666)(-0.6, 8.893583333333332)(-0.5, 6.1771666666666665)(-0.4, 3.93775)(-0.3, 2.2185499999999996)(-0.2, 0.6443066666666666)(-0.1, -1.1354883333333332)(0., -3.1404666666666667)(0.1, -5.102483333333333)(0.2, -6.7215)(0.3, -7.853616666666666)(0.4, -8.502949999999998)(0.5, -8.715499999999999)(0.6, -8.529966666666667)(0.7, -8.01405)(0.8, -7.29505)(0.9, -6.517366666666666)(1., -5.768883333333333)(1.1, -5.062049999999999)(1.2, -4.385183333333333)(1.3, -3.7552166666666666)(1.4, -3.2112)(1.5, -2.768133333333333)(1.6, -2.398133333333333)(1.7, -2.0633999999999997)(1.8, -1.7568333333333332)(1.9, -1.499753333333333)(2., -1.303675)(2.1, -1.1483183333333333)(2.2, -1.0027283333333332)(2.3, -0.8594916666666667)(2.4, -0.7373583333333333)(2.5, -0.6509383333333333)(2.6, -0.5883683333333333)(2.7, -0.5249783333333333)(2.8, -0.45312166666666664)(2.9, -0.3888133333333333)(3., -0.3478866666666667)(3.1, -0.32400666666666667)(3.2, -0.2966933333333333)(3.3, -0.25738666666666665)(3.4, -0.21898666666666666)(3.5, -0.19734833333333335)(3.6, -0.18996833333333332)(3.7, -0.17951166666666665)(3.8, -0.156453)(3.9, -0.13066216666666666)(4., -0.11743699999999999)(4.1, -0.11714416666666666)(4.2, -0.11510533333333334)(4.3, -0.10103116666666667)(4.4, -0.08201599999999999)(4.5, -0.07249233333333333)(4.6, -0.07506633333333333)(4.7, -0.07744933333333333)(4.8, -0.06886583333333332)(4.9, -0.05391666666666667)(5., -0.045976166666666665)};
 
\legend{{\rm Re},{\rm Im}}
	\end{axis}
\end{tikzpicture}
}
\caption{\it Newman-Penrose scalar $\Psi_4^0$ in the time domain at tree level as a function of the rescaled retarded time $u/\sqrt{-b^2}$.}
\label{Figure-2}
\end{figure}

In Figure~\ref{Figure-1} we plot the quantities 
$\omega^2   W_{\pm}^{(0)} (\omega, \hat{\mathbf{n}})$  at 
$\theta \to \frac{\pi }{4}$, 
$\phi  \to \frac{\pi }{2}$,  
$y \to 2$, where we stripped off an overall dimensionful factor of $ \frac{\kappa^3 (m_1+ m_2)^2\chi(1-\chi)}{ (4\pi)^2 (-b^2)}$;  these appear in the integrand of the   Newman-Penrose scalar  \eqref{tdwf2}. 
In Figure~\ref{Figure-2} we then plot the Newman-Penrose scalar in the time domain  given in  \eqref{tdwf2} (up to a factor of $\frac{\kappa^4 (m_1+ m_2)^2\chi(1-\chi)}{(4\pi)^4 (-b^2)^{3/2}}$),
%the additional (4\pi)^2 is 2\pi (FT) and 8\pi definition
as a function of $u/\sqrt{-b^2}$, where $u$ is as usual the retarded~time.

At tree level the waveforms depend on the mass ratio $\chi$ only through the prefactor $\chi(1-\chi)$ in  \eqref{eq: treeLevelWaveform}. Thus, they are maximised when both masses are equal, for a given total mass. As such we have only plotted the equal-mass case in Figures~\ref{Figure-1} and \ref{Figure-2}.

Finally, we mention that tree-level  waveforms for non-spinning objects were derived  in \cite{Kovacs:1978eu,Jakobsen:2021smu} in the time domain, see also  \cite{DiVecchia:2021bdo} for a derivation in the frequency domain and \cite{Mougiakakos:2021ckm} for a one-parameter integral representation of the time-domain waveform.

\subsubsection{One loop}

In the present section we  evaluate numerically the following quantity,
\begin{equation}
\begin{split}
    \widehat{W}^{(1)}(b, k^h) &\coloneqq -i \int\!d\mu^{(4)} \ e^{i q_1\Cdot b}\  \Big[\cM_{\bar{m}_1^3 \bar{m}_2^2}^{(1)} + \cM_{\bar{m}_1^2 \bar{m}_2^3}^{(1)} \\
    &- i G \left(\bar{m}_1 w_1 \log\frac{w_1^2}{\mu_{\rm IR}^2} +\bar{m}_2 w_2 \log\frac{w_2^2}{\mu_{\rm IR}^2} \right) \cM_{\bar{m}_1^2 \bar{m}_2^2}^{(0)}\Big]_{\rm fin}\, ,
\end{split}
\end{equation}
where the subscript ``fin'' means that we are dropping all infrared divergences in the corresponding amplitudes, as discussed near \eqref{intwf2}.   This contribution is just the part of the waveform which comes directly from the amplitude (the first term in the KMOC formula \eqref{eq: KMOCkernel}), and  we plot  it purely for illustrative purposes. 
This corresponds to isolating the new contributions of the one-loop amplitude from the lower order in the PM expansion and the tails      \cite{Blanchet:1993ec,Porto:2012as}:
\begin{equation}
    e^{i \theta_{\rm tail} (\mu_{\rm IR},\omega)} \cM_{\bar{m}_1^2 \bar{m}_2^2}^{(0)}\, , 
\end{equation}
with%
\footnote{{From the field theory  viewpoint such exponentiation is  natural. Indeed, we know from \cite{Weinberg:1965nx} that the infrared divergences exponentiate as per   \eqref{eq:classicalWeinberg} and we expect them  to be accompanied by   an infrared-running logarithm, \textit{i.e.} schematically $\frac{1}{\epsilon}\rightarrow \frac{1}{\epsilon} - \log\left( -\frac{\omega^2}{\mu_{\rm IR}^2}\right)$.}}
\begin{equation}
    \theta_{\rm tail} (\mu_{\rm IR},\omega) = G \Big( \bar{m}_1 w_1 \log\frac{w_1^2}{\mu_{\rm IR}^2} + \bar{m}_2 w_2 \log\frac{w_2^2}{\mu_{\rm IR}^2}\Big) \ .
\end{equation}
We now  present the result of the waveform%
\footnote{Recall that $d\mu^{(4)}$ is proportional to $(m_1 m_2)^{-1}$.} 
\begin{align}\label{eq: 1-loopWaveform}
\begin{split}
    \widehat{W}_{\pm}^{(1)}&\coloneqq m_1^2m_2\widehat{W}^{(1)}_{\pm, m_1^2m_2}+m_1m_2^2\widehat{W}^{(1)}_{\pm, m_1m_2^2}\\%&=(m_1^2 m_2+m_1 m_2^2)\Big[{(1-\chi)\over m_1^2m_2}\widehat{W}^{(m_1^2m_2)}_{\pm}+{\chi \over m_1m_2^2}\widehat{W}^{(m_1m_2^2)}_{\pm}\Big]\nn\\ &
    &= (m_1+m_2)^3 \, \chi (1-\chi)\Big[(1-\chi)\widehat{W}^{(1)}_{\pm, m_1^2m_2}+\chi\widehat{W}^{(1)}_{\pm, m_1m_2^2}\Big]\, .
\end{split}
\end{align}
As before,  $W_{\pm}$ is shorthand for  $W(b, h^{\pm})$ and the subscripts $m_1^2m_2$, $m_1 m_2^2$ indicate the mass dependence of the corresponding terms. 
In the plots  displayed below
we  will show results for several values of $\chi$.

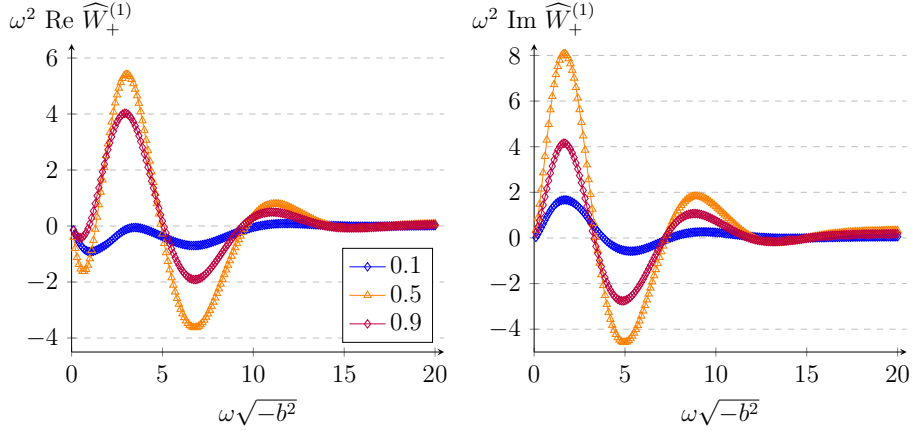
\begin{figure}[h]
    \centering
\scalebox{0.8}{
\begin{tikzpicture}
	\begin{axis}[width=0.5\textwidth,
			title={},
			xmin=0, xmax=20.5,
			ymin=-4.5, ymax=6.5,
			axis lines=left,
			xtick={0,5,10,15,20},
			ytick={-4,-2,0,2,4,6},
			compat=newest,
			xlabel=$\omega \sqrt{-b^2}$, %xlabel style={at={(1,0)}, anchor=west},
			ylabel=$\omega^2$ Re  $\widehat{W}_{+}^{(1)}$, ylabel style={rotate=-90,at={(0,1)},anchor=south},
			legend pos=south east,
			ymajorgrids=true,
			grid style=dashed,
		]
		\addplot[
			color=blue,
			mark=diamond,
		]
		coordinates {(0.1, -0.13705327455528726)(0.2, -0.30499920045433626)(0.3, -0.4525339948835804)(0.4, -0.5786935948851213)(0.5, -0.6834045706022147)(0.6, -0.7672140929028574)(0.7, -0.8310838560398909)(0.8, -0.8762597988729091)(0.9, -0.9041489322053314)(1., -0.9162435402226023)(1.1, -0.9142680402056803)(1.2, -0.9007000976434147)(1.3, -0.8781666064431997)(1.4, -0.8492376115910786)(1.5, -0.816390778575901)(1.6, -0.7815589707235766)(1.7, -0.7447113948683736)(1.8, -0.7052535060411703)(1.9, -0.6624983797756513)(2., -0.6156809243386439)(2.1, -0.5643771415250771)(2.2, -0.5098021770288135)(2.3, -0.4535406945324916)(2.4, -0.3970991904518938)(2.5, -0.34193204969156477)(2.6, -0.28936100967289824)(2.7, -0.24036908299424273)(2.8, -0.19582771624579687)(2.9, -0.1565619293986565)(3., -0.12336997605678177)(3.1, -0.09684924363536174)(3.2, -0.07692298609057358)(3.3, -0.06332543471510474)(3.4, -0.05577921414686703)(3.5, -0.05398041952714209)(3.6, -0.057567112723333175)(3.7, -0.06596890955789252)(3.8, -0.07855455013332646)(3.9, -0.09468566843697253)(4., -0.11372129288060638)(4.1, -0.13506403604903888)(4.2, -0.15833916880236948)(4.3, -0.18322123106586785)(4.4, -0.20938642085834303)(4.5, -0.23650525130646907)(4.6, -0.2642835292422583)(4.7, -0.29254454926851364)(4.8, -0.3211395567077018)(4.9, -0.3499287979615034)(5., -0.3787677820214864)(5.1, -0.4075072804691066)(5.2, -0.43597674654031293)(5.3, -0.46398668383060454)(5.4, -0.49136654557593057)(5.5, -0.5179363101920152)(5.6, -0.5435088499794137)(5.7, -0.5678496631375564)(5.8, -0.5907124043405919)(5.9, -0.6118625717879503)(6., -0.6310561888588369)(6.1, -0.6480753346880752)(6.2, -0.6628015740228527)(6.3, -0.6751283151356373)(6.4, -0.6849631785292348)(6.5, -0.6922066785912823)(6.6, -0.6967830167599793)(6.7, -0.6986874556252136)(6.8, -0.6979294700072101)(6.9, -0.6945280095464186)(7., -0.6884925490630643)(7.1, -0.6798728313633289)(7.2, -0.668810978750588)(7.3, -0.6554846441040612)(7.4, -0.6400738490080242)(7.5, -0.6227515089315842)(7.6, -0.6037000141640735)(7.7, -0.5831064924049364)(7.8, -0.5611628087637304)(7.9, -0.5380631970550687)(8., -0.514001891093565)(8.1, -0.4891660185786637)(8.2, -0.46371902015924693)(8.3, -0.4378243364841969)(8.4, -0.41164777690745163)(8.5, -0.3853480446677808)(8.6, -0.35908147429889764)(8.7, -0.33295702623339013)(8.8, -0.30708366090384637)(8.9, -0.2815656013327418)(9., -0.2565047018374957)(9.1, -0.23200400108805547)(9.2, -0.2081352708476257)(9.3, -0.1849667298218269)(9.4, -0.16256375427021177)(9.5, -0.14099598412143444)(9.6, -0.12032429509544079)(9.7, -0.10057948035796227)(9.8, -0.08178712192360646)(9.9, -0.063974696771026)(10., -0.047162575763704764)(10.1, -0.031373972211194104)(10.2, -0.016611610124308708)(10.3, -0.002869046625295551)(10.4, 0.00985765033623876)(10.5, 0.021568196515687473)(10.6, 0.032270527074989064)(10.7, 0.041984112769706)(10.8, 0.05072273946326572)(10.9, 0.058509857335725224)(11., 0.06536394228652334)(11.1, 0.0713041808266158)(11.2, 0.07636373482679025)(11.3, 0.08056984439519368)(11.4, 0.08396064568323196)(11.5, 0.08656290505804079)(11.6, 0.08841475867102597)(11.7, 0.08956594932836905)(11.8, 0.0900664567067571)(11.9, 0.08996886605843911)(12., 0.08932884195223725)(12.1, 0.08819517971231056)(12.2, 0.08661098977068289)(12.3, 0.08462554119252455)(12.4, 0.08228099692783698)(12.5, 0.07962354672521725)(12.6, 0.07669464292314997)(12.7, 0.07353550098961407)(12.8, 0.07018378333500414)(12.9, 0.06667786298123157)(13., 0.0630558760797022)(13.1, 0.05935430068828243)(13.2, 0.05560321936118681)(13.3, 0.05182560853746106)(13.4, 0.04805178764182532)(13.5, 0.04430662807737034)(13.6, 0.040614053765164373)(13.7, 0.03699514618020814)(13.8, 0.0334676706104236)(13.9, 0.030045365545137058)(14., 0.026749549329854884)(14.1, 0.023594055202105634)(14.2, 0.02058940021233912)(14.3, 0.01774051126707235)(14.4, 0.015054991909535936)(14.5, 0.012542435395207732)(14.6, 0.010206418468722693)(14.7, 0.008038840158788406)(14.8, 0.006036336904224981)(14.9, 0.0041891970143017425)(15., 0.0024897695716869683)(15.1, 0.0009297688460938553)(15.2, -0.0004906014538427582)(15.3, -0.0017763416644966396)(15.4, -0.002926061355999764)(15.5, -0.003946381058984382)(15.6, -0.004835924555812804)(15.7, -0.005599689814153746)(15.8, -0.0062421773736141)(15.9, -0.006768622072368207)(16., -0.007179284467972254)(16.1, -0.007481270675595026)(16.2, -0.007679578662905234)(16.3, -0.007778708969509774)(16.4, -0.007783635876026799)(16.5, -0.007701276001220589)(16.6, -0.007536840496214919)(16.7, -0.007296701177611133)(16.8, -0.0069839847360834975)(16.9, -0.006607668563795241)(17., -0.006172537444960014)(17.1, -0.005685673807696036)(17.2, -0.005149683227565196)(17.3, -0.00457359047083185)(17.4, -0.003962369818114146)(17.5, -0.0033207586795246085)(17.6, -0.0026558868572825836)(17.7, -0.0019724893927955374)(17.8, -0.0012752918526507122)(17.9, -0.000571651434752856)(18., 0.00013369113459842978)(18.1, 0.0008364679226327732)(18.2, 0.0015373890671260412)(18.3, 0.002231002756520743)(18.4, 0.0029206133343703624)(18.5, 0.0036033925668377243)(18.6, 0.004277905018658734)(18.7, 0.004944174376883953)(18.8, 0.0055978895983109875)(18.9, 0.006240495593024155)(19., 0.006870097396978447)(19.1, 0.007484326305117602)(19.2, 0.008081974277862928)(19.3, 0.008666618059849378)(19.4, 0.009239915744616337)(19.5, 0.009798764328540098)(19.6, 0.010346503685749997)(19.7, 0.010882660075234772)(19.8, 0.011408631032977628)(19.9, 0.011925364041001061)(20., 0.01243451719284446)};
    \addplot[
			color=orange,
			mark=triangle,
		]
		coordinates {(0.1, -0.40000980258713775)(0.2, -0.7969047580252917)(0.3, -1.1206277823810227)(0.4, -1.3664204103857518)(0.5, -1.5320126330272292)(0.6, -1.616832013250191)(0.7, -1.6214115096922967)(0.8, -1.5469946925080829)(0.9, -1.395226495764398)(1., -1.1679163469347753)(1.1, -0.8691502622412658)(1.2, -0.5118271566624663)(1.3, -0.1109198780484561)(1.4, 0.3188211208365232)(1.5, 0.7629269999404746)(1.6, 1.2089607417707722)(1.7, 1.6518756886103254)(1.8, 2.088678060457471)(1.9, 2.5166570059700426)(2., 2.9333451240477677)(2.1, 3.335788113106587)(2.2, 3.7183142404840086)(2.3, 4.074699075671078)(2.4, 4.398941899191933)(2.5, 4.685186745768594)(2.6, 4.928268522431157)(2.7, 5.125239507331832)(2.8, 5.273809952249563)(2.9, 5.371788805007311)(3., 5.417077129735776)(3.1, 5.408490573906828)(3.2, 5.348259668115114)(3.3, 5.23949662761511)(3.4, 5.085353146078896)(3.5, 4.889040134804961)(3.6, 4.654064593223826)(3.7, 4.38499285830506)(3.8, 4.086674195677733)(3.9, 3.763984189915983)(4., 3.4217984255939475)(4.1, 3.0647621965164062)(4.2, 2.696487777894158)(4.3, 2.320343994696108)(4.4, 1.939686512418626)(4.5, 1.5578907357668856)(4.6, 1.1780623002590953)(4.7, 0.8023396201821582)(4.8, 0.432603842134922)(4.9, 0.07071768945468547)(5., -0.2814482188377315)(5.1, -0.6221436728952024)(5.2, -0.9500612791213963)(5.3, -1.2640068153837816)(5.4, -1.5628123784948953)(5.5, -1.84528703619034)(5.6, -2.1102859143595896)(5.7, -2.3566246604745125)(5.8, -2.583151820688315)(5.9, -2.7887291006267394)(6., -2.9722050464429906)(6.1, -3.1326782342684356)(6.2, -3.2701552438393433)(6.3, -3.3848926848701404)(6.4, -3.477134007602722)(6.5, -3.5471489812240495)(6.6, -3.5952271141298904)(6.7, -3.6217697702325533)(6.8, -3.627211212125687)(6.9, -3.611992282139207)(7., -3.5765472428667615)(7.1, -3.5215274881988217)(7.2, -3.4483608209054126)(7.3, -3.358685595317115)(7.4, -3.254140165764509)(7.5, -3.136362886578176)(7.6, -3.0069657931436273)(7.7, -2.8674951234836996)(7.8, -2.7194707966761618)(7.9, -2.56441273179878)(8., -2.4038474276655917)(8.1, -2.239255324936758)(8.2, -2.071899732975619)(8.3, -1.903024221936713)(8.4, -1.7338657822383086)(8.5, -1.5656416650898741)(8.6, -1.3995164838107383)(8.7, -1.2362600675441855)(8.8, -1.0765698683345597)(8.9, -0.9211433382262046)(9., -0.7706713495271962)(9.1, -0.6257987163917392)(9.2, -0.48697944062228365)(9.3, -0.3546115962630069)(9.4, -0.22909259938445942)(9.5, -0.11082644579345913)(9.6, -0.0001443554658648399)(9.7, 0.10288733601322286)(9.8, 0.19827311470799108)(9.9, 0.28601192391279523)(10., 0.3661110013375295)(10.1, 0.43860258768990407)(10.2, 0.5036597300337515)(10.3, 0.5614975857450153)(10.4, 0.6123240744897452)(10.5, 0.6563438260658573)(10.6, 0.6937673920339079)(10.7, 0.7248434864248046)(10.8, 0.7497806868782236)(10.9, 0.7688290233723261)(11., 0.7822122069402034)(11.1, 0.7901736878237487)(11.2, 0.7929634960011234)(11.3, 0.7908250817142206)(11.4, 0.7840282141500037)(11.5, 0.7728097638140988)(11.6, 0.757472398574806)(11.7, 0.7384043228719014)(11.8, 0.7160463790353003)(11.9, 0.6908262499223833)(12., 0.6631913575993331)(12.1, 0.6335516196356082)(12.2, 0.6022162836357763)(12.3, 0.5695024928879255)(12.4, 0.5357017297087014)(12.5, 0.5011206098081645)(12.6, 0.46604469374031965)(12.7, 0.43072927527234167)(12.8, 0.3954079350417231)(12.9, 0.3603274131584911)(13., 0.3257265540491521)(13.1, 0.2918264368522903)(13.2, 0.25878036942293625)(13.3, 0.2267199464864376)(13.4, 0.1957833425044096)(13.5, 0.16610149422857345)(13.6, 0.13779875867438288)(13.7, 0.11096133038694068)(13.8, 0.08567014012233584)(13.9, 0.061991840608957016)(14., 0.040010915660477535)(14.1, 0.01978651710594127)(14.2, 0.0013350350683902878)(14.3, -0.015353682985235997)(14.4, -0.030279814707816804)(14.5, -0.043440208405680046)(14.6, -0.054864670284117026)(14.7, -0.06469144060210565)(14.8, -0.07307191909115865)(14.9, -0.0801681646555419)(15., -0.08613730139732073)(15.1, -0.09112079364624412)(15.2, -0.09516930537157098)(15.3, -0.09832428891178588)(15.4, -0.10061930092185252)(15.5, -0.10209447779300201)(15.6, -0.10278206023294457)(15.7, -0.10272547450104498)(15.8, -0.10196025117314719)(15.9, -0.10053310637674968)(16., -0.09847364892505571)(16.1, -0.095827936971937)(16.2, -0.09263873880313167)(16.3, -0.0889415849944836)(16.4, -0.08478055977898433)(16.5, -0.08019777346874515)(16.6, -0.07523928421763787)(16.7, -0.06994193854875984)(16.8, -0.06434764938213437)(16.9, -0.058503067047897275)(17., -0.052447538368927564)(17.1, -0.046227187296459665)(17.2, -0.03987820237996422)(17.3, -0.03344670757067568)(17.4, -0.026975471154332086)(17.5, -0.020503971548537794)(17.6, -0.014078332704527227)(17.7, -0.007737967242542078)(17.8, -0.0015224522762307075)(17.9, 0.004518890491346506)(18., 0.010346569483112668)(18.1, 0.01592763337984068)(18.2, 0.021272024163030565)(18.3, 0.026382900106090675)(18.4, 0.03128006621518587)(18.5, 0.03597668196285093)(18.6, 0.04047932708535324)(18.7, 0.044787935785330135)(18.8, 0.048895928326514224)(18.9, 0.05279679077000079)(19., 0.056490457318427126)(19.1, 0.05997028243816322)(19.2, 0.06324626732833548)(19.3, 0.06634150686324244)(19.4, 0.06927870113300663)(19.5, 0.07206785133675449)(19.6, 0.07472869668328817)(19.7, 0.07728097638140988)(19.8, 0.07974508761354852)(19.9, 0.08213287390498542)(20., 0.0844686802799099)};

 \addplot[
			color=purple,
			mark=diamond,
		]
		coordinates {(0.1, -0.15095378330745193)(0.2, -0.26877459402893)(0.3, -0.35431800843075584)(0.4, -0.4051291005926201)(0.5, -0.41964689388244647)(0.6, -0.3969073253423366)(0.7, -0.3363324309385627)(0.8, -0.23757874843796675)(0.9, -0.10041580283857458)(1., 0.07534234920653031)(1.1, 0.28847985139196897)(1.2, 0.5321840711054279)(1.3, 0.7983057154713449)(1.4, 1.0787888185933754)(1.5, 1.3656982185330429)(1.6, 1.6520107047985328)(1.7, 1.934061890667808)(1.8, 2.20910170957055)(1.9, 2.4744914240740816)(2., 2.7276823075378682)(2.1, 2.966139845551708)(2.2, 3.1869979049975248)(2.3, 3.3873192916055554)(2.4, 3.5643326204599735)(2.5, 3.715266506644953)(2.6, 3.8377048710031065)(2.7, 3.9305344221579923)(2.8, 3.9929734874410445)(2.9, 4.024240394183695)(3., 4.023671904970192)(3.1, 3.990960088143223)(3.2, 3.927668289039916)(3.3, 3.835762532856972)(3.4, 3.7172325318416495)(3.5, 3.5740916852917706)(3.6, 3.408495514808533)(3.7, 3.2231680312066375)(3.8, 3.0209516805535976)(3.9, 2.804759970068616)(4., 2.5774116587686415)(4.1, 2.3416923437998403)(4.2, 2.099809658274647)(4.3, 1.8538693809880769)(4.4, 1.6059630785048105)(4.5, 1.358187054799638)(4.6, 1.1124907541338633)(4.7, 0.8702290757996676)(4.8, 0.6326124280808005)(4.9, 0.4008441131458432)(5., 0.17612435384680292)(5.1, -0.04043497966291101)(5.2, -0.24806737442709245)(5.3, -0.4460958545406618)(5.4, -0.6338559982353378)(5.5, -0.810672724570086)(5.6, -0.9758970083594906)(5.7, -1.1289200924040925)(5.8, -1.2691569065549952)(5.9, -1.3960247493683582)(6., -1.5089338132851726)(6.1, -1.6074529939851987)(6.2, -1.6917125702465305)(6.3, -1.761992049265808)(6.4, -1.8185733069447252)(6.5, -1.8617405878900335)(6.6, -1.8917805054135415)(6.7, -1.9089867789422243)(6.8, -1.913660234018228)(6.9, -1.9061064335938105)(7., -1.8866214658010039)(7.1, -1.855629328844879)(7.2, -1.8140111810063653)(7.3, -1.7627689845242618)(7.4, -1.7029070703424223)(7.5, -1.6354274006996465)(7.6, -1.5613153568993383)(7.7, -1.4814899965033275)(7.8, -1.3968537961380498)(7.9, -1.3083139698400532)(8., -1.216770625530717)(8.1, -1.1230954466707457)(8.2, -1.028048787583199)(8.3, -0.9323507346051803)(8.4, -0.8367332175990744)(8.5, -0.7419139541969288)(8.6, -0.648570394044834)(8.7, -0.5571478536933673)(8.8, -0.4680466442970367)(8.9, -0.38165760219012546)(9., -0.29837630111702923)(9.1, -0.21857107471399678)(9.2, -0.1424896895299129)(9.3, -0.07035338261703246)(9.4, -0.0023829172865990145)(9.5, 0.061200943150143856)(9.6, 0.12022054581397654)(9.7, 0.17465859915798834)(9.8, 0.22454376451336003)(9.9, 0.2699044663407667)(10., 0.3107622598562204)(10.1, 0.3471668878659025)(10.2, 0.3792462604428513)(10.3, 0.4071472373005549)(10.4, 0.4310166781525015)(10.5, 0.4509990740071231)(10.6, 0.467243653282964)(10.7, 0.4799020131036252)(10.8, 0.48911864447753844)(10.9, 0.4950451445283046)(11., 0.4978283729694118)(11.1, 0.4976199269244608)(11.2, 0.4945690348119961)(11.3, 0.488824925050562)(11.4, 0.4805391947637595)(11.5, 0.46986344107518924)(11.6, 0.45696347333878934)(11.7, 0.4420832681753549)(11.8, 0.42548338314107476)(11.9, 0.40742674449719374)(12., 0.3881715410948444)(12.1, 0.3679617495548218)(12.2, 0.34698449757657046)(12.3, 0.3254150693342537)(12.4, 0.3034240115919224)(12.5, 0.28118423981868373)(12.6, 0.25885682595836335)(12.7, 0.23658934033596637)(12.8, 0.21450992989503656)(12.9, 0.19275787449288204)(13., 0.17146747970619294)(13.1, 0.15076073384536662)(13.2, 0.13071840975282165)(13.3, 0.111412752932774)(13.4, 0.09291221896134959)(13.5, 0.07528644776720254)(13.6, 0.05860128935089692)(13.7, 0.04289677482788353)(13.8, 0.028215067148163835)(13.9, 0.014588807067413119)(14., 0.0020582696777029894)(14.1, -0.009347739198777358)(14.2, -0.01962815601345766)(14.3, -0.028795163016442266)(14.4, -0.03685634006391122)(14.5, -0.04381938544729736)(14.6, -0.049708933699185824)(14.7, -0.05461665370525391)(14.8, -0.05864818971101089)(14.9, -0.061910133443988526)(15., -0.06450836602020171)(15.1, -0.06653692503038401)(15.2, -0.06803110417987374)(15.3, -0.06901719609479537)(15.4, -0.06952007217823969)(15.5, -0.06956176138722989)(15.6, -0.06916737199536235)(15.7, -0.06836272288775033)(15.8, -0.06716936928040582)(15.9, -0.06561526189299269)(16., -0.06372171907101729)(16.1, -0.0615147965700985)(16.2, -0.05902007640484394)(16.3, -0.05625935066177123)(16.4, -0.0532584382259936)(16.5, -0.05004102614807368)(16.6, -0.0466355388886866)(16.7, -0.043061399829293696)(16.8, -0.03934632281905325)(16.9, -0.03551446864953911)(17., -0.031589761241819524)(17.1, -0.027598019481007728)(17.2, -0.023562646173059604)(17.3, -0.019508062667105204)(17.4, -0.015459969413004954)(17.5, -0.011442100835422603)(17.6, -0.007480489002926459)(17.7, -0.003598844653129702)(17.8, 0.00017912431880055804)(17.9, 0.0038252454824071716)(18., 0.007315840314465726)(18.1, 0.010631435216967683)(18.2, 0.013778449380615518)(18.3, 0.01676467584504432)(18.4, 0.019601034340563463)(18.5, 0.022299818446414942)(18.6, 0.02486713942164391)(18.7, 0.02730311570150318)(18.8, 0.02960715510972869)(18.9, 0.031773099013174155)(19., 0.033803079246390214)(19.1, 0.03569425336330936)(19.2, 0.037455385572639735)(19.3, 0.039099266881685184)(19.4, 0.040640583261794504)(19.5, 0.04208999388572088)(19.6, 0.043458157926217496)(19.7, 0.04475976135463316)(19.8, 0.04600759517827168)(19.9, 0.047210423605841254)(20., 0.04838269573818504)};
\legend{0.1,0.5,0.9}
	\end{axis}
\end{tikzpicture}
\begin{tikzpicture}
	\begin{axis}[width=0.5\textwidth,
			title={},
			axis lines=left,
			xmin=0, xmax=20.5,
			ymin=-5, ymax=8.5,
			xtick={0,5,10,15,20},
			ytick={-4,-2,0,2,4,6,8},
			compat=newest,
			xlabel=$\omega \sqrt{-b^2}$, %xlabel style={at={(1,0)}, anchor=west},
			ylabel=$\omega^2$ Im $\widehat{W}_{+}^{(1)}$, ylabel style={rotate=-90,at={(0,1)},anchor=south},
			legend pos=north west,
			ymajorgrids=true,
			grid style=dashed,
		]
		\addplot[
			color=blue,
			mark=diamond,
		]
		coordinates {(0.1, 0.025697133673357922)(0.2, 0.1554538491733374)(0.3, 0.30050339825755207)(0.4, 0.45522247512243713)(0.5, 0.6142762822402807)(0.6, 0.7727616001446215)(0.7, 0.9262607939055321)(0.8, 1.0708465505397307)(0.9, 1.203062929370132)(1., 1.3199040436803404)(1.1, 1.4191646290629762)(1.2, 1.5006717700489327)(1.3, 1.5649679000960934)(1.4, 1.612865485038756)(1.5, 1.6453972802814507)(1.6, 1.6636860520208454)(1.7, 1.6685537409114628)(1.8, 1.6608412372482753)(1.9, 1.6414889169386193)(2., 1.611498742221293)(2.1, 1.571758977492395)(2.2, 1.522473331386763)(2.3, 1.4636868093004665)(2.4, 1.3954538914497991)(2.5, 1.3178219519358867)(2.6, 1.2311060285312114)(2.7, 1.1366657579380677)(2.8, 1.0361165970048254)(2.9, 0.9310408407090677)(3., 0.8229923595677017)(3.1, 0.7134729125863976)(3.2, 0.6038587174028429)(3.3, 0.4954738801434879)(3.4, 0.38960223894882634)(3.5, 0.28750631361384554)(3.6, 0.1902612593635825)(3.7, 0.0983663992238972)(3.8, 0.012148614492553705)(3.9, -0.06809671730993218)(4., -0.142095537008551)(4.1, -0.2096673492780156)(4.2, -0.2709016911694528)(4.3, -0.3259646089073064)(4.4, -0.37505365249326855)(4.5, -0.4183772679722904)(4.6, -0.45615100750449167)(4.7, -0.48854068044381066)(4.8, -0.5157049900290169)(4.9, -0.5378286952544988)(5., -0.555098923819701)(5.1, -0.5676980660139557)(5.2, -0.5757611380254697)(5.3, -0.5794184186323373)(5.4, -0.5788049240227655)(5.5, -0.574060407795074)(5.6, -0.5653459418930881)(5.7, -0.5528747097718716)(5.8, -0.536878844526938)(5.9, -0.5175857418436886)(6., -0.4952322722277493)(6.1, -0.47007188712014014)(6.2, -0.4424054120630065)(6.3, -0.4125620970591685)(6.4, -0.3808522424709966)(6.5, -0.34760509830131087)(6.6, -0.31314280843776315)(6.7, -0.2778017289983423)(6.8, -0.2419158473959815)(6.9, -0.2058255465472652)(7., -0.16986576134714895)(7.1, -0.13435271392064357)(7.2, -0.09951687926972339)(7.3, -0.06557309894299124)(7.4, -0.032731950819948785)(7.5, -0.0012044012477269198)(7.6, 0.028824061218128656)(7.7, 0.05724449509467036)(7.8, 0.08397604226609764)(7.9, 0.10893627179645264)(8., 0.13204488458432803)(8.1, 0.15323650437017108)(8.2, 0.17250970993094794)(8.3, 0.1898853458711538)(8.4, 0.2053754925865755)(8.5, 0.21900217903423636)(8.6, 0.23079264532228333)(8.7, 0.24080018731448233)(8.8, 0.24909302371645367)(8.9, 0.2557301352840982)(9., 0.2607825831691039)(9.1, 0.26431669111304595)(9.2, 0.2664153637928936)(9.3, 0.2671591371805597)(9.4, 0.2666332846580696)(9.5, 0.2649207109023926)(9.6, 0.2621090580006103)(9.7, 0.2582788619246355)(9.8, 0.25351539605649376)(9.9, 0.24789919636809787)(10., 0.24151553624147326)(10.1, 0.2344454253895441)(10.2, 0.22675992496399824)(10.3, 0.21852559557691656)(10.4, 0.20981184028644762)(10.5, 0.2006821403880991)(10.6, 0.19120779390406456)(10.7, 0.18145701953996432)(10.8, 0.17149495668484555)(10.9, 0.16138935030331736)(11., 0.1512060503959438)(11.1, 0.14101493376188454)(11.2, 0.1308745074160293)(11.3, 0.12084469959630144)(11.4, 0.11099041282124258)(11.5, 0.10136660104815794)(11.6, 0.09203224503294846)(11.7, 0.08301624297730052)(11.8, 0.0743474930829005)(11.9, 0.06605015614132229)(12., 0.058150998519701624)(12.1, 0.05067252291607297)(12.2, 0.04362609911470643)(12.3, 0.03701954384228763)(12.4, 0.030856883897412226)(12.5, 0.025148304711822146)(12.6, 0.01989700403734333)(12.7, 0.01509698905018345)(12.8, 0.010739732412139958)(12.9, 0.006814338079954053)(13., 0.0033122550283726307)(13.1, 0.00022376825047794004)(13.2, -0.0024574130606184372)(13.3, -0.004742976569405106)(13.4, -0.0066404514417739)(13.5, -0.008157109602247541)(13.6, -0.009310337345938988)(13.7, -0.010140189475349623)(13.8, -0.010696882537672188)(13.9, -0.011032433295942183)(14., -0.011190544358447635)(14.1, -0.011208949196734787)(14.2, -0.01106445818830284)(14.3, -0.010721090703347178)(14.4, -0.010138744565265303)(14.5, -0.009279449432005338)(14.6, -0.00812366348685313)(14.7, -0.006717007989192269)(14.8, -0.005116402921524724)(14.9, -0.0033818002088244663)(15., -0.0015767379955206565)(15.1, 0.00025021578991312154)(15.2, 0.0020860427444876553)(15.3, 0.003922480824966703)(15.4, 0.005753395085254551)(15.5, 0.007576492618856735)(15.6, 0.009383790889733656)(15.7, 0.01117109728993573)(15.8, 0.012936374733114573)(15.9, 0.014673559334326153)(16., 0.01638089825182884)(16.1, 0.018052351287729168)(16.2, 0.019687800006774323)(16.3, 0.021283383419722593)(16.4, 0.022839693702838053)(16.5, 0.02435218294241267)(16.6, 0.025819595724766636)(16.7, 0.02724413494560227)(16.8, 0.028621536935818304)(16.9, 0.029952749177437236)(17., 0.03123942976399845)(17.1, 0.032480394342973826)(17.2, 0.0336761166553746)(17.3, 0.034828254794740166)(17.4, 0.035939651207138046)(17.5, 0.037008884669534454)(17.6, 0.038036902663951926)(17.7, 0.03902417893140169)(17.8, 0.03996526545025436)(17.9, 0.04085518793989179)(18., 0.04169465701183085)(18.1, 0.04248011960848714)(18.2, 0.043213707564411306)(18.3, 0.04390157951274964)(18.4, 0.04454444606501899)(18.5, 0.04515083455942192)(18.6, 0.04572145560747531)(18.7, 0.046263415324347926)(18.8, 0.046774345004983515)(18.9, 0.04726300885809026)(19., 0.04772964375417377)(19.1, 0.04817780275081843)(19.2, 0.048607722718529874)(19.3, 0.04902129862135312)(19.4, 0.049420188552827544)(19.5, 0.04980178693739127)(19.6, 0.05017059431465118)(19.7, 0.05052353136803414)(19.8, 0.0508674673422033)(19.9, 0.05119671734502364)(20., 0.05151412382256267)}   ;
   \addplot[
			color=orange,
			mark=triangle,
		]
		coordinates  {(0.1, 0.3494254481370478)(0.2, 0.8709662694510661)(0.3, 1.4797629473278622)(0.4, 2.1502578121840012)(0.5, 2.8575465622474012)(0.6, 3.5783566703402947)(0.7, 4.291534284363012)(0.8, 4.978169242283066)(0.9, 5.6216477100252895)(1., 6.207514007010221)(1.1, 6.723964086374156)(1.2, 7.163687861124026)(1.3, 7.5213623246195045)(1.4, 7.793434419276201)(1.5, 7.977403845312506)(1.6, 8.07241523701366)(1.7, 8.081561070425336)(1.8, 8.009183971484013)(1.9, 7.860152945027565)(2., 7.639929172157923)(2.1, 7.353776441889006)(2.2, 7.005642595981247)(2.3, 6.599212286744389)(2.4, 6.138157007015633)(2.5, 5.626293003830068)(2.6, 5.068397165717824)(2.7, 4.473135005343056)(2.8, 3.8500931984473543)(2.9, 3.2087334057832306)(3., 2.5584251717954527)(3.1, 1.908044560708734)(3.2, 1.2649477176700186)(3.3, 0.6360045453160924)(3.4, 0.02792394013727677)(3.5, -0.5527353509577282)(3.6, -1.100237179688372)(3.7, -1.611640601335218)(3.8, -2.084822335004779)(3.9, -2.517801880080569)(4., -2.9087105914626488)(4.1, -3.2562325218975396)(4.2, -3.560729556879949)(4.3, -3.8230833810697056)(4.4, -4.044274375170653)(4.5, -4.225388195666911)(4.6, -4.367635514031678)(4.7, -4.472503350661386)(4.8, -4.541590581469012)(4.9, -4.576568459466473)(5., -4.579174035028359)(5.1, -4.5511509382655335)(5.2, -4.494098045090969)(5.3, -4.409620811153912)(5.4, -4.299370750257476)(5.5, -4.165019115413581)(5.6, -4.008309536733083)(5.7, -3.831189616151134)(5.8, -3.6356661732292865)(5.9, -3.423739447792828)(6., -3.197462317557186)(6.1, -2.9588810805015195)(6.2, -2.7100617738137895)(6.3, -2.453070434681958)(6.4, -2.1899796800302522)(6.5, -1.9228752862554375)(6.6, -1.6538430297542766)(6.7, -1.3849884261323346)(6.8, -1.1184169909951787)(6.9, -0.8562408196846408)(7., -0.6005627959117804)(7.1, -0.35332920566449205)(7.2, -0.1158027003324884)(7.3, 0.11090211276053416)(7.4, 0.3256916814469349)(7.5, 0.5274612680074185)(7.6, 0.7152633904194805)(7.7, 0.8886723397466204)(7.8, 1.0474150569337417)(7.9, 1.1912283525301486)(8., 1.3198292978763428)(8.1, 1.433086298246977)(8.2, 1.5312822823015484)(8.3, 1.6148449328974384)(8.4, 1.6841887734194927)(8.5, 1.7397480664613578)(8.6, 1.7820031327705554)(8.7, 1.811592206765021)(8.8, 1.829239059434169)(8.9, 1.8356345630860746)(9., 1.8315024887101523)(9.1, 1.817553447823279)(9.2, 1.7945835885138104)(9.3, 1.7633693196612987)(9.4, 1.7247002096178305)(9.5, 1.6793724064717608)(9.6, 1.628182058311444)(9.7, 1.571793718499887)(9.8, 1.510891679608898)(9.9, 1.4461536544740192)(10., 1.378244196458257)(10.1, 1.307794960243281)(10.2, 1.2352599476315418)(10.3, 1.1610339427990826)(10.4, 1.0855248893944813)(10.5, 1.0091144121212476)(10.6, 0.9322301938367615)(10.7, 0.8552670187170668)(10.8, 0.7786328304107415)(10.9, 0.7027355725663643)(11., 0.6279864787006473)(11.1, 0.5547954663830489)(11.2, 0.4835619256049999)(11.3, 0.41468919419969136)(11.4, 0.3485865317629552)(11.5, 0.2856493804444619)(11.6, 0.22623238802902376)(11.7, 0.17044741003318647)(11.8, 0.1183569539514903)(11.9, 0.0700268171466092)(12., 0.02551457231088284)(12.1, -0.015135498930609247)(12.2, -0.051961624845580597)(12.3, -0.08501743028461047)(12.4, -0.11435910619542239)(12.5, -0.1400450806360708)(12.6, -0.1621437828637366)(12.7, -0.1808039149180629)(12.8, -0.19618536439034728)(12.9, -0.20845262468727463)(13., -0.21776689934739601)(13.1, -0.224278206357608)(13.2, -0.22806616052674614)(13.3, -0.22920313895375163)(13.4, -0.22776151873756587)(13.5, -0.2238070972408627)(13.6, -0.21749515623955273)(13.7, -0.20932378176907745)(13.8, -0.19987922833086166)(13.9, -0.18974577650544983)(14., -0.17949981118986563)(14.1, -0.1695288788502584)(14.2, -0.15940134878748727)(14.3, -0.14848096050449527)(14.4, -0.13612947958334531)(14.5, -0.121709329579727)(14.6, -0.10481585671319571)(14.7, -0.08594648904556634)(14.8, -0.06581118012009057)(14.9, -0.045134490494533726)(15., -0.02463854622424215)(15.1, -0.00491685268000723)(15.2, 0.013930749220049606)(15.3, 0.031922906448510154)(15.4, 0.04908160848398005)(15.5, 0.06542896324031772)(15.6, 0.08098734182083231)(15.7, 0.09577793097630481)(15.8, 0.1098276418280689)(15.9, 0.12315884547943366)(16., 0.13579917682272216)(16.1, 0.14777232290849704)(16.2, 0.15910328673457436)(16.3, 0.1698183872460237)(16.4, 0.179953155018689)(16.5, 0.18952338141961197)(16.6, 0.19856591297189005)(16.7, 0.20711167443597994)(16.8, 0.21518172096793736)(16.9, 0.2228102671963527)(17., 0.23003152774981642)(17.1, 0.23687379549427828)(17.2, 0.2433699691110753)(17.3, 0.24954965741341073)(17.4, 0.2554483909771284)(17.5, 0.2610997264571922)(17.6, 0.26651945522064374)(17.7, 0.27168389021692047)(17.8, 0.27654763126577725)(17.9, 0.28106396223971575)(18., 0.285197352562892)(18.1, 0.2889168774748492)(18.2, 0.2922593834986848)(18.3, 0.29527750852453793)(18.4, 0.2980133628645199)(18.5, 0.30052945401317094)(18.6, 0.30286591836172216)(18.7, 0.30506486622228485)(18.8, 0.30714603680366126)(18.9, 0.309155488259723)(19., 0.3111195395355398)(19.1, 0.31306450957618115)(19.2, 0.31499039838164705)(19.3, 0.31689983784644443)(19.4, 0.3187901960760665)(19.5, 0.3206509454924852)(19.6, 0.32248932380559475)(19.7, 0.3242921715428604)(19.8, 0.32607199020319017)(19.9, 0.3278090405777819)(20., 0.3295105603765297)};
   
   \addplot[
			color=purple,
			mark=diamond,
		]
		coordinates  {(0.1, 0.2258894258558221)(0.2, 0.4716399698673853)(0.3, 0.7649259238185087)(0.4, 1.0929631496500438)(0.5, 1.4431572425778485)(0.6, 1.8036552025003905)(0.7, 2.163643890835836)(0.8, 2.5134329351990203)(0.9, 2.8445305279632453)(1., 3.149501303956906)(1.1, 3.4221155688820364)(1.2, 3.6571621716148592)(1.3, 3.8504248171552304)(1.4, 3.998421509070446)(1.5, 4.098309801292991)(1.6, 4.148478974384608)(1.7, 4.150184442025116)(1.8, 4.1057712222202145)(1.9, 4.017821203481227)(2., 3.889224205976793)(2.1, 3.7229647980778027)(2.2, 3.5216011812450168)(2.3, 3.2877389310403244)(2.4, 3.024007310076175)(2.5, 2.7331066421167067)(2.6, 2.418139930785622)(2.7, 2.0839938146139887)(2.8, 1.7359505058772695)(2.9, 1.3792472114548584)(3., 1.0190737641250243)(3.1, 0.6603168024188344)(3.2, 0.3069036393195705)(3.3, -0.03755084438640707)(3.4, -0.36949903913633547)(3.5, -0.6854748188213873)(3.6, -0.982432265609716)(3.7, -1.2587488165377825)(3.8, -1.5132235381420618)(3.9, -1.7447218207006057)(4., -1.9521777469380956)(4.1, -2.1348191190061905)(4.2, -2.2928235897841103)(4.3, -2.4266435819376007)(4.4, -2.5368357411548836)(4.5, -2.623909339023054)(4.6, -2.6885513000084287)(4.7, -2.731661732032387)(4.8, -2.7542354912185587)(4.9, -2.757291120741136)(5., -2.741918224925999)(5.1, -2.709135346947341)(5.2, -2.659984717029916)(5.3, -2.5955085653984793)(5.4, -2.5167491222777865)(5.5, -2.4247486178925914)(5.6, -2.320639293259789)(5.7, -2.2055818138569454)(5.8, -2.080798431493092)(5.9, -1.9475066605671478)(6., -1.8069429651184812)(6.1, -1.66032485954601)(6.2, -1.508839065082922)(6.3, -1.3536509846168974)(6.4, -1.1959354958558412)(6.5, -1.0368651078026039)(6.6, -0.8776241729853158)(6.7, -0.7193899378169386)(6.8, -0.563342017415491)(6.9, -0.4106671330141594)(7., -0.26254016232084987)(7.1, -0.12004407728728507)(7.2, 0.01613881659507892)(7.3, 0.14542262013057583)(7.4, 0.26723019833224754)(7.5, 0.38097541513392214)(7.6, 0.48616486927238045)(7.7, 0.5826019582279527)(7.8, 0.6701635093377134)(7.9, 0.7487452995791868)(8., 0.8182336311096725)(8.1, 0.8785834985331016)(8.2, 0.9300128227146501)(8.3, 0.9728034795560131)(8.4, 1.0072397136639422)(8.5, 1.0336176131704695)(8.6, 1.0522474784379658)(8.7, 1.0635485702613892)(8.8, 1.0679614677812044)(8.9, 1.0659267501378757)(9., 1.0578968399971493)(9.1, 1.0443217913197151)(9.2, 1.0256848199370499)(9.3, 1.0024644042705189)(9.4, 0.9751508662667684)(9.5, 0.9442297904623315)(9.6, 0.9101820239836295)(9.7, 0.8734126153952829)(9.8, 0.834328981966969)(9.9, 0.793331434853196)(10., 0.7508179165034155)(10.1, 0.7071674197266294)(10.2, 0.6626286585537456)(10.3, 0.617421922554996)(10.4, 0.5717651325955566)(10.5, 0.5258809469507161)(10.6, 0.47999676130587565)(10.7, 0.43433523393632373)(10.8, 0.389119023117349)(10.9, 0.34458026194446506)(11., 0.3009439773980167)(11.1, 0.25843756516340505)(11.2, 0.21729031589007625)(11.3, 0.17773152022747632)(11.4, 0.13999165317757958)(11.5, 0.10430095287185462)(11.6, 0.07085531121845426)(11.7, 0.03970589224659374)(11.8, 0.010869632197425337)(11.9, -0.01563089516986478)(12., -0.03978050645586597)(12.1, -0.06157022426841502)(12.2, -0.08103861112582783)(12.3, -0.09823209364720718)(12.4, -0.11319544035811635)(12.5, -0.12598076276979311)(12.6, -0.13664040926398088)(12.7, -0.14527599728759324)(12.8, -0.15199317108613944)(12.9, -0.15690018048069065)(13., -0.16010432781029552)(13.1, -0.16170438807580012)(13.2, -0.16175010408338597)(13.3, -0.16028340191254886)(13.4, -0.15734762886581846)(13.5, -0.15298376354066798)(13.6, -0.14728631726884234)(13.7, -0.14057317026889177)(13.8, -0.1332159723641437)(13.9, -0.1255844784138806)(14., -0.1180493907594073)(14.1, -0.11085184357545128)(14.2, -0.10370427606818236)(14.3, -0.09618529560809166)(14.4, -0.08787422017718712)(14.5, -0.07835131523949923)(14.6, -0.06734370597254666)(14.7, -0.05516453518476719)(14.8, -0.0422676467649405)(14.9, -0.029114938199037564)(15., -0.016163024760753913)(15.1, -0.0037903544569284405)(15.2, 0.007944115643588512)(15.3, 0.01906201181796061)(15.4, 0.029585363023211086)(15.5, 0.039532266165969776)(15.6, 0.0489267872896083)(15.7, 0.057789297257610485)(15.8, 0.06613969319244897)(15.9, 0.07400071466266384)(16., 0.08139462749578392)(16.1, 0.08834369751933813)(16.2, 0.09486663750327091)(16.3, 0.10098571327511109)(16.4, 0.10672650684946634)(16.5, 0.11210488855021358)(16.6, 0.11714762474448856)(16.7, 0.12187627064830328)(16.8, 0.12630906529059097)(16.9, 0.13047088007444232)(17., 0.13438303334536372)(17.1, 0.1380685015424009)(17.2, 0.1415502611045996)(17.3, 0.14484749854291556)(17.4, 0.14798342716690008)(17.5, 0.15098291837964392)(17.6, 0.15385686822440595)(17.7, 0.15658822202478104)(17.8, 0.15914902906110526)(17.9, 0.16151110174320918)(18., 0.16364719996294574)(18.1, 0.16554003217340427)(18.2, 0.16721304855464178)(18.3, 0.16869822662491765)(18.4, 0.17002517519743532)(18.5, 0.17123060920056676)(18.6, 0.17234200561296467)(18.7, 0.1733832883556972)(18.8, 0.1743710383641582)(18.9, 0.1753289426889103)(19., 0.1762766615819205)(19.1, 0.177228644144032)(19.2, 0.178185364116256)(19.3, 0.1791465846280869)(19.4, 0.18010875262194032)(19.5, 0.18106689381719807)(19.6, 0.18202171882537704)(19.7, 0.18296659527231976)(19.8, 0.18390436560409365)(19.9, 0.18482579187097936)(20., 0.18573347964853873)};
  
  %\legend{0.1,0.3,0.5,0.7,0.9}
	\end{axis}
\end{tikzpicture}
}
\caption{\it Spectral version of the amplitude contribution to the Newman-Penrose  scalar at  one loop for positive-helicity graviton, and for mass ratios $\chi = 0.1, 0.5, 0.9$.}
\label{Figure-3}
\end{figure}

We begin by plotting the quantity  $\omega^2 W^{\pm}$ which is the spectral version of the Newman-Penrose scalar \eqref{tdwf2}, for various choices of $\chi$. 
For the positive-helicity waveform  at $\theta \to \frac{\pi }{4}$, $\phi \to \frac{\pi }{2}$, and $y \to 2$ this is shown in Figure~\ref{Figure-3}, where we stripped off a dimensionful factor of $\frac{\kappa^5 (m_1 + m_2)^3} {(4\pi)^4 (-b^2)^{3/2}}$.

The corresponding negative-helicity waveform is  shown in Figure~\ref{Figure-4}. 
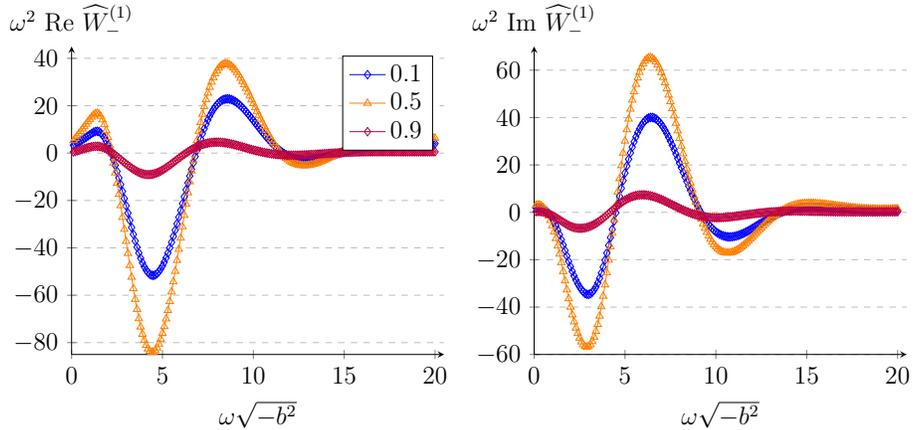
\begin{figure}
    \centering
\scalebox{0.8}{
\begin{tikzpicture}
	\begin{axis}[width=0.5\textwidth,
			title={},
			xmin=0.0, xmax=20.5,
			ymin=-85.0, ymax=45.0,
			axis lines=left,
			xtick={0,5,10,15,20},
			ytick={-80,-60,-40,-20,0,20,40},
			compat=newest,
			xlabel=$\omega \sqrt{-b^2}$, %xlabel style={at={(1,0)}, anchor=west},
			ylabel=$\omega^2$ Re $\widehat{W}_{-}^{(1)}$, ylabel style={rotate=-90,at={(0,1)},anchor=south},
			legend pos=north east,
			ymajorgrids=true,
			grid style=dashed,
		]
		\addplot[
			color=blue,
			mark=diamond,
		]
		coordinates {(0.1, 3.3112364851984375)(0.2, 3.7566477839778396)(0.3, 4.2179530936847565)(0.4, 4.69264158695955)(0.5, 5.178865673858337)(0.6, 5.6754173148024245)(0.7, 6.181799081729996)(0.8, 6.698034661691615)(0.9, 7.224692543900787)(1., 7.762720210380011)(1.1, 8.300487319303047)(1.2, 8.774678384516024)(1.3, 9.109186912561267)(1.4, 9.228190654587841)(1.5, 9.05593842289651)(1.6, 8.534846997569554)(1.7, 7.679247044197358)(1.8, 6.521471386797888)(1.9, 5.093947597591365)(2., 3.4291032487980075)(2.1, 1.556037882033988)(2.2, -0.5092668496860983)(2.3, -2.754164430066871)(2.4, -5.1660746665545245)(2.5, -7.732377098609301)(2.6, -10.438835808843068)(2.7, -13.264487757507908)(2.8, -16.18664075016484)(2.9, -19.182697340577132)(3., -22.230083769558608)(3.1, -25.303454893007267)(3.2, -28.366427401258946)(3.3, -31.379893973834754)(3.4, -34.304297236295135)(3.5, -37.1007904257174)(3.6, -39.73194800221262)(3.7, -42.16673992954375)(3.8, -44.375794265013184)(3.9, -46.32973906592325)(4., -47.999439260081935)(4.1, -49.36002344439851)(4.2, -50.40580672673794)(4.3, -51.13465727254959)(4.4, -51.545153858799694)(4.5, -51.63563839194888)(4.6, -51.40753209503091)(4.7, -50.8738628458552)(4.8, -50.0512115798156)(4.9, -48.95521175028343)(5., -47.6022074221469)(5.1, -46.00925327181107)(5.2, -44.19600955124293)(5.3, -42.182610253420705)(5.4, -39.98966311233386)(5.5, -37.63777586197187)(5.6, -35.147082495312965)(5.7, -32.53748013482972)(5.8, -29.828392161983512)(5.9, -27.039478828741284)(6., -24.190400387070017)(6.1, -21.300485470228796)(6.2, -18.38794942010029)(6.3, -15.470723333960382)(6.4, -12.56680937023666)(6.5, -9.694138626205032)(6.6, -6.87012108402901)(6.7, -4.110034891321485)(6.8, -1.4285518072009404)(6.9, 1.159582979357397)(7., 3.6396574412490867)(7.1, 5.998058630515796)(7.2, 8.225328307867871)(7.3, 10.313152318557833)(7.4, 12.253169133737082)(7.5, 14.03704091160758)(7.6, 15.657732598152222)(7.7, 17.11320710702263)(7.8, 18.40280120080305)(7.9, 19.525756893875478)(8., 20.48141094882416)(8.1, 21.27007129730641)(8.2, 21.896333227131365)(8.3, 22.365810569282385)(8.4, 22.6841171547428)(8.5, 22.85691418859707)(8.6, 22.89033661694092)(8.7, 22.79281702977464)(8.8, 22.573285445160327)(8.9, 22.240695568210658)(9., 21.80397741698773)(9.1, 21.27217944480648)(9.2, 20.65461066253799)(9.3, 19.960627455154512)(9.4, 19.199586207628276)(9.5, 18.380890679032632)(9.6, 17.51366038383419)(9.7, 16.606162102679306)(9.8, 15.666330997506451)(9.9, 14.702220665506916)(10., 13.721789955669744)(10.1, 12.732713472377215)(10.2, 11.741126161725045)(10.3, 10.752855038151646)(10.4, 9.773703429044852)(10.5, 8.80952203589363)(10.6, 7.865711506226258)(10.7, 6.946274951587815)(10.8, 6.054836490714383)(10.9, 5.194996555291477)(11., 4.370355577004618)(11.1, 3.5842297429325694)(11.2, 2.8385140171203407)(11.3, 2.1348072754809095)(11.4, 1.4747297122727572)(11.5, 0.85986835988358)(11.6, 0.29153784961960244)(11.7, -0.23011566941720382)(11.8, -0.7052203441703826)(11.9, -1.1339133226626916)(12., -1.5163265417657645)(12.1, -1.8527987263026466)(12.2, -2.144471592110456)(12.3, -2.3926763514308114)(12.4, -2.5988010654266827)(12.5, -2.764136678353732)(12.6, -2.8901517873468405)(12.7, -2.9788361046532685)(12.8, -3.0322030295708395)(12.9, -3.052337022549062)(13., -3.041298856986883)(13.1, -3.0011729933338143)(13.2, -2.934186014342741)(13.3, -2.84258818981711)(13.4, -2.728653476610935)(13.5, -2.594584770426537)(13.6, -2.442656028117928)(13.7, -2.2748759115728174)(13.8, -2.093328881240715)(13.9, -1.900042548649781)(14., -1.6970492627382878)(14.1, -1.4863103112928195)(14.2, -1.2694601008021966)(14.3, -1.0480525017833269)(14.4, -0.82364612216323)(14.5, -0.5977948324588136)(14.6, -0.3719695985100162)(14.7, -0.14729745018260676)(14.8, 0.0751774873346145)(14.9, 0.2944110888528476)(15., 0.5093663352984613)(15.1, 0.7190843748646809)(15.2, 0.9229659251722724)(15.3, 1.1205040833391955)(15.4, 1.311182471663186)(15.5, 1.4944941872622028)(15.6, 1.6699939135856694)(15.7, 1.8374992603442541)(15.8, 1.9969036358104244)(15.9, 2.1480909734364237)(16., 2.290952312789664)(16.1, 2.425411855308345)(16.2, 2.551616460705955)(16.3, 2.669672720710025)(16.4, 2.77979381877562)(16.5, 2.882098190155552)(16.6, 2.976822705355447)(16.7, 3.0642042348809317)(16.8, 3.1445980844904455)(16.9, 3.2182884987907387)(17., 3.2856307835402516)(17.1, 3.346909183345735)(17.2, 3.402526378066754)(17.3, 3.452837673461747)(17.4, 3.498245749390279)(17.5, 3.539082224560226)(17.6, 3.575773465881716)(17.7, 3.608627405012063)(17.8, 3.638023034760267)(17.9, 3.6643156608847685)(18., 3.687813215042883)(18.1, 3.708918377094172)(18.2, 3.7278917045948257)(18.3, 3.745088503303284)(18.4, 3.760793017826297)(18.5, 3.7753131798211803)(18.6, 3.7889569209452465)(18.7, 3.8019611117041214)(18.8, 3.8145626226034324)(18.9, 3.826998324148805)(19., 3.8394813997953023)(19.1, 3.852248720048552)(19.2, 3.865442407211928)(19.3, 3.8791808965382444)(19.4, 3.8936063103308767)(19.5, 3.9088607708932)(19.6, 3.9250627134780287)(19.7, 3.942354260388737)(19.8, 3.96085384687814)(19.9, 3.9807035952496097)(20., 4.002021940755963)};
    \addplot[
			color=orange,
			mark=triangle,
		]
		coordinates  {(0.1, 5.297700974636467)(0.2, 6.213468668332212)(0.3, 7.1714519499195495)(0.4, 8.159070363655223)(0.5, 9.16577001256634)(0.6, 10.182602455329238)(0.7, 11.20285636095118)(0.8, 12.221202143055582)(0.9, 13.234020946695372)(1., 14.238878269451616)(1.1, 15.213600400103202)(1.2, 16.05238517947045)(1.3, 16.62929645539546)(1.4, 16.81905604934707)(1.5, 16.49684636433284)(1.6, 15.568313982278353)(1.7, 14.059383064077135)(1.8, 12.026178760090056)(1.9, 9.525089410128667)(2., 6.612569151367196)(2.1, 3.3402228553508104)(2.2, -0.2589442048559943)(2.3, -4.156807604551874)(2.4, -8.325274501769568)(2.5, -12.736329695429772)(2.6, -17.360239357340138)(2.7, -22.159696382864542)(2.8, -27.095617138574678)(2.9, -32.129115383130255)(3., -37.22143646991631)(3.1, -42.3296147211068)(3.2, -47.394037726119116)(3.3, -52.35094784052223)(3.4, -57.13658741988511)(3.5, -61.687462009227424)(3.6, -65.94277484543844)(3.7, -69.8524541355233)(3.8, -73.36866519681807)(3.9, -76.44403392819751)(4., -79.03184420216314)(4.1, -81.09195962748386)(4.2, -82.61582654701206)(4.3, -83.60212901349425)(4.4, -84.05086702693045)(4.5, -83.96138261369391)(4.6, -83.33696564191831)(4.7, -82.20393505667325)(4.8, -80.59058372390851)(4.9, -78.52849437770759)(5., -76.04661785764698)(5.1, -73.1765368978102)(5.2, -69.95378207404114)(5.3, -66.41519990943723)(5.4, -62.59730794028251)(5.5, -58.53688689231167)(5.6, -54.27032270708343)(5.7, -49.831764215825515)(5.8, -45.25483387086434)(5.9, -40.573285719251615)(6., -35.82067641595104)(6.1, -31.029904642299595)(6.2, -26.229789643148443)(6.3, -21.448361094996695)(6.4, -16.713648674343435)(6.5, -12.053616260325084)(6.6, -7.495438163725975)(6.7, -3.0629396095703387)(6.8, 1.2208766441510208)(6.9, 5.333014419183565)(7., 9.25051701569036)(7.1, 12.952276639725607)(7.2, 16.42499564429291)(7.3, 19.657225288287)(7.4, 22.637648425327967)(7.5, 25.354816314310536)(7.6, 27.79931993237235)(7.7, 29.969382750721206)(7.8, 31.86526795880779)(7.9, 33.4870413539948)(8., 34.835031923095585)(8.1, 35.91108199226502)(8.2, 36.72374521865072)(8.3, 37.283417585555156)(8.4, 37.600363481555476)(8.5, 37.6848472952288)(8.6, 37.54798878086703)(8.7, 37.20439495298377)(8.8, 36.6692650023567)(8.9, 35.958127106576875)(9., 35.08618045642197)(9.1, 34.068887432120356)(9.2, 32.92177621126307)(9.3, 31.66050656616653)(9.4, 30.30054087705909)(9.5, 28.85753891625715)(9.6, 27.34669987453839)(9.7, 25.78157800861364)(9.8, 24.175332791017684)(9.9, 22.54112369428531)(10., 20.892175988313973)(10.1, 19.240859577286376)(10.2, 17.596649281427563)(10.3, 15.968296149973169)(10.4, 14.364551232158819)(10.5, 12.794099779857481)(10.6, 11.265034868678045)(10.7, 9.783344058623838)(10.8, 8.354356936071444)(10.9, 6.983337290034786)(11., 5.6757134029344565)(11.1, 4.436216105146714)(11.2, 3.2676088859038623)(11.3, 2.1721288555368155)(11.4, 1.1519933851676845)(11.5, 0.20942445173396856)(11.6, -0.6538882684908662)(11.7, -1.4384224643598325)(11.8, -2.145178255785574)(11.9, -2.775161684443375)(12., -3.329386029718414)(12.1, -3.8091014415014963)(12.2, -4.216538450387267)(12.3, -4.554177616948534)(12.4, -4.824492922021836)(12.5, -5.02997150591625)(12.6, -5.173265002347532)(12.7, -5.25776855522966)(12.8, -5.287081280300895)(12.9, -5.264769394618166)(13., -5.194418854447201)(13.1, -5.079628775526264)(13.2, -4.924083810165095)(13.3, -4.7314883498822375)(13.4, -4.5055336267236985)(13.5, -4.249910872735484)(13.6, -3.968271841545998)(13.7, -3.6639656189153436)(13.8, -3.3403083919222865)(13.9, -3.0005966084367905)(14., -2.6481332960650876)(14.1, -2.2860372497979213)(14.2, -1.9167363923179597)(14.3, -1.542474413692384)(14.4, -1.165495003988375)(14.5, -0.7880484330093809)(14.6, -0.4122981180401208)(14.7, -0.040131324834170826)(14.8, 0.32663060981229186)(14.9, 0.6861743763813366)(15., 1.0366900868178914)(15.1, 1.3765597839738044)(15.2, 1.7049215223980474)(15.3, 2.0211304879364156)(15.4, 2.3245221272259027)(15.5, 2.6144384666397684)(15.6, 2.8903268083315528)(15.7, 3.152042398103373)(15.8, 3.3995391778013575)(15.9, 3.6327842487441684)(16., 3.8517249730416676)(16.1, 4.056400829111459)(16.2, 4.247107905085574)(16.3, 4.424234405403792)(16.4, 4.588161954769618)(16.5, 4.739265598150296)(16.6, 4.877959858930672)(16.7, 5.004830333638541)(16.8, 5.120502097219308)(16.9, 5.225587065145841)(17., 5.320703732627272)(17.1, 5.406490334081541)(17.2, 5.48364432155299)(17.3, 5.552876306558499)(17.4, 5.614896900614944)(17.5, 5.67042987471174)(17.6, 5.720133202475627)(17.7, 5.764500364126657)(17.8, 5.803972201994747)(17.9, 5.839002717882346)(18., 5.870019594646838)(18.1, 5.897503153035737)(18.2, 5.921966612477903)(18.3, 5.943956091083531)(18.4, 5.964024286699079)(18.5, 5.982717317434742)(18.6, 6.000535243246842)(18.7, 6.017866268575156)(18.8, 6.035052539705585)(18.9, 6.052423043451502)(19., 6.070333085571345)(19.1, 6.089091913669684)(19.2, 6.108916659043339)(19.3, 6.129991554307797)(19.4, 6.152507411814816)(19.5, 6.176655043916148)(19.6, 6.202612103491012)(19.7, 6.230575982627434)(19.8, 6.260724334204628)(19.9, 6.293254550310618)(20., 6.328350863560892)};

 \addplot[
			color=purple,
			mark=diamond,
		]
		coordinates {(0.1, 0.5031176913600436)(0.2, 0.7170496572213523)(0.3, 0.9454757293219248)(0.4, 1.1818985496924361)(0.5, 1.4204721542540326)(0.6, 1.6560659278548522)(0.7, 1.8842456546295736)(0.8, 2.101219037783122)(0.9, 2.3037930628996572)(1., 2.489295830675716)(1.1, 2.653304968771258)(1.2, 2.7830626317532605)(1.3, 2.8639302223740266)(1.4, 2.8815533879926116)(1.5, 2.821814646473698)(1.6, 2.67433906967086)(1.7, 2.443508761938182)(1.8, 2.137384426582122)(1.9, 1.764128621226556)(2., 1.3319370653661484)(2.1, 0.8489225738185958)(2.2, 0.32282607470775987)(2.3, -0.23872757039025358)(2.4, -0.8281277121296765)(2.5, -1.4378016004456413)(2.6, -2.060538897146888)(2.7, -2.6904936381545634)(2.8, -3.3222272766594907)(2.9, -3.9502894223272134)(3., -4.569350488781142)(3.1, -5.1738203320885)(3.2, -5.757208700395132)(3.3, -6.312812158391817)(3.4, -6.833974644870461)(3.5, -7.3140874727240925)(3.6, -7.7468498865030595)(3.7, -8.127074422134147)(3.8, -8.44983417310033)(3.9, -8.710296981086838)(4., -8.903607000728336)(4.1, -9.025974303934802)(4.2, -9.077422577756801)(4.3, -9.05894667831796)(4.4, -8.971588835843038)(4.5, -8.81641496760735)(4.6, -8.595225289453658)(4.7, -8.312709837393355)(4.8, -7.974292946005282)(4.9, -7.585398949868278)(5., -7.151428496510619)(5.1, -6.6779954169156435)(5.2, -6.170974099622882)(5.3, -5.636404742525797)(5.4, -5.0803038564672995)(5.5, -4.508735326391411)(5.6, -3.9274551055848534)(5.7, -3.3413664135140833)(5.8, -2.7550882250388127)(5.9, -2.1732513585440327)(6., -1.6004890011197899)(6.1, -1.0410363974066803)(6.2, -0.49750386037670397)(6.3, 0.027908793584389236)(6.4, 0.532991799529613)(6.5, 1.0155420248861384)(6.6, 1.473415080966532)(6.7, 1.9047088976106163)(6.8, 2.3075806222846196)(6.9, 2.6801897711598257)(7., 3.020691122997409)(7.1, 3.3275805500866413)(7.2, 3.600644868972461)(7.3, 3.8400735760593685)(7.4, 4.04593773249905)(7.5, 4.218426834696008)(7.6, 4.35777775315587)(7.7, 4.464748473496637)(7.8, 4.540168042488001)(7.9, 4.584889193950217)(8., 4.599788348754101)(8.1, 4.585884050073847)(8.2, 4.544763330297148)(8.3, 4.478250092317326)(8.4, 4.388120864926579)(8.5, 4.2761521769171)(8.6, 4.144215305283338)(8.7, 3.9943236493231136)(8.8, 3.828609043587063)(8.9, 3.6491559485246956)(9., 3.4580725116360846)(9.1, 3.2574195063201774)(9.2, 3.0490918966219835)(9.3, 2.8349372724853867)(9.4, 2.61680322385427)(9.5, 2.3965373406725177)(9.6, 2.175946944898058)(9.7, 1.956566957407348)(9.8, 1.73988492497572)(9.9, 1.527383656968393)(10., 1.320553068865756)(10.1, 1.1207077919740343)(10.2, 0.9284850079533615)(10.3, 0.7443513516998219)(10.4, 0.5687758268145546)(10.5, 0.40222269948858735)(10.6, 0.24512307404215947)(10.7, 0.09774106108904418)(10.8, -0.03970020735445871)(10.9, -0.16697191437991435)(11., -0.28384903300697784)(11.1, -0.39014704111176646)(11.2, -0.48582140703922255)(11.3, -0.570872130789346)(11.4, -0.6452921062469681)(11.5, -0.7090837021171451)(11.6, -0.7623369291920149)(11.7, -0.8055492155333924)(11.8, -0.8393079999952305)(11.9, -0.8642054588415946)(12., -0.8808313996314938)(12.1, -0.8897519428733744)(12.2, -0.8914360921683762)(12.3, -0.8863196892468514)(12.4, -0.8748433132492648)(12.5, -0.8574499120211371)(12.6, -0.8345871708181016)(12.7, -0.8067619925221984)(12.8, -0.7744931235407487)(12.9, -0.7382945728709612)(13., -0.6986827182151011)(13.1, -0.6561573563400394)(13.2, -0.6111495915660157)(13.3, -0.5640715785728194)(13.4, -0.5153331033351839)(13.5, -0.46534632053289854)(13.6, -0.41451154132047174)(13.7, -0.36317933404622993)(13.8, -0.31169316094333116)(13.9, -0.2603870094247082)(14., -0.20960481546453033)(14.1, -0.15963650856168385)(14.2, -0.11059175976027376)(14.3, -0.06252978668670651)(14.4, -0.015510707075310073)(14.5, 0.03040209252661003)(14.6, 0.07511613787365734)(14.7, 0.1184028015538015)(14.8, 0.1599965517302356)(14.9, 0.19963351465969215)(15., 0.2370528959154767)(15.1, 0.2720386695964583)(15.2, 0.304579939659378)(15.3, 0.3347098679750236)(15.4, 0.362473459939464)(15.5, 0.3879015087184306)(15.6, 0.4110437571181049)(15.7, 0.4319712662901748)(15.8, 0.4507669409116094)(15.9, 0.4675113169543215)(16., 0.4822896678003367)(16.1, 0.4951872668316803)(16.2, 0.506310705775884)(16.3, 0.5157736824756485)(16.4, 0.5236851573635618)(16.5, 0.5301635656924367)(16.6, 0.5353154991898055)(16.7, 0.5392759740438746)(16.8, 0.5421705316226261)(16.9, 0.544127081999098)(17., 0.5452782726564411)(17.1, 0.545756751077806)(17.2, 0.5456904273362306)(17.3, 0.5452214237350907)(17.4, 0.5444847564625935)(17.5, 0.5436154417069455)(17.6, 0.5427248086057912)(17.7, 0.5418081197490181)(17.8, 0.5408322132658384)(17.9, 0.5397757708107458)(18., 0.5385961556927276)(18.1, 0.5372909992067275)(18.2, 0.5359195189791521)(18.3, 0.5345646196869706)(18.4, 0.5333139434172645)(18.5, 0.5322480261419469)(18.6, 0.5314379290127055)(18.7, 0.5309073390801029)(18.8, 0.5306752059845893)(18.9, 0.5307533732514459)(19., 0.5311584218160667)(19.1, 0.5318974577936202)(19.2, 0.5329799560043316)(19.3, 0.5344106538583135)(19.4, 0.5361871826505097)(19.5, 0.538319017201145)(19.6, 0.5408061575102194)(19.7, 0.5436533409878457)(19.8, 0.54686293633908)(19.9, 0.550442049679091)(20., 0.5543883123028226)};
\legend{0.1,0.5,0.9}
	\end{axis}
\end{tikzpicture}
\begin{tikzpicture}
	\begin{axis}[width=0.5\textwidth,
			title={},
			axis lines=left,
			xmin=0, xmax=20.5,
			ymin=-60, ymax=70,
			xtick={0,5,10,15,20},
			ytick={-60,-40,-20,0,20,40,60},
			compat=newest,
			xlabel=$\omega \sqrt{-b^2}$, %xlabel style={at={(1,0)}, anchor=west},
			ylabel=$\omega^2$ Im $\widehat{W}_{-}^{(1)}$, ylabel style={rotate=-90,at={(0,1)},anchor=south},
			legend pos=south east,
			ymajorgrids=true,
			grid style=dashed,
		]
		\addplot[
			color=blue,
			mark=diamond,
		]
		coordinates {(0.1, 1.869817872131885)(0.2, 1.9406824213000589)(0.3, 1.8973090630148555)(0.4, 1.7265112162280676)(0.5, 1.415490767520715)(0.6, 0.9515704074316856)(0.7, 0.3220538768594187)(0.8, -0.4858853620757416)(0.9, -1.485253868837727)(1., -2.6892619115253074)(1.1, -4.105960718624715)(1.2, -5.72203343030965)(1.3, -7.518980460090708)(1.4, -9.478539091984114)(1.5, -11.582565045258905)(1.6, -13.807063337695153)(1.7, -16.103759760246252)(1.8, -18.418458341224937)(1.9, -20.69708154419676)(2., -22.88562289387896)(2.1, -24.937253091309245)(2.2, -26.832690877329657)(2.3, -28.560187474861127)(2.4, -30.107899358622358)(2.5, -31.46398300333204)(2.6, -32.613515567135714)(2.7, -33.53233625845953)(2.8, -34.192494357639596)(2.9, -34.566512886023276)(3., -34.62715173546357)(3.1, -34.351197596409115)(3.2, -33.73438679975863)(3.3, -32.775771863489624)(3.4, -31.475826528613347)(3.5, -29.833840183612914)(3.6, -27.855260850117727)(3.7, -25.567091765769167)(3.8, -23.00228161789984)(3.9, -20.19332903988164)(4., -17.172874787389855)(4.1, -13.974791342729027)(4.2, -10.637570163063401)(4.3, -7.200958119237046)(4.4, -3.7046310209423385)(4.5, -0.18825543992194027)(4.6, 3.31062062188381)(4.7, 6.763055615485992)(4.8, 10.14222656169801)(4.9, 13.42134340633354)(5., 16.573639782256837)(5.1, 19.574670653287274)(5.2, 22.4093947423176)(5.3, 25.065163164347364)(5.4, 27.52909016387052)(5.5, 29.788834787543944)(5.6, 31.832790380591945)(5.7, 33.65669327391326)(5.8, 35.256516668912234)(5.9, 36.629418119521375)(6., 37.77184456815627)(6.1, 38.681664180266296)(6.2, 39.3612456609077)(6.3, 39.81414206766488)(6.4, 40.04390645812225)(6.5, 40.054091889864175)(6.6, 39.84943577300319)(6.7, 39.43893918675307)(6.8, 38.83278756285577)(6.9, 38.04116633305319)(7., 37.074260929087274)(7.1, 35.9429673942168)(7.2, 34.66126108827373)(7.3, 33.24382798260689)(7.4, 31.705590919070705)(7.5, 30.061472739519637)(7.6, 28.32592254479687)(7.7, 26.514810658779368)(7.8, 24.643296793827208)(7.9, 22.72725127381732)(8., 20.781857498160356)(8.1, 18.82229886626695)(8.2, 16.861437446592603)(8.3, 14.911732627733242)(8.4, 12.985620111234248)(8.5, 11.095464537489303)(8.6, 9.252872561274085)(8.7, 7.466087276184389)(8.8, 5.742499041995748)(8.9, 4.089474531433136)(9., 2.51442779132265)(9.1, 1.0239059224397975)(9.2, -0.37874883238103646)(9.3, -1.6910469641257213)(9.4, -2.910498963780127)(9.5, -4.034591635279558)(9.6, -5.061448964219458)(9.7, -5.991592065712202)(9.8, -6.82618160523536)(9.9, -7.566354561215936)(10., -8.213247912080938)(10.1, -8.768353942015807)(10.2, -9.234633532340872)(10.3, -9.6154028701349)(10.4, -9.913954455426094)(10.5, -10.13362816234378)(10.6, -10.277905987320663)(10.7, -10.350956851255768)(10.8, -10.35713917145261)(10.9, -10.300811365214711)(11., -10.18635553689616)(11.1, -10.018153790851034)(11.2, -9.800730353736796)(11.3, -9.538633139261467)(11.4, -9.236481122284756)(11.5, -8.898774842413562)(11.6, -8.530038526305344)(11.7, -8.134417407808556)(11.8, -7.715914598468284)(11.9, -7.2786279580318585)(12., -6.826584285094925)(12.1, -6.363691943000313)(12.2, -5.893361867029039)(12.3, -5.418791809007058)(12.4, -4.943203207810885)(12.5, -4.469841189367599)(12.6, -4.0017376961492115)(12.7, -3.541190372060299)(12.8, -3.09028367755037)(12.9, -2.651102073068936)(13., -2.2257347564756174)(13.1, -1.8161145910963257)(13.2, -1.4234022424086281)(13.3, -1.0485925665361544)(13.4, -0.6926733134873657)(13.5, -0.3566346019757795)(13.6, -0.04131353236827838)(13.7, 0.25307955432614165)(13.8, 0.5264897041501753)(13.9, 0.778853909549326)(14., 1.0101153216022434)(14.1, 1.2203094708847717)(14.2, 1.4098342998463627)(14.3, 1.5791848678437745)(14.4, 1.7288443907084854)(14.5, 1.859305559092197)(14.6, 1.9711487057336936)(14.7, 2.0653094691301983)(14.8, 2.1428087611609605)(14.9, 2.2046627562951153)(15., 2.2519018412321374)(15.1, 2.2855303469158814)(15.2, 2.3065147050093016)(15.3, 2.3158047662399586)(15.4, 2.3143480126303584)(15.5, 2.3030966636131165)(15.6, 2.2829863576854565)(15.7, 2.2548958844232523)(15.8, 2.2196921898770947)(15.9, 2.1782303765722943)(16., 2.1313821279695553)(16.1, 2.0799788595436257)(16.2, 2.02473355151644)(16.3, 1.9663331283543142)(16.4, 1.9054645145235634)(16.5, 1.8428122657854482)(16.6, 1.779025407325384)(16.7, 1.7146179481405792)(16.8, 1.6500612605372296)(16.9, 1.585838560346813)(17., 1.5224212198755254)(17.1, 1.4602569243789998)(17.2, 1.399646499399382)(17.3, 1.3408765582484794)(17.4, 1.2842147645976492)(17.5, 1.2299406256435306)(17.6, 1.1783075928271438)(17.7, 1.1294696319771451)(17.8, 1.0835759715120796)(17.9, 1.0407545215049854)(18., 1.0011450355541813)(18.1, 0.9648635802074248)(18.2, 0.9319101554647156)(18.3, 0.9022729178007723)(18.4, 0.8759305488700888)(18.5, 0.8528664677372712)(18.6, 0.8330498812365879)(18.7, 0.8164002533961262)(18.8, 0.802827573423748)(18.9, 0.7922418305273156)(19., 0.7845506452096348)(19.1, 0.7796545318583424)(19.2, 0.7774255804004004)(19.3, 0.7777240372374893)(19.4, 0.7804148861814023)(19.5, 0.7853607423388762)(19.6, 0.7924218521115916)(19.7, 0.8014631993113416)(19.8, 0.8123473990448628)(19.9, 0.8249370664188924)(20., 0.8390924478351107)};
   \addplot[
			color=orange,
			mark=triangle,
		]
		coordinates {(0.1, 2.9761397287308915)(0.2, 3.153141214060028)(0.3, 3.073861971774211)(0.4, 2.7265966510537485)(0.5, 2.1008571522884156)(0.6, 1.1860698392965126)(0.7, -0.029155535171807377)(0.8, -1.5574235744919007)(0.9, -3.412797346380154)(1., -5.610850362810497)(1.1, -8.159136161017898)(1.2, -11.028361755139922)(1.3, -14.181108185023904)(1.4, -17.581252698561862)(1.5, -21.193659504085918)(1.6, -24.973783787065816)(1.7, -28.836286368113473)(1.8, -32.686024260802384)(1.9, -36.428709844420816)(2., -39.970318687707724)(2.1, -43.23070960292626)(2.2, -46.182839874017425)(2.3, -48.813484231083784)(2.4, -51.10928580950252)(2.5, -53.05734832618954)(2.6, -54.640564466849625)(2.7, -55.82531177915639)(2.8, -56.57401996902302)(2.9, -56.848987147637374)(3., -56.61237983146193)(3.1, -55.83327326003993)(3.2, -54.50666683380817)(3.3, -52.6344686762842)(3.4, -50.21825792417217)(3.5, -47.25954791681361)(3.6, -43.76873463751097)(3.7, -39.791020874421264)(3.8, -35.38049205966245)(3.9, -30.591036233264486)(4., -25.476409840531957)(4.1, -20.09247484237504)(4.2, -14.503515262126161)(4.3, -8.775657449272613)(4.4, -2.975027753301702)(4.5, 2.8322803749806127)(4.6, 8.583923934440783)(4.7, 14.231969546370852)(4.8, 19.732365876460626)(4.9, 25.04102869171857)(5., 30.113807961790474)(5.1, 34.91089628225862)(5.2, 39.409067184099115)(5.3, 43.589305229499175)(5.4, 47.43259498064604)(5.5, 50.919920999726955)(5.6, 54.03522873024947)(5.7, 56.773385977925024)(5.8, 59.13222142978538)(5.9, 61.10962957022497)(6., 62.703307491550206)(6.1, 63.91299200431039)(6.2, 64.74493385795955)(6.3, 65.2074235201946)(6.4, 65.30855406662444)(6.5, 65.05661596494596)(6.6, 64.46134722483492)(6.7, 63.5394603764105)(6.8, 62.309378681221396)(6.9, 60.78939380609096)(7., 58.997994809930574)(7.1, 56.95446032000369)(7.2, 54.682082602696866)(7.3, 52.20533827692483)(7.4, 49.54844077215158)(7.5, 46.73580090992913)(7.6, 43.791829511809524)(7.7, 40.74152957560885)(7.8, 37.60990409914319)(7.9, 34.42208767495401)(8., 31.203149098220045)(8.1, 27.977236001042655)(8.2, 24.76487716057609)(8.3, 21.585745988259863)(8.4, 18.459450098170795)(8.5, 15.405597104385725)(8.6, 12.442478673728004)(8.7, 9.582727899831026)(8.8, 6.837530334349359)(8.9, 4.218157065509046)(9., 1.7358265436459916)(9.1, -0.5995843891288191)(9.2, -2.7835903266019053)(9.3, -4.813063920125375)(9.4, -6.684880452945844)(9.5, -8.395875071918693)(9.6, -9.944021218273571)(9.7, -11.33109542080267)(9.8, -12.559992763463642)(9.9, -13.633608330214143)(10., -14.554837205011824)(10.1, -15.327100850715729)(10.2, -15.955794651065121)(10.3, -16.446906166063325)(10.4, -16.806357158351005)(10.5, -17.040200985294145)(10.6, -17.154556801621435)(10.7, -17.15685970931502)(10.8, -17.05467640508241)(10.9, -16.855507788268426)(11., -16.566986352943246)(11.1, -16.19681039053972)(11.2, -15.752809787216052)(11.3, -15.2428802264931)(11.4, -14.674851594529068)(11.5, -14.056816966932853)(11.6, -13.396211445686607)(11.7, -12.699417374969697)(11.8, -11.972356517422783)(11.9, -11.221016433049186)(12., -10.45145047921491)(12.1, -9.669448823835266)(12.2, -8.880406850649509)(12.3, -8.089456753946207)(12.4, -7.301928120101951)(12.5, -6.522992621822909)(12.6, -5.757506104474418)(12.7, -5.008857132234186)(12.8, -4.280065804048945)(12.9, -3.574145639129163)(13., -2.894129895894106)(13.1, -2.242742585158476)(13.2, -1.621602322044051)(13.3, -1.0320579524856466)(13.4, -0.4754464788927974)(13.5, 0.046888713980783335)(13.6, 0.5338699311318592)(13.7, 0.9854734197125048)(13.8, 1.4019115078120694)(13.9, 1.7834309355405795)(14., 2.130242254458592)(14.1, 2.4425560161266637)(14.2, 2.7204314381712007)(14.3, 2.963888259801006)(14.4, 3.1729593796974154)(14.5, 3.347664537069232)(14.6, 3.4885959081805202)(14.7, 3.5985104025273187)(14.8, 3.680743946397195)(14.9, 3.738606147132649)(15., 3.7754263512849793)(15.1, 3.794191759119584)(15.2, 3.7965538844395783)(15.3, 3.7838418339709747)(15.4, 3.7573715549672535)(15.5, 3.718458994681892)(15.6, 3.6684135206321016)(15.7, 3.608439224554815)(15.8, 3.53972703871443)(15.9, 3.46346131563908)(16., 3.3808264078568913)(16.1, 3.2929671894783943)(16.2, 3.200837722262359)(16.3, 3.1053591692862206)(16.4, 3.007432954418612)(16.5, 2.907973661000701)(16.6, 2.8078432344835154)(16.7, 2.7076864890212606)(16.8, 2.6080824414054673)(16.9, 2.509636427372734)(17., 2.412914304242059)(17.1, 2.3184819293324357)(17.2, 2.2266419705121656)(17.3, 2.137703675385816)(17.4, 2.0519236536678145)(17.5, 1.9695914137539268)(17.6, 1.891016203248721)(17.7, 1.8166059658007745)(17.8, 1.7468212829488055)(17.9, 1.682102997022729)(18., 1.622905109824995)(18.1, 1.5695631879052407)(18.2, 1.5220377528458617)(18.3, 1.4801840504489754)(18.4, 1.4438441670441642)(18.5, 1.4128865079060806)(18.6, 1.3871531593643072)(18.7, 1.3664993672209607)(18.8, 1.3507672178056243)(18.9, 1.3398185366566824)(19., 1.333495410103718)(19.1, 1.3316267650037783)(19.2, 1.3339823105875048)(19.3, 1.3403185966130042)(19.4, 1.3503790133658482)(19.5, 1.3639201106041425)(19.6, 1.3806852786134598)(19.7, 1.4004310671519058)(19.8, 1.4229074462413203)(19.9, 1.4478709656398088)(20., 1.475058435896676)};
  
   \addplot[
			color=purple,
			mark=diamond,
		]
		coordinates  {(0.1, 0.2730027325543567)(0.2, 0.32957688411810504)(0.3, 0.31586918795752006)(0.4, 0.23663789878961972)(0.5, 0.09712756647947246)(0.6, -0.09760154436123036)(0.7, -0.3430453410680076)(0.8, -0.6354596115584268)(0.9, -0.9719625892610404)(1., -1.3505479809931946)(1.1, -1.7685959989626647)(1.2, -2.2183704524557006)(1.3, -2.6914174331265053)(1.4, -3.179962850980428)(1.5, -3.676846110632392)(1.6, -4.174060988992232)(1.7, -4.658342737744885)(1.8, -5.115502813603343)(1.9, -5.531589543786229)(2., -5.893030248321163)(2.1, -6.188952570999906)(2.2, -6.419001206064017)(2.3, -6.5854737974180715)(2.4, -6.690786424219456)(2.5, -6.7374262267772425)(2.6, -6.7275724137431965)(2.7, -6.661935596634191)(2.8, -6.5408710812086674)(2.9, -6.3647341732250675)(3., -6.133809117290139)(3.1, -5.848640715566824)(3.2, -5.510531755836065)(3.3, -5.120950835232744)(3.4, -4.6813428638411825)(3.5, -4.193058003543449)(3.6, -3.658275462991302)(3.7, -3.0823958897130193)(3.8, -2.4716489780065682)(3.9, -1.8322028358384534)(4., -1.1701260855628166)(4.1, -0.49181659953662027)(4.2, 0.19504770167077)(4.3, 0.8824989679911018)(4.4, 1.562608669860057)(4.5, 2.227497073037476)(4.6, 2.8697809238629923)(4.7, 3.4839861449515848)(4.8, 4.0651005564042055)(4.9, 4.608173564653267)(5., 5.108278263181746)(5.1, 5.561150982888373)(5.2, 5.965133630233762)(5.3, 6.319183974993161)(5.4, 6.6223308480935)(5.5, 6.873603080461714)(5.6, 7.072455869934863)(5.7, 7.220215691344451)(5.8, 7.318635386432113)(5.9, 7.369491483990047)(6., 7.374560512810447)(6.1, 7.335737436938322)(6.2, 7.255201465025434)(6.3, 7.13532130212804)(6.4, 6.978370905100158)(6.5, 6.786695291947481)(6.6, 6.562757915928524)(6.7, 6.309472284262488)(6.8, 6.0299177135225115)(6.9, 5.727173520281737)(7., 5.404271647012177)(7.1, 5.064244036185846)(7.2, 4.709885759769134)(7.3, 4.343968202677866)(7.4, 3.9691916886761796)(7.5, 3.5883276026799016)(7.6, 3.204099955503733)(7.7, 2.819043261557872)(7.8, 2.435715722303082)(7.9, 2.0566636956748443)(8., 1.6843909029176267)(8.1, 1.3213015796635348)(8.2, 0.9692575280867897)(8.3, 0.6299736906481254)(8.4, 0.3051721159234434)(8.5, -0.0034314482668037064)(8.6, -0.29429028489497827)(8.7, -0.5665374005363876)(8.8, -0.819470095149041)(8.9, -1.0523967068565103)(9., -1.2646208363722546)(9.1, -1.4556071563535584)(9.2, -1.6254385714773916)(9.3, -1.7743614270696053)(9.4, -1.9026196997509934)(9.5, -2.0104573661423526)(9.6, -2.098239206822345)(9.7, -2.1667800563303263)(9.8, -2.2170108157534067)(9.9, -2.249862386178697)(10., -2.2662680373983632)(10.1, -2.2671681453197428)(10.2, -2.253538616426014)(10.3, -2.2263600946104707)(10.4, -2.1866155924714596)(10.5, -2.1352904913123867)(10.6, -2.0733891220771064)(10.7, -2.0019963516813863)(10.8, -1.9222136279763882)(10.9, -1.8351400301082177)(11., -1.741891218158373)(11.1, -1.643568639978017)(11.2, -1.541297430468873)(11.3, -1.4362121993528898)(11.4, -1.3294404502368489)(11.5, -1.222114424137643)(11.6, -1.11521713365362)(11.7, -1.009151258644345)(11.8, -0.9041655131407254)(11.9, -0.8005204546989495)(12., -0.6984647973499252)(12.1, -0.5982993666357973)(12.2, -0.500512115798156)(12.3, -0.40562889735949126)(12.4, -0.31418503866251823)(12.5, -0.22671065589882797)(12.6, -0.14366006669821169)(12.7, -0.06518747375983187)(12.8, 0.008627108059510937)(12.9, 0.07770394814762298)(13., 0.14196217891388346)(13.1, 0.2013418247462679)(13.2, 0.25585093924196756)(13.3, 0.30551084074648877)(13.4, 0.3503527961665742)(13.5, 0.39039575514267394)(13.6, 0.42569893530119446)(13.7, 0.4564589391618056)(13.8, 0.4828889501795708)(13.9, 0.505213995334835)(14., 0.5236614703129991)(14.1, 0.5383308607264263)(14.2, 0.548873966931846)(14.3, 0.5548146792129497)(14.4, 0.5556863626736539)(14.5, 0.5510152763027063)(14.6, 0.5406379794512249)(14.7, 0.525618020689471)(14.8, 0.5073268802450202)(14.9, 0.48713130093533513)(15., 0.4664051316930474)(15.1, 0.4462853509451627)(15.2, 0.4270040917871946)(15.3, 0.40856372292419924)(15.4, 0.3909571382410079)(15.5, 0.3741938125578456)(15.6, 0.3582713771696562)(15.7, 0.3431827259612708)(15.8, 0.3289136467023519)(15.9, 0.3154594019827868)(16., 0.302812885687407)(16.1, 0.29095751688081845)(16.2, 0.2798696085124586)(16.3, 0.2695231048267086)(16.4, 0.25988721265783704)(16.5, 0.25093113884011253)(16.6, 0.24262409020780365)(16.7, 0.23491608708422326)(16.8, 0.22775833414521243)(16.9, 0.2210972986564996)(17., 0.21487802666077957)(17.1, 0.2090491172583314)(17.2, 0.20353714059241104)(17.3, 0.19827056177031935)(17.4, 0.19316979230216594)(17.5, 0.18816471851828548)(17.6, 0.18322312602991286)(17.7, 0.17848595278789556)(17.8, 0.17413582595207147)(17.9, 0.17036058383340197)(18., 0.16734380107374738)(18.1, 0.1652242837894046)(18.2, 0.1639565528432935)(18.3, 0.1634584141699617)(18.4, 0.16364009384777697)(18.5, 0.1644130023076351)(18.6, 0.16570228846975818)(18.7, 0.16747952787347117)(18.8, 0.16972719210135762)(18.9, 0.17242846334751818)(19., 0.17556462884200832)(19.1, 0.17911437023932175)(19.2, 0.1830423938341201)(19.3, 0.18730369423033447)(19.4, 0.1918577665715027)(19.5, 0.19666031607307277)(19.6, 0.2016703641375714)(19.7, 0.20684764277904202)(19.8, 0.21214738347192139)(19.9, 0.21752908135974738)(20., 0.22295009975150734)};
  
  %\legend{$\mu$=0.1,$\mu$=0.3,$\mu$=0.5,$\mu$=0.7,$\mu$=0.9}
	\end{axis}
\end{tikzpicture}
}
\caption{\it Spectral version of the amplitude contribution to the Newman-Penrose  scalar at one loop for negative-helicity graviton, and  for mass ratios $\chi = 0.1, 0.5, 0.9$.}
    \label{Figure-4}
\end{figure}
In the frequency domain, the most interesting part of the spectrum is contained in the region $\omega\sqrt{-b^2}\in[0,20]$. Beyond that, the amplitude is very small and tends to zero as $\omega\rightarrow\infty$. At one loop, the dependence on the mass ratio $\chi$ follows the same pattern as at tree level, in that the equal-mass case has the largest waveform due to the prefactor $\chi(1-\chi)$ in \eqref{eq: 1-loopWaveform}. 
However, due to the two terms in 
\eqref{eq: 1-loopWaveform} the $\chi$-dependence of the waveform is not as simple, as  Figures~\ref{Figure-3} and \ref{Figure-4} show. This is a new feature at one loop.

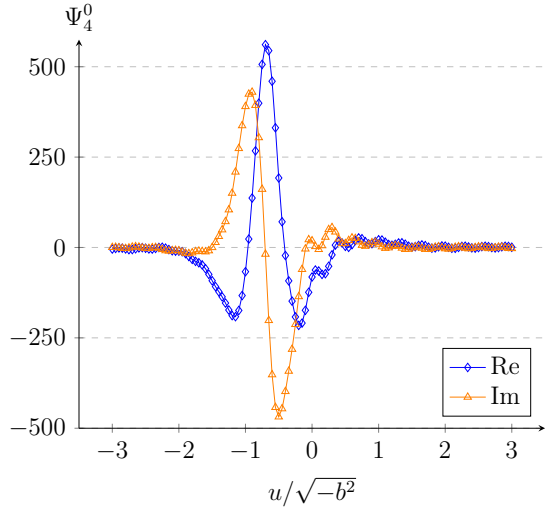
\begin{figure}[h]
\centering
\scalebox{0.8}{
\begin{tikzpicture}
	\begin{axis}[width=0.6\textwidth,
			xmin=-3.5, xmax=3.5,
    			ymin=-500, ymax=575,
			axis lines=left,
			xtick={-3,-2,-1,0,1,2,3},
			ytick={-500,-250,0,250,500},
			compat=newest,
			xlabel=$u/\sqrt{-b^2}$, %xlabel style={at={(1,0)}, anchor=west},
			ylabel=$\Psi_4^0$, ylabel style={rotate=-90,at={(0,1)},anchor=south},
			legend pos=south east,
			ymajorgrids=true,
			grid style=dashed,
		]
		\addplot[
			color=blue,
			mark=diamond,
		]
		coordinates {(-3., -4.300870930126441)(-2.95, -2.5534114127596994)(-2.9, -1.2053682057687758)(-2.85, -1.7706070295554308)(-2.8, -3.9639028966644494)(-2.75, -5.891824840636977)(-2.7, -5.758723355683886)(-2.65, -3.546030426058593)(-2.6, -1.094729940432564)(-2.55, -0.4593465222201137)(-2.5, -2.013971734877492)(-2.45, -4.075008323275646)(-2.4, -4.449086069285735)(-2.35, -2.5264871319535276)(-2.3, 0.022878006191175825)(-2.25, 0.659421168718117)(-2.2, -1.7471042116083033)(-2.15, -5.869230026294749)(-2.1, -9.133595102218788)(-2.05, -10.143979403439642)(-2., -10.086209319011932)(-1.95, -11.750027228947575)(-1.9, -16.986905121528263)(-1.85, -24.938253211221895)(-1.8, -32.759585712271836)(-1.75, -38.298012917987144)(-1.7, -42.30855956505114)(-1.65, -48.089252660131855)(-1.6, -58.718882397467766)(-1.55, -74.3339125072446)(-1.5, -91.9695796247378)(-1.45, -108.21626441618436)(-1.4, -122.28110866136343)(-1.35, -136.52492173301562)(-1.3, -153.66579268320083)(-1.25, -172.91875897521925)(-1.2, -188.4186437003167)(-1.15, -191.38018299427029)(-1.1, -174.1840422594389)(-1.05, -132.97383806950364)(-1., -67.03698498670587)(-0.95, 23.318980115816498)(-0.9, 136.78153144744394)(-0.85, 267.27809881227415)(-0.8, 399.11035263938516)(-0.75, 506.47717310344916)(-0.7, 560.948177766688)(-0.65, 544.4330397355318)(-0.6, 459.76696736760005)(-0.55, 331.0574563729938)(-0.5, 192.2941083618111)(-0.45, 71.03417476914706)(-0.4, -22.210491946848816)(-0.35, -92.35515217000703)(-0.3, -148.4066094846737)(-0.25, -192.06908138146628)(-0.2, -215.42846107796458)(-0.15, -209.13954915359045)(-0.1, -173.72806653610857)(-0.05, -123.94512396338709)(0., -81.67953007616204)(0.05, -62.32345931683362)(0.1, -65.01272912404245)(0.15, -74.37075903034201)(0.2, -72.47645295895958)(0.25, -51.939056350183435)(0.3, -20.35737502450028)(0.35, 6.589934858607364)(0.4, 17.401560101099363)(0.45, 12.690403136283368)(0.5, 3.2018510016475377)(0.55, 0.434769233340521)(0.6, 7.879760559104396)(0.65, 19.61518077353836)(0.7, 26.506928134731034)(0.75, 24.17875425387673)(0.8, 16.11265556367977)(0.85, 9.896054940884275)(0.9, 10.459214568010434)(0.95, 16.25135640419641)(1., 21.266233995115257)(1.05, 20.639250926195388)(1.1, 14.728015863569604)(1.15, 8.335604687709377)(1.2, 6.174411353848967)(1.25, 8.804016112585076)(1.3, 12.518343032891046)(1.35, 12.964317557094935)(1.4, 9.043650107443526)(1.45, 3.659873022957025)(1.5, 0.8776052233448657)(1.55, 2.2436834874447134)(1.6, 5.575826426659165)(1.65, 7.137566308142476)(1.7, 5.065600777804182)(1.75, 0.93064447739632)(1.8, -1.8531169223485378)(1.85, -1.1835958584599726)(1.9, 1.904932345454257)(1.95, 4.310582620857114)(2., 3.730493332314953)(2.05, 0.6791143193664239)(2.1, -2.1051800390160924)(2.15, -2.2142523271206644)(2.2, 0.2873634017421177)(2.25, 3.0284617915291996)(2.3, 3.5532154980625856)(2.35, 1.5185373331516085)(2.4, -1.0999739902376762)(2.45, -1.8900358225448792)(2.5, -0.1925730891795486)(2.55, 2.381028902290273)(2.6, 3.4987747601861767)(2.65, 2.2435453129830982)(2.7, -0.11509669463099714)(2.75, -1.356070485237143)(2.8, -0.3564098381848854)(2.85, 1.9283167281485716)(2.9, 3.4017894476048056)(2.95, 2.76717388461476)(3., 0.6782918523329998)};
    \addplot[
			color=orange,
			mark=triangle,
		]
		coordinates {(-3., 0.2853315791827468)(-2.95, 0.7771458100143108)(-2.9, -0.570071640075055)(-2.85, -2.3918788873952037)(-2.8, -2.805428471273382)(-2.75, -1.1667846322967839)(-2.7, 1.2633027836031703)(-2.65, 2.4058476674908786)(-2.6, 1.2486036527818145)(-2.55, -1.2026376152178078)(-2.5, -2.8255756237241396)(-2.45, -2.354479666451342)(-2.4, -0.6137163466839256)(-2.35, 0.17165610758550653)(-2.3, -1.5859927893649206)(-2.25, -5.309564239126577)(-2.2, -8.692489582852767)(-2.15, -9.764328620811071)(-2.1, -8.712557778468314)(-2.05, -7.739546379246251)(-2., -9.010225047205171)(-1.95, -12.544727775323292)(-1.9, -16.044752685400272)(-1.85, -16.910053801925113)(-1.8, -14.642808278906864)(-1.75, -11.375179653794202)(-1.7, -9.898160456489842)(-1.65, -10.887950183193759)(-1.6, -11.873726270774561)(-1.55, -8.995289045878188)(-1.5, -0.03668064794609929)(-1.45, 14.058067116823658)(-1.4, 30.71045844651767)(-1.35, 49.01627170284484)(-1.3, 71.79808214979138)(-1.25, 104.29671552169842)(-1.2, 150.30815326595027)(-1.15, 208.84609291606472)(-1.1, 273.9117889170597)(-1.05, 337.07857303128503)(-1., 390.00794548707375)(-0.95, 424.3725920644134)(-0.9, 429.7692917509291)(-0.85, 393.27478454383436)(-0.8, 303.90749061285044)(-0.75, 161.1173440060233)(-0.7, -18.361543622511988)(-0.65, -201.82156647699605)(-0.6, -351.88100571203887)(-0.55, -442.274738500736)(-0.5, -468.4370858471437)(-0.45, -445.96070675772955)(-0.4, -398.212218638886)(-0.35, -342.0291665988714)(-0.3, -281.4284796289293)(-0.25, -212.77222154681803)(-0.2, -135.80509858536283)(-0.15, -60.56647234135169)(-0.1, -3.9216346708827174)(-0.05, 21.76425423319423)(0., 18.514061909190158)(0.05, 2.7447040852616134)(0.1, -5.300365767824761)(0.15, 5.035998544337181)(0.2, 28.600797607103477)(0.25, 49.81820995911468)(0.3, 55.157665940104025)(0.35, 42.99364170521209)(0.4, 23.718764891423294)(0.45, 10.887226412204345)(0.5, 10.951773624987467)(0.55, 19.570043782744044)(0.6, 26.605624178741927)(0.65, 24.94647788155613)(0.7, 15.754717910733591)(0.75, 6.3872658220991285)(0.8, 3.521191921649185)(0.85, 7.648088045129491)(0.9, 13.336533237741353)(0.95, 14.439823415057795)(1., 9.379611441256607)(1.05, 2.0374087554619456)(1.1, -2.0881780005011494)(1.15, -0.7465631958434685)(1.2, 3.4372344868772515)(1.25, 5.69208378676773)(1.3, 3.4584738755483957)(1.35, -1.5455274113204542)(1.4, -5.174107208589759)(1.45, -4.704281140416568)(1.5, -1.1572045362914598)(1.55, 1.8526366016010183)(1.6, 1.494343642896405)(1.65, -1.8097959387640232)(1.7, -4.960739520910741)(1.75, -5.115810745260658)(1.8, -2.233642809900672)(1.85, 1.019438018324787)(1.9, 1.783134847408547)(1.95, -0.4288612881460289)(2., -3.4050200981120953)(2.05, -4.341349467643443)(2.1, -2.382469864532832)(2.15, 0.6934449849568056)(2.2, 2.175964841780705)(2.25, 0.883671740183402)(2.3, -1.848767716675791)(2.35, -3.475811480612976)(2.4, -2.570308175494364)(2.45, -0.04711328037958946)(2.5, 1.766409157816834)(2.55, 1.2851080295933102)(2.6, -0.9418102898420856)(2.65, -2.820305254973958)(2.7, -2.66889894372498)(2.75, -0.7227708695005758)(2.8, 1.1555727616971463)(2.85, 1.228699950572951)(2.9, -0.5408174746565994)(2.95, -2.51328160126487)(3., -2.916751029181403)};
 
\legend{{\rm Re},{\rm Im}}
	\end{axis}
\end{tikzpicture} 
}
\caption{\it The one-loop amplitude contribution to the Newman-Penrose scalar $\Psi_4^0$ in the time domain for equal masses as a function of the rescaled retarded time $u/\sqrt{-b^2}$.}
\label{Figure-5}
\end{figure}

The time-domain waveform is obtained using  \eqref{tdwf2}  by performing a numerical Fourier transform.  For example, in the equal-mass case, corresponding to $\chi=0.5$, the result is shown in Figure~\ref{Figure-5} (up to an overall factor of $\frac{\kappa^6 (m_1 + m_2)^3}{(4\pi)^6 (-b^2)^{2}}$). The small oscillations are due to the use of a finite frequency domain in the numerical computations and  vanish when  the range of  frequencies  is enlarged. 

%\FloatBarrier

\section{Conclusions}
\label{sec: theend}

The HEFT approach provides a powerful method to compute classical effects both in the conservative sector \cite{Brandhuber:2021eyq} and in the presence of radiation. It has the clear advantage of computing directly classical quantities, leading to integrals with linearised propagators  that can either be computed directly or using differential equations. It  holds the promise to be efficiently applicable to higher PM orders and to other problems.

Several issues are left to investigate. Having computed the graviton emission amplitude in a scattering process, it is important to determine the corresponding one for  bound states. It would be remarkable if an analytic continuation  similar to that of \cite{Kalin:2019rwq,Kalin:2019inp,Cho:2021arx} or  a generalisation of the Bethe-Salpeter equation approach of \cite{Adamo:2022ooq}
could be applicable also to  waveforms.

A number of concrete problems can also be tackled with our method and results:
one could  determine the one-loop waveforms fully analytically at one loop, or study the radiated energy, power and angular momentum. Going in a different direction, there are  intriguing differences and complementarities between the HEFT approach initiated in \cite{Brandhuber:2021eyq} and pursued in this paper,   and the eikonal approach \cite{Levy:1969cr,Amati:1987wq,Amati:1987uf,Muzinich:1987in,Amati:1990xe,Kabat:1992tb,Ciafaloni:2018uwe,Addazi:2019mjh,DiVecchia:2021bdo,DiVecchia:2022nna,DiVecchia:2022piu,Cristofoli:2021jas}.
In the former, which  appears to be intimately  related to the $N$ operator discussed in \cite{Damgaard:2021ipf},  experience so far indicates that 
 classical contributions can be computed directly without pollution either from quantum or hyper-classical terms, which can be discarded at the diagrammatic level. Clarifying the relationship between these approaches would be highly desirable.
We hope to come back to some of these questions in the near future.

\newpage

\section*{Acknowledgements}

We would like to thank Asaad Elkhidir, Aidan Herderschee, Donal O'Connell, Radu Roiban, Matteo Sergola,  Fei Teng and  Ingrid Vazquez-Holm
for  coordination on ongoing work. 
We thank  Congkao Wen for  initial collaboration on this project and interesting discussions on related topics, Fabian Bautista, Brando Bellazzini, Donato Bini, Alessandra Buonanno, Thibault Damour, Riccardo Gonzo, Pavel Novichkov, Jan Plefka and Rodolfo Russo for several interesting conversations, Zheyong Fan for aid on numerical integrations, Roman Lee for help with the LiteRed2 package, and Alex Owen  for computer assistance. 
AB, GB, JG and  GT would like to thank the Kavli Institute for Theoretical Physics at the University of California, Santa Barbara, where their research was supported in part by the National Science Foundation under Grant
No.~PHY-1748958.
This work was supported by the Science and Technology Facilities Council (STFC) Consolidated Grants ST/P000754/1 \textit{``String theory, gauge theory \& duality''} and  ST/T000686/1 \textit{``Amplitudes, strings  \& duality''}, and  by the European Union's Horizon 2020 research and innovation programme under the Marie Sk\l{}odowska-Curie grant agreement No.~764850 {\it ``\href{https://sagex.org}{SAGEX}''}.
The work of GRB and JG  is supported by an STFC quota studentship. GC has received funding from the European Union's Horizon 2020 research and innovation program under the Marie Sk\l{}odowska-Curie grant agreement No.~847523 ``INTERACTIONS''. SDA's research is supported by the European Research Council, under grant ERC–AdG–885414. This research utilised Queen Mary's Apocrita HPC facility, supported by \href{http://doi.org/10.5281/zenodo.438045}{QMUL Research-IT.}
No new data were generated or analysed during this study.

\newpage

\appendix

\section{Integrals from differential equations}
\label{sec:diffEq}

\subsection{The differential equation for \texorpdfstring{$j_{a_1,1,a_3,a_4,0}$}{ja1aa0} with respect to \texorpdfstring{$w_1$}{w1}}

The subset of MIs which appear in our basis as sub-topologies of  $j_{1,1,1,1,0}$ are
\begin{equation}
	\vec{j}_1 = \begin{pmatrix} j_{0,1,0,1,0} \\ j_{1,1,0,1,0} \\ j_{1,1,1,0,0} \\ j_{1,1,1,1,0} \end{pmatrix}\ ,
\end{equation}
and the differential equation with respect to $w_1$ in $D=4-2\epsilon$ looks like
\begin{equation}
	\frac{\partial \vec{j}_1}{\partial w_1} = ( A_0 + \epsilon A_1) \vec{j}_1\ ,
\end{equation}
where
\begin{align}
	A_0 & = \begin{pmatrix}
		        \frac{1}{w_1}                                                             & 0                        & 0              & 0 \\
		        -\frac{1}{w_1 \left(w_1^2-q_2^2\right)}                                   & -\frac{w_1}{w_1^2-q_2^2} & 0              & 0 \\
		        0                                                                         & 0                        & 0              & 0 \\
		        -\frac{2}{w_1 \left[ 4 w_1^2 q_2^2 + \left(q_1^2 - q_2^2\right)^2\right]} &
		        0                                                                         & 0                        & -\frac{1}{w_1}
	        \end{pmatrix}\ ,                                                                                                                \\
	A_1 & = \begin{pmatrix}
		        -\frac{2}{w_1}                                                                            & 0                                                                                                    & 0 & 0 \\
		        \frac{2}{w_1 \left(w_1^2-q_2^2\right)}                                                    & \frac{2 w_1}{w_1^2-q_2^2}                                                                            & 0 & 0 \\
		        0                                                                                         & 0                                                                                                    & 0 & 0 \\
		        \frac{4}{w_1 \left[ 4 w_1^2 q_2^2 + \left(q_1^2 - q_2^2\right)^2\right]}                  & \frac{4 \left(2
		        w_1^2+q_1^2-q_2^2\right)}{w_1 \left[ 4 w_1^2 q_2^2 + \left(q_1^2 - q_2^2\right)^2\right]} & -\frac{4
		        \left(q_1^2-q_2^2\right)}{w_1 \left[ 4 w_1^2 q_2^2 + \left(q_1^2 - q_2^2\right)^2\right]} & -\frac{2 \left(q_1^2-q_2^2\right)^2}{w_1 \left[ 4 w_1^2 q_2^2 + \left(q_1^2 - q_2^2\right)^2\right]}
	        \end{pmatrix}\ .
\end{align}

If we normalise the integrals in the basis through the transformation
\begin{equation}
	\vec{j}_1 \rightarrow \vec{j}^\prime_1 =  S_1^{-1} \vec{j}_1\ ,
\end{equation}
with
\begin{equation}
	S_1 = \begin{pmatrix}
		\frac{4w_1 \epsilon }{2 \epsilon -1} & 0                            & 0                       & 0                    \\
		0                                    & \frac{1}{\sqrt{w_1^2-q_2^2}} & 0                       & 0                    \\
		0                                    & 0                            & \frac{1}{\sqrt{-q_2^2}} & 0                    \\
		0                                    & 0                            & 0                       & -\frac{1}{q_2^2 w_1} \\
	\end{pmatrix}\ ,
\end{equation}
the system of differential equations takes the canonical form
\begin{equation}
	\frac{\partial \vec{j}^\prime_1}{\partial w_1} = \epsilon A_{\rm res}  \vec{j}^\prime_1\ ,
\end{equation}
with
\begin{equation}
	A_{\rm res} = \begin{pmatrix}
		-\frac{2}{w_1}                                                                                                         & 0                                                                                              & 0                                                                                                    & 0 \\
		\frac{4}{\sqrt{w_1^2-q_2^2}}                                                                                           & \frac{2 w_1}{w_1^2-q_2^2}                                                                      & 0                                                                                                    & 0 \\
		0                                                                                                                      & 0                                                                                              & 0                                                                                                    & 0 \\
		- \frac{8 q_2^2 w_1}{ 4 w_1^2 q_2^2 + \left(q_1^2 - q_2^2\right)^2 }                                                   & - \frac{4
		q_2^2 \left(2w_1^2+q_1^2-q_2^2\right)}{ \sqrt{w_1^2-q_2^2} \left[ 4 w_1^2 q_2^2 + \left(q_1^2 - q_2^2\right)^2\right]} & -\frac{4 \sqrt{-q_2^2} \left(q_1^2-q_2^2\right)}{4 w_1^2 q_2^2 + \left(q_1^2 - q_2^2\right)^2} & -\frac{2 \left(q_1^2-q_2^2\right)^2}{w_1 \left[ 4 w_1^2 q_2^2 + \left(q_1^2 - q_2^2\right)^2\right]}
	\end{pmatrix}\ .
\end{equation}
In order to make the singularities of the integrals manifest, we need to rationalise the system of differential equations and write it in $d$log forms. We define
\begin{equation}
	\begin{split}
		w_1 &= \sqrt{-q_2^2} \frac{\alpha^2-1}{2\alpha}\ ,\\
		(-q_1^2) &= (-q_2^2)\, \beta\ ,
	\end{split}
\end{equation}
with $\alpha\geq 1$. Then, the system of differential equation can be written in term of forms~as
\begin{equation}
	d \vec{j}^\prime_1 (\alpha,\epsilon) = \epsilon\ d A^\prime_{\rm res} (\alpha) \ \vec{j}^\prime_1 (\alpha,\epsilon)\ ,
\end{equation}
where we omitted any dependence on the kinematic variables other than $\alpha$, and
\begin{equation}
	\begin{split}
		d A_{\rm res}^\prime (\alpha) &= \begin{pmatrix}
			-2 & 0 & 0 & 0  \\
			0  & 0 & 0 & 0  \\
			0  & 0 & 0 & 0  \\
			0  & 0 & 0 & -2
		\end{pmatrix} d{\rm log} (\alpha^2 -1) + \begin{pmatrix}
			0 & 0 & 0 & 0 \\
			0 & 2 & 0 & 0 \\
			0 & 0 & 0 & 0 \\
			0 & 0 & 0 & 0
		\end{pmatrix} d{\rm log} (\alpha^2 +1)+ \begin{pmatrix}
			2                  & 0                   & 0 & 0 \\
			4                  & -2                  & 0 & 0 \\
			0                  & 0                   & 0 & 0 \\
			\frac{2}{\beta ^2} & -\frac{2}{\beta ^2} & 0 & 0
		\end{pmatrix} d{\rm log} (\alpha)\\
		&+ \begin{pmatrix}
			0                   & 0                  & 0                & 0 \\
			0                   & 0                  & 0                & 0 \\
			0                   & 0                  & 0                & 0 \\
			-\frac{1}{\beta ^2} & \frac{1}{\beta ^2} & \frac{1}{\beta } & 1 \\
		\end{pmatrix} d{\rm log} (\alpha+\beta)+ \begin{pmatrix}
			0                   & 0                   & 0                 & 0 \\
			0                   & 0                   & 0                 & 0 \\
			0                   & 0                   & 0                 & 0 \\
			-\frac{1}{\beta ^2} & -\frac{1}{\beta ^2} & -\frac{1}{\beta } & 1 \\
		\end{pmatrix} d{\rm log} \left(\alpha+\frac{1}{\beta}\right)\\
		&+ \begin{pmatrix}
			0                   & 0                  & 0                 & 0 \\
			0                   & 0                  & 0                 & 0 \\
			0                   & 0                  & 0                 & 0 \\
			-\frac{1}{\beta ^2} & \frac{1}{\beta ^2} & -\frac{1}{\beta } & 1 \\
		\end{pmatrix} d{\rm log} (\alpha-\beta)+ \begin{pmatrix}
			0                   & 0                   & 0                & 0 \\
			0                   & 0                   & 0                & 0 \\
			0                   & 0                   & 0                & 0 \\
			-\frac{1}{\beta ^2} & -\frac{1}{\beta ^2} & \frac{1}{\beta } & 1 \\
		\end{pmatrix} d{\rm log} \left(\alpha-\frac{1}{\beta}\right)\ .
	\end{split}
\end{equation}
The solution to the differential equations are given by
\begin{equation}
	\label{eq:pathorderedsolution1}
	\vec{j}_1 (\alpha,\epsilon) = S_1 (\alpha ) \cdot \mathbb{P} e^{\epsilon \int_{\alpha_0}^\alpha d A_{\rm res}^\prime(\alpha)} \cdot S_1^{-1}(\alpha_0 )\cdot \vec{j}_1 (\alpha_0,\epsilon)\ ,
\end{equation}
where $\vec{j}_1 (\alpha_0,\epsilon)$ is a chosen boundary value of the Feynman integral which needs to be fixed, $\mathbb{P} e^{\epsilon \int_{\alpha_0}^\alpha d A_{\rm res}^\prime(\alpha)}$ is a path-ordered exponential. %We notice from the $d$log representation of $d A_{\rm res}^\prime$ that there are two singularities on the physical region which are not expected from the leading singularities of the Feynman integrals we are considering, {\it i.e.} $\alpha = \beta$ and $\alpha=\frac{1}{\beta}$. Then, the requirement that such singularities do not appear in the $\epsilon$ expansion of the solution \eqref{eq:pathorderedsolution1} fixes $j_{1,1,1,1,0}(\alpha_0,\epsilon)$ and $j_{1,1,0,1,0}(\alpha_0,\epsilon)$ in terms of $j_{1,1,1,0,0}(\alpha_0,\epsilon)$ and $j_{0,1,0,1,0}(\alpha_0,\epsilon)$. In particular, the latter can easily be computed by integrating directly over Feynman parameters. We want to evaluate the integrals up to $\mathcal{O}(\epsilon^0)$:

Finally, we are left with the evaluation of the boundary value of the $\mathcal{I}_5$ integral and we choose to compute its asymptotic behaviour near the singular point $\alpha\sim 1$ ($w_1 \sim 0$), using the geometric approach to the method of regions \cite{Pak:2010pt} implemented in the \texttt{Mathematica} package \texttt{asy2.1.m} \cite{Jantzen:2012mw}:

\begin{equation}
	\label{eq::box1Integral}
	\begin{split}
		j_{1,1,1,1,0} &= \frac{i \, \Gamma\left[4-\frac{D}{2}\right]}{(4\pi)^{D/2}} \int_{-\infty}^{+\infty} \!d x_2 \int_0^{+\infty} d^3 x_{1,3,4}\,\delta \left(1-\sum_l x_l\right)\, \left(x_1+x_3+x_4\right)^{4-D}\\
		&\hspace{6cm}\left(x_2^2 - q_1^2 x_1 x_3 - q_1^2 x_1 x_4 - 2w_1 x_2 x_4 - i \varepsilon\right)^{\frac{D}{2}-4}\\
		&\overset{\alpha \sim 1}{\sim} i \frac{\sqrt{\pi} \,\Gamma\left[7-\frac{D}{2}\right]}{(4\pi)^{D/2}} (-q_2^2)^{\frac{D-7}{2}} (\alpha -1)^{D-5} \int_0^{+\infty} d^3 x_{1,3,4}\,\delta \left(1-x_4\right) \left(x_1+x_3+x_4\right)^{4-D} \\
		&\hspace{9cm} \left(x_1 x_4 + x_1 x_3 \beta^2 -x_4^2 - i \varepsilon\right)^{\frac{D-7}{2}}\ \\
		& = - i \frac{2 \sqrt{\pi} \,\Gamma\left[7-\frac{D}{2}\right]}{(4\pi)^{D/2} (D-5)^2} e^{- i \pi \frac{D-5}{2}} (-q_2^2)^{\frac{D-7}{2}} (\alpha -1)^{D-5}\bigg[\pFq{2}{1}{1,1}{6-D}{\beta^2}\\
			& \hspace{8.2cm} -\pi (D-5) \frac{\beta ^{2 (D-5)} \left(1-\beta ^2\right)^{4-D}}{\sin \pi D}\Bigg]\ ,
	\end{split}
\end{equation}
where in the second step we performed the $x_2$ integration, redefined $(x_1,x_3,x_4)\rightarrow\left(x_1,\frac{x_3}{(\alpha-1)^2},\frac{x_4}{(\alpha-1)^2}\right)$ keeping only the leading term in the $\alpha\sim 1$ expansion and used the Cheng-Wu theorem \cite{Cheng:1987ga} to make the integration on $x_4$ trivial. The $\epsilon$-expansion in $D=4-2\epsilon$ gives us the boundary value of the integral we are after.

%\subsection{The DEs for $j_{a_1,1,a_3,0,a_5}$, $j_{a_1,1,0,a_4,a_5}$, $j_{0,1,a_3,a_4,a_5}$ with respect to $y$}
\subsection{The DEs for $j_{0,1,a_3,a_4,a_5}$ with respect to $y$}

The subsets of MIs which appear in our basis as sub-topologies of %$j_{1,1,1,0,1}$, $j_{1,1,0,1,1}$, 
$j_{0,1,1,1,1}$ respectively are
\begin{equation}
	\vec{j}_2 %= \begin{pmatrix} j_{0,1,1,0,1} \\ j_{1,1,1,0,0} \\ j_{1,1,1,0,1} \end{pmatrix}\ , \qquad \vec{j}_3 = \begin{pmatrix} j_{0,1,0,1,0} \\ j_{0,1,0,1,1} \\ j_{1,1,0,1,0} \\ j_{1,1,0,1,1} \end{pmatrix}\ ,\qquad \vec{j}_4 = 
	\begin{pmatrix} j_{0,1,0,1,0} \\ j_{0,1,0,1,1} \\ j_{0,1,1,0,1} \\ j_{0,1,1,1,1} \end{pmatrix}\ ,
\end{equation}
and the differential equations with respect to $y$ in $D=4-2\epsilon$ are
\begin{equation}
	\frac{\partial \vec{j}_2}{\partial y} = ( B_0 + \epsilon B_1) \vec{j}_2\ , %\qquad \frac{\partial \vec{j}_3}{\partial y} = ( C_0 + \epsilon C_1) \vec{j}_2\ , \qquad \frac{\partial \vec{j}_3}{\partial y} = ( D_0 + \epsilon D_1) \vec{j}_3\ , 
\end{equation}
where
\begin{align}
	%     B_0 &= \begin{pmatrix}
	%  \frac{y}{1-y^2} & 0 & 0 \\
	%  0 & 0 & 0 \\
	%  0 & 0 & \frac{y}{1-y^2}
	% \end{pmatrix}\ ,\\
	% B_1 &= \begin{pmatrix}
	%  \frac{2 y}{y^2-1} & 0 & 0 \\
	%  0 & 0 & 0 \\
	%  \frac{2 y}{w_2^2+q_1^2 y^2-q_1^2} & -\frac{4 w_2 y}{\left(y^2-1\right) \left(w_2^2+q_1^2 y^2-q_1^2\right)} & -\frac{2 w_2^2 y}{\left(y^2-1\right)
	%    \left(w_2^2+q_1^2 y^2-q_1^2\right)}
	% \end{pmatrix}\ ,\\
	% C_0 &= \begin{pmatrix}
	%  0 & 0 & 0 & 0 \\
	%  \frac{1}{w_1-w_1 y^2} & \frac{y}{1-y^2} & 0 & 0 \\
	%  0 & 0 & 0 & 0 \\
	%  -\frac{1}{-q_2^2 w_1+q_2^2 w_1 y^2} & 0 & 0 & \frac{y}{1-y^2}
	% \end{pmatrix}\ ,\\
	% C_1 &= \begin{pmatrix}
	%  0 & 0 & 0 & 0 \\
	%  -\frac{2}{w_1-w_1 y^2} & \frac{2 y}{y^2-1} & 0 & 0 \\
	%  0 & 0 & 0 & 0 \\
	%  \frac{2}{-q_2^2 w_1+q_2^2 w_1 y^2} & \frac{2 y}{w_1^2+q_2^2 y^2-q_2^2} & -\frac{4 w_1
	%    \left(w_1^2-q_2^2\right)}{-q_2^2 \left(y^2-1\right) \left(w_1^2+q_2^2 y^2-q_2^2\right)} & -\frac{2 w_1^2 y}{\left(y^2-1\right)
	%    \left(w_1^2+q_2^2 y^2-q_2^2\right)}
	% \end{pmatrix}\ ,\\
	%D_0 
	B_0 & = \begin{pmatrix}
		        0                                                      & 0                       & 0                       & 0 \\
		        \frac{1}{w_1-w_1 y^2}                                  & \frac{1}{\frac{1}{y}-y} & 0                       & 0 \\
		        0                                                      & 0                       & \frac{1}{\frac{1}{y}-y} & 0 \\
		        \frac{1}{4 w_1^2 w_2 y-2 w_1 \left(w_1^2+w_2^2\right)} & 0                       & 0                       & 0
	        \end{pmatrix}\ ,                                                                                         \\
	%D_1 
	B_1 & = \begin{pmatrix}
		        0                                         & 0                                                           & 0                                                           & 0                                           \\
		        -\frac{2}{w_1-w_1 y^2}                    & \frac{2 y}{y^2-1}                                           & 0                                                           & 0                                           \\
		        0                                         & 0                                                           & \frac{2 y}{y^2-1}                                           & 0                                           \\
		        \frac{1}{w_1^3-2 w_1^2 w_2 y +w_1 w_2^2 } & \frac{w_1 y-w_2}{w_1 \left(w_1^2-2 w_1 w_2 y +w_2^2\right)} & \frac{w_2 y-w_1}{w_2 \left(w_1^2-2 w_1 w_2 y +w_2^2\right)} & -\frac{2 w_1 w_2}{w_1^2-2 w_1 w_2 y +w_2^2}
	        \end{pmatrix}\ .
\end{align}
If we perform the change of basis with
\begin{align}
	S_2 & = % \begin{pmatrix}
	%  \frac{1}{\sqrt{y^2-1}} & 0 & 0 \\
	%  0 & \frac{1}{\sqrt{-q_1^2}} & 0 \\
	%  0 & 0 & \frac{1}{-q_1^2 \sqrt{y^2-1}}
	% \end{pmatrix}\ ,\\
	% S_3 &= \begin{pmatrix}
	%  \frac{\epsilon }{2 \epsilon -1} & 0 & 0 & 0 \\
	%  0 & \frac{1}{\sqrt{-q_2^2} \sqrt{y^2-1}} & 0 & 0 \\
	%  0 & 0 & \frac{1}{-q_2^2} & 0 \\
	%  0 & 0 & 0 & \frac{1}{\left(-q_2^2\right)^{3/2} \sqrt{y^2-1}}
	% \end{pmatrix}\ ,\\
	% S_4 &= 
	\begin{pmatrix}
		\frac{\epsilon }{2 \epsilon -1} & 0                          & 0                          & 0               \\
		0                               & \frac{1}{w_2 \sqrt{y^2-1}} & 0                          & 0               \\
		0                               & 0                          & \frac{1}{w_2 \sqrt{y^2-1}} & 0               \\
		0                               & 0                          & 0                          & \frac{1}{w_2^3}
	\end{pmatrix}\ ,
\end{align}
the systems of differential equations take the canonical form. Moreover, we can rationalise the system of differential equations and write it in terms of $d$log forms through the following change of variables:
\begin{equation}
	\begin{split}
		y & = \frac{1+x^2}{2x}\ ,\qquad 0< x \leq 1\ ,\\
		%w_2 & = \sqrt{-q_1^2} \gamma\ ,\\
		%w_1 & = \sqrt{-q_2^2} \delta\ ,\\
		\alpha^\prime & = %\frac{\delta}{\beta \gamma}
		\frac{w_1}{w_2}\ .
	\end{split}
\end{equation}
Then, we find
\begin{equation}
	d \vec{j}^\prime_2 %= \epsilon\ d B^\prime_{\rm res} (x) \ \vec{j}^\prime_2\ ,\qquad d \vec{j}^\prime_3  = \epsilon\ d C^\prime_{\rm res} (x) \ \vec{j}^\prime_3 \ ,\qquad d \vec{j}^\prime_4 
	= \epsilon\ d D^\prime_{\rm res} (x) \ \vec{j}^\prime_4 \ ,
\end{equation}
where we omitted any dependence on the kinematic variables other than $x$ and
\begin{equation}
	\begin{split}
		d B_{\rm res}^\prime (x)
		%     &= \begin{pmatrix}
		%     2 & 0 & 0 \\
		%  0 & 0 & 0 \\
		%  0 & 0 & -2
		%     \end{pmatrix} d{\rm log} (x+1)(1-x) + \begin{pmatrix}
		%     -2 & 0 & 0 \\
		%  0 & 0 & 0 \\
		%  2 & 0 & 0
		%     \end{pmatrix} d{\rm log} (x)\\
		%     &+ \begin{pmatrix}
		%     0 & 0 & 0 \\
		%  0 & 0 & 0 \\
		%  -1 & -2 & 1
		%     \end{pmatrix} d{\rm log} \left(x+\sqrt{\gamma ^2+1}-\gamma\right) \left(\sqrt{\gamma ^2+1}+\gamma -x\right)\\
		%     &+ \begin{pmatrix}
		%     0 & 0 & 0 \\
		%  0 & 0 & 0 \\
		%  -1 & 2 & 1
		%     \end{pmatrix} d{\rm log} \left(x-\sqrt{\gamma ^2+1}+\gamma\right) \left(x+\sqrt{\gamma ^2+1}+\gamma \right)\ ,
		% \end{split}
		% \end{equation}
		% \begin{equation}
		% \begin{split}
		%     d C_{\rm res}^\prime (x) &= \begin{pmatrix}
		%     0 & 0 & 0 & 0 \\
		%  0 & 2 & 0 & 0 \\
		%  0 & 0 & 0 & 0 \\
		%  0 & 0 & 0 & -2
		%     \end{pmatrix} d{\rm log} (x+1)(1-x) + \begin{pmatrix}
		%     0 & 0 & 0 & 0 \\
		%  -\frac{1}{\delta } & -2 & 0 & 0 \\
		%  0 & 0 & 0 & 0 \\
		%  \frac{1}{\delta } & 2 & 0 & 0
		%     \end{pmatrix} d{\rm log} (x)\\
		%     &+ \begin{pmatrix}
		%     0 & 0 & 0 & 0 \\
		%  0 & 0 & 0 & 0 \\
		%  0 & 0 & 0 & 0 \\
		%  0 & -1 & 2 \sqrt{\delta ^2+1} & 1
		%     \end{pmatrix} d{\rm log} \left(x-\delta+\sqrt{\delta ^2+1}\right) \left(x+\delta-\sqrt{\delta ^2+1}\right)\\
		%     &+ \begin{pmatrix}
		%     0 & 0 & 0 & 0 \\
		%  0 & 0 & 0 & 0 \\
		%  0 & 0 & 0 & 0 \\
		%  0 & -1 & -2 \sqrt{\delta ^2+1} & 1
		%     \end{pmatrix} d{\rm log} \left(\delta+\sqrt{\delta ^2+1}-x\right) \left(x+\delta+\sqrt{\delta ^2+1}\right)\ ,
		% \end{split}
		% \end{equation}
		% \begin{equation}
		% \begin{split}
		%     d D_{\rm res}^\prime (x) 
		&= \begin{pmatrix}
			0 & 0 & 0 & 0 \\
			0 & 2 & 0 & 0 \\
			0 & 0 & 2 & 0 \\
			0 & 0 & 0 & 0
		\end{pmatrix} d{\rm log} (x+1)(1-x) + \begin{pmatrix}
			0                               & 0                         & 0                         & 0  \\
			-\frac{1}{\alpha^\prime}        & -2                        & 0                         & 0  \\
			0                               & 0                         & -2                        & 0  \\
			\frac{1}{4 \alpha^{\prime\, 2}} & \frac{1}{2 \alpha^\prime} & \frac{1}{2 \alpha^\prime} & -1
		\end{pmatrix} d{\rm log} (x)\\
		&+ \begin{pmatrix}
			0                                & 0                         & 0                          & 0 \\
			0                                & 0                         & 0                          & 0 \\
			0                                & 0                         & 0                          & 0 \\
			-\frac{1}{4 \alpha^{\prime\, 2}} & \frac{1}{2 \alpha^\prime} & -\frac{1}{2 \alpha^\prime} & 1
		\end{pmatrix} d{\rm log} \left(x-\alpha^\prime\right)+ \begin{pmatrix}
			0                          & 0            & 0           & 0      \\
			0                          & 0            & 0           & 0      \\
			0                          & 0            & 0           & 0      \\
			-\frac{1}{4 \alpha^\prime} & -\frac{1}{2} & \frac{1}{2} & \alpha
		\end{pmatrix} d{\rm log} \left(x-\frac{1}{\alpha^\prime}\right)\ .
	\end{split}
\end{equation}
At this point, we only need a boundary value of the $\widetilde{\mathcal{I}}_4$ integral. We choose to compute its value around the regular point $x\sim 1$ ($y \sim 1$):
\begin{equation}
	\label{eq::box4Integral}
	\begin{split}
		j_{0,1,1,1,1} &= \frac{i \, \Gamma\left[4-\frac{D}{2}\right]}{(4\pi)^{D/2}} \int_{-\infty}^{+\infty} \!d x_2 \left(\int_0^{+\infty} d x_5 - \int_{-\infty}^0 d x_5\right) \int_0^{+\infty} d^2 x_{3,4}\,\delta \left(1-\sum_l x_l\right)\\
		&\hspace{3cm}\left(x_3+x_4\right)^{4-D} \left(x_2^2 + x_5^2 - 2 w_1 x_2 x_4 +2 w_2 x_3 x_5 + 2 y x_2 x_5 - i \varepsilon\right)^{\frac{D}{2}-4}\\
		&\overset{x \sim 1}{\sim} i \frac{\sqrt{\pi} \,\Gamma\left[7-\frac{D}{2}\right]}{(4\pi)^{D/2}} w_1^{D-6} \left(\int_0^{+\infty} d x_5 - \int_{-\infty}^0 d x_5\right) \int_0^{+\infty} d^2 x_{3,4}\,\delta \left(1-x_4\right) \left(x_3+x_4\right)^{4-D} \\
		&\hspace{6cm} \left(2 x_4 x_5 + 2 x_3 x_5 \alpha -x_4^2 - i \varepsilon\right)^{\frac{D-7}{2}}\ \\
		& = - i \frac{2 \sqrt{\pi} \,\Gamma\left[7-\frac{D}{2}\right]}{(4\pi)^{D/2} (D-5)^2} e^{- i \pi \frac{D-5}{2}} w_1^{D-6}\bigg[\pFq{2}{1}{1,1}{6-D}{\alpha}\\
			& \hspace{6.5cm} -\pi (D-5) \frac{\alpha ^{(D-5)} \left(1-\alpha\right)^{4-D}}{\sin \pi D}\Bigg]\ .
	\end{split}
\end{equation}

\section{Infrared divergences and  heavy-mass expansion}
\label{sec:Weinberg-resurrection}

In this appendix, we review the classic result of \cite{Weinberg:1965nx} for the infrared divergences in gravity and consider its limit in the large-mass expansion.

\subsection{Weinberg's formula for infrared divergences of gravitational amplitudes}

In \cite{Weinberg:1965nx}, Weinberg presented a compact formula for the resummation of infrared divergences in gravitational amplitudes arising from the exchange of virtual soft gravitons, in addition to re-deriving similar formulae for photons in electrodynamics, reproducing (and in part upgrading) the work of \cite{Bloch:1937pw,Jauch:1954rnc}.

There are two types of contribution:
first, each pair of particles in the initial or final state contributes an infrared-divergent phase $e^{- i W_{ij} \log \left(\frac{\lambda}{\Lambda}\right)}$, where
\begin{align}
	\label{Wij}
	W_{ij} = G \frac{m_i m_j(1 + \beta^2_{ij}) }{\beta_{ij} \sqrt{ 1 - \beta_{ij}^2}}\ = \ G (p_i\Cdot p_j) \frac{1 + \beta^2_{ij} }{\beta_{ij} }\, .
\end{align}
Here \begin{align}
	\beta_{ij} \coloneqq \sqrt{1 - \frac{m_i^2 m_j^2}{(p_i \Cdot p_j)^2}}
\end{align}
is the relative velocity of any one particle in the rest frame of the other, and $\lambda$ is an infrared cutoff (see below for a translation to dimensional regularisation).
This phase is usually discarded in the computation of observables such as cross sections but will be important for us.
In addition there is a divergent contribution of the type $e^{\sum_{i, j} B_{ij} \, \log \left(\frac{\lambda}{\Lambda}\right)}$, where
\begin{align}
	\begin{split}
		\label{IRphaseW}
		B_{ij}  & = \frac{G}{2\pi} \eta_i \eta_j \frac{m_i m_j }{\sqrt{1 - \beta_{ij}^2}} \frac{1 + \beta_{ij}^2}{\beta_{ij} }\log\frac{1+\beta_{ij} }{1 - \beta_{ij} }
		\\ &
		= \frac{G}{2\pi}\sum_{i, j} \eta_i \eta_j (p_i\Cdot p_j) \frac{1 + \beta_{ij}^2}{\beta_{ij} }\log\frac{1+\beta_{ij} }{1 - \beta_{ij} }  \ , \end{split}
\end{align}
where $\eta_{i} = \pm 1$ depending on whether the particle is outgoing $(+)$ or incoming $(-)$. In this case the sum is over all pairs of particles, including the case $i=j$.

If  one of two particles in a pair is massless, say particle $\hat{i}$, then $\beta_{\hat{i} j} \to 1$. In this case Weinberg's formulae simplify to
\begin{align}
	\label{We-massless}
	W_{\hat{i} j} \to 2 G (p_{\hat{i}}\Cdot p_j)  \, , \qquad
	B_{\hat{i}j} \to \frac{2 G}{\pi} \eta_{\hat{i} }\eta_j (p_{\hat{i}}\Cdot p_j) \log \left( \frac{2 (p_{\hat{i}}\Cdot p_j)}{\mu^2}\right)\, ,
\end{align}
with the result being independent of the choice of $\mu$ after summing over $j$; indeed, a shift in $\mu^2$ changes $\sum_j B_{\hat{i}j}$ by an amount proportional to $p_{\hat{i}}\Cdot \sum_j \eta_j p_j= - \eta_{\hat{i}} p_{\hat{i}}^2=0 $.
Note that we can combine $W_{\hat{i}j}$ and $B_{\hat{i}j}$ into one quantity valid for all kinematic regimes, by replacing the exponent by
\begin{align}
	\label{combinedIRmassless}
	\frac{2G}{\pi} \eta_i \eta_j (p_i \Cdot p_j) \log \big(\!-\eta_i \eta_i (p_i \Cdot p_j) + i \varepsilon \big)
	\ ,
\end{align}
where as usual $\log\big[\!-(s+ i \varepsilon)\big] = \log (s) - i \pi $ for $s>0$.%
\footnote{Note that \eqref{combinedIRmassless} is equivalent to the known universal form of infrared divergences for massless gravitons found in \cite{Dunbar:1995ed} after replacing  $2 (p_{\hat{i}}\Cdot p_j) \eta_i \eta_j \to s_{ij}$. Note that in the conventions that we are using, all energies are positive. }

We also comment that to express Weinberg's formula in dimensional regularisation. we need to make the replacement
\begin{align}
	\label{negeps}
	\log \left( \frac{\lambda}{\Lambda}\right) \to \frac{(4 \pi)^\eps}{\Gamma(1 - \eps)}
	\frac{\ \Lambda^{- 2 \eps} }{2 \eps}\ .
\end{align}
Note that both sides of \eqref{negeps} are negative.

% Internal note: this factor is obtained by multiplying together the missing angular contribution  
% \begin{align}
% \frac{\Omega (2-2 \eps)}{( 2\pi)^{1- 2 \eps} }  = \frac{(4 \pi)^\eps}{\Gamma(1 - \eps)}
% \, , 
% \end{align}
% and  
% \begin {align}
% \label{6.8}
% \int_0^\Lambda \frac{dk}{k } k^{- 2 \eps} = - \frac{\Lambda^{- 2 \eps} }{2 \eps} \ , 
% \end{align}
% where $\Omega (D) = \dfrac{2 \pi^{\frac{D}{2}}}{\Gamma\left( \frac{D}{2}\right)}$. 
% The  contribution obtained in this way replaces 
% \begin {align}
% \int_\lambda^\Lambda \frac{dk}{k } = \log \left( \frac{\Lambda}{\lambda}\right) \ ,  
% \end{align}
% which is a positive quantity, as \eqref{6.8}. 
% Note that officially speaking there is also another adjustment  to perform  in \eqref{IRphaseW}, namely 
% \begin{align}  
% \begin{split}
% \frac{1}{\beta} \log \frac{1 + \beta}{1-\beta}\to \frac{2^{1-2 \epsilon} \Gamma^2 (1-\epsilon ) \, \mbox{}_2F_1\Big(1, 1-\epsilon , 2-2 \epsilon
%    ; \frac{2 \beta }{\beta +1}\Big)}{(\beta +1) \Gamma (2-2 \epsilon )} = 
%    \frac{1}{\beta} \log \frac{1 + \beta}{1-\beta} + \cO (\eps)
% \, . \end{split}
% \end{align}
% This will shift slightly finite terms, but we are using Weinberg's formula only for the divergent part.  I just mention it in case someone wants to check this. 
% 

\subsection{The large-\texorpdfstring{$\bar{m}$}{mbar} expansion of Weinberg's formula}

We now apply Weinberg's formulae to the process we are describing. We will first discuss the  phase and then the real contribution.

The phase contribution arises from pairs of particles either in the initial or final states. Thus we have to consider the pairs $(1,2), (1^\prime, 2^\prime), (1^\prime, k), (2^\prime, k)$.

We will use \eqref{IRphaseW}, and compute the quantities $\beta_{ij}$ for the various cases. Because we want to perform a HEFT expansion, in order to be able to compare to the result of our calculation we need to expand $p_1\Cdot p_2$ and $p_1^\prime\Cdot p_2^\prime$ around $\bar{p}_1\Cdot \bar{p}_2$, that is an expansion in $\mb_i$.
We begin with the contributions from the pairs $(1,2), (1^\prime, 2^\prime)$.
These take the form
\begin{align}
	W_{12} = W_{\bar{1} \bar{2}  } + \Delta\, , \qquad
	W_{1^\prime 2^\prime} = W_{\bar{1} \bar{2}  } - \Delta\, ,
\end{align}
hence the contribution from pairs
$(1,2), (1^\prime, 2^\prime)$ is simply
\begin{align}
	e^{-2 i W_{\bar{1} \bar{2}}\log \left( \frac{\lambda}{\Lambda}\right)} \ ,
\end{align}
where
$W_{\bar{1} \bar{2}} $ is obtained from \eqref{Wij} by replacing $p_1, p_2$ with $\bar{p}_1, \bar{p}_2$.
When we expand the exponential in powers of $G$, its contribution will be of order $\bar{m}_1^3 \bar{m}_2^3$ in the large mass expansion. Then, there is no need to compute it --  as we have explained in Appendix~\ref{sec:factorisation}, 
such hyper-classical contributions are obtained simply from the exponentiation of lower-order amplitudes.

Next we consider the pairs $(1^\prime, k), (2^\prime, k)$. Because one of the particles is massless we can use  the first of \eqref{We-massless}. The result is then simply
$2G (p_1^\prime + p_2^\prime)\Cdot k = 2G(\mb_1 \bar{w}_1 + \mb_2 \bar{w}_2)$, with the corresponding contribution to the phase being
\begin{align}
	\label{Wpred}
	e^{- 2 i G (\mb_1 \bar{w}_1 + \mb_2 \bar{w}_2) \log \left( \frac{\lambda}{\Lambda}\right)}\ ,
\end{align}
which, once expanded, will appear in the large-mass expansion of the amplitude at the order we are considering.
Translating to dimensional regularisation, the expected infrared divergence at one loop is
\begin{align}
	\label{Wphase}
	-\frac{ i G (\bar{m}_1 \bar{w}_1 + \bar{m}_2 \bar{w}_2)}{\epsilon}  \cM_5^{(0)}\ ,
\end{align}
where $\cM_5^{(0)}$ is the classical tree-level five-point amplitude.

One can repeat the same calculation for the real part of the exponent. In this case one has to sum over all  pairs of particles, and a short calculation shows that  contributions cancel in pairs, both at order $\bar{m}_1^3 \bar{m}_2^3$, $\bar{m}_1^3 \bar{m}_2^2$ and $\bar{m}_1^2 \bar{m}_2^3$.

\section{Factorisation in impact parameter space}
\label{sec:factorisation}

The purpose of this section is to show that also in the presence of radiation, HEFT diagrams that are hyper-classical factorise in impact parameter space.%
\footnote{See also Section~4 of \cite{Cristofoli:2021jas} for a related discussion.}
To this end,  consider the one-loop five-point diagram with two cut massive lines shown below:
\begin{equation}\label{eq: 2mprdiagram}
	\begin{tikzpicture}[baseline={([yshift=-0.4ex]current bounding box.center)}]\tikzstyle{every node}=[font=\small]
		\begin{feynman}
			\vertex (p1);
			\vertex [above=2.35cm of p1](p2){};
			\vertex [left=1cm of p2](f1){$p_2=\pb_2+\frac{q_2}{2}$};
			\vertex [below=2.3cm of f1] (g1){$p_{1}=\pb_1+\frac{q_1}{2}$};
			\vertex [right=1.0cm of p2] (u1) [HV]{$~~~$};
			\vertex [below=1.25cm of u1] (m1) [GR2]{H};
			\vertex [below=1.25cm of m1] (b0) [HV]{$~~~$};
			\vertex [right=3cm of u1] (p3){};
			\vertex [right=1cm of p3] (f3){$p_{2'}=\pb_2-\frac{q_2}{2}$};
			\vertex[left=1cm of p3](u2)[HV]{$~~~$};
			\vertex [below=1.176cm of u2] (m2) [GR2]{H};
			\vertex [below=1.176cm of m2] (bx) [HV]{$~~~$};
			\vertex [right=1.0cm of p1] (b1) []{};
			\vertex [right=1.0cm of b1] (b2) []{};
			\vertex [right=0.85cm of b2] (b3) [HV]{$~~~$};
			\vertex [right=1.0cm of b3](p4);
			\vertex [above=1.25cm of p1] (cutL);
			\vertex [right=4.3cm of cutL] (cutR){$k$};
			\vertex [right=0.05cm of p4] (g3){$p_{1'}=\pb_1-\frac{q_1}{2}$};
			\vertex [right=0.82cm of b1] (cut1);
			\vertex [above=0.19cm of cut1] (cut1u);
			\vertex [below=0.19cm of cut1] (cut1b);
			\vertex [right=1.0cm of u1] (cut3);
			\vertex [above=0.17cm of cut3] (cut3u);
			\vertex [below=0.18cm of cut3] (cut3b);
			\vertex [above=0.4cm of cut1u] (cutd1);
			\vertex [right=0.4cm of cutd1] (cutd2);
			\vertex [right=0.5cm of b2] (cut2);
			\vertex [above=0.19cm of cut2] (cut2u);
			\vertex [below=0.19cm of cut2] (cut2b);
			\diagram* {
			(p2) -- [thick] (u1) -- [thick,
			edge label=$\delta(\pb_2 \cdot \ell)$] (u2)--[thick] (p3),
			(p1) -- [thick] (b0)-- [thick, edge label'=$\delta(\pb_1 \cdot \ell)$] (b3)-- [thick] (p4),  (cut1u)--[ red,thick] (cut1b),(cutR)-- [photon,ultra thick] (m2),(cut3u)--[ red,thick] (cut3b),
			};
		\end{feynman}
	\end{tikzpicture}\, \,,
\end{equation}
In impact parameter space, it becomes
\begin{align}
	\begin{split}
		\label{nonmanff}
		\widetilde{\cM}_5^{2{\rm MPR}} (\vec{b})  = &\int\!d^Dq_1 d^Dq_2 \, \delta^{(D)} (q_1 + q_2-k) \delta (\bar{p}_{1}\Cdot q_1 ) \delta (\bar{p}_{2}\Cdot q_2 )\, e^{i( q_1\Cdot b_1 + q_2 \Cdot b_2) }  \\ &
		\int\!d^D\ell \, \delta (\bar{p}_1\Cdot \ell)
		\delta (\bar{p}_2\Cdot \ell)\, \cM_4(\ell) \cM_5 (q_{1R}, q_{2R})\, ,
	\end{split}
\end{align}
where we have identified the four-point tree amplitude with momentum transfer $\ell$ and the five-point tree amplitude with shifted momentum transfers
\begin{align}
	q_{1R} = q_1 - \ell\, , \qquad
	q_{2R} = q_2 + \ell \ .
\end{align}
As usual, we adopt the parameterisation of the  kinematics
introduced in \eqref{barredv} (and in the four-point case we set $k=0$).
As a first step we define what we mean by    Fourier transforms to impact parameter space for four- and five-point tree-level amplitudes.
At four points we then define
\begin{align}
	\begin{split}
		\label{4pIPS}
		\widetilde{\cM}_4  (\vec{b}) & \coloneqq \int\!d^Dq_1 d^Dq_2 \, \delta^{(D)} (q_1 + q_2) \delta (\bar{p}_{1}\Cdot q_1 ) \delta (\bar{p}_{2}\Cdot q_2 )\, e^{i( q_1\Cdot b_1 + q_2 \Cdot b_2) }\cM_4 (q_1) \\ & =
		\int\!d^Dq  \, \delta (\bar{p}_1\Cdot q ) \delta (\bar{p}_2\Cdot q )\, e^{iq\Cdot (  b_1 -  b_2) }\cM_4 (q) \\
		&= \int\!d^Dq  \ \delta (\bar{p}_1\Cdot q ) \delta (\bar{p}_2\Cdot q )\, e^{iq\Cdot (  b_1 -  b_2) } \begin{tikzpicture}[baseline={([yshift=-0.8ex]current bounding box.center)}]\tikzstyle{every node}=[font=\small]	\begin{feynman}
				\vertex (a) {\(\pb_1+\frac{q}{2}\)};
				\vertex [right=1.15cm of a] (m);
				\vertex [right=2.4cm of a] (d){$\pb_{1}-\frac{q}{2}$};
				\vertex [above=0.25cm of m](x)[HV]{ $\mathcal{M}_4$};
				\vertex [above=1.6cm of a] (b){$\pb_2-\frac{q}{2}$};
				\vertex [right=2.4cm of b] (c){$\pb_{2}+\frac{q}{2}$};
				\diagram* {
				(a) -- [thick] (x) --  [thick] (c),
				(b)--[thick](x)--[thick](d)
				};
			\end{feynman}
		\end{tikzpicture}\, ,
	\end{split}
\end{align}
while at five points
\begin{align}
	\label{5pIPS}
	\begin{split}
		\widetilde{\cM}_5(\vec{b}) & \coloneqq \int\!d^Dq_1 d^Dq_2 \, \delta^{(D)} (q_1 + q_2-k) \delta (\bar{p}_{1}\Cdot q_1 ) \delta (\bar{p}_{2}\Cdot q_2 )\, e^{i( q_1\Cdot b_1 + q_2 \Cdot b_2) }\cM_5 (q_1, q_2) \\
		&=\int\!d^Dq_1 d^Dq_2 \, \delta^{(D)} (q_1 + q_2-k) \delta (\bar{p}_{1}\Cdot q_1 ) \delta (\bar{p}_{2}\Cdot q_2 )\, e^{i( q_1\Cdot b_1 + q_2 \Cdot b_2) }\begin{tikzpicture}[baseline={([yshift=-0.8ex]current bounding box.center)}]\tikzstyle{every node}=[font=\small]	\begin{feynman}
				\vertex (a) {\(\pb_1+\frac{q_1}{2}\)};
				\vertex [right=1.15cm of a] (m);
				\vertex [right=2.4cm of a] (d){$\pb_{1}-\frac{q_1}{2}$};
				\vertex [above=0.25cm of m](x)[HV]{$\mathcal{M}_5$};
				\vertex [above=1.6cm of a] (b){$\pb_2+\frac{q_2}{2}$};
				\vertex [right=2.4cm of b] (c){$\pb_{2}-\frac{q_2}{2}$};
				\vertex [above=0.8cm of d] (k){$k$};
				\diagram* {
				(a) -- [thick] (x) --  [thick] (c),
				(b)--[thick](x)--[thick](d), (k)--[photon, ultra thick](x)
				};
			\end{feynman}
		\end{tikzpicture}\,.
	\end{split}
\end{align}
We  now  turn to exposing  the factorised structure of \eqref{nonmanff}. In order to do so we change the order of integrations and observe that on the support of $\delta (\bar{p}_1\Cdot \ell)
	\delta (\bar{p}_2\Cdot \ell)$
we can rewrite
\begin{align}\begin{split}
		\bar{p}_{1}\Cdot q_1 \to \bar{p}_{1}\Cdot (q_1 - \ell) & = \bar{p}_{1}\Cdot q_{1R}\, , \\
		\bar{p}_{2}\Cdot q_2 \to \bar{p}_{2}\Cdot (q_2 + \ell) &= \bar{p}_{2}\Cdot q_{2R}\, .
	\end{split}
\end{align}
Then, changing also integration variables from $(q_1,  q_2) \to (q_{1R}, q_{2R})$, where $q_1 + q_2 = q_{1R} + q_{2R} = k$,
and using
\begin{align}
	q_1\Cdot b_1 +  q_2\Cdot b_2 = q_{1R}\Cdot b_1  +  q_{2R}\Cdot b_2 + \ell\Cdot (b_1 - b_2)\, ,
\end{align}
we can rewrite \eqref{nonmanff} as
\begin{align}
	\begin{split}
		\label{manff}
		\widetilde{\cM}_5^{2{\rm MPR}} (\vec{b})  &=
		\int\!d^D\ell \, \delta (\bar{p}_1\Cdot \ell)
		\delta (\bar{p}_2\Cdot \ell)\, e^{i \ell\Cdot ( b_1 - b_2)}\cM_4(\ell) \\
		&
		\int\!d^Dq_{1R} d^Dq_{2R} \, \delta^{(D)} (q_{1R} {+} q_{2R}{-}k) \delta (\bar{p}_{1}\Cdot q_{1R} ) \delta (\bar{p}_{2}\Cdot q_{2R} )\, e^{i( q_{1R}\Cdot b_{1} {+} q_{2R} \Cdot b_{2}) }\cM_5 (q_{1R}, q_{2R})   \\
		&=
		\widetilde{\cM}_4(\vec{b}) \, \widetilde{\cM}_5(\vec{b}) \, ,
	\end{split}
\end{align}
where we have used \eqref{4pIPS} and \eqref{5pIPS}.

In conclusion, when transformed to impact parameter space,
the particular two massive particle reducible diagram considered in
\eqref{eq: 2mprdiagram} is a product of two tree-level amplitudes in impact parameter space. This shows that  factorisation is made manifest by the HEFT expansion in impact parameter space also in the radiative case.

A short comment is in order here. In the four-point case \eqref{4pIPS}, the delta functions impose that $q$ lives in the $(D{-}2)$-dimensional subspace orthogonal to $\bar{p}_1$ and $\bar{p}_2$, hence $\widetilde{\cM}_4 (\vec{b}) $ will   depend only on the projection $\vec{b}_{\perp}$ of $\vec{b}$ living in the same subspace. For the case of $\widetilde{\cM}_5 (\vec{b}) $ in \eqref{5pIPSbis}, the particular orthogonal subspace is slightly different because of the presence of $k$ within the delta functions. This is a general feature of five-point kinematics and beyond.

%-------------------------------------------------------------
\section{Details of the \texorpdfstring{$\cC_4$}{} calculation}
\label{app: c4Details}
Here we present the details of the computation of the cut $\cC_4$ discussed in Section~\ref{sec:second-snail}. 
 Starting from the cut diagram \eqref{eq: cutfour}, if we simply plug in the HEFT  amplitudes as before we obtain
\begin{equation}\label{eq: C4Naive}
	\int\! \frac{d^D \ell}{(2 \pi)^{D}} \delta(\vb_1\Cdot \ell_1)\sum_{h_3} \frac{\cM_6^{h_3}(\ell_3, k, \vb_1, \vb_2)
		\cA_3^{-h_3}(-\ell_3,\vb_1)}{ \ell_3^2  }\, .
\end{equation}
An unpleasant feature of the above integrand
is that it contains divergences of the form
\begin{equation}
    \frac{\delta(\vb_1\Cdot \ell_1)}{\vb_1\Cdot \ell_3}= -\frac{\delta(\vb_1\Cdot \ell_1)}{\vb_1\Cdot \ell_1}\,, \quad \text{and} \quad \frac{\delta(\vb_1\Cdot \ell_1)}{\vb_1\Cdot \ell_1(\vb_1\Cdot \ell_1 - \vb_1\Cdot q_2)}
\end{equation}
which come from the linearised massive propagators present in the six-point HEFT amplitude.
To deal with these divergences we perform the following steps:
\begin{itemize}
    \item[{\bf 1.}] First we consider $\cC_4$ without the delta function $\delta(\vb_!\Cdot \ell_1)$ and then expand the integrand in powers of $\vb_1\Cdot \ell_1$.
    \item[{\bf 2.}] Then we only keep the $(\vb_1\Cdot \ell_1)^0$ term in the expansion which has no divergences. 
    \item[{\bf 3.}] Finally, we reinstate the $\delta(\vb_1\Cdot \ell_1)$.
\end{itemize}
Despite this procedure being rather ad hoc, the resulting expression for $\cC_4$ merges exactly with the other cut diagrams. In addition, the resulting amplitude built using $\cC_4$ passes many nontrivial checks (see Section \ref{sec:finalresult}) including the cancellation of spurious poles involving the new contribution from this cut diagram.

To make the calculation of $\cC_4$ more rigorous here we outline how to calculate $\cC_4$ using the forward limit, which will turn out to be equivalent to the simplified procedure above. First, we find the cut tree-level amplitude with three pairs of massive particles 
\begin{align}
    \mathcal{E}(q_1,q_2,q_3,\pb_{1},\pb_{2},\pb_{3})&=  \begin{tikzpicture}[baseline={([yshift=-0.4ex]current bounding box.center)}]\tikzstyle{every node}=[font=\small]
		      \begin{feynman}
			\vertex (p1) {\(p_1\)};
			\vertex [above=2.5cm of p1](p2){$p_2$};
			\vertex [right=1.5cm of p2] (u1) [HV]{$~~~$};
                \vertex [right=1.5cm of u1] (u2) ;
			\vertex [below=1.25cm of u1] (m1) [GR2]{H};
			\vertex [below=1.25cm of m1] (b0) [HV]{$~~~$};
			\vertex [right=1.5cm of u1] (p3){$p_{2'}$};
			\vertex [right=1.0cm of p1] (b1) []{};
			\vertex [right=1.0cm of b1] (b2) []{};
			\vertex [right=1.0cm of b2] (p4){$p_{1'}$};
   			\vertex [above=2.0cm of p4] (k1){};
                \vertex [right=3.0cm of m1] (h3)[HV]{H};
                \vertex [right=1.5cm of m1] (cut){};
                \vertex [right=0.3cm of h3] (hr){};
                \vertex [above=1.3cm of hr] (hu){$p_{3'}$};
                \vertex [below=1.3cm of hr] (hd){$p_3$};
                \vertex [right=0.3cm of k1] (k1l){$k_1$};
                \vertex [above=0.5cm of cut] (cutu){};
                \vertex [below=0.5cm of cut] (cutd){};
			\diagram* {
			(p2) -- [thick] (u1)-- [thick] (p3), (p1) -- [thick] (b0)--  [thick] (p4), (m1)--[photon,ultra thick] (k1),
    (m1)--[photon,ultra thick] (h3), (h3)--[thick] (hu), (h3)--[thick] (hd), (cutu)--[dashed, red,thick] (cutd)
			};
		\end{feynman}
	      \end{tikzpicture}\\
    &=\sum_{h_3}\frac{\cM_6^{h_3}(-q_3, k, \pb_{1}, \pb_{2})
		\cA_3^{-h_3}(q_3,\pb_{3})}{ q_3^2  }\,  
\end{align}
For each massive line, we define a momentum transfer $ q_i\coloneqq p_i-p_{i'}$ and hence momentum conservation for this amplitude can be written as $q_1+q_2+q_3=k$. Both massive particles $p_1$ and $p_3$ have the same mass $p_{1}^2=p_{1'}^2=p_{3}^2=p_{3'}^2=m_1^2$, and we can define barred masses in the usual way: $\mb_{i}\coloneqq \sqrt{m_i^2-q_i^2/4}$. Thus the cut we are considering above is a cut in $q_3$, and is homogeneous in the masses with scaling $\mb_{1}^2 \mb_2^2 \mb_{3}^2$.

Once we have this cut amplitude we take the forward limit on the two massive lines $p_{1'}$ and $p_{3}$ to form the loop diagram in \eqref{eq: cutfour}. This method was also used in the worldline formalism \cite{Shen:2018ebu} to  compute classical radiation in dilaton gravity and Yang-Mills theory. The forward limit process is most clearly described by writing the amplitude in impact parameter space, which we obtain by performing a Fourier transform with respect to each $q_i$
\begin{equation}
\begin{split}
    \int\!\left(\prod_{i=1}^3 d^D q_i \, e^{iq_i\Cdot b_i} \right)& \delta (\bar{p}_{1}\Cdot q_1) \delta (\bar{p}_{2}\Cdot q_2) \delta (\bar{p}_{3}\Cdot q_3) \\
    &\delta^{(D)} (q_1 + q_2+q_3-k) \,  \mathcal{E}( q_1,q_2,q_3,\pb_1,\pb_2,\pb_3)\, .
\end{split}
\end{equation}
The first step is to reparameterise the momentum transfer as $q_1\rightarrow q_1-q_3$ which also shifts 
\begin{equation}
    \pb_1=p_1-\frac{q_1}{2}\rightarrow p_1-\frac{q_1}{2}+\frac{q_3}{2}= \pb_1+\frac{q_3}{2}\, .
\end{equation}
Applying this to the cut amplitude in impact parameter space we have
\begin{equation}
\begin{split}
    \int\!\left(\prod_{i=1}^2 d^D q_i \, e^{iq_i\Cdot b_i} \right)
    \int\! d^D q_3 \, e^{iq_3\Cdot (b_3-b_1)} & \delta ((\bar{p}_{1}+\frac{q_3}{2})\Cdot (q_1-q_3)) \delta (\bar{p}_{2}\Cdot q_2) \delta (\bar{p}_{3}\Cdot q_3) \\
    &\delta^{(D)} (q_1 + q_2-k) \,  \mathcal{E}( q_1-q_3,q_2,q_3,\pb_1+\frac{q_3}{2},\pb_2,\pb_3)\, .
\end{split}
\end{equation}
Note that this removes the appearance of $q_3$ in the momentum-conserving delta function. Next we take the forward limit by taking $p_3\rightarrow p_{1'}$ which amounts to the replacement
\begin{equation}
    \pb_3= p_3-\frac{q_3}{2}\rightarrow p_1+\frac{q_3}{2}-q_1= \pb_1-\frac{q_1}{2}+\frac{q_3}{2}\, .
\end{equation}
Since we are identifying particles 1 and 3 we will also identify their impact parameters $b_3\rightarrow b_1$. Applying the forward limit we find
\begin{equation}\label{eq: forward}
\begin{split}
    \int\!\left(\prod_{i=1}^2 d^D q_i \, e^{iq_i\Cdot b_i} \right)&\delta (\pb_1\Cdot q_1) \delta (\bar{p}_{2}\Cdot q_2)
    \int\! d^D q_3 \,   \delta (\pb_{1}\Cdot q_3- \frac{q_3}{2}(q_1-q_3)) \\
    &\delta^{(D)} (q_1 + q_2-k) \,  \mathcal{E}\Big( q_1-q_3,q_2,q_3,\pb_1+\frac{q_3}{2},\pb_2,\pb_1-\frac{q_1}{2}+\frac{q_3}{2}\Big)\,,
\end{split}
\end{equation}
which crucially is not singular. However, the above expression is no longer homogeneous in the mass $\mb_1=\sqrt{m_1^2-q_1^2/4}$ and must be re-expanded in the large-$\mb_1$ limit. Before performing this expansion we note that now the integrals over $q_1$ and $q_2$ simply transform the HEFT amplitude with four scalars to impact parameter space, and hence can be stripped off, leaving us with%
\footnote{We note that this expression for the forward limit is similar to expressions found in  \textit{e.g.}~\cite{Brandhuber:2005kd,Catani:2008xa,Caron-Huot:2010fvq} and it would be interesting to investigate this further.}
\begin{equation}\label{eq: forwardBefExp}
    \int\! d^D \ell_1 \,   \delta \Big(\pb_{1}\Cdot \ell_1 + \frac{\ell_1}{2}\Cdot(\ell_1+q_1)\Big) \\
    \delta^{(D)} (q_1 + q_2-k) \,  \mathcal{E}\Big( -\ell_1,q_2,q_3,\pb_1+\frac{q_1}{2}+\frac{\ell_1}{2},\pb_2,\pb_1+\frac{\ell_1}{2}\Big)\,,
\end{equation}
where we have reparameterised the momentum as $q_3=-\ell_3=\ell_1+q_1$ to make contact with our usual loop integration variables in \eqref{eq: C4Naive}. To recover $\cC_4$ the final step is to perform the large-$\mb_1$ expansion on  \eqref{eq: forwardBefExp}. However, we will perform this expansion in a specific way in order to make contact with the simplified procedure for computing $\cC_4$ described above:
\begin{itemize}
    \item[{\bf 1.}] First we only expand $\mathcal{E}$ as $\mb_1\rightarrow\infty$ and  never use the delta function in \eqref{eq: forwardBefExp} to simplify the expression. To leading order, this yields the term $\mathcal{E}( -\ell_1,q_2,q_3,\pb_1,\pb_2,\pb_1)$ which is $\cO(\mb_1^4\mb_2^2)$. This is exactly the naive integrand in \eqref{eq: C4Naive}, without the delta function.
    \item[{\bf 2.}] Next, we use the delta function in \eqref{eq: forwardBefExp} to replace all instances of $\pb_1\Cdot \ell_1$ in $\mathcal{E}$ with $\ell_1\Cdot(\ell_1+q_1)/2$, which results in many terms with differing powers of $\mb_1$. When this replacement happens to a denominator, for example,
    \begin{equation}
         \frac{1}{\pb_1\Cdot \ell_1}= \frac{2}{\ell_1\Cdot(\ell_1+q_1)}\,,
    \end{equation}
    the power of $\mb_1$ is increased and a spurious massless pole is introduced. These poles must cancel, which is why expanding $\mathcal{E}$ to only the leading order is justified.
    \item[{\bf 3.}] Now expanding the delta function in \eqref{eq: forwardBefExp} we can write 
    \begin{equation}
        \delta \Big(\pb_{1}\Cdot \ell_1 + \frac{\ell_1}{2}\Cdot(\ell_1+q_1)\Big)= \delta (\pb_{1}\Cdot \ell_1) +\cdots
    \end{equation}
    where $+\cdots$ are terms involving derivatives of the delta function $\delta'(\pb_1\Cdot \ell_1)$. These derivative terms can be dropped since, by construction, the expression for $\mathcal{\mathcal{E}}$ no longer depends on $\pb_1\Cdot \ell_1$.
    \item[{\bf 4.}] Finally, we re-expand the expression $\delta(\pb_1\Cdot \ell_1)\mathcal{E}$ in powers of $\mb_1$ to find that the leading term is of $\cO(\mb_{1}^3\mb_{2}^2)$, as expected.
\end{itemize}
The result of this process is exactly the same as the naive method described at the start of this appendix. As was the case for $\cC_2$, we now need to perform additional loop momentum reparameterisations in order to write  $\cC_4$ in the topology \eqref{eq: masterGraph}. 

As a final remark, we note that it would be possible to compute the entirety of $\cM_{\bar{m}_1^3 \bar{m}_2^2}^{(1)}$ in one fell swoop following the above method if we had instead started from the \textit{full} tree-level six-scalar one-graviton amplitude at order $\mb_1^2 \mb_2^2 \mb_3^2$. This would bypass the use of generalised unitarity completely. One can compute this tree-level amplitude using the BCFW method presented in Section \ref{sec: BCFWtwosources} and generalising it to six scalars. However, unsurprisingly, we found this amplitude contains a very large number of terms, so we find it more practical to split up the calculation using multiple cut diagrams and generalised unitarity.

\newpage

\bibliographystyle{JHEP}
\bibliography{ScatEq}

\providecommand{\href}[2]{#2}\begingroup\raggedright\begin{thebibliography}{100}

\bibitem{LIGOScientific:2016dsl}
{\scshape LIGO Scientific, Virgo} collaboration, \emph{{Binary Black Hole
  Mergers in the first Advanced LIGO Observing Run}},
  \href{https://doi.org/10.1103/PhysRevX.6.041015}{\emph{Phys. Rev. X}
  {\bfseries 6} (2016) 041015}
  [\href{https://arxiv.org/abs/1606.04856}{{\ttfamily 1606.04856}}].

\bibitem{LIGOScientific:2016aoc}
{\scshape LIGO Scientific, Virgo} collaboration, \emph{{Observation of
  Gravitational Waves from a Binary Black Hole Merger}},
  \href{https://doi.org/10.1103/PhysRevLett.116.061102}{\emph{Phys. Rev. Lett.}
  {\bfseries 116} (2016) 061102}
  [\href{https://arxiv.org/abs/1602.03837}{{\ttfamily 1602.03837}}].

\bibitem{LIGOScientific:2016sjg}
{\scshape LIGO Scientific, Virgo} collaboration, \emph{{GW151226: Observation
  of Gravitational Waves from a 22-Solar-Mass Binary Black Hole Coalescence}},
  \href{https://doi.org/10.1103/PhysRevLett.116.241103}{\emph{Phys. Rev. Lett.}
  {\bfseries 116} (2016) 241103}
  [\href{https://arxiv.org/abs/1606.04855}{{\ttfamily 1606.04855}}].

\bibitem{LIGOScientific:2017bnn}
{\scshape LIGO Scientific, VIRGO} collaboration, \emph{{GW170104: Observation
  of a 50-Solar-Mass Binary Black Hole Coalescence at Redshift 0.2}},
  \href{https://doi.org/10.1103/PhysRevLett.118.221101}{\emph{Phys. Rev. Lett.}
  {\bfseries 118} (2017) 221101}
  [\href{https://arxiv.org/abs/1706.01812}{{\ttfamily 1706.01812}}].

\bibitem{LIGOScientific:2017vwq}
{\scshape LIGO Scientific, Virgo} collaboration, \emph{{GW170817: Observation
  of Gravitational Waves from a Binary Neutron Star Inspiral}},
  \href{https://doi.org/10.1103/PhysRevLett.119.161101}{\emph{Phys. Rev. Lett.}
  {\bfseries 119} (2017) 161101}
  [\href{https://arxiv.org/abs/1710.05832}{{\ttfamily 1710.05832}}].

\bibitem{Travaglini:2022uwo}
G.~Travaglini et~al., \emph{{The SAGEX review on scattering amplitudes}},
  \href{https://doi.org/10.1088/1751-8121/ac8380}{\emph{J. Phys. A} {\bfseries
  55} (2022) 443001} [\href{https://arxiv.org/abs/2203.13011}{{\ttfamily
  2203.13011}}].

\bibitem{Iwasaki:1971iy}
Y.~Iwasaki, \emph{{Fourth-order gravitational potential based on quantum field
  theory}}, \href{https://doi.org/10.1007/BF02770190}{\emph{Lett. Nuovo Cim.}
  {\bfseries 1S2} (1971) 783}.

\bibitem{Iwasaki:1971vb}
Y.~Iwasaki, \emph{{Quantum theory of gravitation vs. classical theory.
  Fourth-order potential}},
  \href{https://doi.org/10.1143/PTP.46.1587}{\emph{Prog. Theor. Phys.}
  {\bfseries 46} (1971) 1587}.

\bibitem{Bjerrum-Bohr:2002gqz}
N.E.J.~Bjerrum-Bohr, J.F.~Donoghue and B.R.~Holstein, \emph{{Quantum
  gravitational corrections to the nonrelativistic scattering potential of two
  masses}}, \href{https://doi.org/10.1103/PhysRevD.71.069903}{\emph{Phys. Rev.
  D} {\bfseries 67} (2003) 084033}
  [\href{https://arxiv.org/abs/hep-th/0211072}{{\ttfamily hep-th/0211072}}].

\bibitem{Holstein:2004dn}
B.R.~Holstein and J.F.~Donoghue, \emph{{Classical physics and quantum loops}},
  \href{https://doi.org/10.1103/PhysRevLett.93.201602}{\emph{Phys. Rev. Lett.}
  {\bfseries 93} (2004) 201602}
  [\href{https://arxiv.org/abs/hep-th/0405239}{{\ttfamily hep-th/0405239}}].

\bibitem{Neill:2013wsa}
D.~Neill and I.Z.~Rothstein, \emph{{Classical Space-Times from the S Matrix}},
  \href{https://doi.org/10.1016/j.nuclphysb.2013.09.007}{\emph{Nucl. Phys. B}
  {\bfseries 877} (2013) 177}
  [\href{https://arxiv.org/abs/1304.7263}{{\ttfamily 1304.7263}}].

\bibitem{Bjerrum-Bohr:2013bxa}
N.~Bjerrum-Bohr, J.F.~Donoghue and P.~Vanhove, \emph{{On-shell Techniques and
  Universal Results in Quantum Gravity}},
  \href{https://doi.org/10.1007/JHEP02(2014)111}{\emph{JHEP} {\bfseries 02}
  (2014) 111} [\href{https://arxiv.org/abs/1309.0804}{{\ttfamily 1309.0804}}].

\bibitem{Bjerrum-Bohr:2014zsa}
N.E.J.~Bjerrum-Bohr, J.F.~Donoghue, B.R.~Holstein, L.~Plant\'e and P.~Vanhove,
  \emph{{Bending of Light in Quantum Gravity}},
  \href{https://doi.org/10.1103/PhysRevLett.114.061301}{\emph{Phys. Rev. Lett.}
  {\bfseries 114} (2015) 061301}
  [\href{https://arxiv.org/abs/1410.7590}{{\ttfamily 1410.7590}}].

\bibitem{Bjerrum-Bohr:2016hpa}
N.~Bjerrum-Bohr, J.F.~Donoghue, B.R.~Holstein, L.~Plante and P.~Vanhove,
  \emph{{Light-like Scattering in Quantum Gravity}},
  \href{https://doi.org/10.1007/JHEP11(2016)117}{\emph{JHEP} {\bfseries 11}
  (2016) 117} [\href{https://arxiv.org/abs/1609.07477}{{\ttfamily
  1609.07477}}].

\bibitem{Bern:1994zx}
Z.~Bern, L.J.~Dixon, D.C.~Dunbar and D.A.~Kosower, \emph{{One loop n point
  gauge theory amplitudes, unitarity and collinear limits}},
  \href{https://doi.org/10.1016/0550-3213(94)90179-1}{\emph{Nucl. Phys. B}
  {\bfseries 425} (1994) 217}
  [\href{https://arxiv.org/abs/hep-ph/9403226}{{\ttfamily hep-ph/9403226}}].

\bibitem{Bern:1994cg}
Z.~Bern, L.J.~Dixon, D.C.~Dunbar and D.A.~Kosower, \emph{{Fusing gauge theory
  tree amplitudes into loop amplitudes}},
  \href{https://doi.org/10.1016/0550-3213(94)00488-Z}{\emph{Nucl. Phys. B}
  {\bfseries 435} (1995) 59}
  [\href{https://arxiv.org/abs/hep-ph/9409265}{{\ttfamily hep-ph/9409265}}].

\bibitem{Bern:2019nnu}
Z.~Bern, C.~Cheung, R.~Roiban, C.-H.~Shen, M.P.~Solon and M.~Zeng,
  \emph{{Scattering Amplitudes and the Conservative Hamiltonian for Binary
  Systems at Third Post-Minkowskian Order}},
  \href{https://doi.org/10.1103/PhysRevLett.122.201603}{\emph{Phys. Rev. Lett.}
  {\bfseries 122} (2019) 201603}
  [\href{https://arxiv.org/abs/1901.04424}{{\ttfamily 1901.04424}}].

\bibitem{Bern:2019crd}
Z.~Bern, C.~Cheung, R.~Roiban, C.-H.~Shen, M.P.~Solon and M.~Zeng, \emph{{Black
  Hole Binary Dynamics from the Double Copy and Effective Theory}},
  \href{https://doi.org/10.1007/JHEP10(2019)206}{\emph{JHEP} {\bfseries 10}
  (2019) 206} [\href{https://arxiv.org/abs/1908.01493}{{\ttfamily
  1908.01493}}].

\bibitem{Parra-Martinez:2020dzs}
J.~Parra-Martinez, M.S.~Ruf and M.~Zeng, \emph{{Extremal black hole scattering
  at $\mathcal{O}(G^3)$: graviton dominance, eikonal exponentiation, and
  differential equations}},
  \href{https://doi.org/10.1007/JHEP11(2020)023}{\emph{JHEP} {\bfseries 11}
  (2020) 023} [\href{https://arxiv.org/abs/2005.04236}{{\ttfamily
  2005.04236}}].

\bibitem{Cheung:2020gyp}
C.~Cheung and M.P.~Solon, \emph{{Classical gravitational scattering at $
  \mathcal{O} $(G$^{3}$) from Feynman diagrams}},
  \href{https://doi.org/10.1007/JHEP06(2020)144}{\emph{JHEP} {\bfseries 06}
  (2020) 144} [\href{https://arxiv.org/abs/2003.08351}{{\ttfamily
  2003.08351}}].

\bibitem{Bjerrum-Bohr:2021din}
N.E.J.~Bjerrum-Bohr, P.H.~Damgaard, L.~Plant\'e and P.~Vanhove, \emph{{The
  amplitude for classical gravitational scattering at third Post-Minkowskian
  order}}, \href{https://doi.org/10.1007/JHEP08(2021)172}{\emph{JHEP}
  {\bfseries 08} (2021) 172}
  [\href{https://arxiv.org/abs/2105.05218}{{\ttfamily 2105.05218}}].

\bibitem{DiVecchia:2021bdo}
P.~Di~Vecchia, C.~Heissenberg, R.~Russo and G.~Veneziano, \emph{{The eikonal
  approach to gravitational scattering and radiation at $ \mathcal{O}
  $(G$^{3}$)}}, \href{https://doi.org/10.1007/JHEP07(2021)169}{\emph{JHEP}
  {\bfseries 07} (2021) 169}
  [\href{https://arxiv.org/abs/2104.03256}{{\ttfamily 2104.03256}}].

\bibitem{Brandhuber:2021eyq}
A.~Brandhuber, G.~Chen, G.~Travaglini and C.~Wen, \emph{{Classical
  gravitational scattering from a gauge-invariant double copy}},
  \href{https://doi.org/10.1007/JHEP10(2021)118}{\emph{JHEP} {\bfseries 10}
  (2021) 118} [\href{https://arxiv.org/abs/2108.04216}{{\ttfamily
  2108.04216}}].

\bibitem{Bern:2021dqo}
Z.~Bern, J.~Parra-Martinez, R.~Roiban, M.S.~Ruf, C.-H.~Shen, M.P.~Solon et~al.,
  \emph{{Scattering Amplitudes and Conservative Binary Dynamics at ${\cal
  O}(G^4)$}}, \href{https://doi.org/10.1103/PhysRevLett.126.171601}{\emph{Phys.
  Rev. Lett.} {\bfseries 126} (2021) 171601}
  [\href{https://arxiv.org/abs/2101.07254}{{\ttfamily 2101.07254}}].

\bibitem{Bern:2021yeh}
Z.~Bern, J.~Parra-Martinez, R.~Roiban, M.S.~Ruf, C.-H.~Shen, M.P.~Solon et~al.,
  \emph{{Scattering Amplitudes, the Tail Effect, and Conservative Binary
  Dynamics at O(G4)}},
  \href{https://doi.org/10.1103/PhysRevLett.128.161103}{\emph{Phys. Rev. Lett.}
  {\bfseries 128} (2022) 161103}
  [\href{https://arxiv.org/abs/2112.10750}{{\ttfamily 2112.10750}}].

\bibitem{Bern:2022jvn}
Z.~Bern, J.~Parra-Martinez, R.~Roiban, M.S.~Ruf, C.-H.~Shen, M.P.~Solon et~al.,
  \emph{{Scattering amplitudes and conservative dynamics at the fourth
  post-Minkowskian order}},
  \href{https://doi.org/10.22323/1.416.0051}{\emph{PoS} {\bfseries LL2022}
  (2022) 051}.

\bibitem{Luna:2017dtq}
A.~Luna, I.~Nicholson, D.~O'Connell and C.D.~White, \emph{{Inelastic Black Hole
  Scattering from Charged Scalar Amplitudes}},
  \href{https://doi.org/10.1007/JHEP03(2018)044}{\emph{JHEP} {\bfseries 03}
  (2018) 044} [\href{https://arxiv.org/abs/1711.03901}{{\ttfamily
  1711.03901}}].

\bibitem{Shen:2018ebu}
C.-H.~Shen, \emph{{Gravitational Radiation from Color-Kinematics Duality}},
  \href{https://doi.org/10.1007/JHEP11(2018)162}{\emph{JHEP} {\bfseries 11}
  (2018) 162} [\href{https://arxiv.org/abs/1806.07388}{{\ttfamily
  1806.07388}}].

\bibitem{Bautista:2019tdr}
Y.F.~Bautista and A.~Guevara, \emph{{From Scattering Amplitudes to Classical
  Physics: Universality, Double Copy and Soft Theorems}},
  \href{https://arxiv.org/abs/1903.12419}{{\ttfamily 1903.12419}}.

\bibitem{Herrmann:2021lqe}
E.~Herrmann, J.~Parra-Martinez, M.S.~Ruf and M.~Zeng, \emph{{Gravitational
  Bremsstrahlung from Reverse Unitarity}},
  \href{https://doi.org/10.1103/PhysRevLett.126.201602}{\emph{Phys. Rev. Lett.}
  {\bfseries 126} (2021) 201602}
  [\href{https://arxiv.org/abs/2101.07255}{{\ttfamily 2101.07255}}].

\bibitem{Herrmann:2021tct}
E.~Herrmann, J.~Parra-Martinez, M.S.~Ruf and M.~Zeng, \emph{{Radiative
  classical gravitational observables at $ \mathcal{O} $(G$^{3}$) from
  scattering amplitudes}},
  \href{https://doi.org/10.1007/JHEP10(2021)148}{\emph{JHEP} {\bfseries 10}
  (2021) 148} [\href{https://arxiv.org/abs/2104.03957}{{\ttfamily
  2104.03957}}].

\bibitem{Guevara:2017csg}
A.~Guevara, \emph{{Holomorphic Classical Limit for Spin Effects in
  Gravitational and Electromagnetic Scattering}},
  \href{https://doi.org/10.1007/JHEP04(2019)033}{\emph{JHEP} {\bfseries 04}
  (2019) 033} [\href{https://arxiv.org/abs/1706.02314}{{\ttfamily
  1706.02314}}].

\bibitem{Arkani-Hamed:2017jhn}
N.~Arkani-Hamed, T.-C.~Huang and Y.-t.~Huang, \emph{{Scattering Amplitudes For
  All Masses and Spins}},  \href{https://arxiv.org/abs/1709.04891}{{\ttfamily
  1709.04891}}.

\bibitem{Guevara:2018wpp}
A.~Guevara, A.~Ochirov and J.~Vines, \emph{{Scattering of Spinning Black Holes
  from Exponentiated Soft Factors}},
  \href{https://doi.org/10.1007/JHEP09(2019)056}{\emph{JHEP} {\bfseries 09}
  (2019) 056} [\href{https://arxiv.org/abs/1812.06895}{{\ttfamily
  1812.06895}}].

\bibitem{Chung:2018kqs}
M.-Z.~Chung, Y.-T.~Huang, J.-W.~Kim and S.~Lee, \emph{{The simplest massive
  S-matrix: from minimal coupling to Black Holes}},
  \href{https://doi.org/10.1007/JHEP04(2019)156}{\emph{JHEP} {\bfseries 04}
  (2019) 156} [\href{https://arxiv.org/abs/1812.08752}{{\ttfamily
  1812.08752}}].

\bibitem{Guevara:2019fsj}
A.~Guevara, A.~Ochirov and J.~Vines, \emph{{Black-hole scattering with general
  spin directions from minimal-coupling amplitudes}},
  \href{https://doi.org/10.1103/PhysRevD.100.104024}{\emph{Phys. Rev.}
  {\bfseries D100} (2019) 104024}
  [\href{https://arxiv.org/abs/1906.10071}{{\ttfamily 1906.10071}}].

\bibitem{Arkani-Hamed:2019ymq}
N.~Arkani-Hamed, Y.-t.~Huang and D.~O'Connell, \emph{{Kerr black holes as
  elementary particles}},
  \href{https://doi.org/10.1007/JHEP01(2020)046}{\emph{JHEP} {\bfseries 01}
  (2020) 046} [\href{https://arxiv.org/abs/1906.10100}{{\ttfamily
  1906.10100}}].

\bibitem{Aoude:2020onz}
R.~Aoude, K.~Haddad and A.~Helset, \emph{{On-shell heavy particle effective
  theories}}, \href{https://doi.org/10.1007/JHEP05(2020)051}{\emph{JHEP}
  {\bfseries 05} (2020) 051}
  [\href{https://arxiv.org/abs/2001.09164}{{\ttfamily 2001.09164}}].

\bibitem{Chung:2020rrz}
M.-Z.~Chung, Y.-t.~Huang, J.-W.~Kim and S.~Lee, \emph{{Complete Hamiltonian for
  spinning binary systems at first post-Minkowskian order}},
  \href{https://doi.org/10.1007/JHEP05(2020)105}{\emph{JHEP} {\bfseries 05}
  (2020) 105} [\href{https://arxiv.org/abs/2003.06600}{{\ttfamily
  2003.06600}}].

\bibitem{Guevara:2020xjx}
A.~Guevara, B.~Maybee, A.~Ochirov, D.~O'Connell and J.~Vines, \emph{{A
  worldsheet for Kerr}},
  \href{https://doi.org/10.1007/JHEP03(2021)201}{\emph{JHEP} {\bfseries 03}
  (2021) 201} [\href{https://arxiv.org/abs/2012.11570}{{\ttfamily
  2012.11570}}].

\bibitem{Chen:2021kxt}
W.-M.~Chen, M.-Z.~Chung, Y.-t.~Huang and J.-W.~Kim, \emph{{The 2PM Hamiltonian
  for binary Kerr to quartic in spin}},
  \href{https://doi.org/10.1007/JHEP08(2022)148}{\emph{JHEP} {\bfseries 08}
  (2022) 148} [\href{https://arxiv.org/abs/2111.13639}{{\ttfamily
  2111.13639}}].

\bibitem{Kosmopoulos:2021zoq}
D.~Kosmopoulos and A.~Luna, \emph{{Quadratic-in-spin Hamiltonian at $
  \mathcal{O} $(G$^{2}$) from scattering amplitudes}},
  \href{https://doi.org/10.1007/JHEP07(2021)037}{\emph{JHEP} {\bfseries 07}
  (2021) 037} [\href{https://arxiv.org/abs/2102.10137}{{\ttfamily
  2102.10137}}].

\bibitem{Chiodaroli:2021eug}
M.~Chiodaroli, H.~Johansson and P.~Pichini, \emph{{Compton black-hole
  scattering for s \ensuremath{\leq} 5/2}},
  \href{https://doi.org/10.1007/JHEP02(2022)156}{\emph{JHEP} {\bfseries 02}
  (2022) 156} [\href{https://arxiv.org/abs/2107.14779}{{\ttfamily
  2107.14779}}].

\bibitem{Bautista:2021wfy}
Y.F.~Bautista, A.~Guevara, C.~Kavanagh and J.~Vines, \emph{{Scattering in Black
  Hole Backgrounds and Higher-Spin Amplitudes: Part I}},
  \href{https://arxiv.org/abs/2107.10179}{{\ttfamily 2107.10179}}.

\bibitem{Cangemi:2022bew}
L.~Cangemi, M.~Chiodaroli, H.~Johansson, A.~Ochirov, P.~Pichini and
  E.~Skvortsov, \emph{{Kerr Black Holes Enjoy Massive Higher-Spin Gauge
  Symmetry}},  \href{https://arxiv.org/abs/2212.06120}{{\ttfamily 2212.06120}}.

\bibitem{Ochirov_2022}
A.~Ochirov and E.~Skvortsov, \emph{Chiral approach to massive higher spins},
  \href{https://doi.org/10.1103/physrevlett.129.241601}{\emph{Physical Review
  Letters} {\bfseries 129} (2022) }.

\bibitem{Damgaard:2019lfh}
P.H.~Damgaard, K.~Haddad and A.~Helset, \emph{{Heavy Black Hole Effective
  Theory}}, \href{https://doi.org/10.1007/JHEP11(2019)070}{\emph{JHEP}
  {\bfseries 11} (2019) 070}
  [\href{https://arxiv.org/abs/1908.10308}{{\ttfamily 1908.10308}}].

\bibitem{Bern:2020buy}
Z.~Bern, A.~Luna, R.~Roiban, C.-H.~Shen and M.~Zeng, \emph{{Spinning black hole
  binary dynamics, scattering amplitudes, and effective field theory}},
  \href{https://doi.org/10.1103/PhysRevD.104.065014}{\emph{Phys. Rev. D}
  {\bfseries 104} (2021) 065014}
  [\href{https://arxiv.org/abs/2005.03071}{{\ttfamily 2005.03071}}].

\bibitem{Comberiati:2022ldk}
F.~Comberiati and L.~de~la Cruz, \emph{{Classical off-shell currents}},
  \href{https://arxiv.org/abs/2212.09259}{{\ttfamily 2212.09259}}.

\bibitem{Maybee:2019jus}
B.~Maybee, D.~O'Connell and J.~Vines, \emph{{Observables and amplitudes for
  spinning particles and black holes}},
  \href{https://doi.org/10.1007/JHEP12(2019)156}{\emph{JHEP} {\bfseries 12}
  (2019) 156} [\href{https://arxiv.org/abs/1906.09260}{{\ttfamily
  1906.09260}}].

\bibitem{Haddad:2021znf}
K.~Haddad, \emph{{Exponentiation of the leading eikonal phase with spin}},
  \href{https://doi.org/10.1103/PhysRevD.105.026004}{\emph{Phys. Rev. D}
  {\bfseries 105} (2022) 026004}
  [\href{https://arxiv.org/abs/2109.04427}{{\ttfamily 2109.04427}}].

\bibitem{Chen:2022clh}
W.-M.~Chen, M.-Z.~Chung, Y.-t.~Huang and J.-W.~Kim, \emph{{Gravitational
  Faraday effect from on-shell amplitudes}},
  \href{https://doi.org/10.1007/JHEP12(2022)058}{\emph{JHEP} {\bfseries 12}
  (2022) 058} [\href{https://arxiv.org/abs/2205.07305}{{\ttfamily
  2205.07305}}].

\bibitem{Menezes:2022tcs}
G.~Menezes and M.~Sergola, \emph{{NLO deflections for spinning particles and
  Kerr black holes}},
  \href{https://doi.org/10.1007/JHEP10(2022)105}{\emph{JHEP} {\bfseries 10}
  (2022) 105} [\href{https://arxiv.org/abs/2205.11701}{{\ttfamily
  2205.11701}}].

\bibitem{FebresCordero:2022jts}
F.~Febres~Cordero, M.~Kraus, G.~Lin, M.S.~Ruf and M.~Zeng, \emph{{Conservative
  Binary Dynamics with a Spinning Black Hole at O(G3) from Scattering
  Amplitudes}},
  \href{https://doi.org/10.1103/PhysRevLett.130.021601}{\emph{Phys. Rev. Lett.}
  {\bfseries 130} (2023) 021601}
  [\href{https://arxiv.org/abs/2205.07357}{{\ttfamily 2205.07357}}].

\bibitem{Alessio:2022kwv}
F.~Alessio and P.~Di~Vecchia, \emph{{Radiation reaction for spinning black-hole
  scattering}},
  \href{https://doi.org/10.1016/j.physletb.2022.137258}{\emph{Phys. Lett. B}
  {\bfseries 832} (2022) 137258}
  [\href{https://arxiv.org/abs/2203.13272}{{\ttfamily 2203.13272}}].

\bibitem{Bern:2022kto}
Z.~Bern, D.~Kosmopoulos, A.~Luna, R.~Roiban and F.~Teng, \emph{{Binary Dynamics
  Through the Fifth Power of Spin at $\mathcal{O}(G^2)$}},
  \href{https://arxiv.org/abs/2203.06202}{{\ttfamily 2203.06202}}.

\bibitem{Aoude:2022thd}
R.~Aoude, K.~Haddad and A.~Helset, \emph{{Classical Gravitational
  Spinning-Spinless Scattering at O(G2S\ensuremath{\infty})}},
  \href{https://doi.org/10.1103/PhysRevLett.129.141102}{\emph{Phys. Rev. Lett.}
  {\bfseries 129} (2022) 141102}
  [\href{https://arxiv.org/abs/2205.02809}{{\ttfamily 2205.02809}}].

\bibitem{Aoude:2022trd}
R.~Aoude, K.~Haddad and A.~Helset, \emph{{Searching for Kerr in the 2PM
  amplitude}}, \href{https://doi.org/10.1007/JHEP07(2022)072}{\emph{JHEP}
  {\bfseries 07} (2022) 072}
  [\href{https://arxiv.org/abs/2203.06197}{{\ttfamily 2203.06197}}].

\bibitem{Bjerrum-Bohr:2023jau}
N.E.J.~Bjerrum-Bohr, G.~Chen and M.~Skowronek, \emph{{Classical Spin
  Gravitational Compton Scattering}},
  \href{https://arxiv.org/abs/2302.00498}{{\ttfamily 2302.00498}}.

\bibitem{Brandhuber:2019qpg}
A.~Brandhuber and G.~Travaglini, \emph{{On higher-derivative effects on the
  gravitational potential and particle bending}},
  \href{https://doi.org/10.1007/JHEP01(2020)010}{\emph{JHEP} {\bfseries 01}
  (2020) 010} [\href{https://arxiv.org/abs/1905.05657}{{\ttfamily
  1905.05657}}].

\bibitem{Emond:2019crr}
W.T.~Emond and N.~Moynihan, \emph{{Scattering Amplitudes, Black Holes and
  Leading Singularities in Cubic Theories of Gravity}},
  \href{https://doi.org/10.1007/JHEP12(2019)019}{\emph{JHEP} {\bfseries 12}
  (2019) 019} [\href{https://arxiv.org/abs/1905.08213}{{\ttfamily
  1905.08213}}].

\bibitem{AccettulliHuber:2019jqo}
M.~Accettulli~Huber, A.~Brandhuber, S.~De~Angelis and G.~Travaglini,
  \emph{{Note on the absence of $R^2$ corrections to Newton\textquoteright{}s
  potential}}, \href{https://doi.org/10.1103/PhysRevD.101.046011}{\emph{Phys.
  Rev. D} {\bfseries 101} (2020) 046011}
  [\href{https://arxiv.org/abs/1911.10108}{{\ttfamily 1911.10108}}].

\bibitem{AccettulliHuber:2020oou}
M.~Accettulli~Huber, A.~Brandhuber, S.~De~Angelis and G.~Travaglini,
  \emph{{Eikonal phase matrix, deflection angle and time delay in effective
  field theories of gravity}},
  \href{https://doi.org/10.1103/PhysRevD.102.046014}{\emph{Phys. Rev. D}
  {\bfseries 102} (2020) 046014}
  [\href{https://arxiv.org/abs/2006.02375}{{\ttfamily 2006.02375}}].

\bibitem{AccettulliHuber:2020dal}
M.~Accettulli~Huber, A.~Brandhuber, S.~De~Angelis and G.~Travaglini,
  \emph{{From amplitudes to gravitational radiation with cubic interactions and
  tidal effects}},
  \href{https://doi.org/10.1103/PhysRevD.103.045015}{\emph{Phys. Rev. D}
  {\bfseries 103} (2021) 045015}
  [\href{https://arxiv.org/abs/2012.06548}{{\ttfamily 2012.06548}}].

\bibitem{Carrillo-Gonzalez:2021mqj}
M.~Carrillo-Gonz\'alez, C.~de~Rham and A.J.~Tolley, \emph{{Scattering
  amplitudes for binary systems beyond GR}},
  \href{https://doi.org/10.1007/JHEP11(2021)087}{\emph{JHEP} {\bfseries 11}
  (2021) 087} [\href{https://arxiv.org/abs/2107.11384}{{\ttfamily
  2107.11384}}].

\bibitem{Bellazzini:2021shn}
B.~Bellazzini, G.~Isabella, M.~Lewandowski and F.~Sgarlata,
  \emph{{Gravitational causality and the self-stress of photons}},
  \href{https://doi.org/10.1007/JHEP05(2022)154}{\emph{JHEP} {\bfseries 05}
  (2022) 154} [\href{https://arxiv.org/abs/2108.05896}{{\ttfamily
  2108.05896}}].

\bibitem{Donoghue:1994dn}
J.F.~Donoghue, \emph{{General relativity as an effective field theory: The
  leading quantum corrections}},
  \href{https://doi.org/10.1103/PhysRevD.50.3874}{\emph{Phys. Rev. D}
  {\bfseries 50} (1994) 3874}
  [\href{https://arxiv.org/abs/gr-qc/9405057}{{\ttfamily gr-qc/9405057}}].

\bibitem{Buonanno:1998gg}
A.~Buonanno and T.~Damour, \emph{{Effective one-body approach to general
  relativistic two-body dynamics}},
  \href{https://doi.org/10.1103/PhysRevD.59.084006}{\emph{Phys. Rev. D}
  {\bfseries 59} (1999) 084006}
  [\href{https://arxiv.org/abs/gr-qc/9811091}{{\ttfamily gr-qc/9811091}}].

\bibitem{Damour:2016gwp}
T.~Damour, \emph{{Gravitational scattering, post-Minkowskian approximation and
  Effective One-Body theory}},
  \href{https://doi.org/10.1103/PhysRevD.94.104015}{\emph{Phys. Rev. D}
  {\bfseries 94} (2016) 104015}
  [\href{https://arxiv.org/abs/1609.00354}{{\ttfamily 1609.00354}}].

\bibitem{Damour:2017zjx}
T.~Damour, \emph{{High-energy gravitational scattering and the general
  relativistic two-body problem}},
  \href{https://doi.org/10.1103/PhysRevD.97.044038}{\emph{Phys. Rev. D}
  {\bfseries 97} (2018) 044038}
  [\href{https://arxiv.org/abs/1710.10599}{{\ttfamily 1710.10599}}].

\bibitem{Vines:2017hyw}
J.~Vines, \emph{{Scattering of two spinning black holes in post-Minkowskian
  gravity, to all orders in spin, and effective-one-body mappings}},
  \href{https://doi.org/10.1088/1361-6382/aaa3a8}{\emph{Class. Quant. Grav.}
  {\bfseries 35} (2018) 084002}
  [\href{https://arxiv.org/abs/1709.06016}{{\ttfamily 1709.06016}}].

\bibitem{Vines:2018gqi}
J.~Vines, J.~Steinhoff and A.~Buonanno, \emph{{Spinning-black-hole scattering
  and the test-black-hole limit at second post-Minkowskian order}},
  \href{https://doi.org/10.1103/PhysRevD.99.064054}{\emph{Phys. Rev. D}
  {\bfseries 99} (2019) 064054}
  [\href{https://arxiv.org/abs/1812.00956}{{\ttfamily 1812.00956}}].

\bibitem{Damour:2019lcq}
T.~Damour, \emph{{Classical and quantum scattering in post-Minkowskian
  gravity}}, \href{https://doi.org/10.1103/PhysRevD.102.024060}{\emph{Phys.
  Rev. D} {\bfseries 102} (2020) 024060}
  [\href{https://arxiv.org/abs/1912.02139}{{\ttfamily 1912.02139}}].

\bibitem{Goldberger:2004jt}
W.D.~Goldberger and I.Z.~Rothstein, \emph{{An Effective field theory of gravity
  for extended objects}},
  \href{https://doi.org/10.1103/PhysRevD.73.104029}{\emph{Phys. Rev. D}
  {\bfseries 73} (2006) 104029}
  [\href{https://arxiv.org/abs/hep-th/0409156}{{\ttfamily hep-th/0409156}}].

\bibitem{Goldberger:2009qd}
W.D.~Goldberger and A.~Ross, \emph{{Gravitational radiative corrections from
  effective field theory}},
  \href{https://doi.org/10.1103/PhysRevD.81.124015}{\emph{Phys. Rev. D}
  {\bfseries 81} (2010) 124015}
  [\href{https://arxiv.org/abs/0912.4254}{{\ttfamily 0912.4254}}].

\bibitem{Kalin:2020mvi}
G.~K\"alin and R.A.~Porto, \emph{{Post-Minkowskian Effective Field Theory for
  Conservative Binary Dynamics}},
  \href{https://doi.org/10.1007/JHEP11(2020)106}{\emph{JHEP} {\bfseries 11}
  (2020) 106} [\href{https://arxiv.org/abs/2006.01184}{{\ttfamily
  2006.01184}}].

\bibitem{Kalin:2020fhe}
G.~K\"alin, Z.~Liu and R.A.~Porto, \emph{{Conservative Dynamics of Binary
  Systems to Third Post-Minkowskian Order from the Effective Field Theory
  Approach}}, \href{https://doi.org/10.1103/PhysRevLett.125.261103}{\emph{Phys.
  Rev. Lett.} {\bfseries 125} (2020) 261103}
  [\href{https://arxiv.org/abs/2007.04977}{{\ttfamily 2007.04977}}].

\bibitem{Mogull:2020sak}
G.~Mogull, J.~Plefka and J.~Steinhoff, \emph{{Classical black hole scattering
  from a worldline quantum field theory}},
  \href{https://doi.org/10.1007/JHEP02(2021)048}{\emph{JHEP} {\bfseries 02}
  (2021) 048} [\href{https://arxiv.org/abs/2010.02865}{{\ttfamily
  2010.02865}}].

\bibitem{Jakobsen:2021smu}
G.U.~Jakobsen, G.~Mogull, J.~Plefka and J.~Steinhoff, \emph{{Classical
  Gravitational Bremsstrahlung from a Worldline Quantum Field Theory}},
  \href{https://doi.org/10.1103/PhysRevLett.126.201103}{\emph{Phys. Rev. Lett.}
  {\bfseries 126} (2021) 201103}
  [\href{https://arxiv.org/abs/2101.12688}{{\ttfamily 2101.12688}}].

\bibitem{Mougiakakos:2021ckm}
S.~Mougiakakos, M.M.~Riva and F.~Vernizzi, \emph{{Gravitational Bremsstrahlung
  in the post-Minkowskian effective field theory}},
  \href{https://doi.org/10.1103/PhysRevD.104.024041}{\emph{Phys. Rev. D}
  {\bfseries 104} (2021) 024041}
  [\href{https://arxiv.org/abs/2102.08339}{{\ttfamily 2102.08339}}].

\bibitem{Liu:2021zxr}
Z.~Liu, R.A.~Porto and Z.~Yang, \emph{{Spin Effects in the Effective Field
  Theory Approach to Post-Minkowskian Conservative Dynamics}},
  \href{https://doi.org/10.1007/JHEP06(2021)012}{\emph{JHEP} {\bfseries 06}
  (2021) 012} [\href{https://arxiv.org/abs/2102.10059}{{\ttfamily
  2102.10059}}].

\bibitem{Dlapa:2021npj}
C.~Dlapa, G.~K\"alin, Z.~Liu and R.A.~Porto, \emph{{Dynamics of Binary Systems
  to Fourth Post-Minkowskian Order from the Effective Field Theory Approach}},
  \href{https://arxiv.org/abs/2106.08276}{{\ttfamily 2106.08276}}.

\bibitem{Jakobsen:2021lvp}
G.U.~Jakobsen, G.~Mogull, J.~Plefka and J.~Steinhoff, \emph{{Gravitational
  Bremsstrahlung and Hidden Supersymmetry of Spinning Bodies}},
  \href{https://doi.org/10.1103/PhysRevLett.128.011101}{\emph{Phys. Rev. Lett.}
  {\bfseries 128} (2022) 011101}
  [\href{https://arxiv.org/abs/2106.10256}{{\ttfamily 2106.10256}}].

\bibitem{Dlapa:2021vgp}
C.~Dlapa, G.~K\"alin, Z.~Liu and R.A.~Porto, \emph{{Conservative Dynamics of
  Binary Systems at Fourth Post-Minkowskian Order in the Large-Eccentricity
  Expansion}},
  \href{https://doi.org/10.1103/PhysRevLett.128.161104}{\emph{Phys. Rev. Lett.}
  {\bfseries 128} (2022) 161104}
  [\href{https://arxiv.org/abs/2112.11296}{{\ttfamily 2112.11296}}].

\bibitem{Jakobsen:2022fcj}
G.U.~Jakobsen and G.~Mogull, \emph{{Conservative and Radiative Dynamics of
  Spinning Bodies at Third Post-Minkowskian Order Using Worldline Quantum Field
  Theory}}, \href{https://doi.org/10.1103/PhysRevLett.128.141102}{\emph{Phys.
  Rev. Lett.} {\bfseries 128} (2022) 141102}
  [\href{https://arxiv.org/abs/2201.07778}{{\ttfamily 2201.07778}}].

\bibitem{Riva:2022fru}
M.M.~Riva, F.~Vernizzi and L.K.~Wong, \emph{{Gravitational bremsstrahlung from
  spinning binaries in the post-Minkowskian expansion}},
  \href{https://doi.org/10.1103/PhysRevD.106.044013}{\emph{Phys. Rev. D}
  {\bfseries 106} (2022) 044013}
  [\href{https://arxiv.org/abs/2205.15295}{{\ttfamily 2205.15295}}].

\bibitem{Jakobsen:2022psy}
G.U.~Jakobsen, G.~Mogull, J.~Plefka and B.~Sauer, \emph{{All things retarded:
  radiation-reaction in worldline quantum field theory}},
  \href{https://doi.org/10.1007/JHEP10(2022)128}{\emph{JHEP} {\bfseries 10}
  (2022) 128} [\href{https://arxiv.org/abs/2207.00569}{{\ttfamily
  2207.00569}}].

\bibitem{Dlapa:2022lmu}
C.~Dlapa, G.~K\"alin, Z.~Liu, J.~Neef and R.A.~Porto, \emph{{Radiation Reaction
  and Gravitational Waves at Fourth Post-Minkowskian Order}},
  \href{https://arxiv.org/abs/2210.05541}{{\ttfamily 2210.05541}}.

\bibitem{Damour:1985mt}
T.~Damour and G.~Sch\"afer, \emph{{Lagrangians for$n$ point masses at the
  second post-Newtonian approximation of general relativity}},
  \href{https://doi.org/10.1007/BF00773685}{\emph{Gen. Rel. Grav.} {\bfseries
  17} (1985) 879}.

\bibitem{Gilmore:2008gq}
J.B.~Gilmore and A.~Ross, \emph{{Effective field theory calculation of second
  post-Newtonian binary dynamics}},
  \href{https://doi.org/10.1103/PhysRevD.78.124021}{\emph{Phys. Rev. D}
  {\bfseries 78} (2008) 124021}
  [\href{https://arxiv.org/abs/0810.1328}{{\ttfamily 0810.1328}}].

\bibitem{Damour:2001bu}
T.~Damour, P.~Jaranowski and G.~Schaefer, \emph{{Dimensional regularization of
  the gravitational interaction of point masses}},
  \href{https://doi.org/10.1016/S0370-2693(01)00642-6}{\emph{Phys. Lett. B}
  {\bfseries 513} (2001) 147}
  [\href{https://arxiv.org/abs/gr-qc/0105038}{{\ttfamily gr-qc/0105038}}].

\bibitem{Blanchet:2003gy}
L.~Blanchet, T.~Damour and G.~Esposito-Farese, \emph{{Dimensional
  regularization of the third postNewtonian dynamics of point particles in
  harmonic coordinates}},
  \href{https://doi.org/10.1103/PhysRevD.69.124007}{\emph{Phys. Rev. D}
  {\bfseries 69} (2004) 124007}
  [\href{https://arxiv.org/abs/gr-qc/0311052}{{\ttfamily gr-qc/0311052}}].

\bibitem{Itoh:2003fy}
Y.~Itoh and T.~Futamase, \emph{{New derivation of a third postNewtonian
  equation of motion for relativistic compact binaries without ambiguity}},
  \href{https://doi.org/10.1103/PhysRevD.68.121501}{\emph{Phys. Rev. D}
  {\bfseries 68} (2003) 121501}
  [\href{https://arxiv.org/abs/gr-qc/0310028}{{\ttfamily gr-qc/0310028}}].

\bibitem{Foffa:2011ub}
S.~Foffa and R.~Sturani, \emph{{Effective field theory calculation of
  conservative binary dynamics at third post-Newtonian order}},
  \href{https://doi.org/10.1103/PhysRevD.84.044031}{\emph{Phys. Rev. D}
  {\bfseries 84} (2011) 044031}
  [\href{https://arxiv.org/abs/1104.1122}{{\ttfamily 1104.1122}}].

\bibitem{Jaranowski:2012eb}
P.~Jaranowski and G.~Schafer, \emph{{Towards the 4th post-Newtonian Hamiltonian
  for two-point-mass systems}},
  \href{https://doi.org/10.1103/PhysRevD.86.061503}{\emph{Phys. Rev. D}
  {\bfseries 86} (2012) 061503}
  [\href{https://arxiv.org/abs/1207.5448}{{\ttfamily 1207.5448}}].

\bibitem{Damour:2014jta}
T.~Damour, P.~Jaranowski and G.~Sch\"afer, \emph{{Nonlocal-in-time action for
  the fourth post-Newtonian conservative dynamics of two-body systems}},
  \href{https://doi.org/10.1103/PhysRevD.89.064058}{\emph{Phys. Rev. D}
  {\bfseries 89} (2014) 064058}
  [\href{https://arxiv.org/abs/1401.4548}{{\ttfamily 1401.4548}}].

\bibitem{Galley:2015kus}
C.R.~Galley, A.K.~Leibovich, R.A.~Porto and A.~Ross, \emph{{Tail effect in
  gravitational radiation reaction: Time nonlocality and renormalization group
  evolution}}, \href{https://doi.org/10.1103/PhysRevD.93.124010}{\emph{Phys.
  Rev. D} {\bfseries 93} (2016) 124010}
  [\href{https://arxiv.org/abs/1511.07379}{{\ttfamily 1511.07379}}].

\bibitem{Damour:2015isa}
T.~Damour, P.~Jaranowski and G.~Sch\"afer, \emph{{Fourth post-Newtonian
  effective one-body dynamics}},
  \href{https://doi.org/10.1103/PhysRevD.91.084024}{\emph{Phys. Rev. D}
  {\bfseries 91} (2015) 084024}
  [\href{https://arxiv.org/abs/1502.07245}{{\ttfamily 1502.07245}}].

\bibitem{Damour:2016abl}
T.~Damour, P.~Jaranowski and G.~Sch\"afer, \emph{{Conservative dynamics of
  two-body systems at the fourth post-Newtonian approximation of general
  relativity}}, \href{https://doi.org/10.1103/PhysRevD.93.084014}{\emph{Phys.
  Rev. D} {\bfseries 93} (2016) 084014}
  [\href{https://arxiv.org/abs/1601.01283}{{\ttfamily 1601.01283}}].

\bibitem{Bernard:2015njp}
L.~Bernard, L.~Blanchet, A.~Boh\'e, G.~Faye and S.~Marsat, \emph{{Fokker action
  of nonspinning compact binaries at the fourth post-Newtonian approximation}},
  \href{https://doi.org/10.1103/PhysRevD.93.084037}{\emph{Phys. Rev. D}
  {\bfseries 93} (2016) 084037}
  [\href{https://arxiv.org/abs/1512.02876}{{\ttfamily 1512.02876}}].

\bibitem{Bernard:2016wrg}
L.~Bernard, L.~Blanchet, A.~Boh\'e, G.~Faye and S.~Marsat, \emph{{Energy and
  periastron advance of compact binaries on circular orbits at the fourth
  post-Newtonian order}},
  \href{https://doi.org/10.1103/PhysRevD.95.044026}{\emph{Phys. Rev. D}
  {\bfseries 95} (2017) 044026}
  [\href{https://arxiv.org/abs/1610.07934}{{\ttfamily 1610.07934}}].

\bibitem{Foffa:2012rn}
S.~Foffa and R.~Sturani, \emph{{Dynamics of the gravitational two-body problem
  at fourth post-Newtonian order and at quadratic order in the Newton
  constant}}, \href{https://doi.org/10.1103/PhysRevD.87.064011}{\emph{Phys.
  Rev. D} {\bfseries 87} (2013) 064011}
  [\href{https://arxiv.org/abs/1206.7087}{{\ttfamily 1206.7087}}].

\bibitem{Foffa:2016rgu}
S.~Foffa, P.~Mastrolia, R.~Sturani and C.~Sturm, \emph{{Effective field theory
  approach to the gravitational two-body dynamics, at fourth post-Newtonian
  order and quintic in the Newton constant}},
  \href{https://doi.org/10.1103/PhysRevD.95.104009}{\emph{Phys. Rev. D}
  {\bfseries 95} (2017) 104009}
  [\href{https://arxiv.org/abs/1612.00482}{{\ttfamily 1612.00482}}].

\bibitem{Porto:2017dgs}
R.A.~Porto and I.Z.~Rothstein, \emph{{Apparent ambiguities in the
  post-Newtonian expansion for binary systems}},
  \href{https://doi.org/10.1103/PhysRevD.96.024062}{\emph{Phys. Rev. D}
  {\bfseries 96} (2017) 024062}
  [\href{https://arxiv.org/abs/1703.06433}{{\ttfamily 1703.06433}}].

\bibitem{Porto:2017shd}
R.A.~Porto, \emph{{Lamb shift and the gravitational binding energy for binary
  black holes}}, \href{https://doi.org/10.1103/PhysRevD.96.024063}{\emph{Phys.
  Rev. D} {\bfseries 96} (2017) 024063}
  [\href{https://arxiv.org/abs/1703.06434}{{\ttfamily 1703.06434}}].

\bibitem{Foffa:2019yfl}
S.~Foffa, R.A.~Porto, I.~Rothstein and R.~Sturani, \emph{{Conservative dynamics
  of binary systems to fourth Post-Newtonian order in the EFT approach II:
  Renormalized Lagrangian}},
  \href{https://doi.org/10.1103/PhysRevD.100.024048}{\emph{Phys. Rev. D}
  {\bfseries 100} (2019) 024048}
  [\href{https://arxiv.org/abs/1903.05118}{{\ttfamily 1903.05118}}].

\bibitem{Blumlein:2020pog}
J.~Bl\"umlein, A.~Maier, P.~Marquard and G.~Sch\"afer, \emph{{Fourth
  post-Newtonian Hamiltonian dynamics of two-body systems from an effective
  field theory approach}},
  \href{https://doi.org/10.1016/j.nuclphysb.2020.115041}{\emph{Nucl. Phys. B}
  {\bfseries 955} (2020) 115041}
  [\href{https://arxiv.org/abs/2003.01692}{{\ttfamily 2003.01692}}].

\bibitem{Foffa:2019hrb}
S.~Foffa, P.~Mastrolia, R.~Sturani, C.~Sturm and W.J.~Torres~Bobadilla,
  \emph{{Static two-body potential at fifth post-Newtonian order}},
  \href{https://doi.org/10.1103/PhysRevLett.122.241605}{\emph{Phys. Rev. Lett.}
  {\bfseries 122} (2019) 241605}
  [\href{https://arxiv.org/abs/1902.10571}{{\ttfamily 1902.10571}}].

\bibitem{Blumlein:2019zku}
J.~Bl\"umlein, A.~Maier and P.~Marquard, \emph{{Five-Loop Static Contribution
  to the Gravitational Interaction Potential of Two Point Masses}},
  \href{https://doi.org/10.1016/j.physletb.2019.135100}{\emph{Phys. Lett. B}
  {\bfseries 800} (2020) 135100}
  [\href{https://arxiv.org/abs/1902.11180}{{\ttfamily 1902.11180}}].

\bibitem{Bini:2020wpo}
D.~Bini, T.~Damour and A.~Geralico, \emph{{Binary dynamics at the fifth and
  fifth-and-a-half post-Newtonian orders}},
  \href{https://doi.org/10.1103/PhysRevD.102.024062}{\emph{Phys. Rev. D}
  {\bfseries 102} (2020) 024062}
  [\href{https://arxiv.org/abs/2003.11891}{{\ttfamily 2003.11891}}].

\bibitem{Blumlein:2020pyo}
J.~Bl\"umlein, A.~Maier, P.~Marquard and G.~Sch\"afer, \emph{{The fifth-order
  post-Newtonian Hamiltonian dynamics of two-body systems from an effective
  field theory approach: potential contributions}},
  \href{https://doi.org/10.1016/j.nuclphysb.2021.115352}{\emph{Nucl. Phys. B}
  {\bfseries 965} (2021) 115352}
  [\href{https://arxiv.org/abs/2010.13672}{{\ttfamily 2010.13672}}].

\bibitem{Blumlein:2020znm}
J.~Bl\"umlein, A.~Maier, P.~Marquard and G.~Sch\"afer, \emph{{Testing binary
  dynamics in gravity at the sixth post-Newtonian level}},
  \href{https://doi.org/10.1016/j.physletb.2020.135496}{\emph{Phys. Lett. B}
  {\bfseries 807} (2020) 135496}
  [\href{https://arxiv.org/abs/2003.07145}{{\ttfamily 2003.07145}}].

\bibitem{Bini:2020uiq}
D.~Bini, T.~Damour, A.~Geralico, S.~Laporta and P.~Mastrolia,
  \emph{{Gravitational dynamics at $O(G^6)$: perturbative gravitational
  scattering meets experimental mathematics}},
  \href{https://arxiv.org/abs/2008.09389}{{\ttfamily 2008.09389}}.

\bibitem{Blumlein:2021txj}
J.~Bl\"umlein, A.~Maier, P.~Marquard and G.~Sch\"afer, \emph{{The 6th
  post-Newtonian potential terms at $O(G_N^4)$}},
  \href{https://doi.org/10.1016/j.physletb.2021.136260}{\emph{Phys. Lett. B}
  {\bfseries 816} (2021) 136260}
  [\href{https://arxiv.org/abs/2101.08630}{{\ttfamily 2101.08630}}].

\bibitem{Porto:2005ac}
R.A.~Porto, \emph{{Post-Newtonian corrections to the motion of spinning bodies
  in NRGR}}, \href{https://doi.org/10.1103/PhysRevD.73.104031}{\emph{Phys. Rev.
  D} {\bfseries 73} (2006) 104031}
  [\href{https://arxiv.org/abs/gr-qc/0511061}{{\ttfamily gr-qc/0511061}}].

\bibitem{Steinhoff:2010zz}
J.~Steinhoff, \emph{{Canonical formulation of spin in general relativity}},
  \href{https://doi.org/10.1002/andp.201000178}{\emph{Annalen Phys.} {\bfseries
  523} (2011) 296} [\href{https://arxiv.org/abs/1106.4203}{{\ttfamily
  1106.4203}}].

\bibitem{Levi:2014gsa}
M.~Levi and J.~Steinhoff, \emph{{Leading order finite size effects with spins
  for inspiralling compact binaries}},
  \href{https://doi.org/10.1007/JHEP06(2015)059}{\emph{JHEP} {\bfseries 06}
  (2015) 059} [\href{https://arxiv.org/abs/1410.2601}{{\ttfamily 1410.2601}}].

\bibitem{Levi:2015msa}
M.~Levi and J.~Steinhoff, \emph{{Spinning gravitating objects in the effective
  field theory in the post-Newtonian scheme}},
  \href{https://doi.org/10.1007/JHEP09(2015)219}{\emph{JHEP} {\bfseries 09}
  (2015) 219} [\href{https://arxiv.org/abs/1501.04956}{{\ttfamily
  1501.04956}}].

\bibitem{Maia:2017yok}
N.T.~Maia, C.R.~Galley, A.K.~Leibovich and R.A.~Porto, \emph{{Radiation
  reaction for spinning bodies in effective field theory II: Spin-spin
  effects}}, \href{https://doi.org/10.1103/PhysRevD.96.084065}{\emph{Phys. Rev.
  D} {\bfseries 96} (2017) 084065}
  [\href{https://arxiv.org/abs/1705.07938}{{\ttfamily 1705.07938}}].

\bibitem{Levi:2018nxp}
M.~Levi, \emph{{Effective Field Theories of Post-Newtonian Gravity: A
  comprehensive review}},
  \href{https://doi.org/10.1088/1361-6633/ab12bc}{\emph{Rept. Prog. Phys.}
  {\bfseries 83} (2020) 075901}
  [\href{https://arxiv.org/abs/1807.01699}{{\ttfamily 1807.01699}}].

\bibitem{Levi:2020uwu}
M.~Levi, A.J.~Mcleod and M.~Von~Hippel, \emph{{N$^{3}$LO gravitational
  quadratic-in-spin interactions at G$^{4}$}},
  \href{https://doi.org/10.1007/JHEP07(2021)116}{\emph{JHEP} {\bfseries 07}
  (2021) 116} [\href{https://arxiv.org/abs/2003.07890}{{\ttfamily
  2003.07890}}].

\bibitem{Levy:1969cr}
M.~Levy and J.~Sucher, \emph{{Eikonal approximation in quantum field theory}},
  \href{https://doi.org/10.1103/PhysRev.186.1656}{\emph{Phys. Rev.} {\bfseries
  186} (1969) 1656}.

\bibitem{Amati:1987wq}
D.~Amati, M.~Ciafaloni and G.~Veneziano, \emph{{Superstring Collisions at
  Planckian Energies}},
  \href{https://doi.org/10.1016/0370-2693(87)90346-7}{\emph{Phys.\ Lett.\ B}
  {\bfseries 197} (1987) 81}.

\bibitem{Amati:1987uf}
D.~Amati, M.~Ciafaloni and G.~Veneziano, \emph{{Classical and Quantum Gravity
  Effects from Planckian Energy Superstring Collisions}},
  \href{https://doi.org/10.1142/S0217751X88000710}{\emph{Int. J. Mod. Phys. A}
  {\bfseries 3} (1988) 1615}.

\bibitem{Amati:1990xe}
D.~Amati, M.~Ciafaloni and G.~Veneziano, \emph{{Higher Order Gravitational
  Deflection and Soft Bremsstrahlung in Planckian Energy Superstring
  Collisions}}, \href{https://doi.org/10.1016/0550-3213(90)90375-N}{\emph{Nucl.
  Phys. B} {\bfseries 347} (1990) 550}.

\bibitem{Kabat:1992tb}
D.N.~Kabat and M.~Ortiz, \emph{{Eikonal quantum gravity and Planckian
  scattering}}, \href{https://doi.org/10.1016/0550-3213(92)90627-N}{\emph{Nucl.
  Phys. B} {\bfseries 388} (1992) 570}
  [\href{https://arxiv.org/abs/hep-th/9203082}{{\ttfamily hep-th/9203082}}].

\bibitem{Bellazzini:2022wzv}
B.~Bellazzini, G.~Isabella and M.M.~Riva, \emph{{Classical vs Quantum Eikonal
  Scattering and its Causal Structure}},
  \href{https://arxiv.org/abs/2211.00085}{{\ttfamily 2211.00085}}.

\bibitem{Damgaard:2021ipf}
P.H.~Damgaard, L.~Plante and P.~Vanhove, \emph{{On an exponential
  representation of the gravitational S-matrix}},
  \href{https://doi.org/10.1007/JHEP11(2021)213}{\emph{JHEP} {\bfseries 11}
  (2021) 213} [\href{https://arxiv.org/abs/2107.12891}{{\ttfamily
  2107.12891}}].

\bibitem{Brandhuber:2021kpo}
A.~Brandhuber, G.~Chen, G.~Travaglini and C.~Wen, \emph{{A new gauge-invariant
  double copy for heavy-mass effective theory}},
  \href{https://doi.org/10.1007/JHEP07(2021)047}{\emph{JHEP} {\bfseries 07}
  (2021) 047} [\href{https://arxiv.org/abs/2104.11206}{{\ttfamily
  2104.11206}}].

\bibitem{Georgi:1990um}
H.~Georgi, \emph{{An Effective Field Theory for Heavy Quarks at Low-energies}},
  \href{https://doi.org/10.1016/0370-2693(90)91128-X}{\emph{Phys. Lett. B}
  {\bfseries 240} (1990) 447}.

\bibitem{Luke:1992cs}
M.E.~Luke and A.V.~Manohar, \emph{{Reparametrization invariance constraints on
  heavy particle effective field theories}},
  \href{https://doi.org/10.1016/0370-2693(92)91786-9}{\emph{Phys. Lett. B}
  {\bfseries 286} (1992) 348}
  [\href{https://arxiv.org/abs/hep-ph/9205228}{{\ttfamily hep-ph/9205228}}].

\bibitem{Neubert:1993mb}
M.~Neubert, \emph{{Heavy quark symmetry}},
  \href{https://doi.org/10.1016/0370-1573(94)90091-4}{\emph{Phys. Rept.}
  {\bfseries 245} (1994) 259}
  [\href{https://arxiv.org/abs/hep-ph/9306320}{{\ttfamily hep-ph/9306320}}].

\bibitem{Manohar:2000dt}
A.V.~Manohar and M.B.~Wise, \emph{{Heavy quark physics}}, vol.~10, Cambridge
  University Press (2000).

\bibitem{Bern:2008qj}
Z.~Bern, J.J.M.~Carrasco and H.~Johansson, \emph{{New Relations for
  Gauge-Theory Amplitudes}},
  \href{https://doi.org/10.1103/PhysRevD.78.085011}{\emph{Phys. Rev.}
  {\bfseries D78} (2008) 085011}
  [\href{https://arxiv.org/abs/0805.3993}{{\ttfamily 0805.3993}}].

\bibitem{Bern:2010ue}
Z.~Bern, J.J.M.~Carrasco and H.~Johansson, \emph{{Perturbative Quantum Gravity
  as a Double Copy of Gauge Theory}},
  \href{https://doi.org/10.1103/PhysRevLett.105.061602}{\emph{Phys. Rev. Lett.}
  {\bfseries 105} (2010) 061602}
  [\href{https://arxiv.org/abs/1004.0476}{{\ttfamily 1004.0476}}].

\bibitem{Bern:2019prr}
Z.~Bern, J.J.~Carrasco, M.~Chiodaroli, H.~Johansson and R.~Roiban, \emph{{The
  Duality Between Color and Kinematics and its Applications}},
  \href{https://arxiv.org/abs/1909.01358}{{\ttfamily 1909.01358}}.

\bibitem{Brandhuber:2021bsf}
A.~Brandhuber, G.~Chen, H.~Johansson, G.~Travaglini and C.~Wen,
  \emph{{Kinematic Hopf Algebra for Bern-Carrasco-Johansson Numerators in
  Heavy-Mass Effective Field Theory and Yang-Mills Theory}},
  \href{https://doi.org/10.1103/PhysRevLett.128.121601}{\emph{Phys. Rev. Lett.}
  {\bfseries 128} (2022) 121601}
  [\href{https://arxiv.org/abs/2111.15649}{{\ttfamily 2111.15649}}].

\bibitem{Chen:2022nei}
G.~Chen, G.~Lin and C.~Wen, \emph{{Kinematic Hopf algebra for amplitudes and
  form factors}},  \href{https://arxiv.org/abs/2208.05519}{{\ttfamily
  2208.05519}}.

\bibitem{Brandhuber:2022enp}
A.~Brandhuber, G.R.~Brown, G.~Chen, J.~Gowdy, G.~Travaglini and C.~Wen,
  \emph{{Amplitudes, Hopf algebras and the colour-kinematics duality}},
  \href{https://doi.org/10.1007/JHEP12(2022)101}{\emph{JHEP} {\bfseries 12}
  (2022) 101} [\href{https://arxiv.org/abs/2208.05886}{{\ttfamily
  2208.05886}}].

\bibitem{Cao:2022vou}
Q.~Cao, J.~Dong, S.~He and Y.-Q.~Zhang, \emph{{Covariant color-kinematics
  duality, Hopf algebras, and permutohedra}},
  \href{https://doi.org/10.1103/PhysRevD.107.026022}{\emph{Phys. Rev. D}
  {\bfseries 107} (2023) 026022}
  [\href{https://arxiv.org/abs/2211.05404}{{\ttfamily 2211.05404}}].

\bibitem{DEath:1976bbo}
P.D.~D'Eath, \emph{{High Speed Black Hole Encounters and Gravitational
  Radiation}}, \href{https://doi.org/10.1103/PhysRevD.18.990}{\emph{Phys. Rev.
  D} {\bfseries 18} (1978) 990}.

\bibitem{Kovacs:1977uw}
S.J.~Kovacs and K.S.~Thorne, \emph{{The Generation of Gravitational Waves. 3.
  Derivation of Bremsstrahlung Formulas}},
  \href{https://doi.org/10.1086/155576}{\emph{Astrophys. J.} {\bfseries 217}
  (1977) 252}.

\bibitem{Kovacs:1978eu}
S.J.~Kovacs and K.S.~Thorne, \emph{{The Generation of Gravitational Waves. 4.
  Bremsstrahlung}}, \href{https://doi.org/10.1086/156350}{\emph{Astrophys. J.}
  {\bfseries 224} (1978) 62}.

\bibitem{DiVecchia:2022nna}
P.~Di~Vecchia, C.~Heissenberg, R.~Russo and G.~Veneziano, \emph{{The eikonal
  operator at arbitrary velocities I: the soft-radiation limit}},
  \href{https://doi.org/10.1007/JHEP07(2022)039}{\emph{JHEP} {\bfseries 07}
  (2022) 039} [\href{https://arxiv.org/abs/2204.02378}{{\ttfamily
  2204.02378}}].

\bibitem{Blanchet:1993ec}
L.~Blanchet and G.~Schaefer, \emph{{Gravitational wave tails and binary star
  systems}}, \href{https://doi.org/10.1088/0264-9381/10/12/026}{\emph{Class.
  Quant. Grav.} {\bfseries 10} (1993) 2699}.

\bibitem{Ross:2012fc}
A.~Ross, \emph{{Multipole expansion at the level of the action}},
  \href{https://doi.org/10.1103/PhysRevD.85.125033}{\emph{Phys. Rev. D}
  {\bfseries 85} (2012) 125033}
  [\href{https://arxiv.org/abs/1202.4750}{{\ttfamily 1202.4750}}].

\bibitem{Galley:2009px}
C.R.~Galley and M.~Tiglio, \emph{{Radiation reaction and gravitational waves in
  the effective field theory approach}},
  \href{https://doi.org/10.1103/PhysRevD.79.124027}{\emph{Phys. Rev. D}
  {\bfseries 79} (2009) 124027}
  [\href{https://arxiv.org/abs/0903.1122}{{\ttfamily 0903.1122}}].

\bibitem{Porto:2010zg}
R.A.~Porto, A.~Ross and I.Z.~Rothstein, \emph{{Spin induced multipole moments
  for the gravitational wave flux from binary inspirals to third Post-Newtonian
  order}}, \href{https://doi.org/10.1088/1475-7516/2011/03/009}{\emph{JCAP}
  {\bfseries 03} (2011) 009} [\href{https://arxiv.org/abs/1007.1312}{{\ttfamily
  1007.1312}}].

\bibitem{Porto:2012as}
R.A.~Porto, A.~Ross and I.Z.~Rothstein, \emph{{Spin induced multipole moments
  for the gravitational wave amplitude from binary inspirals to 2.5
  Post-Newtonian order}},
  \href{https://doi.org/10.1088/1475-7516/2012/09/028}{\emph{JCAP} {\bfseries
  09} (2012) 028} [\href{https://arxiv.org/abs/1203.2962}{{\ttfamily
  1203.2962}}].

\bibitem{Cristofoli:2021vyo}
A.~Cristofoli, R.~Gonzo, D.A.~Kosower and D.~O'Connell, \emph{{Waveforms from
  amplitudes}}, \href{https://doi.org/10.1103/PhysRevD.106.056007}{\emph{Phys.
  Rev. D} {\bfseries 106} (2022) 056007}
  [\href{https://arxiv.org/abs/2107.10193}{{\ttfamily 2107.10193}}].

\bibitem{Cristofoli:2021jas}
A.~Cristofoli, R.~Gonzo, N.~Moynihan, D.~O'Connell, A.~Ross, M.~Sergola et~al.,
  \emph{{The Uncertainty Principle and Classical Amplitudes}},
  \href{https://arxiv.org/abs/2112.07556}{{\ttfamily 2112.07556}}.

\bibitem{Kosower:2018adc}
D.A.~Kosower, B.~Maybee and D.~O'Connell, \emph{{Amplitudes, Observables, and
  Classical Scattering}},
  \href{https://doi.org/10.1007/JHEP02(2019)137}{\emph{JHEP} {\bfseries 02}
  (2019) 137} [\href{https://arxiv.org/abs/1811.10950}{{\ttfamily
  1811.10950}}].

\bibitem{Caron-Huot:2023vxl}
S.~Caron-Huot, M.~Giroux, H.S.~Hannesdottir and S.~Mizera, \emph{{What can be
  measured asymptotically?}},
  \href{https://arxiv.org/abs/2308.02125}{{\ttfamily 2308.02125}}.

\bibitem{Carrasco:2020ywq}
J.J.M.~Carrasco and I.A.~Vazquez-Holm, \emph{{Loop-Level Double-Copy for
  Massive Quantum Particles}},
  \href{https://doi.org/10.1103/PhysRevD.103.045002}{\emph{Phys. Rev. D}
  {\bfseries 103} (2021) 045002}
  [\href{https://arxiv.org/abs/2010.13435}{{\ttfamily 2010.13435}}].

\bibitem{Carrasco:2021bmu}
J.J.M.~Carrasco and I.A.~Vazquez-Holm, \emph{{Extracting Einstein from the
  loop-level double-copy}},
  \href{https://doi.org/10.1007/JHEP11(2021)088}{\emph{JHEP} {\bfseries 11}
  (2021) 088} [\href{https://arxiv.org/abs/2108.06798}{{\ttfamily
  2108.06798}}].

\bibitem{Lee:2012cn}
R.N.~Lee, \emph{{Presenting LiteRed: a tool for the Loop InTEgrals REDuction}},
   \href{https://arxiv.org/abs/1212.2685}{{\ttfamily 1212.2685}}.

\bibitem{Lee:2013mka}
R.N.~Lee, \emph{{LiteRed 1.4: a powerful tool for reduction of multiloop
  integrals}}, \href{https://doi.org/10.1088/1742-6596/523/1/012059}{\emph{J.
  Phys. Conf. Ser.} {\bfseries 523} (2014) 012059}
  [\href{https://arxiv.org/abs/1310.1145}{{\ttfamily 1310.1145}}].

\bibitem{Kotikov:1990kg}
A.V.~Kotikov, \emph{{Differential equations method: New technique for massive
  Feynman diagrams calculation}},
  \href{https://doi.org/10.1016/0370-2693(91)90413-K}{\emph{Phys. Lett. B}
  {\bfseries 254} (1991) 158}.

\bibitem{Bern:1993kr}
Z.~Bern, L.J.~Dixon and D.A.~Kosower, \emph{{Dimensionally regulated pentagon
  integrals}}, \href{https://doi.org/10.1016/0550-3213(94)90398-0}{\emph{Nucl.
  Phys. B} {\bfseries 412} (1994) 751}
  [\href{https://arxiv.org/abs/hep-ph/9306240}{{\ttfamily hep-ph/9306240}}].

\bibitem{Remiddi:1997ny}
E.~Remiddi, \emph{{Differential equations for Feynman graph amplitudes}},
  \href{https://doi.org/10.1007/BF03185566}{\emph{Nuovo Cim. A} {\bfseries 110}
  (1997) 1435} [\href{https://arxiv.org/abs/hep-th/9711188}{{\ttfamily
  hep-th/9711188}}].

\bibitem{Gehrmann:1999as}
T.~Gehrmann and E.~Remiddi, \emph{{Differential equations for two loop four
  point functions}},
  \href{https://doi.org/10.1016/S0550-3213(00)00223-6}{\emph{Nucl. Phys. B}
  {\bfseries 580} (2000) 485}
  [\href{https://arxiv.org/abs/hep-ph/9912329}{{\ttfamily hep-ph/9912329}}].

\bibitem{Henn:2013pwa}
J.M.~Henn, \emph{{Multiloop integrals in dimensional regularization made
  simple}}, \href{https://doi.org/10.1103/PhysRevLett.110.251601}{\emph{Phys.
  Rev. Lett.} {\bfseries 110} (2013) 251601}
  [\href{https://arxiv.org/abs/1304.1806}{{\ttfamily 1304.1806}}].

\bibitem{Weinberg:1965nx}
S.~Weinberg, \emph{{Infrared photons and gravitons}},
  \href{https://doi.org/10.1103/PhysRev.140.B516}{\emph{Phys. Rev.} {\bfseries
  140} (1965) B516}.

\bibitem{Newman:1961qr}
E.~Newman and R.~Penrose, \emph{{An Approach to gravitational radiation by a
  method of spin coefficients}},
  \href{https://doi.org/10.1063/1.1724257}{\emph{J. Math. Phys.} {\bfseries 3}
  (1962) 566}.

\bibitem{Herderschee:2023fxh}
A.~Herderschee, R.~Roiban and F.~Teng, \emph{{The Sub-Leading Scattering
  Waveform from Amplitudes}},
  \href{https://arxiv.org/abs/2303.06112}{{\ttfamily 2303.06112}}.

\bibitem{Elkhidir:2023dco}
A.~Elkhidir, D.~O'Connell, M.~Sergola and I.A.~Vazquez-Holm, \emph{{Radiation
  and Reaction at One Loop}},
  \href{https://arxiv.org/abs/2303.06211}{{\ttfamily 2303.06211}}.

\bibitem{Landshoff:1969yyn}
P.V.~Landshoff and J.C.~Polkinghorne, \emph{{Iterations of Regge cuts}},
  \href{https://doi.org/10.1103/PhysRev.181.1989}{\emph{Phys. Rev.} {\bfseries
  181} (1969) 1989}.

\bibitem{Bautista:2019evw}
Y.F.~Bautista and A.~Guevara, \emph{{On the Double Copy for Spinning Matter}},
  \href{https://arxiv.org/abs/1908.11349}{{\ttfamily 1908.11349}}.

\bibitem{Britto:2004ap}
R.~Britto, F.~Cachazo and B.~Feng, \emph{{New recursion relations for tree
  amplitudes of gluons}},
  \href{https://doi.org/10.1016/j.nuclphysb.2005.02.030}{\emph{Nucl. Phys. B}
  {\bfseries 715} (2005) 499}
  [\href{https://arxiv.org/abs/hep-th/0412308}{{\ttfamily hep-th/0412308}}].

\bibitem{Britto:2005fq}
R.~Britto, F.~Cachazo, B.~Feng and E.~Witten, \emph{{Direct proof of tree-level
  recursion relation in Yang-Mills theory}},
  \href{https://doi.org/10.1103/PhysRevLett.94.181602}{\emph{Phys. Rev. Lett.}
  {\bfseries 94} (2005) 181602}
  [\href{https://arxiv.org/abs/hep-th/0501052}{{\ttfamily hep-th/0501052}}].

\bibitem{Britto:2021pud}
R.~Britto, R.~Gonzo and G.R.~Jehu, \emph{{Graviton particle statistics and
  coherent states from classical scattering amplitudes}},
  \href{https://doi.org/10.1007/JHEP03(2022)214}{\emph{JHEP} {\bfseries 03}
  (2022) 214} [\href{https://arxiv.org/abs/2112.07036}{{\ttfamily
  2112.07036}}].

\bibitem{Kosmopoulos:2020pcd}
D.~Kosmopoulos, \emph{{Simplifying D-dimensional physical-state sums in gauge
  theory and gravity}},
  \href{https://doi.org/10.1103/PhysRevD.105.056025}{\emph{Phys. Rev. D}
  {\bfseries 105} (2022) 056025}
  [\href{https://arxiv.org/abs/2009.00141}{{\ttfamily 2009.00141}}].

\bibitem{Bautista:2021inx}
Y.F.~Bautista and N.~Siemonsen, \emph{{Post-Newtonian waveforms from spinning
  scattering amplitudes}},
  \href{https://doi.org/10.1007/JHEP01(2022)006}{\emph{JHEP} {\bfseries 01}
  (2022) 006} [\href{https://arxiv.org/abs/2110.12537}{{\ttfamily
  2110.12537}}].

\bibitem{Feng:2020jck}
B.~Feng, X.-D.~Li and R.~Huang, \emph{{Expansion of EYM Amplitudes in Gauge
  Invariant Vector Space}},
  \href{https://doi.org/10.1088/1674-1137/abb4ce}{\emph{Chin. Phys. C}
  {\bfseries 44} (2020) 123104}
  [\href{https://arxiv.org/abs/2005.06287}{{\ttfamily 2005.06287}}].

\bibitem{Beneke:1997zp}
M.~Beneke and V.A.~Smirnov, \emph{{Asymptotic expansion of Feynman integrals
  near threshold}},
  \href{https://doi.org/10.1016/S0550-3213(98)00138-2}{\emph{Nucl. Phys. B}
  {\bfseries 522} (1998) 321}
  [\href{https://arxiv.org/abs/hep-ph/9711391}{{\ttfamily hep-ph/9711391}}].

\bibitem{Pak:2010pt}
A.~Pak and A.~Smirnov, \emph{{Geometric approach to asymptotic expansion of
  Feynman integrals}},
  \href{https://doi.org/10.1140/epjc/s10052-011-1626-1}{\emph{Eur. Phys. J. C}
  {\bfseries 71} (2011) 1626}
  [\href{https://arxiv.org/abs/1011.4863}{{\ttfamily 1011.4863}}].

\bibitem{Jantzen:2012mw}
B.~Jantzen, A.V.~Smirnov and V.A.~Smirnov, \emph{{Expansion by regions:
  revealing potential and Glauber regions automatically}},
  \href{https://doi.org/10.1140/epjc/s10052-012-2139-2}{\emph{Eur. Phys. J. C}
  {\bfseries 72} (2012) 2139}
  [\href{https://arxiv.org/abs/1206.0546}{{\ttfamily 1206.0546}}].

\bibitem{vanNeerven:1983vr}
W.L.~van Neerven and J.A.M.~Vermaseren, \emph{{Large loop integrals}},
  \href{https://doi.org/10.1016/0370-2693(84)90237-5}{\emph{Phys. Lett. B}
  {\bfseries 137} (1984) 241}.

\bibitem{Ellis:2011cr}
R.K.~Ellis, Z.~Kunszt, K.~Melnikov and G.~Zanderighi, \emph{{One-loop
  calculations in quantum field theory: from Feynman diagrams to unitarity
  cuts}}, \href{https://doi.org/10.1016/j.physrep.2012.01.008}{\emph{Phys.
  Rept.} {\bfseries 518} (2012) 141}
  [\href{https://arxiv.org/abs/1105.4319}{{\ttfamily 1105.4319}}].

\bibitem{Caron-Huot:2016cwu}
S.~Caron-Huot and M.~Wilhelm, \emph{{Renormalization group coefficients and the
  S-matrix}}, \href{https://doi.org/10.1007/JHEP12(2016)010}{\emph{JHEP}
  {\bfseries 12} (2016) 010}
  [\href{https://arxiv.org/abs/1607.06448}{{\ttfamily 1607.06448}}].

\bibitem{Bern:2007dw}
Z.~Bern, L.J.~Dixon and D.A.~Kosower, \emph{{On-Shell Methods in Perturbative
  QCD}}, \href{https://doi.org/10.1016/j.aop.2007.04.014}{\emph{Annals Phys.}
  {\bfseries 322} (2007) 1587}
  [\href{https://arxiv.org/abs/0704.2798}{{\ttfamily 0704.2798}}].

\bibitem{Huang:2013vha}
Y.-t.~Huang and D.~McGady, \emph{{Consistency Conditions for Gauge Theory S
  Matrices from Requirements of Generalized Unitarity}},
  \href{https://doi.org/10.1103/PhysRevLett.112.241601}{\emph{Phys. Rev. Lett.}
  {\bfseries 112} (2014) 241601}
  [\href{https://arxiv.org/abs/1307.4065}{{\ttfamily 1307.4065}}].

\bibitem{Ciafaloni:2018uwe}
M.~Ciafaloni, D.~Colferai and G.~Veneziano, \emph{{Infrared features of
  gravitational scattering and radiation in the eikonal approach}},
  \href{https://doi.org/10.1103/PhysRevD.99.066008}{\emph{Phys. Rev. D}
  {\bfseries 99} (2019) 066008}
  [\href{https://arxiv.org/abs/1812.08137}{{\ttfamily 1812.08137}}].

\bibitem{DiVecchia:2022piu}
P.~Di~Vecchia, C.~Heissenberg, R.~Russo and G.~Veneziano, \emph{{Classical
  Gravitational Observables from the Eikonal Operator}},
  \href{https://arxiv.org/abs/2210.12118}{{\ttfamily 2210.12118}}.

\bibitem{He:2014laa}
T.~He, V.~Lysov, P.~Mitra and A.~Strominger, \emph{{BMS supertranslations and
  Weinberg\textquoteright{}s soft graviton theorem}},
  \href{https://doi.org/10.1007/JHEP05(2015)151}{\emph{JHEP} {\bfseries 05}
  (2015) 151} [\href{https://arxiv.org/abs/1401.7026}{{\ttfamily 1401.7026}}].

\bibitem{Pretorius:2005gq}
F.~Pretorius, \emph{{Evolution of binary black hole spacetimes}},
  \href{https://doi.org/10.1103/PhysRevLett.95.121101}{\emph{Phys. Rev. Lett.}
  {\bfseries 95} (2005) 121101}
  [\href{https://arxiv.org/abs/gr-qc/0507014}{{\ttfamily gr-qc/0507014}}].

\bibitem{Pretorius:2007nq}
F.~Pretorius, \emph{{Binary Black Hole Coalescence}},
  \href{https://arxiv.org/abs/0710.1338}{{\ttfamily 0710.1338}}.

\bibitem{CalderonBustillo:2022dph}
J.~Calderon~Bustillo, I.C.F.~Wong, N.~Sanchis-Gual, S.H.W.~Leong,
  A.~Torres-Forne, K.~Chandra et~al., \emph{{Gravitational-wave parameter
  inference with the Newman-Penrose scalar}},
  \href{https://arxiv.org/abs/2205.15029}{{\ttfamily 2205.15029}}.

\bibitem{GRCHombo}
T.~Andrade, L.~Saló, J.~Aurrekoetxea, J.~Bamber, K.~Clough, R.~Croft et~al.,
  \emph{Grchombo: An adaptable numerical relativity code for fundamental
  physics}, \href{https://doi.org/10.21105/joss.03703}{\emph{Journal of Open
  Source Software} {\bfseries 6} (2021) 3703}.

\bibitem{Clough:2015sqa}
K.~Clough, P.~Figueras, H.~Finkel, M.~Kunesch, E.A.~Lim and S.~Tunyasuvunakool,
  \emph{{GRChombo: Numerical Relativity with Adaptive Mesh Refinement}},
  \href{https://doi.org/10.1088/0264-9381/32/24/245011}{\emph{Class. Quant.
  Grav.} {\bfseries 32} (2015) 245011}
  [\href{https://arxiv.org/abs/1503.03436}{{\ttfamily 1503.03436}}].

\bibitem{Manohar:2022dea}
A.V.~Manohar, A.K.~Ridgway and C.-H.~Shen, \emph{{Radiated Angular Momentum and
  Dissipative Effects in Classical Scattering}},
  \href{https://doi.org/10.1103/PhysRevLett.129.121601}{\emph{Phys. Rev. Lett.}
  {\bfseries 129} (2022) 121601}
  [\href{https://arxiv.org/abs/2203.04283}{{\ttfamily 2203.04283}}].

\bibitem{Kalin:2019rwq}
G.~K\"alin and R.A.~Porto, \emph{{From Boundary Data to Bound States}},
  \href{https://doi.org/10.1007/JHEP01(2020)072}{\emph{JHEP} {\bfseries 01}
  (2020) 072} [\href{https://arxiv.org/abs/1910.03008}{{\ttfamily
  1910.03008}}].

\bibitem{Kalin:2019inp}
G.~K\"alin and R.A.~Porto, \emph{{From boundary data to bound states. Part II.
  Scattering angle to dynamical invariants (with twist)}},
  \href{https://doi.org/10.1007/JHEP02(2020)120}{\emph{JHEP} {\bfseries 02}
  (2020) 120} [\href{https://arxiv.org/abs/1911.09130}{{\ttfamily
  1911.09130}}].

\bibitem{Cho:2021arx}
G.~Cho, G.~K\"alin and R.A.~Porto, \emph{{From boundary data to bound states.
  Part III. Radiative effects}},
  \href{https://doi.org/10.1007/JHEP04(2022)154}{\emph{JHEP} {\bfseries 04}
  (2022) 154} [\href{https://arxiv.org/abs/2112.03976}{{\ttfamily
  2112.03976}}].

\bibitem{Adamo:2022ooq}
T.~Adamo and R.~Gonzo, \emph{{Bethe-Salpeter equation for classical
  gravitational bound states}},
  \href{https://arxiv.org/abs/2212.13269}{{\ttfamily 2212.13269}}.

\bibitem{Muzinich:1987in}
I.J.~Muzinich and M.~Soldate, \emph{{High-Energy Unitarity of Gravitation and
  Strings}}, \href{https://doi.org/10.1103/PhysRevD.37.359}{\emph{Phys. Rev. D}
  {\bfseries 37} (1988) 359}.

\bibitem{Addazi:2019mjh}
A.~Addazi, M.~Bianchi and G.~Veneziano, \emph{{Soft gravitational radiation
  from ultra-relativistic collisions at sub- and sub-sub-leading order}},
  \href{https://doi.org/10.1007/JHEP05(2019)050}{\emph{JHEP} {\bfseries 05}
  (2019) 050} [\href{https://arxiv.org/abs/1901.10986}{{\ttfamily
  1901.10986}}].

\bibitem{Cheng:1987ga}
H.~Cheng and T.~Wu, \emph{{Expanding protons: scattering at high energies}},
  MIT press, Cambridge (1987).

\bibitem{Bloch:1937pw}
F.~Bloch and A.~Nordsieck, \emph{{Note on the Radiation Field of the
  electron}}, \href{https://doi.org/10.1103/PhysRev.52.54}{\emph{Phys. Rev.}
  {\bfseries 52} (1937) 54}.

\bibitem{Jauch:1954rnc}
J.-M.~Jauch and F.~Rohrlich, \emph{{The infrared divergence}},
  \href{https://doi.org/10.5169/seals-112533}{\emph{Helv. Phys. Acta}
  {\bfseries 27} (1954) 613}.

\bibitem{Dunbar:1995ed}
D.C.~Dunbar and P.S.~Norridge, \emph{{Infinities within graviton scattering
  amplitudes}}, \href{https://doi.org/10.1088/0264-9381/14/2/009}{\emph{Class.
  Quant. Grav.} {\bfseries 14} (1997) 351}
  [\href{https://arxiv.org/abs/hep-th/9512084}{{\ttfamily hep-th/9512084}}].

\bibitem{Brandhuber:2005kd}
A.~Brandhuber, B.~Spence and G.~Travaglini, \emph{{From trees to loops and
  back}}, \href{https://doi.org/10.1088/1126-6708/2006/01/142}{\emph{JHEP}
  {\bfseries 01} (2006) 142}
  [\href{https://arxiv.org/abs/hep-th/0510253}{{\ttfamily hep-th/0510253}}].

\bibitem{Catani:2008xa}
S.~Catani, T.~Gleisberg, F.~Krauss, G.~Rodrigo and J.-C.~Winter, \emph{{From
  loops to trees by-passing Feynman's theorem}},
  \href{https://doi.org/10.1088/1126-6708/2008/09/065}{\emph{JHEP} {\bfseries
  09} (2008) 065} [\href{https://arxiv.org/abs/0804.3170}{{\ttfamily
  0804.3170}}].

\bibitem{Caron-Huot:2010fvq}
S.~Caron-Huot, \emph{{Loops and trees}},
  \href{https://doi.org/10.1007/JHEP05(2011)080}{\emph{JHEP} {\bfseries 05}
  (2011) 080} [\href{https://arxiv.org/abs/1007.3224}{{\ttfamily 1007.3224}}].

\end{thebibliography}\endgroup

\end{document}